\documentclass[a4paper,11pt]{article}

\usepackage{jheppub}
\usepackage[table]{xcolor}
\usepackage{slashbox}
\allowdisplaybreaks
\usepackage{graphicx}
\usepackage{epstopdf}

\newcommand{\be}{\begin{equation}}
	\newcommand{\ee}{\end{equation}}
\newcommand{\bea}{\begin{eqnarray}}
	\newcommand{\eea}{\end{eqnarray}}
\newcommand{\nn}{\nonumber}
\newcommand{\bdm}{\begin{displaymath}}
	\newcommand{\edm}{\end{displaymath}}

\title{N$^3$LO Quadratic-in-Spin Interactions for 
	\\Generic Compact Binaries}

\author[a]{Jung-Wook Kim,}
\author[b]{Mich\`ele Levi,}
\author[c]{Zhewei Yin}

\affiliation[a]{Queen Mary University of London, Mile End Road, 
	London E1 4NS, United Kingdom}

\affiliation[b]{Mathematical Institute, University of Oxford, 
	Oxford OX2 6GG, United Kingdom}

\affiliation[c]{Department of Physics and Astronomy, Uppsala University, 
	75108 Uppsala, Sweden}

\emailAdd{jung-wook.kim@qmul.ac.uk}
\emailAdd{levi@maths.ox.ac.uk}
\emailAdd{zhewei.yin@physics.uu.se}

\abstract{
We derive the third subleading (N$^3$LO) corrections of the quadratic-in-spin sectors 
via the EFT of spinning objects in post-Newtonian (PN) gravity.
These corrections consist of contributions from $4$ sectors for generic compact binaries, 
that enter at the fifth PN order. 
One of these contributions is due to a new tidal interaction, that is unique to  
the sectors with spin, and complements the first tidal interaction that also enters 
at this PN order in the simple point-mass sector. 
The evaluation of Feynman graphs is carried out in a generic dimension via advanced multi-loop 
methods, and gives rise to dimensional-regularization poles in conjunction with logarithms.
At these higher-spin sectors the reduction of generalized Lagrangians entails 
redefinitions of the position beyond linear order. 
We provide here the most general Lagrangians and Hamiltonians. 
We then specify the latter to simplified configurations, and derive the consequent gauge-invariant 
relations among the binding energy, angular momentum, and frequency. 
We end with a derivation of all the scattering angles that correspond to an extension of our 
Hamiltonians to the scattering problem in the simplified aligned-spins configuration, as a guide to 
scattering-amplitudes studies. 
}

\begin{document}
	
%\today

\preprint{QMUL-PH-22-29, UUITP-39/22} 
	
\maketitle
	
\flushbottom

\section{Introduction} 
\label{intro}

The inspirals and mergers of compact binaries have been successfully measured through their
emitted gravitational waves (GWs) as of $7$ years now, with the first black-hole (BH) 
binary merger detected \cite{Abbott:2016blz} by the advanced LIGO 
\cite{LIGOScientific:2014pky} in collaboration with Advanced VIRGO 
\cite{VIRGO:2014yos}. By now there is already an operational array of second-generation 
ground-based GW detectors on the northern hemisphere, including also KAGRA in Japan,
and several elaborate plans for diverse ground- and space-based GW experiments that 
will extend the measured bandwidth of frequencies to no less than $16$ orders of magnitude.  

Since these compact-binary sources of GWs spend most of their lifetime with their components 
orbiting in non-relativistic (NR) velocities, they have been analytically studied in the 
post-Newtonian (PN) approximation of General Relativity (GR) \cite{Blanchet:2013haa}. 
With the input from PN theory, supplemented with numerical simulations that cover 
the short-lived phase of the binary in strong gravity, the effective one-body (EOB)
approach \cite{Buonanno:1998gg} enables the generation of complete theoretical 
waveforms to be tested against our real-world measurements. The wealth of data that is being recorded 
\cite{LIGOScientific:2018mvr,LIGOScientific:2020ibl,LIGOScientific:2021djp}, contains coveted 
information on gravity and QCD theory due to the diverse 
sources involved, including rotating BHs in a surprising span of masses
\cite{TheLIGOScientific:2016src,TheLIGOScientific:2016wfe,Abbott:2016izl},
and neutron stars (NSs), whose interiors probe QCD in extreme conditions that only 
nature can generate \cite{LIGOScientific:2017vwq,LIGOScientific:2021qlt}.

These impressive advances in the experimental front of GW science have catalysed a resurgence
of theoretical studies aimed at improving the precision of predicted gravitational waveforms.
Modern HEP approaches have joined traditional GR efforts to better describe such GW sources,
yet they both essentially share a similar perspective of capturing these two-body systems via 
effective descriptions that consider either the single compact object, or the whole binary, as
a point particle \cite{Blanchet:2013haa,Goldberger:2004jt}. The UV physics of their internal 
structure is then captured by some unknown characteristic coefficients, which remain to be 
specified. For the conservative point-mass and spin-orbit sectors of the two-body interactions, 
such UV coefficients show up only as of the high fifth PN ($5$PN) accuracy 
\cite{Goldberger:2004jt,Levi:2020uwu}, which has been reached in the point-mass sector 
via heroic undertakings only recently in both traditional GR 
\cite{Bini:2019nra,Bini:2020wpo,Bini:2020uiq}, and EFT methods \cite{Blumlein:2020pyo}.
Subsequently, the spin-orbit sector at the $4.5$PN accuracy has been approached following a similar 
approach to \cite{Bini:2019nra,Bini:2020wpo} via traditional GR methods, without the need 
to account for any UV coefficients \cite{Antonelli:2020aeb,Antonelli:2020ybz}. 
At the same time, this third subleading spin-orbit sector was approached, completed, and verified in 
\cite{Levi:2020kvb,Kim:2022pou,Levi:2022dqm}, and following those was also derived in 
\cite{Mandal:2022nty}, via the EFT of spinning gravitating objects \cite{Levi:2015msa}, and the public 
\texttt{EFTofPNG} code \cite{Levi:2017kzq}. 

However more generally, namely beyond the spin-orbit sector which requires only the minimal coupling to 
gravity, much lower PN orders than the present precision frontier necessitate to take into 
account finite-size effects related with such unknown UV coefficients, due to the rotation of the 
individual compact objects. The leading finite-size effects, starting from the 
$2$PN order \cite{Barker:1975ae}, are due to spin-induced multipolar deformations, which have been 
uniquely formulated within the EFT of spinning gravitating objects to all orders in spin
\cite{Levi:2014gsa,Levi:2015msa}. Indeed the EFT presented in \cite{Levi:2015msa}, has granted unique 
access to higher-spin sectors, and thus the completion of the current state of the art at the $4$PN 
order in sectors with spins \cite{Levi:2011eq,Levi:2014gsa,Levi:2015uxa,Levi:2015ixa,Levi:2016ofk}. At 
the present $5$PN order however, another type of finite-size effects comes into play due to the tidal 
response to the gravitational field. The most simple unknown UV coefficients of this type arise in the 
point-mass sector, and are referred to as ``Love numbers'' in traditional GR. These have been studied 
for decades, see e.g.~\cite{Hinderer:2007mb,Damour:2009vw,Binnington:2009bb,Kol:2011vg}, with a recent 
booming activity, that notably considers also the real case of rotating compact objects, 
e.g.~\cite{LeTiec:2020spy,LeTiec:2020bos,Chia:2020yla,Charalambous:2021mea,Charalambous:2021kcz,
	Castro:2021wyc}. Interestingly almost all studies to date 
indicate that such simple ``Love numbers'' vanish for BHs in $4$-dimensional GR.
On the other hand, the UV coefficients of tidal effects can be significantly large for NSs, and thus 
these effects are very important for GW astronomy \cite{Blanchet:2000ub}.

In a recent letter by the present authors, which reported on the completion of all quadratic-in-spin 
interactions at the $5$PN order \cite{Kim:2021rfj}, a new finite-size feature was found, related with 
these tidal effects at the point-mass sector. This new feature, unique to spinning objects, comes with a 
new unknown UV coefficient, which thus provides a new unique probe for gravity and QCD. In the present 
paper we provide in full detail the derivation of all these quadratic-in-spin sectors at the 
third subleading order (N$^3$LO), for generic compact binaries, and in the most general settings. These 
results constitute the PN state of the art, and we provide the consequent useful GW quantities and 
observables. The derivations are carried out via the EFT of spinning gravitating objects 
\cite{Levi:2015msa}, and see \cite{Levi:2018nxp,Levi:2017kzq}, which provides a unique independent 
conceptual framework to tackle general sectors in PN theory, such as those in the present challenging 
overlap of high-loop and higher-spin orders. The overall derivation builds on the public code 
\texttt{EFTofPNG} introduced in \cite{Levi:2017kzq}, and on extensions of this framework in 
\cite{Levi:2011eq,Levi:2014sba,Levi:2014gsa,Levi:2015uxa,Levi:2015ixa,Levi:2016ofk, 
	Levi:2019kgk,Levi:2020kvb,Levi:2020uwu,Levi:2020lfn,Kim:2022pou}. 

In section \ref{eftform} we present the formulation of the EFT required for the present 
higher-spin sectors, which importantly entails non-minimal coupling of spin to gravity. 
Moreover, in section \ref{tidaltheory} we present the extension of the effective theory of a spinning 
particle to couplings that are quadratic in the curvature, which is required at the present $5$PN 
order, where we point out the relation to the point-mass sector.
In section \ref{eftcomput} we present in full detail the evaluation of the diagrammatic expansion, 
which consists of $4$ independent sectors of quadratic-in-spin corrections at the $5$PN order. 
The Feynman graphs are evaluated via advanced multi-loop methods with a computational load of 
graphs, and a generic dimensional expansion, due to the appearance of dimensional-regularization 
(DimReg) poles across all loop orders, which we handled with independent development of the 
\texttt{EFTofPNG} code, similar to \cite{Kim:2022pou}. In section \ref{bulkyaction} we provide the full 
generalized Lagrangians for all of the $4$ independent contributions included in the present 
quadratic-in-spin sectors.

We proceed in section \ref{minimalism} to reduce these effective actions via a formal procedure that 
involves redefinitions of the position and rotational variables through lower-order sectors, which 
requires in higher-spin sectors such as the present ones, an intricate implementation of the 
redefinitions beyond linear order. We provide the full redefinitions in section \ref{redefsinhere}, and 
present in section \ref{normalaction} the consequent reduced actions of all $4$ subsectors for the 
first time. 
In section \ref{thebeast} we present the full general Hamiltonians of the relevant $4$ subsectors for 
the first time, with the consequent useful simplified Hamiltonians in section \ref{constrainedhams}. 
Finally, in section \ref{seeforreal} we proceed to discuss the consequent gauge-invariant 
binding-energy relations to the angular momentum and orbital frequency, noting in particular the 
binding energy associated to the new tidal effect with spins, in comparison to that from the point-mass 
sector, which also contributes at the present $5$PN order.

We end in section \ref{imaginaryevents} with a derivation of all the extrapolated scattering angles 
that correspond to an extension of our Hamiltonians to the scattering problem in the simplified 
aligned-spins configuration, as a guide to scattering-amplitudes studies.

\section{EFT of Spinning Gravitating Objects}
\label{eftform}

We start by reviewing the EFT of spinning gravitating objects \cite{Goldberger:2004jt, 
Levi:2015msa,Levi:2018nxp}. At the orbital scale the compact-binary inspiral is 
described as a two-particle system with an effective action that consists of the purely 
gravitational action with the relevant weak field modes, augmented by a 
point-particle action, $S_{\text{pp}}$, for each of the compact objects in the binary 
\cite{Goldberger:2004jt}:
\be 
\label{fullorbital}
S_{\text{eff}}=S_{\text{gr}}[g_{\mu\nu}]+\sum_{a=1}^{2}S_{\text{pp}}(\lambda_a).
\ee
$S_{\text{gr}}$, the action of the field in some theory of gravity, 
must be supplemented by $S_{\text{pp}}$, a new infinite tower of interactions between the 
weak field modes and the new worldline degrees of freedom at this scale. These 
interactions are localized on the worldline of each object in the binary, parametrized by 
$\lambda_a$ for the $a$-th object.

For the gravitational action we consider here the theory of GR, which we supplement with 
the fully-harmonic gauge-fixing term:
\be 
\label{puregr}
S_{\text{gr}}[g_{\mu\nu}]=S_{\text{EH}} + S_{\text{GF}} 
= -\frac{1}{16\pi G_d} \int d^{D}x \sqrt{g} \,R 
+ \frac{1}{32\pi G_d} \int d^{D}x\sqrt{g}
\,g_{\mu\nu}\Gamma^\mu\Gamma^\nu, 
\ee
with $\Gamma^\mu\equiv\Gamma^\mu_{\rho\sigma}g^{\rho\sigma}$.
As of third subleading orders in PN theory, such as in this work, the dimension must be 
kept generic throughout the whole evaluation of two-body interactions, including a modified 
minimal subtraction ($\overline{\text{MS}}$) (see e.g.~\cite{Peskin:1995ev}), for the 
$d$-dimensional coupling constant of the NR theory:
\be
G_d\equiv G_N \left(\sqrt{4\pi e^\gamma} \,R_0 \right)^{d-3},
\ee
with $d\equiv D-1$ the number of spatial dimensions, $G_N\equiv G$ Newton's gravitational 
constant in three-dimensional space, 
$\gamma$ 
%\equiv \lim_{z\to 0$}\tfrac{1}{z}-\Gamma(z)$ 
Euler's constant, and $R_0$ some renormalization scale. 

For the NR gravitational field it is beneficial to use a $d+1$ Kaluza-Klein reduction 
over the time dimension \cite{Kol:2007bc,Kol:2010ze}: 
\begin{align}
\label{kkred}
ds^2=g_{\mu\nu}dx^{\mu}dx^{\nu}\equiv 
e^{2\phi}\left(dt-A_idx^i\right)^2-e^{-\frac{2}{d-2}\phi}\gamma_{ij}dx^idx^j,
\end{align}
which defines the set of fields: $\phi$, $A_i$, and 
$\gamma_{ij}\equiv\delta_{ij}+\sigma_{ij}$.
This decomposition has been useful also in sectors with spin since its application 
in \cite{Levi:2008nh}, and as applied later all across the public 
\texttt{EFTofPNG} code \cite{Levi:2017kzq}.
The propagators in a NR approximation are instantaneous, leaving the 
momenta integrals as purely spatial, namely Euclidean in $d$ dimensions. The relativistic 
corrections in this approximation are then obtained as quadratic insertions on the 
propagators with two time derivatives. All the Feynman rules which are required for 
the present sectors in $d$ dimensions, involving only the gravitational field, 
including the propagators, their relativistic perturbative corrections, and the 
self-interaction vertices, are similar to those provided in \cite{Kim:2022pou}. 

For rotating objects one then needs to specify the effective action of a spinning 
particle as the point-particle action of each object.
This effective action starts with the leading couplings of mass 
and spin, whose associated position and rotational degrees of freedom (DOFs), 
respectively, are minimally coupled to gravity 
\cite{Levi:2010zu,Levi:2015msa,Levi:2018nxp}:
\begin{align} 
\label{mcwspin}
S_{\text{pp}}(\lambda)=&\int 
d\lambda\left[-m \sqrt{u^2}-\frac{1}{2} \hat{S}_{\mu\nu} \hat{\Omega}^{\mu\nu}
-\frac{\hat{S}^{\mu\nu} p_{\nu}}{p^2} \frac{D p_{\mu}}{D \lambda}\right].
\end{align}
Here we introduced the mass $m$, the $4$-velocity $u^{\mu}$, the linear momentum 
$p_{\mu}$, as well as generic angular velocity and spin variables, $\hat{\Omega}^{\mu\nu}$ and 
$\hat{S}_{\mu\nu}$, respectively. The latter generic rotational variables, together with 
the extra term in eq.~\eqref{mcwspin} with the covariant derivative, enable to switch the 
rotational gauge along the worldline \cite{Levi:2015msa}, which was not accounted for in past 
formulations of spin in relativity \cite{Hanson:1974qy,Bailey:1975fe,Yee:1993ya,Porto:2005ac}.

Next, we must consider the part of the point-particle action that accounts for 
finite-size effects, of which we have in the present sectors two types: spin-induced 
multipolar, and tidal deformations, that are both uniquely due to the presence of spin. 
Tidal effects enter (only) as of the present high $5$PN order, and thus we need to 
consider them properly for the present sectors, which we do in section \ref{tidaltheory} below.
For the spin-induced multipolar deformations, we consider the leading non-minimal 
couplings to gravity to all orders in spin, that are linear in the curvature, which we introduced in 
\cite{Levi:2014gsa,Levi:2015msa}:
\begin{align}
\label{spindeform}
L_{\text{NMC(RS$^\infty$)}}&
=\sum_{n=1}^\infty \frac{(-1)^n}{(2n)!}\frac{C_{ES^{2n}}}{m^{2n-1}}
D_{\mu_{2n}}\cdots D_{\mu_{3}}\frac{E_{\mu_{1}\mu_{2}}}{\sqrt{u^2}}
\bullet S^{\mu_1}S^{\mu_2}\cdots S^{\mu_{2n-1}}S^{\mu_{2n}}\nn\\
&+\sum_{n=1}^\infty \frac{(-1)^n}{(2n+1)!}\frac{C_{BS^{2n+1}}}{m^{2n}}
D_{\mu_{2n+1}}\cdots D_{\mu_{3}}\frac{B_{\mu_{1}\mu_{2}}}{\sqrt{u^2}}
\bullet S^{\mu_1}S^{\mu_2}\cdots 
S^{\mu_{2n}}S^{\mu_{2n+1}},
\end{align}
where $\bullet$ stands for tensor contraction of the spin-induced multipole tensors with 
the definite-parity electric and magnetic components of the curvature:
\bea
E_{\mu\nu}&\equiv& R_{\mu\alpha\nu\beta}u^{\alpha}u^{\beta}, \label{elec}\\
B_{\mu\nu}&\equiv& \frac{1}{2} \epsilon_{\alpha\beta\gamma\mu} 
R^{\alpha\beta}_{\,\,\,\,\,\,\,\delta\nu}u^{\gamma}u^{\delta}\label{mag},
\eea
and their covariant derivatives, $D_{\mu}$. This infinite series of operators in 
eq.~\eqref{spindeform} is preceded by a corresponding series of Wilson coefficients, 
that would be referred to as ``multipolar deformation parameters'' in traditional GR. From 
this series only the leading term is needed in the present sectors, with the electric 
component and no further covariant derivatives, and whose coefficient corresponds to the 
quadrupolar deformation constant, as in \cite{Barker:1975ae}.

From the effective action of a spinning particle in eqs.~\eqref{mcwspin}, 
\eqref{spindeform}, all the necessary worldline couplings for the present sectors can be 
extracted using the \texttt{EFTofPNG} code, including both mass and spin, but excluding 
new couplings that are beyond linear in the curvature, which are discussed in the next section 
\ref{tidaltheory}. 
We remind that in our spin couplings we implement the generalized canonical gauge that we
introduced in \cite{Levi:2015msa}, and so all the indices in the Feynman rules are 
Euclidean. All of these rules are provided in both human- and machine-readable formats in 
the ancillary files to this publication.

\subsection{New Physics}
\label{tidaltheory}

At such high perturbative orders the point-particle effective action needs to be 
further extended to include non-minimal coupling to gravity that is quadratic in the 
curvature. In the EFT formulation of the point-mass sector, defined as that whose 
worldline DOFs are only the position coordinates, two leading 
tidal operators were presented already in \cite{Goldberger:2004jt}: 
\begin{align}
\label{ordinarylove}
L_{\text{NMC(R$^2$S$^0$)}}^{\text{LO}}&
= C_{\text{E$^2$}} \frac{E_{\alpha\beta}E^{\alpha\beta}}{\sqrt{u^2}^{\,3}}
+ C_{\text{B$^2$}} \frac{B_{\alpha\beta}B^{\alpha\beta}}{\sqrt{u^2}^{\,3}},
\end{align}
where the coefficients in this expression contain all dependence in the mass and coupling constant, and
numerical symmetry factors, in contrast to the new coefficients we introduced in 
eq.~\eqref{spindeform}. 
Initial power-counting indicates that these operators may enter at the $5$PN order, notably 
being the leading finite-size effects to affect the point-mass sector, where this high PN 
accuracy has been obtained only recently in this simplest of all sectors
\cite{Bini:2019nra,Bini:2020wpo,Bini:2020uiq,Blumlein:2020pyo}. 

An extension of the effective action of a spinning particle to include such general
operators was outlined in \cite{Levi:2015msa}, and approached in 
\cite{Levi:2020uwu,Levi:2020lfn}. There it was shown that quadratic-in-curvature operators, 
which are linear in the spin may enter only as of the $6.5$PN order, notably making also the 
spin-orbit sector affected solely by minimal coupling to gravity to a very high PN 
order. Following \cite{Levi:2020uwu} and then \cite{Kim:2021rfj}, we consider 
up to quadratic order in the spins and at the present $5$PN accuracy, the following 
extension to the non-minimal coupling of spin to gravity:
\begin{subequations}
\label{nmcwspin2}
\begin{align}
L_{\text{NMC(R$^2$S$^2$)}}^{\text{LO}}&
= \frac{1}{2} C_{\text{E$^2$}} G^4 m^5 \frac{E_{\alpha\beta}E^{\alpha\beta}}{\sqrt{u^2}^{\,3}}
+ \frac{1}{2} C_{\text{B$^2$}} G^4 m^5 \frac{B_{\alpha\beta}B^{\alpha\beta}}{\sqrt{u^2}^{\,3}}
\label{ordLove}\\
& + \frac{1}{2} C_{\text{E$^2$S$^2$}} G^2 m 
\frac{E_{\mu\alpha}E_{\nu}^{\,\,\,\alpha}}{\sqrt{u^2}^{\,3}}
 S^{\mu} S^{\nu}
+ \frac{1}{2} C_{\text{B$^2$S$^2$}} G^2 m 
\frac{B_{\mu\alpha}B_{\nu}^{\,\,\,\alpha}}{\sqrt{u^2}^{\,3}}
S^{\mu} S^{\nu}.
\label{spinLove}
\end{align}
\end{subequations}
As detailed in \cite{Levi:2015msa} we construct the effective action of a spinning 
particle by considering the contraction of $SO(3)$ tensors in the local frame, 
and specifically for rotating objects -- in the body-fixed frame. Accordingly, the 
$SO(3)$ building blocks are symmetrized and their traces are removed.

Note that in eq.~\eqref{nmcwspin2} we already defined the dimensionless normalized Wilson 
coefficients of the new operators, similar to eq.~\eqref{spindeform}. Note that these are 
the leading operators to carry an additional scaling in the gravitational coupling-constant $G$, in 
contrast to all operators encountered up to this PN order, 
including the operator that is quadratic in curvature and quartic in spin at the $5$PN 
order \cite{Levi:2020lfn}. 
The signs of new operators are fixed according to the leading non-minimal 
coupling, and the number of derivatives. 
The Wilson coefficients in eq.~\eqref{ordLove} correspond to the long-studied ``Love 
numbers'' in traditional GR.
Trace terms of the new spin operators in eq.~\eqref{spinLove} are 
absorbed in the usual ``Love numbers'' from the point-mass sector, 
e.g.~$C_{\text{E$^2$}} \supset \tfrac{C_{\text{E$^2$S$^2$}}}{3} \tfrac{S^2}{G^2m^4}$
(recall that $[S^2/(G^2m^4)]\le1$ for Kerr black holes). 
This is similar to that the basic constant scalar spin length, S$^2$, which is the trace of the square
of the spin tensor, is absorbed in the mass, and other Wilson coefficients, and is therefore
omitted from our traceless tensors \cite{Levi:2015msa}.
For this reason such trace terms cannot be matched independently of the mass.
Yet, as we shall clearly see the new operators in eq.~\eqref{spinLove} represent a 
new type of tidal effects that are only relevant to spinning objects, and are thus 
preceded by new Wilson coefficients \cite{Kim:2021rfj}.
Their intrinsic $G$ scaling gives further indication that the operators in 
eq.~\eqref{spinLove} belong to a different type of effects than spin-induced 
multipolar deformations as in eq.~\eqref{spindeform}, and that the new Wilson 
coefficients may correspond to a generic concept of ``Love numbers'' in gravity \cite{Levi:2022rrq}. 
% reconsider symmetry factor of spin "Love" coefficeints

Proceeding to derive the consequent Feynman rules, one finds that the operators in 
eq.~\eqref{nmcwspin2}, which contain the magnetic curvature component, actually enter 
only as of the 6PN order, and thus we are left with the contributions:
\begin{align}
\label{nmcwspin2at5pn}
L_{\text{NMC(R$^2$S$^2$)}}^{5\text{PN}}&
= \frac{1}{2} C_{\text{E$^2$}} G^4 m^5 \frac{E_{ij}E_{ij}}{\sqrt{u^2}^{\,3}} 
- \frac{1}{2} C_{\text{E$^2$S$^2$}} G^2 m  
\Bigg[ 
\frac{E_{ik}E_{jk}}{\sqrt{u^2}^{\,3}} S^{i} S^{j}
- \frac{S^2}{3} \frac{E_{ij}E_{ij}}{\sqrt{u^2}^{\,3}}\Bigg].
\end{align} 
From eq.~\eqref{nmcwspin2at5pn} we can now derive the Feynman rules for the point-mass 
tidal two-graviton coupling as 
(using JaxoDraw \cite{Vermaseren:1994je,Binosi:2003yf,Binosi:2008ig}):
\begin{align}
\label{frmE2} 
\parbox{12mm}{\includegraphics[scale=0.6]{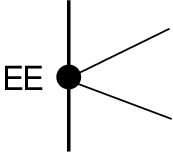}} \qquad = 
& \quad \frac{C_{\text{E}^2}\,G^4m^5}{2} \int dt \, \phi_{,ij}\phi_{,ij}\,,
\end{align}
and for the quadratic-in-spin new two-graviton coupling as:
\begin{align}
\label{frs2E2} \parbox{12mm}{\includegraphics[scale=0.6]{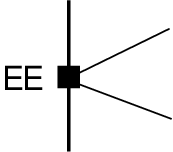}} \qquad = 
& \quad -\frac{C_{\text{E}^2\text{S}^2}\,G^2m}{2} \int dt \, 
\Big[S_iS_j\phi_{,ik}\phi_{,jk} - \frac{S^2}{3} \phi_{,ij}\phi_{,ij} \Big].
\end{align}
% fix 2nd Feynman rule reg trace removal 
Let us highlight here again that these types of tidal operators for the point-mass and for 
the quadratic-in-spin sectors, will enter the potentials at orders $G^6$ and $G^4$, respectively, though 
they start to contribute through the graph topology of two-graviton exchange at 
order $G^2$. This is in contrast with all other (non-tidal) PN contributions up to the present $5$PN 
order, where the order in $G$ has always been identical to the order of the graph topology.

\section{Diagrammatic Expansion}
\label{eftcomput}

\begin{table}[t]
\begin{center}
\begin{tabular}{|l|r|r|r|r||r|}
\hline
Order in $G$ & 1 & 2 & 3 & 4 & Total\\
\hline
S$_1$S$_2$ & 10 & 94 & 294 & 73 & 471\\
\hline
S$_1^2$ & 0 & 50 & 181 & 52 & 283\\
\hline
C$_{\text{E}}$S$_1^2$ & 9 & 107 & 213 & 38 & 367\\
\hline
C$_{\text{E}^2}$S$_1^2$ & 0 & 1 & 0 & 0 & 1\\
\hline
Total in S$^2$ & 19 & 252 & 688 & 163 & 1122\\
\hline
\end{tabular}
\caption{The number of graphs across topology orders in $G$ in the N$^3$LO 
quadratic-in-spin sectors.}
\label{graphbreakdown}
\end{center}
\end{table}

\begin{figure}[t]
\centering
\includegraphics[scale=0.5]{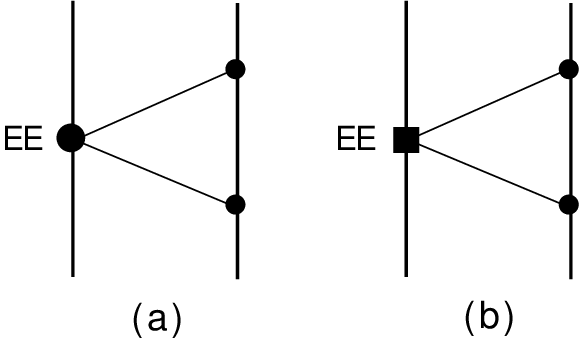}
\caption{The leading Feynman graphs of two-graviton exchange due to quadratic-in-curvature operators 
entering first at the $5$PN order.
(a) The tidal interaction in the point-mass sector.
(b) The new interaction in the present quadratic-in-spin sectors, due to the new 
quadratic-in-curvature coupling to spin.}
\label{5pne2}
\end{figure}

The Wick contractions and corresponding Feynman graphs in the diagrammatic expansion 
for the present sectors were then generated using the \texttt{EFTofPNG} code. The 
distribution of graphs among the $4$ topology orders in $G$, and among the $4$ quadratic-in-spin 
sectors at the N$^3$LO, is 
displayed in table \ref{graphbreakdown}. There are $1122$ graphs altogether, with $1$ 
graph originating from the new coupling beyond linear in curvature from section 
\ref{tidaltheory}, as shown in figure \ref{5pne2}(b). The topologies at order $G^4$, and their 
corresponding integral structure, were studied  in \cite{Levi:2020kvb,Levi:2020uwu}, where $163$ 
graphs at this order in the present sectors were evaluated. The analysis of all graphs in 
the present sectors builds on the N$^2$LO quadratic-in-spin sectors completed via our EFT in 
\cite{Levi:2011eq,Levi:2015ixa}.

We provide all the graphs and their values enumerated in $3$ separate lists, which 
correspond to the interactions of spin$_1$ with spin$_2$, of each spin non-linearly with 
itself, and of each spin-induced quadrupole. These lists are included in the ancillary 
files to this publication in both human visual and machine-readable formats. The graphs 
are indexed as $(n_1,n_2,n_3)$, with $n_1$ for order in $G$, $n_2$ for topology type (for 
the topologies at each order in $G$, see \cite{Levi:2018nxp,Levi:2020kvb}), and $n_3$ as 
a graph counter within each topology. Only unique graphs with spin on worldline ``1'' are 
presented in each list, while a copy of each list from the exchange of worldline labels, 
$1\leftrightarrow2$, is suppressed.

Dimensional regularization (DimReg) is used to evaluate all integrals, and as in all 
third and higher subleading sectors \cite{Levi:2020kvb}, the dimension needs to be 
kept generic across the whole derivation. An expansion in the dimensional parameter, 
$\epsilon_d\equiv d-3$, is imperative due to the appearance of DimReg poles (in 
conjunction with logarithms) in the values of individual graphs across all loop orders, 
and in their summation. 
This dimensional expansion is the most time-consuming and demanding task in the 
evaluation of the present sectors. Similar to the spin-orbit sector in \cite{Kim:2022pou}, 
there is also a concentration of graphs in the highest-loop of each order in $G$ (see 
\cite{Levi:2020kvb} for the definition of loop order in the worldline picture). 
At order $G^2$ where one-loop is the highest loop order, $192$ out of $252$ graphs are in 
the one-loop topology, which in the present sectors already contain DimReg poles and 
logarithms, and at order $G^3$, $384$ of the $688$ graphs are of two-loop order.
%, and the majority of these topologies yield such DimReg poles with logarithms. 
% 5 out of 9 toplopgies at 2-loop: 2 1-loops + 2 nested + 1 IBP

In the present sectors similar to the spin-orbit sector, the topologies at order $G^3$, 
whose evaluation is the most demanding, are of the nested type, with $214$ graphs, and of the 
rank-two topology (see \cite{Levi:2020kvb} for the definition of such rank), which 
requires the use of integration-by-parts (IBP) methods, with $87$ graphs overall.
We also recall that at the N$^3$LO the graphs at order $G^1$ can carry up to $3$ 
relativistic corrections (with $2$ time derivatives each) on their propagators, and that 
these insertions are also allowed on two-loop graphs at order $G^3$. Therefore we implemented 
upgrades of the \texttt{EFTofPNG} code, which included improvements of the IBP and 
projection methods, see e.g.~\cite{Smirnov:2006ry,Laporta:2001dd} and  
\cite{Boels:2018nrr,Chen:2019wyb}, respectively, for reviews of such modern multi-loop 
integration methods. 
Independent development, implementation and evaluation, were carried out in parallel to 
verify the validity of all results. 

We note that similar to the spin-orbit sector \cite{Kim:2022pou}, factorizable 
topologies gave rise either to zeros due to contact-interaction terms 
\cite{Levi:2011eq,Levi:2020kvb},
% 20 out of 36 and 14 out of 47 in the 2 factorizable toplogies
or to factors of the transcendental Riemann zeta value $\zeta(2)\propto\pi^2$, which  
also appeared in the single rank-two topology up to order $G^3$.

Finally, let us examine the contributions from the quadratic-in-curvature operators in 
eq.~\eqref{nmcwspin2at5pn} to our present sectors. It is clear that there are no explicit 
contributions to the present sectors from contractions/graphs that contain the point-mass 
tidal coupling in eq.~\eqref{frmE2}. Yet let us evaluate the related leading contribution 
as a point of reference. It gives rise to a graph of two-graviton exchange, 
shown in figure \ref{5pne2}(a), which equals:
\begin{align}
\text{Fig.~1}(a) & =
3 C_{1(\text{E}^2)}\, \frac{G^6 m_1^5 m_2^2}{r^6}.
\end{align}
The new quadratic-in-spin coupling in eq.~\eqref{frs2E2} however, gives rise to a graph 
of two-graviton exchange, shown in figure \ref{5pne2}(b), that contributes to the present 
sectors:
\begin{align}
\text{Fig.~1}(b) & = 
\frac{C_{1(\text{E}^2\text{S}^2)}}{2}\frac{G^4\,m_1m_2^2}{r^6}
\left[S_1^2-3\big(\vec{S}_1\cdot\vec{n}\big)^2\right].
\end{align}
Thus this contribution introduces a new Wilson coefficient, possibly of a ``Love number'' 
type, appearing first in the present sectors.

\subsection{Effective Actions}
\label{bulkyaction}

From the sum of all graphs an effective action is obtained, which contains higher-order 
time derivatives up to the sixth order in total, beyond the standard velocity and spin 
variables. In the present quadratic-in-spin sectors this effective action consists of 
$4$ independent parts for the potential, 
$-V^{\text{N}^3\text{LO}}_{\text{S$^2$}}\subset L$:
\bea
V^{\text{N}^3\text{LO}}_{\text{S}^2} =V^{\text{N}^3\text{LO}}_{\text{S}_1 \text{S}_2} 
+V^{\text{N}^3\text{LO}}_{\text{S}_1^2} + C_{1\text{ES}^2} V^{\text{N}^3\text{LO}}_{C_{1\text{ES}^2}} 
+ C_{1\text{E$^2$S$^2$}} V^{\text{N}^3\text{LO}}_{C_{1\text{E$^2$S$^2$}}} + (1 \leftrightarrow 2),
\eea
where
\bea
V^{\text{N}^3\text{LO}}_{\text{S}_1 \text{S}_2} &= & \sum_{i=0}^6 
\stackrel{(i)}{V}_{3,\text{S}_1 \text{S}_2},\\
V^{\text{N}^3\text{LO}}_{\text{S}_1^2} &= & \sum_{i=0}^4 
\stackrel{(i)}{V}_{3,\text{S}_1^2},\\
V^{\text{N}^3\text{LO}}_{C_{1\text{ES}^2}} &= & \sum_{i=0}^5 \stackrel{(i)}{V}_{3,\text{E}},\\
V^{\text{N}^3\text{LO}}_{C_{1\text{E$^2$S$^2$}}} &=& 	- \frac{G^4 m_{1} m_{2}{}^2}{2 r{}^6} 
\Big[ S_{1}^2 - 3 ( \vec{S}_{1}\cdot\vec{n})^{2} \Big],
\eea
where the superscript $(i)$ stands for the total number of higher-order time 
derivatives in each of these pieces, and the subscript contains the 
subleading/maximal-loop counter $n$ (see \cite{Levi:2019kgk} where the notation of sectors, $(n,l)$, 
was introduced), and a notation for the respective specific sector within the present
quadratic-in-spin sectors. Notice that the part of the potential that is due to the 
non-linear interaction of the spin with itself contains only up to $4$ higher-order time 
derivatives in total, since it enters only as of the NLO at order $G^2$.
Since this unreduced action of all the present sectors is very lengthy, we provided it  
in a separate tex file, on top of two machine-readable files, which cover separately the 
part bilinear in the spins, and the part that is square in each spin, in the ancillary 
files to this publication.
Notice that all in all, at this stage there are DimReg poles with logarithms at orders 
$G^2$ and $G^3$, and transcendental $\zeta(2)$ factors at orders $G^3$ and $G^4$.

\section{Reduction of Effective Actions}
\label{minimalism}

We press on to reduce the action obtained directly from the EFT computation, which 
contains many higher-order time derivatives beyond velocity and spin, to our final action. 
The formal reduction procedure we carry out 
in the present sectors 
builds on that which was presented in detail for the N$^3$LO spin-orbit sector \cite{Kim:2022pou}. 
It proceeds with gradual subleading redefinitions of the position and rotational variables, 
where the formulation to handle rotational variables was first introduced in \cite{Levi:2014sba}.  
Table \ref{redefsectors} shows all the sectors that are involved in the reduction 
process for the present sectors. Note that redefinitions are applied on all these 
sectors, some of which do not require themselves any redefinitions in order to be 
fixed, as they yield contributions to the present sector. 

\begin{table}[t]
\begin{center}
\begin{tabular}{|l|c|c|c|c|}
\hline
\backslashbox{\quad\boldmath{$l$}}{\boldmath{$n$}} & (N\boldmath{$^{0}$})LO
& N\boldmath{$^{(1)}$}LO & \boldmath{N$^2$LO}
& \boldmath{N$^3$LO} 
\\
\hline
\boldmath{S$^0$} & %\cellcolor{green!25} 
& %\cellcolor{green!25} 
& %\cellcolor{yellow!25}
+ 
& %\cellcolor{yellow!25}
+ 
\\
\hline
\boldmath{S$^1$} & %\cellcolor{yellow!25}
+ 
& %\cellcolor{yellow!25}
++ & %\cellcolor{yellow!25}
++ 
& \cellcolor{gray}
%++
\\
\hline
\boldmath{S$^2$} & %\cellcolor{yellow!25} 
& %\cellcolor{yellow!25}
++ & %\cellcolor{yellow!25}
++ 
& %\cellcolor{yellow!25}
++ 
\\
\hline
\end{tabular}
\caption{The notation $(n,l)$ of sectors was introduced in \cite{Levi:2019kgk}. 
The $11$ sectors that contribute to the present $(3,2)$ one through redefinition of 
variables. 
%include all sectors in the table apart from the N$^3$LO spin-orbit sector which is gray. 
"+" marks sectors that require only position shifts to be fixed, 
whereas "++" marks sectors that require redefinition of both position and rotational 
variables to be fixed.}
\label{redefsectors}
\end{center}
\end{table}

First, as pointed out in \cite{Levi:2015msa} at higher-spin sectors, 
such as the present ones, position shifts need to be applied beyond linear order 
already at the NLO due to the leading position shift that is required already at the LO 
spin-orbit sector. Since position shifts scale like the sector they originated from in 
their spin and PN orders \cite{Levi:2014sba}, we only need to apply here such shifts 
originating from the spin-orbit sector, to quadratic order, while all other position 
shifts can be applied at linear order.
Second, for redefinitions of the rotational variables, that are clearly required only 
in sectors with spins, their power counting is $S^{-1}v^{-1}$ relative to their 
originating sector \cite{Levi:2014sba}. This power counting, together with the extension 
of the formal procedure of redefinition for rotational variables beyond linear order, 
that was carried out in \cite{Kim:2022pou}, show that in the present sectors such 
redefinitions are only needed to be applied at linear order.

\subsection{Redefinition of Degrees of Freedom}
\label{redefsinhere}

As noted we follow here closely the reduction procedure which was presented in detail 
for the N$^3$LO spin-orbit sector in \cite{Kim:2022pou}. As table \ref{redefsectors} shows, we 
proceed through the application of redefinitions fixed in $8$ relevant sectors including 
the present ones, according to their increasing PN order, see \cite{Levi:2019kgk} for our 
general PN-counting formula for any sector, with or without spins. Yet, the redefinitions 
in $5$ of these sectors, which are below quadratic in the spin, are not modified with 
respect to \cite{Kim:2022pou}, and thus we need to consider here in addition only the $3$ 
sectors that are quadratic in the spin, as summarized in tables 
\ref{nlos2redef}--\ref{n3los2redef}. 

The algorithm we use for the reduction procedure is similar to what we described in 
\cite{Kim:2022pou}, with the tables here understood in a similar manner to what was 
explained therein. However, as can be inferred from table \ref{redefsectors}, and tables 
\ref{nlos2redef}--\ref{n3los2redef}, the reduction procedure in the present sectors is 
more laborious and computationally demanding, mainly due to the application of position 
shifts to quadratic order. Nevertheless, our streamlined code following the algorithm we 
described in \cite{Kim:2022pou} still completes the necessary reduction for the present sectors 
efficiently and rapidly.

Let us then go over all the relevant sectors in order, also reviewing those which were 
covered in \cite{Kim:2022pou}. We remind that all of our unreduced actions are computed 
directly with the \texttt{EFTofPNG} code.
First, we note that for the LO spin-orbit, $2$PN, NLO spin-orbit, and $3$PN sectors our 
unreduced actions and redefinitions are identical to those presented in 
\cite{Kim:2022pou}. We remind that in the latter sector, namely at $3$PN, there are DimReg 
poles and logarithms showing up for the first time, and so we also add in advance a total time 
derivative (TTD) to the unreduced action in order to land on a final action 
without these unphysical features. Moreover, since redefinitions in the 3PN sector also 
contain such poles, they should be applied on LO potentials, expanded to linear order in 
their DimReg zeros and containing further logarithms. For the present sectors it means 
considering beyond the Newtonian potential, also the following piece of 
the LO quadratic-in-spin potentials:
\bea
L^{\,\text{LO}}_{\,\text{S$^2$}}|_{{\cal{O}}(\epsilon^1)} &=&\epsilon \left[
- 	\frac{G}{r{}^3} 
\left( \left[2 - \frac{3}{2} \log 
\left( \frac{r}{R_0} \right) \right] \vec{S}_{1}\cdot\vec{n} \vec{S}_{2}\cdot\vec{n} 
-  \left[1 - \frac{1}{2} \log \left( \frac{r}{R_0} \right) \right]
\vec{S}_{1}\cdot\vec{S}_{2} \right) \right.\nn\\
&&\left.+ 	\frac{G C_{1(\text{ES}^2)} m_{2}}{4 m_{1} r{}^3} 
\left( \left[5 - 2 \log \left( \frac{r}{R_0} \right) \right] S_{1}^2 -  
\left[11 - 6 \log \left( \frac{r}{R_0} \right) \right]
 ( \vec{S}_{1}\cdot\vec{n})^{2} \right)\right]\nn\\
&&+ \left( 1 \leftrightarrow 2 \right).
\eea

\begin{table}[t]
\begin{center}
\begin{tabular}{|l|c|c|}
\hline
\backslashbox{from}{\boldmath{to}} 
& (0P)N & LO S$^1$
\\
\hline
LO S$^1$ & $(\Delta \vec{x})^2$ & $\Delta \vec{x}$
\\
\hline
NLO S$^2$ & $\Delta \vec{x}$, $\Delta \vec{S}$ &
\\
\hline
\end{tabular}
\caption{Contributions to the NLO S$^2$ sectors from position shifts and spin 
redefinitions in lower-order sectors.}
\label{nlos2redef}
\end{center}
\end{table}

At this point we proceed to the NLO quadratic-in-spin sectors captured in table 
\ref{nlos2redef}. The unreduced potentials are identical to those presented in our
\cite{Levi:2015msa}. The new redefinitions applied in these sectors for the position and 
spin variables can be written as: 
\bea
\left(\Delta \vec{x}_1\right)^{\text{NLO}}_{\text{S}^2} &=&  	\frac{2 G}{m_{1} r{}^2} 
\Big[ \vec{S}_{1}\cdot\vec{S}_{2} \vec{n} - \vec{S}_{1}\cdot\vec{n} \vec{S}_{2} \Big] 
- 	\frac{G m_{2}}{8 m_{1}{}^2 r{}^2} \Big[ S_{1}^2 \vec{n} - \vec{S}_{1}\cdot\vec{n} 
\vec{S}_{1} \Big] + 	\frac{G C_{1(\text{ES}^2)} m_{2}}{m_{1}{}^2 r{}^2} 
\vec{S}_{1}\cdot\vec{n} \vec{S}_{1} \nn\\
&&- 	\frac{1}{4 m_{1}{}^2} \Big[ \dot{\vec{S}}_{1}\cdot\vec{v}_{1} \vec{S}_{1} - 
\vec{S}_{1}\cdot\vec{v}_{1} \dot{\vec{S}}_{1} \Big] - 	\frac{3}{8 m_{1}{}^2} \Big[ 
S_{1}^2 \vec{a}_{1} - \vec{S}_{1}\cdot\vec{a}_{1} \vec{S}_{1} \Big],
\eea
and
\bea
(\omega^{ij}_1)^{\text{NLO}}_{\text{S}^2} &=&- 	\frac{G}{2 r{}^2} \Big[ 3 S_{2}^{ki} 
n^k \vec{v}_{2}\cdot\vec{n} n^j - 3 S_{2}^{ki} v_{2}^k n^j - S_{2}^{kj} n^k \big( 4 
v_{1}^i - v_{2}^i \big) + S_{2}^{ij} \vec{v}_{2}\cdot\vec{n} \Big] \nn\\ 
&& - 	\frac{G m_{2}}{4 m_{1} r{}^2} \Big[ S_{1}^{ki} v_{1}^k n^j + S_{1}^{kj} n^k 
v_{1}^i \Big] - 	\frac{G C_{1(\text{ES}^2)} m_{2}}{2 m_{1} r{}^2} \Big[ 3 S_{1}^{ki} 
n^k \vec{v}_{2}\cdot\vec{n} n^j - 2 S_{1}^{ki} v_{1}^k n^j \nn\\ 
&& + 3 S_{1}^{ki} v_{2}^k n^j + S_{1}^{kj} n^k \big( 2 v_{1}^i - 3 v_{2}^i \big) - 
S_{1}^{ij} \big( 2 \vec{v}_{1}\cdot\vec{n} - 3 \vec{v}_{2}\cdot\vec{n} \big) \Big]\nn\\
&&- 	\frac{1}{4 m_{1}} \Big[ S_{1}^{kj} v_{1}^k a_{1}^i + S_{1}^{kj} a_{1}^k v_{1}^i 
\Big] - 	\frac{1}{8 m_{1}} v_{1}^i \dot{S}_{1}^{kj} v_{1}^k \nn\\ && - 	\frac{G}{4 
r} \Big[ \dot{S}_{2}^{ki} n^k n^j + \dot{S}_{2}^{ij} \Big] - (i \leftrightarrow j). 
\eea
Note that these redefinitions are different than those we used in \cite{Levi:2015msa}. 
Next, the N$^2$LO spin-orbit sector is unaffected by the redefinitions in the latter 
sectors, that are quadratic in the spin, and thus its unreduced action and redefinitions 
are again identical to those presented in \cite{Kim:2022pou}. 

\begin{table}[t]
\begin{center}
\begin{tabular}{|l|c|c|c|c|c|}
\hline
\backslashbox{from}{\boldmath{to}} 
& (0P)N & 1PN & LO S$^1$ & LO S$^2$ &  NLO S$^1$ 
\\
\hline
LO S$^1$ &  & $(\Delta \vec{x})^2$ &  &  & $\Delta \vec{x}$
\\
\hline
2PN &  &  & & $\Delta \vec{x}$ &  
\\
\hline
NLO S$^1$ &  &  & $\Delta \vec{x}$ & $\Delta \vec{S}$ &
\\
\hline
NLO S$^2$ &  & $\Delta \vec{x}$ & $\Delta \vec{S}$ & &
\\
\hline
N$^2$LO S$^2$ & $\Delta \vec{x}$, $\Delta \vec{S}$ & & & &
\\
\hline
\end{tabular}
\caption{Contributions to the N$^2$LO S$^2$ sector from position shifts and 
spin redefinitions in lower-order sectors.}
\label{n2los2redef}
\end{center}
\end{table}

We proceed then to the N$^2$LO quadratic-in-spin sectors captured in table 
\ref{n2los2redef}. The unreduced potential can be written as the sum:
\bea
V^{\text{N}^2\text{LO}}_{\text{S}^2} =V^{\text{N}^2\text{LO}}_{\text{S}_1 \text{S}_2} 
+V^{\text{N}^2\text{LO}}_{\text{S}_1^2} + C_{1\text{ES}^2} V^{\text{N}^2\text{LO}}_{C_{1\text{ES}^2}} 
+ (1 \leftrightarrow 2),
\eea
where
\bea
V^{\text{N}^2\text{LO}}_{\text{S}_1 \text{S}_2} &= & \sum_{i=0}^4 
\stackrel{(i)}{V}_{2,\text{S}_1 \text{S}_2},\\
V^{\text{N}^2\text{LO}}_{\text{S}_1^2} &= & \sum_{i=0}^2 
\stackrel{(i)}{V}_{2,\text{S}_1^2},\\
V^{\text{N}^2\text{LO}}_{C_{1\text{ES}^2}} &= & \sum_{i=0}^3 \stackrel{(i)}{V}_{2,\text{E}}.
\eea
The new redefinitions that are fixed in these sectors are then written for 
the position shift as:
\bea
\left(\Delta \vec{x}_1\right)^{\text{N}^2\text{LO}}_{\text{S}^2} =
\left(\Delta \vec{x}_1\right)^{\text{N}^2\text{LO}}_{\text{S}_1 \text{S}_2} &&+\left(\Delta \vec{x}_1\right)^{\text{N}^2\text{LO}}_{\text{S}_1^2} 
+ C_{1\text{ES}^2} \left(\Delta \vec{x}_1\right)^{\text{N}^2\text{LO}}_{C_{1\text{ES}^2}}\nn\\
&&  
+\left(\Delta \vec{x}_1\right)^{\text{N}^2\text{LO}}_{\text{S}_2^2}
+ C_{2\text{ES}^2} \left(\Delta \vec{x}_1\right)^{\text{N}^2\text{LO}}_{C_{2\text{ES}^2}} ,
\eea
where
\bea
\left(\Delta \vec{x}_1\right)^{\text{N}^2\text{LO}}_{\text{S}_1 \text{S}_2} &= & 
\sum_{i=0}^2 \stackrel{(i)}{ \Delta \vec{x}_1 }_{(2,\text{S}_1 \text{S}_2)},\\
\left(\Delta \vec{x}_1\right)^{\text{N}^2\text{LO}}_{\text{S}_1^2} &= & \sum_{i=0}^1 
\stackrel{(i)}{\Delta \vec{x}_1}_{(2,\text{S}_1^2)},\\
\left(\Delta \vec{x}_1\right)^{\text{N}^2\text{LO}}_{C_{1\text{ES}^2}} &= & \sum_{i=0}^1 
\stackrel{(i)}{\Delta \vec{x}_1}_{(2,\text{ES}_1^2)},\\
\left(\Delta \vec{x}_1\right)^{\text{N}^2\text{LO}}_{\text{S}_2^2} &= & \sum_{i=0}^1 
\stackrel{(i)}{\Delta \vec{x}_1}_{(2,\text{S}_2^2)},\\
\left(\Delta \vec{x}_1\right)^{\text{N}^2\text{LO}}_{C_{2\text{ES}^2}} &= & \sum_{i=0}^2 
\stackrel{(i)}{\Delta \vec{x}_1}_{(2,\text{ES}_2^2)}.
\eea
The redefinition of spin can be written as:
\bea
\left(\omega^{ij}_1\right)^{\text{N}^2\text{LO}}_{\text{S}^2} 
=\left(\omega^{ij}_1\right)^{\text{N}^2\text{LO}}_{\text{S}_1 \text{S}_2} 
+\left(\omega^{ij}_1\right)^{\text{N}^2\text{LO}}_{\text{S}_1^2} + C_{1\text{ES}^2} 
\left(\omega^{ij}_1\right)^{\text{N}^2\text{LO}}_{C_{1\text{ES}^2}} - (i \leftrightarrow j),
\eea
where
\bea
\left(\omega^{ij}_1\right)^{\text{N}^2\text{LO}}_{\text{S}_1 \text{S}_2} &= & 
\sum_{i=0}^3 \stackrel{(i)}{ \omega^{ij}_1 }_{(2,\text{S}_1 \text{S}_2)},\\
\left(\omega^{ij}_1\right)^{\text{N}^2\text{LO}}_{\text{S}_1^2} &= & \sum_{i=0}^1 
\stackrel{(i)}{\omega^{ij}_1}_{(2,\text{S}_1^2)},\\
\left(\omega^{ij}_1\right)^{\text{N}^2\text{LO}}_{C_{1\text{ES}^2}} &= & \sum_{i=0}^2 
\stackrel{(i)}{\omega^{ij}_1}_{(2,\text{E})}.
\eea
Since the expressions for the unreduced action and for the redefinitions are 
lengthy, they are presented in a separate tex file, on top of machine-readable files for 
the spin$_1$-spin$_2$ and spin-squared parts, in the ancillary files to this publication.

Finally, we arrive at the present N$^3$LO quadratic-in-spin sectors as captured in 
table \ref{n3los2redef}. 
Similar to the N$^3$LO sectors of 3PN and spin-orbit, we also add here a TTD to the 
unreduced potential:
\bea
\Delta V%^{\text{TTD}}_{\text{N$^3$LO:S$^2$}}
&=& \frac{d}{dt}\left[\Bigg(- 	\frac{19 G^3 m_{1}{}^2}{5 r{}^4} \Big[ 5 
\vec{S}_{1}\cdot\vec{n} \vec{S}_{2}\cdot\vec{n} \vec{v}_{1}\cdot\vec{n} - 
\vec{S}_{1}\cdot\vec{S}_{2} \vec{v}_{1}\cdot\vec{n} - \vec{S}_{2}\cdot\vec{n} 
\vec{S}_{1}\cdot\vec{v}_{1} - \vec{S}_{1}\cdot\vec{n} \vec{S}_{2}\cdot\vec{v}_{1} 
\Big]\right. \nn\\ 
&& + 	\frac{G^3 m_{2}{}^2}{5 r{}^4} \Big[ 5 \vec{S}_{1}\cdot\vec{n} 
\vec{S}_{2}\cdot\vec{n} \vec{v}_{1}\cdot\vec{n} - \vec{S}_{1}\cdot\vec{S}_{2} 
\vec{v}_{1}\cdot\vec{n} - \vec{S}_{2}\cdot\vec{n} \vec{S}_{1}\cdot\vec{v}_{1} - 
\vec{S}_{1}\cdot\vec{n} \vec{S}_{2}\cdot\vec{v}_{1} \Big] \nn\\ 
&& + 	\frac{G^3 C_{1(\text{ES}^2)} m_{1} m_{2}}{70 r{}^4} \Big[ S_{1}^2 \big( 73 
\vec{v}_{1}\cdot\vec{n} - 38 \vec{v}_{2}\cdot\vec{n} \big) + 146 
\vec{S}_{1}\cdot\vec{n} \vec{S}_{1}\cdot\vec{v}_{1} - 76 \vec{S}_{1}\cdot\vec{n} 
\vec{S}_{1}\cdot\vec{v}_{2} \nn\\ 
&& - 5 \big( 73 \vec{v}_{1}\cdot\vec{n} - 38 \vec{v}_{2}\cdot\vec{n} \big) ( 
\vec{S}_{1}\cdot\vec{n})^{2} \Big] + 	\frac{G^3 C_{1(\text{ES}^2)} m_{2}{}^3}{2 m_{1} 
r{}^4} \Big[ S_{1}^2 \big( 2 \vec{v}_{1}\cdot\vec{n} - 3 \vec{v}_{2}\cdot\vec{n} \big) 
\nn\\ 
&& + 4 \vec{S}_{1}\cdot\vec{n} \vec{S}_{1}\cdot\vec{v}_{1} - 6 \vec{S}_{1}\cdot\vec{n} 
\vec{S}_{1}\cdot\vec{v}_{2} - 5 \big( 2 \vec{v}_{1}\cdot\vec{n} - 3 
\vec{v}_{2}\cdot\vec{n} \big) ( \vec{S}_{1}\cdot\vec{n})^{2} \Big] \nn\\ 
&& + 	\frac{G^3 m_{1} m_{2}}{70 r{}^4} \Big[ S_{1}^2 \big( 43 \vec{v}_{1}\cdot\vec{n} 
+ 67 \vec{v}_{2}\cdot\vec{n} \big) + 86 \vec{S}_{1}\cdot\vec{n} 
\vec{S}_{1}\cdot\vec{v}_{1} + 134 \vec{S}_{1}\cdot\vec{n} \vec{S}_{1}\cdot\vec{v}_{2} 
\nn\\ 
&&\left. - 5 \big( 43 \vec{v}_{1}\cdot\vec{n} + 67 \vec{v}_{2}\cdot\vec{n} \big) ( 
\vec{S}_{1}\cdot\vec{n})^{2} \Big] \Bigg)\left(\frac{1}{\epsilon} - 3 \log 
\frac{r}{R_0} \right) \right] + (1 \leftrightarrow 2 ),
\label{ttdn3los2}
\eea
in order to land on a reduced potential free of poles and logarithms, after the 
redefinition of variables that we will fix here, will be carried out. 
\begin{table}[t]
\begin{center}
\begin{tabular}{|l|c|c|c|c|c|c|c|c|}
\hline
\backslashbox{from}{\boldmath{to}} 
& (0P)N & 1PN & LO S$^1$
& 2PN & LO S$^2$ 
& NLO S$^1$ & NLO S$^2$ &  N$^2$LO S$^1$ 
\\
\hline
LO S$^1$ & & & & $(\Delta \vec{x})^2$ & & & & $\Delta \vec{x}$
\\
\hline
2PN & & & & & & & $\Delta \vec{x}$ &  
\\
\hline
NLO S$^1$ & $(\Delta \vec{x})^2$ & & & & & $\Delta \vec{x}$ & $\Delta \vec{S}$ & 
\\
\hline
3PN & & & & & $\Delta \vec{x}$ & & & 
\\
\hline
NLO S$^2$ &  & & & $\Delta \vec{x}$ & & $\Delta \vec{S}$ &  & 
\\
\hline
N$^2$LO S$^1$ & & & $\Delta \vec{x}$ & & $\Delta \vec{S}$ & & & 
\\
\hline
N$^2$LO S$^2$ & & $\Delta \vec{x}$ & $\Delta \vec{S}$ & & & & &
\\
\hline
N$^3$LO S$^2$ & $\Delta \vec{x}$, $\Delta \vec{S}$ & & & & & & &
\\
\hline
\end{tabular}
\caption{Contributions to the N$^3$LO S$^2$ sector from position shifts and spin 
redefinitions in lower-order sectors.}
\label{n3los2redef}
\end{center}
\end{table}
The new position shifts fixed in the present sectors can be written as:
\bea
\label{n3los2xredefs}
\left(\Delta \vec{x}_1\right)^{\text{N}^3\text{LO}}_{\text{S}^2} = \left(\Delta 
\vec{x}_1\right)^{\text{N}^3\text{LO}}_{\text{S}_1 \text{S}_2} 
&& +\left(\Delta \vec{x}_1\right)^{\text{N}^3\text{LO}}_{\text{S}_1^2} + C_{1\text{ES}^2} 
\left(\Delta \vec{x}_1\right)^{\text{N}^3\text{LO}}_{C_{1\text{ES}^2}}\nn\\
&& 
+\left(\Delta \vec{x}_1\right)^{\text{N}^3\text{LO}}_{\text{S}_2^2}+ C_{2\text{ES}^2} 
\left(\Delta \vec{x}_1\right)^{\text{N}^3\text{LO}}_{C_{2\text{ES}^2}} ,
\eea
where
\bea
\left(\Delta \vec{x}_1\right)^{\text{N}^3\text{LO}}_{\text{S}_1 \text{S}_2} &= & 
\sum_{i=0}^4 \stackrel{(i)}{ \Delta \vec{x}_1 }_{(3,\text{S}_1 \text{S}_2)},\\
\left(\Delta \vec{x}_1\right)^{\text{N}^3\text{LO}}_{\text{S}_1^2} &= & \sum_{i=0}^3 
\stackrel{(i)}{\Delta \vec{x}_1}_{(3,\text{S}_1^2)},\\
\left(\Delta \vec{x}_1\right)^{\text{N}^3\text{LO}}_{C_{1\text{ES}^2}} &= & \sum_{i=0}^3 
\stackrel{(i)}{\Delta \vec{x}_1}_{(3,\text{ES}_1^2)},\\
\left(\Delta \vec{x}_1\right)^{\text{N}^3\text{LO}}_{\text{S}_2^2} &= & \sum_{i=0}^3 
\stackrel{(i)}{\Delta \vec{x}_1}_{(3,\text{S}_2^2)},\\
\left(\Delta \vec{x}_1\right)^{\text{N}^3\text{LO}}_{C_{2\text{ES}^2}} &= & \sum_{i=0}^4 
\stackrel{(i)}{\Delta \vec{x}_1}_{(3,\text{ES}_2^2)},
\eea
The new redefinitions of spin that are fixed in the present sectors can be written as:
\bea
\label{n3los2spinredefs}
\left(\omega^{ij}_1\right)^{\text{N}^3\text{LO}}_{\text{S}^2} 
=\left(\omega^{ij}_1\right)^{\text{N}^3\text{LO}}_{\text{S}_1 \text{S}_2} 
+\left(\omega^{ij}_1\right)^{\text{N}^3\text{LO}}_{\text{S}_1^2} 
+ C_{1\text{ES}^2} \left(\omega^{ij}_1\right)^{\text{N}^3\text{LO}}_{C_{1\text{ES}^2}} 
- (i \leftrightarrow j),
\eea
where
\bea
\left(\omega^{ij}_1\right)^{\text{N}^3\text{LO}}_{\text{S}_1 \text{S}_2} &= & 
\sum_{i=0}^5 \stackrel{(i)}{ \omega^{ij}_1 }_{(3,\text{S}_1 \text{S}_2)},\\
\left(\omega^{ij}_1\right)^{\text{N}^3\text{LO}}_{\text{S}_1^2} &= & \sum_{i=0}^3 
\stackrel{(i)}{\omega^{ij}_1}_{(3,\text{S}_1^2)},\\
\left(\omega^{ij}_1\right)^{\text{N}^3\text{LO}}_{C_{1\text{ES}^2}} &= & \sum_{i=0}^3 
\stackrel{(i)}{\omega^{ij}_1}_{(3,\text{E})}.
\eea
These redefinitions for the position and spin for the present sectors are extremely 
lengthy, and thus we provide them in separate tex and machine-readable formats in the 
ancillary files to this publication.

\subsection{Reduced Actions}
\label{normalaction}

The action that we obtain after the reduction procedure detailed above is standard in the 
velocity, but with the additional spin variable in the potential, that contains the following $4$ 
sectors:
\bea
\label{greataction}
\hat{V}^{\text{N}^3\text{LO}}_{\text{S}^2} =\hat{V}^{\text{N}^3\text{LO}}_{\text{S}_1 
\text{S}_2} +\hat{V}^{\text{N}^3\text{LO}}_{\text{S}_1^2} + C_{1\text{ES}^2} 
\hat{V}^{\text{N}^3\text{LO}}_{C_{1\text{ES}^2}} + C_{1\text{E$^2$S$^2$}} 
V^{\text{N}^3\text{LO}}_{C_{1\text{E$^2$S$^2$}}} + (1 \leftrightarrow 2),
\eea
where we provide the explicit expressions of all these sectors in Appendix \ref{actionresults} below.
This action is also provided in machine-readable format in the ancillary files to this 
publication.

Similar to the spin-orbit sector, we find the transcendental  
Riemann zeta value $\zeta(2)\propto \pi^2$ at orders $G^3$ and $G^4$.
Similar to the treatment in the spin-orbit sector \cite{Kim:2022pou}, we obtained an 
action without poles and logarithms, thanks to the addition of TTDs we made at the level 
of unreduced actions. Let us reiterate that this removal of the poles at the level of the 
action is not imperative, and one could alternatively carry such poles through to the 
computation of observables, where these unphysical features would vanish anyway. 
The equations of motion (EOMs) for both the position and spin could be obtained directly 
via variations of the unreduced actions, or more readily from the reduced ones that we 
provided here, as discussed for both cases in our \cite{Levi:2015msa}, thanks to the 
generalized canonical gauge that we formulated therein.

\section{Hamiltonians and Observables}
\label{thebeast}

In our formulation \cite{Levi:2015msa} the full general Hamiltonian is directly obtained from the 
reduced action through a straightforward Legendre transform with respect to the position 
variables.
In this Legendre transform all the reduced actions in the sectors that are noted in 
table \ref{redefsectors} must be included. 
Our general Hamiltonian for the present sectors, that contains $4$ parts, can then be written as:
\bea
H^{\text{N}^3\text{LO}}_{\text{S}^2} = H^{\text{N}^3\text{LO}}_{\text{S}_1 \text{S}_2} +H^{\text{N}^3\text{LO}}_{\text{S}_1^2} + C_{1\text{ES}^2} H^{\text{N}^3\text{LO}}_{C_{1\text{ES}^2}} + C_{1\text{E$^2$S$^2$}} H^{\text{N}^3\text{LO}}_{C_{1\text{E$^2$S$^2$}}} + (1 \leftrightarrow 2),
\eea
where we provide the explicit expressions of all these sectors in Appendix \ref{genhamresults} below.
These Hamiltonians are also provided in machine-readable format in the ancillary files to 
this publication. Our general Hamiltonians here have been fully verified in \cite{Levi:2022rrq}, via the 
completion of the Poincar\'e algebra, which constitutes the most stringent check of validity.

\subsection{Specialized Hamiltonians}
\label{constrainedhams}

In this section we derive specialized Hamiltonians by gradually constraining the 
kinematics, and using simplified notation to obtain distilled expressions, see also 
e.g.~\cite{Levi:2014sba,Kim:2022pou}. First we remind various conventional mass 
quantities of the binary: total mass $m \equiv m_1+m_2$, mass ratio $q \equiv m_1/m_2$, 
reduced mass $\mu \equiv m_1 m_2 / m$, and symmetric mass-ratio 
$\nu \equiv m_1 m_2/m^2 = \mu/m = q/(1+q)^2$. All variables are then rescaled to be 
dimensionless, with $Gm$ and $\mu$ for the length and mass units, respectively. 
The dimensionless variables are then all denoted by a tilde. 

The first simplification is to move to the center-of-mass (COM) frame where $\vec{p} 
\equiv \vec{p}_1 = - \vec{p}_2$. With this we define the orbital angular momentum as
$\vec{L} \equiv r \vec{n} \times \vec{p}$. 
The Hamiltonian of the present sectors in the COM frame is then written as:
\bea
\tilde{H}^{\text{N}^3\text{LO}}_{\text{S}^2} =
\tilde{H}^{\text{N}^3\text{LO}}_{\text{S}_1 \text{S}_2} 
+\tilde{H}^{\text{N}^3\text{LO}}_{\text{S}_1^2} 
+ C_{1\text{ES}^2} \tilde{H}^{\text{N}^3\text{LO}}_{C_{1\text{ES}^2}} 
+ C_{1\text{E$^2$S$^2$}} \tilde{H}^{\text{N}^3\text{LO}}_{C_{1\text{E$^2$S$^2$}}} 
+ (1 \leftrightarrow 2),
\eea
where we provide the explicit expressions of all these sectors in Appendix \ref{comhamresults} below.
As can be easily seen, the COM specification is actually a major simplification to the general 
Hamiltonians.

The second common simplification is of the spins being aligned with the orbital angular 
momentum, that is for both spins we have 
$\vec{S}_a \cdot \vec{n} = \vec{S}_a \cdot \vec{p}=0$.
In the spin-orbit sector, where the coupling of the single spin and angular momentum is 
uniquely simple, this constraint does not affect the COM Hamiltonian. 
However, in all other sectors of higher spin -- this is a dramatic simplification, which yields a 
significant loss of physical information on the system. 
Specifically in our present sectors the 
additional aligned-spins constraints yield:
\bea
\label{hs1s2simp}
\tilde{H}^{\text{N}^3\text{LO}}_{\text{S}_1 \text{S}_2} &=& \frac{\nu \tilde{S}_1   \tilde{S}_2}{2\tilde{r}^6}  \left[\left(
\frac{101641}{600}-\frac{81 \pi ^2}{8}\right) \nu +\frac{4537}{100}
+ \left(
-\frac{209 \nu ^2}{8}+\left(\frac{141 \pi ^2}{64}-\frac{256207}{1200}\right) \nu 
\right.\right.\nn\\
&&\left.
-\frac{6337}{100} 
\right) \frac{\tilde{L}^2}{\tilde{r}} 
+\left( 
\frac{483 \nu ^2}{16}+\frac{339 \nu }{16}-\frac{57}{8}
\right) \frac{\tilde{L}^4}{\tilde{r}^2}
+\left( 
-\frac{25 \nu ^3}{32}-\frac{39 \nu ^2}{16}+\frac{69 \nu }{16}-\frac{17}{16}
\right) \frac{\tilde{L}^6}{\tilde{r}^3}\nn\\
&&+ \tilde{p}_r^2 \tilde{r} \left(
\frac{109 \nu ^2}{8}+\left(\frac{306553}{1200}-\frac{141 \pi ^2}{16}\right) \nu +\frac{1112}{25}
+\left(
\frac{215 \nu ^2}{32}-\frac{209 \nu }{32}-\frac{27}{4}
\right) \frac{\tilde{L}^2}{\tilde{r}} \right.\nn\\
&&\left.+\left(
-\frac{237 \nu ^3}{64}-\frac{81 \nu ^2}{8}+\frac{369 \nu }{32}-\frac{15}{16}
\right) \frac{\tilde{L}^4}{\tilde{r}^2}\right)\nn\\
&&+ \tilde{p}_r^4 \tilde{r}^2 \left(
\frac{575 \nu ^2}{32}-\frac{231 \nu }{8}+\frac{3}{8}
+\left(
-\frac{111 \nu ^3}{16}-\frac{237 \nu ^2}{16}-3 \nu +\frac{21}{16}
\right) \frac{\tilde{L}^2}{\tilde{r}} \right)\nn\\
&&\left.+ \tilde{p}_r^6 \tilde{r}^3 \left(
-\frac{187 \nu ^3}{64}+\frac{153 \nu ^2}{8}-\frac{327 \nu }{32}+\frac{19}{16}
\right) \right],%%%%%%%
\eea
\bea
\label{hs1s1simp}
\tilde{H}^{\text{N}^3\text{LO}}_{\text{S}_1^2} &=& \frac{\nu^2 \tilde{S}_1^2}{\tilde{r}^6}  \left[
\left(\frac{45 \pi ^2}{64}-\frac{6679}{3920}\right) \nu +\frac{3987}{2450}
+ \left( 
\left(-\frac{96683}{3920}-\frac{45 \pi ^2}{64}\right) \nu 
\right.\right.\nn\\
&&\left.-\frac{840921}{19600} \right) \frac{\tilde{L}^2}{\tilde{r}} 
+\left(
\frac{19 \nu ^2}{8}+\frac{191 \nu }{32}-\frac{75}{8}
\right) \frac{\tilde{L}^4}{\tilde{r}^2}+\left(
-\frac{119 \nu ^2}{64}+\frac{91 \nu }{32}-\frac{45}{64}
\right)\frac{\tilde{L}^6}{\tilde{r}^3}
\nn\\
&&+ \tilde{p}_r^2 \tilde{r} \left(
\left(\frac{45 \pi ^2}{16}-\frac{143221}{7840}\right) \nu -\frac{54379}{4900}
+\left(
\frac{283 \nu ^2}{32}+\frac{2075 \nu }{16}-\frac{11}{8}
\right) \frac{\tilde{L}^2}{\tilde{r}} \right.\nn\\
&&\left.+\left(
-\frac{897 \nu ^2}{128}+\frac{195 \nu }{32}-\frac{99}{128}
\right) \frac{\tilde{L}^4}{\tilde{r}^2}\right)\nn\\
&&+ \tilde{p}_r^4 \tilde{r}^2 \left(
\frac{177 \nu ^2}{32}+\frac{43 \nu }{32}+\frac{13}{2}
+\left(
-\frac{165 \nu ^2}{16}+\frac{57 \nu }{32}+\frac{9}{16}
\right)\frac{\tilde{L}^2}{\tilde{r}}\right)\nn\\
&&\left.+ \tilde{p}_r^6 \tilde{r}^3 \left(
-\frac{31 \nu ^2}{128}-\frac{47 \nu }{32}+\frac{81}{128}
\right) \right]\nn\\%%%%%%%
&&+\frac{\nu \tilde{S}_1^2}{q\tilde{r}^6}  \left[
\left(\frac{45 \pi ^2}{64}-\frac{6679}{3920}\right) \nu ^2-\frac{11 \nu }{4}+9\right.\nn\\
&&
+ \left( \left(-\frac{37019}{1960}-\frac{45 \pi ^2}{64}\right) \nu ^2  +\left(\frac{63 \pi ^2}{2048}-36\right) \nu +\frac{441}{16} 
\right) \frac{\tilde{L}^2}{\tilde{r}} \nn\\
&&
+\left(
\frac{19 \nu ^3}{8}-\frac{13 \nu ^2}{4}-\frac{1147 \nu }{32}+\frac{31}{4}
\right) \frac{\tilde{L}^4}{\tilde{r}^2}+\left(
-\frac{13 \nu ^3}{8}+\frac{127 \nu ^2}{16}-\frac{19 \nu }{8}+\frac{45}{64}
\right) \frac{\tilde{L}^6}{\tilde{r}^3}
\nn\\
&&+ \tilde{p}_r^2 \tilde{r} \left(
\left(\frac{45 \pi ^2}{16}-\frac{132651}{7840}\right) \nu ^2+\left(-\frac{1835}{32}-\frac{63 \pi ^2}{512}\right) \nu -1
\right.\nn\\
&&
+\left(
\frac{283 \nu ^3}{32}+\frac{323 \nu ^2}{4}-\frac{1177 \nu }{16}+\frac{3}{4}
\right) \frac{\tilde{L}^2}{\tilde{r}} \nn\\
&&\left.+\left(
-\frac{189 \nu ^3}{32}+\frac{537 \nu ^2}{32}-\frac{849 \nu }{64}+\frac{99}{128}
\right) \frac{\tilde{L}^4}{\tilde{r}^2}\right)\nn\\
&&+ \tilde{p}_r^4 \tilde{r}^2 \left(
\frac{177 \nu ^3}{32}+\frac{371 \nu ^2}{8}+\frac{177 \nu }{32}-\frac{11}{2}
+\left(
-\frac{267 \nu ^3}{32}-\frac{3 \nu ^2}{2}-\frac{261 \nu }{32}-\frac{9}{16}
\right)\frac{\tilde{L}^2}{\tilde{r}} \right)\nn\\
&&\left.+ \tilde{p}_r^6 \tilde{r}^3 \left(
-\frac{25 \nu ^3}{32}-\frac{137 \nu ^2}{64}+\frac{175 \nu }{64}-\frac{81}{128}
\right) \right],%%%%%%%
\eea
\bea
\label{hCEsimp}
\tilde{H}^{\text{N}^3\text{LO}}_{C_{1\text{ES}^2}} &=& \frac{\nu^2 \tilde{S}_1^2}{\tilde{r}^6}  \left[
\frac{177 \nu }{7}-\frac{14894}{1225}
+ \left(
\frac{74413}{2450}-\frac{561 \nu }{56}
\right) \frac{\tilde{L}^2}{\tilde{r}} 
+\left(
-\frac{21 \nu ^2}{16}-\frac{223 \nu }{16}-\frac{79}{16}
\right) \frac{\tilde{L}^4}{\tilde{r}^2}\right. 
\nn\\
&&+\left(
\frac{15 \nu ^2}{16}+\frac{\nu }{4}-\frac{9}{16}
\right) \frac{\tilde{L}^6}{\tilde{r}^3}\nn\\
&&+ \tilde{p}_r^2 \tilde{r} \left(
\frac{1445 \nu }{14}-\frac{41779}{4900}
+\left(
-\frac{37 \nu ^2}{8}+\frac{57 \nu }{2}+\frac{65}{4}
\right) \frac{\tilde{L}^2}{\tilde{r}} + 
\frac{27 \nu ^2}{8}
\frac{\tilde{L}^4}{\tilde{r}^2}\right)\nn\\
&&+ \tilde{p}_r^4 \tilde{r}^2 \left(
-\frac{77 \nu ^2}{16}-\frac{269 \nu }{16}+\frac{243}{16}
+\left(
3 \nu ^2-\frac{3 \nu }{4}+\frac{27}{16}
\right)\frac{\tilde{L}^2}{\tilde{r}}\right)\nn\\
&&\left.+ \tilde{p}_r^6 \tilde{r}^3 \left(
-6 \nu ^2-\frac{\nu }{2}+\frac{9}{8}
\right) \right]\nn\\%%%%%%%
&&+\frac{\nu \tilde{S}_1^2}{q\tilde{r}^6}  \left[
\frac{177 \nu ^2}{7}+\left(\frac{1861}{24}-\frac{21 \pi ^2}{4}\right) \nu +\frac{73}{4}
+ \left( 
-\frac{1237 \nu ^2}{112} \right.\right.\nn\\
&&\left.+\left(\frac{2703 \pi ^2}{2048}-\frac{979}{16}\right) \nu -\frac{311}{8}
\right) \frac{\tilde{L}^2}{\tilde{r}} 
+\left(
-\frac{21 \nu ^3}{16}-\frac{35 \nu ^2}{4}+\frac{57 \nu }{16}+\frac{13}{4}
\right) \frac{\tilde{L}^4}{\tilde{r}^2}
\nn\\
&&+\left(
\frac{25 \nu ^3}{32}-\frac{77 \nu ^2}{32}+\frac{3 \nu }{16}+\frac{11}{32}
\right) \frac{\tilde{L}^6}{\tilde{r}^3}\nn\\
&&+ \tilde{p}_r^2 \tilde{r} \left(
\frac{3771 \nu ^2}{28}+\left(\frac{861}{8}-\frac{2703 \pi ^2}{512}\right) \nu +\frac{123}{8}
\right.\nn\\
&&\left.
+\left(
-\frac{37 \nu ^3}{8}+114 \nu ^2+\frac{1557 \nu }{16}-\frac{39}{2}
\right) \frac{\tilde{L}^2}{\tilde{r}} 
+\left(
\frac{21 \nu ^3}{8}-\frac{279 \nu ^2}{32}+\frac{63 \nu }{16}-\frac{21}{32}
\right) \frac{\tilde{L}^4}{\tilde{r}^2}\right)\nn\\
&&+ \tilde{p}_r^4 \tilde{r}^2 \left(
-\frac{77 \nu ^3}{16}-\frac{193 \nu ^2}{4}+\frac{173 \nu }{8}-\frac{67}{4}
+\left(
\frac{3 \nu ^3}{2}-\frac{63 \nu ^2}{8}+\frac{117 \nu }{16}-\frac{75}{32}
\right)\frac{\tilde{L}^2}{\tilde{r}} \right)\nn\\
&&\left.+ \tilde{p}_r^6 \tilde{r}^3 \left(
-8 \nu ^3+5 \nu ^2+\frac{57 \nu }{16}-\frac{43}{32}
\right) \right],%%%%%%%
\eea
\bea
\label{hCE2simp}
\tilde{H}^{\text{N}^3\text{LO}}_{C_{1\text{E$^2$S$^2$}}} &=& 
-\frac{\nu^3 \tilde{S}_1^2   }{2\tilde{r}^6}  \left(1 + \frac{1}{q} \right).
\eea

We note that in \cite{Antonelli:2020ybz} an aligned-spins EOB Hamiltonian was presented
for the simple bilinear in spin, spin$_1$-spin$_2$ sector, via some ansatz, assumptions from the EOB approach, 
and available results from self-force theory, see e.g.~recent review in 
\cite{Barack:2018yvs}. Both that EOB Hamiltonian, and 
our Hamiltonian in eq.~\eqref{hs1s2simp}, are specified to the COM frame and subject to 
the aligned-spins constraints. Yet, the simplified EOB Hamiltonian in 
\cite{Antonelli:2020ybz} is based on some EOB ansatz, and is specified to the so-called 
``quasi-isotropic'' gauge, where dependence in factors of $L^2$ is hidden. 
%(not clearly well-defined).
%there is only dependence in $p_r^2$ whereas dependence in $L^2/r^2$ is hidden
%and on the other hand, it clearly does not satisfy the necessary constraint for 
%circular orbits, in which $p_r\to 0$. 

An additional common assumption in the inspiral phase, where the orbit rapidly gets 
circularized, is that of circular orbits, namely of satisfying the necessary condition: 
$p_r \equiv \vec{p} \cdot \vec{n}=0 \Rightarrow p^2=p_r^2+L^2/r^2\to L^2/r^2$.
After applying this circular-orbit condition our Hamiltonians become:
\bea
\tilde{H}^{\text{N}^3\text{LO}}_{\text{S}_1 \text{S}_2} &=& \frac{\nu \tilde{S}_1   \tilde{S}_2}{2\tilde{r}^6}  \left[
\left(\frac{101641}{600}-\frac{81 \pi ^2}{8}\right) \nu +\frac{4537}{100}
\right.\nn\\
&&
+ \left( 
-\frac{209 \nu ^2}{8}+\left(\frac{141 \pi ^2}{64}-\frac{256207}{1200}\right) \nu -\frac{6337}{100}  
\right) \frac{\tilde{L}^2}{\tilde{r}} 
+\left(
\frac{483 \nu ^2}{16}+\frac{339 \nu }{16}-\frac{57}{8}
\right) \frac{\tilde{L}^4}{\tilde{r}^2}\nn\\
&&\left.
+\left( 
-\frac{25 \nu ^3}{32}-\frac{39 \nu ^2}{16}+\frac{69 \nu }{16}-\frac{17}{16}
\right) \frac{\tilde{L}^6}{\tilde{r}^3}\right],\nn\\
\eea
\bea
\tilde{H}^{\text{N}^3\text{LO}}_{\text{S}_1^2} &=& \frac{\nu^2 \tilde{S}_1^2}{\tilde{r}^6}  \left[
\left(\frac{45 \pi ^2}{64}-\frac{6679}{3920}\right) \nu +\frac{3987}{2450}
+ \left( 
\left(-\frac{96683}{3920}-\frac{45 \pi ^2}{64}\right) \nu -\frac{840921}{19600}
\right) \frac{\tilde{L}^2}{\tilde{r}} \right.\nn\\
&&+\left(
\frac{19 \nu ^2}{8}+\frac{191 \nu }{32}-\frac{75}{8}
\right) \frac{\tilde{L}^4}{\tilde{r}^2}+\left(
-\frac{119 \nu ^2}{64}+\frac{91 \nu }{32}-\frac{45}{64}
\right) \frac{\tilde{L}^6}{\tilde{r}^3}
\nn\\
&&+\frac{\nu \tilde{S}_1^2}{q\tilde{r}^6}  \left[
\left(\frac{45 \pi ^2}{64}-\frac{6679}{3920}\right) \nu ^2-\frac{11 \nu }{4}+9\right.\nn\\
&&
+ \left( 
\left(-\frac{37019}{1960}-\frac{45 \pi ^2}{64}\right) \nu ^2+\left(\frac{63 \pi ^2}{2048}-36\right) \nu +\frac{441}{16}
\right) \frac{\tilde{L}^2}{\tilde{r}} \nn\\
&&+\left(
\frac{19 \nu ^3}{8}-\frac{13 \nu ^2}{4}-\frac{1147 \nu }{32}+\frac{31}{4}
\right) \frac{\tilde{L}^4}{\tilde{r}^2}+\left(
-\frac{13 \nu ^3}{8}+\frac{127 \nu ^2}{16}-\frac{19 \nu }{8}+\frac{45}{64}
\right) \frac{\tilde{L}^6}{\tilde{r}^3},
\nn\\
\eea
\bea
\tilde{H}^{\text{N}^3\text{LO}}_{C_{1\text{ES}^2}}&=& \frac{\nu^2 \tilde{S}_1^2}{\tilde{r}^6}  \left[
\frac{177 \nu }{7}-\frac{14894}{1225}
+ \left( 
\frac{74413}{2450}-\frac{561 \nu }{56}
\right) \frac{\tilde{L}^2}{\tilde{r}} \right.\nn\\
&&+\left(
-\frac{21 \nu ^2}{16}-\frac{223 \nu }{16}-\frac{79}{16}
\right) \frac{\tilde{L}^4}{\tilde{r}^2}+\left(
\frac{15 \nu ^2}{16}+\frac{\nu }{4}-\frac{9}{16}
\right) \frac{\tilde{L}^6}{\tilde{r}^3}
\nn\\
&&+\frac{\nu \tilde{S}_1^2}{q\tilde{r}^6}  \left[
\frac{177 \nu ^2}{7}+\left(\frac{1861}{24}-\frac{21 \pi ^2}{4}\right) \nu +\frac{73}{4}
\right.\nn\\
&&
+ \left( 
-\frac{1237 \nu ^2}{112}+\left(\frac{2703 \pi ^2}{2048}-\frac{979}{16}\right) \nu -\frac{311}{8}
\right) \frac{\tilde{L}^2}{\tilde{r}} \nn\\
&&+\left(
-\frac{21 \nu ^3}{16}-\frac{35 \nu ^2}{4}+\frac{57 \nu }{16}+\frac{13}{4}
\right) \frac{\tilde{L}^4}{\tilde{r}^2}+\left(
\frac{25 \nu ^3}{32}-\frac{77 \nu ^2}{32}+\frac{3 \nu }{16}+\frac{11}{32}
\right) \frac{\tilde{L}^6}{\tilde{r}^3},
\nn\\
\eea
\bea
\tilde{H}^{\text{N}^3\text{LO}}_{C_{1\text{E$^2$S$^2$}}} &=& 
-\frac{\nu^3 \tilde{S}_1^2   }{2\tilde{r}^6}  \left(1 + \frac{1}{q} \right).
\eea

\subsection{GW Observables}
\label{seeforreal}

Despite the various uses of general Lagrangians and Hamiltonians, both formally and 
phenomenologically, it is essential to obtain physical observables that can be 
tested through real-world measurements of GW experiments, such as LIGO, Virgo or KAGRA.
This is especially critical in high-loop and higher-spin sectors such as those we currently tackled, 
where considering gauge-invariant quantities is crucial in order to remove possible 
ambiguities, that multiply when going to higher orders. Yet, it is important 
to bear in mind that the determination of such observables is always possible only under 
some specific kinematic constraints, whereas general Hamiltonians provide the full generic 
information on the system, despite their gauge-dependent ``overload''.
Thus applying kinematic constraints, such as those presented in the previous section, 
enable to define from the related simplified Hamiltonians, the binding energy 
$e\equiv\tilde{H}$ in terms of gauge-invariant quantities. Such gauge-invariant relations 
are regularly used as critical tools in the theoretical construction of gravitational 
waveforms for the LIGO, Virgo, and KAGRA experiments.

Specifically, in order to eliminate the coordinate dependence from the above 
Hamiltonians, we can use the additional condition for circular orbits 
$\dot{p}_r=-\partial\tilde{H}(\tilde{r},\tilde{L})/\partial\tilde{r}=0$, to obtain 
$\tilde{r}(\tilde{L})$. Using this condition one can get the binding energy as a 
function of the gauge-invariant angular momentum, $e(\tilde{L})$. The binding energy can 
also be expressed in terms of the gauge-invariant PN parameter 
$x \equiv \tilde{\omega}^{2/3}$, which stands for the orbital frequency that is 
directly measured through the frequency of emitted GWs, and is inversely related to the 
orbital separation. From Hamilton's equation for the orbital phase, 
$d\phi/d\tilde{t}\equiv\tilde{\omega}
=\partial\tilde{H}(\tilde{r},\tilde{L})/\partial\tilde{L}=0$, one can then express the 
angular momentum as a function of $x$, which is given in our sectors by:
\bea
\label{xtoL}
\frac{1}{\tilde{L}^2} &\supset& x^6 \nu \Bigg[ 
\frac{1}{2} \left(-\frac{403 \nu ^3}{96}+\frac{49 \nu ^2}{2}
+\left(\frac{41 \pi ^2}{4}-\frac{14489}{48}\right) \nu 
+\frac{81}{4} \right) \tilde{S}_1 \tilde{S}_2 \nn \\
&& + \nu \left( -4 \nu  C_{1(\text{E}^2\text{S}^2)}  
+\left(-\frac{14 \nu ^2}{3}-\frac{17 \nu }{2}+\frac{13  }{2}\right) 
C_{1(\text{ES}^2)} -\frac{3607 \nu ^2}{192}-\frac{7099 \nu }{80} 
- \frac{2279 }{64} \right) \tilde{S}_1{}^2\nn \\
&&  + q^{-1} \left(-4 \nu ^2 
C_{1(\text{E}^2\text{S}^2)} +\left(-\frac{14 \nu ^3}{3}-\frac{253 \nu^2}{14}
+\left(\frac{1375 \pi ^2}{256}-\frac{509}{24}\right) \nu \right) 
C_{1(\text{ES}^2)}\right.\nn \\
&& \left. \qquad -\frac{267 \nu ^3}{16}-\frac{65251 \nu ^2}{1120}
+\left(\frac{63 \pi ^2}{256}-\frac{2865}{32}\right) \nu +\frac{1863}{64}\right) 
\tilde{S}_1{}^2 \Bigg] + [1 \leftrightarrow 2] .
\eea

From the previous relations, one can also obtain the binding energy as a function of 
the frequency, $e(x)$. The binding energy of the present sectors as a function of the 
frequency $x$, and as a function of the angular momentum $\tilde{L}$, were already 
presented in our \cite{Kim:2021rfj}. In particular we find binding-energy relations due to the 
new effect discussed in sections \ref{tidaltheory}, \ref{eftcomput} above:
\bea
(e)^{\text{N}^3\text{LO}}_{\text{C$_{1\text{(E$^2$S$^2$)}}$S$_1^2$}} (\tilde{L}) = 
-\frac{1}{2} C_{1(\text{E}^2\text{S}^2)}\tilde{S}_1^2 \frac{\nu}{\tilde{L}^{12}}  
\left[  \nu^2 \left( 1 + q^{-1} \right) \right],
\eea
\bea
\label{es2E2}
(e)^{\text{N}^3\text{LO}}_{\text{C$_{1\text{(E$^2$S$^2$)}}$S$_1^2$}} (x) = 
\frac{3}{2} C_{1(\text{E}^2\text{S}^2)}\tilde{S}_1^2 x^6 \nu^2 
\left[ \nu (1+q^{-1}) \right].
\eea
The effect, quadratic in the curvature and in the spins, and at the $5$PN order, thus represents a 
real new physical effect which is \textit{unique to spinning objects}. 
For reference, we find that the binding energy associated with the point-mass tidal 
effect in fig.~\ref{5pne2}(a) equals:
\bea
\label{eE2}
(e)^{\text{5PN}}_{\text{C$_{1\text{(E$^2$)}}$}} (x) = 
9 C_{1(\text{E}^2)} x^6 \nu^2 
\left[ -\nu + q(1-\nu) \right].
\eea
The new binding energy with spins in eq.~\eqref{es2E2} is always positive, and thus it 
reduces the binding of the compact binary, similar to the point-mass 
tidal effect in eq.~\eqref{eE2}. Most importantly, we found here a new independent 
Wilson coefficient that goes beyond what was presented in \cite{Goldberger:2004jt,Levi:2015msa}, 
which is \textit{unique to spinning objects}, and should provide new 
information on the UV physics of gravity and QCD. 
Such a coefficient can be significantly large for NSs, and thus 
tidal effects are important for GW astronomy \cite{Blanchet:2000ub}.

\subsection{Scattering Observables}
\label{imaginaryevents}

First, let us clarify on the weak-field, or so-called post-Minkowskian (PM) approximation, 
which treats the scattering problem in a perturbative expansion in $G$. It can only be 
useful via its gauge-dependent COM Hamiltonians, or even extrapolated gauge-invariant 
binding-energy relations, if it can provide the N$^n$LO PM results, namely of a similar (or higher) loop 
order, $n$, to the target PN results of real binary inspirals. 
In particular, no PM results for scattering, equivalent to our present ones at the N$^3$LO, exist 
to date, and therefore no comparison whatsoever (even partial) is possible with the above critical 
quantities (gauge-invariant or not) for GW science.

Moreover, a comparison of scattering results in the overlap with the PN approximation, is only 
possible provided that some limited link exists between some kinematic configuration of the scattering 
problem to binary inspirals, which are the events targeted in GW measurements. 
Such a link of the scattering angle, which is already restricted only to aligned spins, to PN 
inspirals, becomes challenging as of the N$^2$LO, once radiation-reaction effects come in, 
and infeasible beyond the N$^3$LO, where physical logarithms show up systematically. 
On the other hand, a unique novel approach was recently devised, which incorporates 
scattering-amplitudes methods directly in the binary-inspiral problem \cite{Edison:2022cdu}. 
This approach fully captures radiation-reaction effects at any loop order, and is 
thus free of all the aforementioned limitations and obstacles of all other  
scattering-amplitudes approaches to the classical gravitational scattering problem.

With all that in mind, we can still assume in the present sectors that our binding energy 
which the Hamiltonian stands for, can be extended to a kinetic energy of scattering, and 
compute the associated scattering angle in the restricted aligned-spins configuration. 
As noted, such a link is possible only as long as the 
Hamiltonians do not contain logarithms, which can still be achieved -- as we have shown 
in the previous sections -- up to the N$^3$LO of PN binary inspirals. 
We then start our computation with our PN aligned-spin Hamiltonians in 
eqs.~\eqref{hs1s2simp}, \eqref{hs1s1simp}, \eqref{hCEsimp}, \eqref{hCE2simp}, which 
depend on $r$, $p_r$, $L$, and $S_{1,2}$. Note that we compute the scattering angles 
with our Hamiltonians not being specified to the ``quasi-isotropic'' gauge, by using the 
integration considerations in \cite{Damour:1988mr}. Thus derivations of scattering 
angles do not require Hamiltonians to be in the ``quasi-isotropic'' gauge, as was 
specified in e.g.~\cite{Antonelli:2020ybz}. 

Our computation first yields the scattering angle $\theta$ as a function of 
$E$, $L$, $S_{1,2}$.
Since we start with a PN Hamiltonian, we arrive at a PN expanded scattering angle 
$\theta$, and unlike the Hamiltonian, each PN order contains terms at all orders in $G$. 
To switch to PM results in the overlapping region of the PN and PM approximations, one 
then needs to expand the scattering angle to a desired order in $G$, and switch to some 
standard scattering variables. Importantly also, due to different natural choices of gauge of 
the rotational DOFs, it is necessary to transform the ``canonical'' $L$ in the Hamiltonian 
to the ``covariant'' one $L_{\text{cov}}$, corresponding to the respective canonical and covariant 
gauges:
\bea
L = L_{\text{cov}} + \Delta L, \quad \Delta L = \left( \sqrt{m_1^2 + p_{\infty}^2} 
- m_1 \right) \frac{S_1}{m_1} + (1 \leftrightarrow 2),
\eea
with
\bea
p_{\infty} = \frac{m_1 m_2}{E} \sqrt{\gamma^2 - 1},\quad E
= \sqrt{m_1^2 + m_2^2 + 2 m_1 m_2 \gamma}, \quad \gamma 
= \frac{1}{\sqrt{1- v_\infty^2}}.
\eea
The relation between $L_{\text{cov}}$ and the impact parameter $b$ is
$L_{\text{cov}} = p_\infty b$.
The above relations enable us to transform the scattering angle $\theta$ from a function 
of $\{ E, L\}$ to a function of $\{v_\infty, b \}$. 

With the following notations:
\bea
\tilde{b} = \frac{v_\infty^2 }{ Gm }b, 
\qquad \tilde{v} = \frac{v_\infty}{c}, 
\qquad \tilde{a}_i = \frac{ S_i }{b m_i c},
\qquad \Gamma = \frac{E}{mc^2} = \sqrt{1+2\nu (\gamma - 1) },
\eea
our consequent scattering angles in the sectors with spin are given by:
\bea
\theta = \left( \theta_{\text{SO}} + \theta_{\text{S}_1 
\text{S}_2} + \theta_{\text{S}_1^2} + C_{1\text{ES}^2}\theta_{C_{1\text{ES}^2}} 
+ C_{1\text{E$^2$S$^2$}}\theta_{C_{1\text{E$^2$S$^2$}}} \right) + (1 \leftrightarrow 2) ,
\eea
with the following pieces:
\bea
\label{thetas1}
\frac{ \theta_{\text{SO}} }{\Gamma} &=&   \tilde{v} \tilde{a}_1 \left[- \frac{4}{ \tilde{b} } + \frac{\pi}{ \tilde{b}^2 } \left(-4 + \nu + \left(\frac{3   \nu }{2}-6  \right) \tilde{v}^2     + \frac{\nu}{q}  \left( 1+ \frac{3}{2} \tilde{v}^2  \right) \right)\right.\nn\\
&&+ \frac{1}{ \tilde{b}^3 } \Big(-12 + 4\nu + \left(50 \nu -120  \right) \tilde{v}^2  + \left( \frac{117 \nu }{2}-60  \right) \tilde{v}^4 +    \frac{177 \nu }{4} \tilde{v}^6\nn\\
&&+ \frac{\nu}{q}  \left( 4+ 40 \tilde{v}^2+20\tilde{v}^4  \right) \Big) + \frac{\pi}{ \tilde{b}^4 } \left( \left(-\frac{3 }{2}   \nu ^2+42   \nu -84   \right) \tilde{v}^2  \right.\nn\\
&&+ \left( -\frac{45}{4}    \nu ^2+\frac{363   \nu }{2}-210   \right) \tilde{v}^4 +    \left(\left(-\frac{257}{48}  -\frac{251 \pi^2 }{128}\right) \nu ^2+\left(\frac{3233  }{12}-\frac{241 \pi ^3}{128}\right) \nu \right.\nn\\
&&\left.-\frac{105  }{2} \right) \tilde{v}^6 + \frac{\nu}{q}  \left( \left(\frac{63}{2}-\frac{3 \nu }{2} \right) \tilde{v}^2+\left(\frac{315}{4}-\frac{45 \nu }{4} \right)\tilde{v}^4 \right.\nn\\
&&\left.\left.\left.+\left( \left(-\frac{257}{48}-\frac{251 \pi ^2}{128}\right) \nu +\frac{315}{16} \right)\tilde{v}^6  \right) \right) \right],
\eea
\bea
\label{thetas1s2}
\frac{ \theta_{\text{S}_1 \text{S}_2} }{\Gamma} &=&\tilde{a}_1 \tilde{a}_2 \left[\frac{1}{ \tilde{b} } \left(2+ 2\tilde{v}^2 \right)+ \frac{\pi}{ \tilde{b}^2 } \left(\frac{3}{2} + \frac{45}{4} \tilde{v}^2     + \frac{9}{4} \tilde{v}^4\right)+ \frac{1}{ \tilde{b}^3 } \Big( 4 + \left(4 \nu +140  \right) \tilde{v}^2 \right.\nn\\
&& + \left(220-33 \nu \right) \tilde{v}^4 +    \left( 20-\frac{1093 \nu }{10} \right) \tilde{v}^6  \Big) + \frac{\pi}{ \tilde{b}^4 } \left(\left( \frac{15 \nu }{4}+\frac{315}{4}   \right) \tilde{v}^2   \right.\nn\\
&&\left.\left.  + \left( \frac{3675}{8}-\frac{495 \nu }{8}   \right) \tilde{v}^4 +  \left(\left(\frac{1845 \pi ^2}{512}-\frac{7995}{16}\right) \nu +\frac{9975}{32} \right) \tilde{v}^6  \right) \right],
\eea
\bea
\label{thetas1s1}
\frac{ \theta_{\text{S}_1^2} }{\Gamma} &=&   \tilde{a}_1^2 \left[ \frac{\pi}{ \tilde{b}^2 } \left(   \left(6-\frac{15 \nu }{4}  \right) \tilde{v}^2 + \left(\frac{15 \nu }{32}+\frac{3}{32}  \right) \tilde{v}^4     + \frac{\nu}{q}  \left( -\frac{15}{4} \tilde{v}^2 +\frac{15}{32}\tilde{v}^4 \right) \right)\right.\nn\\
&&+ \frac{1}{ \tilde{b}^3 } \left( \left(96-72 \nu  \right) \tilde{v}^2  + \left( \frac{1136}{7}-\frac{956 \nu }{7}  \right) \tilde{v}^4 +   \left( \frac{16}{7}-\frac{993 \nu }{35} \right) \tilde{v}^6\right.\nn\\
&&
\left. + \frac{\nu}{q}  \left(-64 \tilde{v}^2-\frac{688}{7}\tilde{v}^4 +\frac{144}{35} \tilde{v}^6 \right) \right) + \frac{\pi}{ \tilde{b}^4 } \left( \left(60-\frac{195 \nu }{4}   \right) \tilde{v}^2  \right.\nn\\
&&+ \left( \frac{165 \nu ^2}{8}-\frac{45555 \nu }{112}+\frac{47295}{112}   \right) \tilde{v}^4 +    \left(\frac{34521 \nu ^2}{448}+\left(-\frac{105303}{224}-\frac{945 \pi ^2}{16384}\right) \nu\right.\nn\\
&&\left. \left.\left.+\frac{8655}{32} \right) \tilde{v}^6 + \frac{\nu}{q}  \left( -\frac{165}{4} \tilde{v}^2+\left(\frac{165 \nu }{8}-\frac{31545}{112} \right)\tilde{v}^4 +\left( \frac{34521 \nu }{448}-165 \right)\tilde{v}^6  \right) \right) \right],\nn\\
\eea
\bea
\label{thetas1cE}
\frac{ \theta_{C_{1\text{ES}^2}} }{\Gamma} &=&   \tilde{a}_1^2 \left[\frac{1}{ \tilde{b} } \left(2+2 \tilde{v}^2 \right) + \frac{\pi}{ \tilde{b}^2 } \left( \frac{3}{2}+  \left(\frac{15}{2}-\frac{3 \nu }{2}  \right) \tilde{v}^2 + \left(\frac{87}{32}-\frac{57 \nu }{32}  \right) \tilde{v}^4 \right.\right.\nn\\
&&\left.    + \frac{\nu}{q}  \left( -\frac{3}{2} \tilde{v}^2 -\frac{57}{32} \tilde{v}^4 \right) \right)+ \frac{1}{ \tilde{b}^3 } \left( 4 + \left(84-20 \nu  \right) \tilde{v}^2  + \left( \frac{964}{7}-\frac{563 \nu }{7}  \right) \tilde{v}^4 \right.\nn\\
&&\left.+   \left( \frac{180}{7}-\frac{6421 \nu }{70} \right) \tilde{v}^6 + \frac{\nu}{q}  \left(-16 \tilde{v}^2 -\frac{432}{7}\tilde{v}^4 -\frac{704}{35} \tilde{v}^6 \right) \right) + \frac{\pi}{ \tilde{b}^4 } \left( \left(45-\frac{45 \nu }{4}   \right) \tilde{v}^2  \right. \nn\\
&&+ \left( \frac{15 \nu ^2}{4}-\frac{14925 \nu }{112}+\frac{26205}{112}  \right) \tilde{v}^4 +    \left(\frac{5925 \nu ^2}{448}+\left(\frac{58095 \pi ^2}{16384}-\frac{89955}{224}\right) \nu \right.\nn\\
&&\left. \left.\left.+\frac{6045}{32}   \right) \tilde{v}^6 + \frac{\nu}{q}  \left( -\frac{15}{2} \tilde{v}^2+\left( \frac{15 \nu }{4}-\frac{9405}{112} \right)\tilde{v}^4 +\left( \frac{5925 \nu }{448}-\frac{3495}{32} \right)\tilde{v}^6  \right) \right) \right],\nn\\
\eea
\bea
\label{thetas1cE2}
\frac{ \theta_{C_{1\text{E$^2$S$^2$}}} }{\Gamma} =   -\tilde{a}_1^2 \frac{\pi}{ \tilde{b}^4 } \frac{15}{16} \tilde{v}^6 \left(\nu^2 - \nu + \frac{\nu^2}{q}\right).
\eea

The above results are organized according to their PM order first, $1/\tilde{b} \sim G$, 
and the $n$th subleading contribution then corresponds to $\tilde{v}^{2n}$ terms in 
even-in-spin sectors, and $\tilde{v}^{2n+1}$ in odd-in-spin sectors. The scattering 
angles due to the present N$^3$LO quadratic-in-spin sectors are thus:
\bea
\label{thetan3los1s2}
\frac{ \theta_{\text{S}_1 \text{S}_2}^{\text{N}^3\text{LO}} }{\Gamma} &=&\tilde{a}_1 \tilde{a}_2 \tilde{v}^6 \left[  \frac{1}{ \tilde{b}^3 }      \left( 20-\frac{1093 \nu }{10} \right)   + \frac{\pi}{ \tilde{b}^4 }   \left(\left(\frac{1845 \pi ^2}{512}-\frac{7995}{16}\right) \nu +\frac{9975}{32} \right) \right],
\eea
\bea
\label{thetan3los1s1}
\frac{ \theta_{\text{S}_1^2}^{\text{N}^3\text{LO}} }{\Gamma} &=&   \tilde{a}_1^2 \tilde{v}^6\left[  \frac{1}{ \tilde{b}^3 } \left(    \frac{16}{7}-\frac{993 \nu }{35} + \frac{144}{35} \frac{\nu}{q}   \right) + \frac{\pi}{ \tilde{b}^4 } \left( \frac{34521 \nu ^2}{448}+\left(-\frac{105303}{224}-\frac{945 \pi ^2}{16384}\right) \nu\right.\right.\nn\\
&&\left.\left.+\frac{8655}{32}  +   \left( \frac{34521 \nu }{448}-165 \right)\frac{\nu}{q}   \right) \right],
\eea
\bea
\label{thetan3los1cE}
\frac{ \theta_{C_{1\text{ES}^2}}^{\text{N}^3\text{LO}} }{\Gamma} &=&   \tilde{a}_1^2 \tilde{v}^6\left[ \frac{1}{ \tilde{b}^3 } \left(     \frac{180}{7}-\frac{6421 \nu }{70} +   \frac{704}{35} \frac{\nu}{q}  \right) + \frac{\pi}{ \tilde{b}^4 } \left(    \frac{5925 \nu ^2}{448}+\left(\frac{58095 \pi ^2}{16384}-\frac{89955}{224}\right) \nu \right.\right.\nn\\
&&\left.\left.+\frac{6045}{32}  +  \left( \frac{5925 \nu }{448}-\frac{3495}{32} \right)\frac{\nu}{q} \right)  \right],
\eea
\bea
\label{thetan3los1cE2}
\frac{ \theta_{C_{1\text{E$^2$S$^2$}}}^{\text{N}^3\text{LO}} }{\Gamma} =  
\frac{ \theta_{C_{1\text{E$^2$S$^2$}}} }{\Gamma}.
%- \tilde{a}_1^2 \frac{\pi}{ \tilde{b}^4 } \frac{15}{16} \tilde{v}^6 
%\left(\nu^2 - \nu + \frac{\nu^2}{q}\right).
\eea

Our results in eqs.~\eqref{thetas1}, and \eqref{thetas1s2}, \eqref{thetan3los1s2}, for 
the spin-orbit, and spin$_1$-spin$_2$ sectors, respectively, agree with 
\cite{Antonelli:2020ybz}, where they were obtained via traditional GR methods, and these 
agreements were already indicated in 
\cite{Kim:2021rfj,Kim:2022pou}. 
Our results in eqs.~\eqref{thetas1s1}, \eqref{thetan3los1s1}, and \eqref{thetas1cE}, 
\eqref{thetan3los1cE}, agree with those obtained at the NLO PM approximation 
for sectors quadratic in the spin of generic objects in \cite{Kosmopoulos:2021zoq} 
via EFT and scattering-amplitudes methods, and in \cite{Liu:2021zxr} via EFT methods. 
Finally, our results in eqs.~\eqref{thetas1s1}, \eqref{thetan3los1s1}, and 
\eqref{thetas1cE}, \eqref{thetan3los1cE}, also agree with those obtained 
at the N$^2$LO PM approximation for sectors quadratic in the spin of generic objects  in 
\cite{Jakobsen:2022fcj} via EFT methods. 

Such agreements with \cite{Kosmopoulos:2021zoq,Jakobsen:2022fcj} are not surprising, as the latter 
approaches have built closely on our theory presented in detail in \cite{Levi:2015msa}, and on our 
comprehensive results along the years, see e.g.~\cite{Levi:2018nxp}, and thus 
\cite{Kosmopoulos:2021zoq,Jakobsen:2022fcj} are clearly dependent on our framework.
Moreover, all the approaches noted above still provide only very partial 
results, that are below the higher orders of either loop or spin of our present new results, 
and are only in the restricted aligned-spins configuration, which is significantly less 
informative at higher-spin sectors. 
In contrast, our results in this work cover all quadratic-in-spin sectors, are 
at the higher N$^3$LO (or three-loop) level, and are not limited to the aligned-spins 
simplification, but rather hold for any generic spin orientations, and moreover -- in general reference 
frames.

\section{Conclusions}

In this paper we presented for the first time the complete N$^3$LO quadratic-in-spin interactions 
that push the state of the art of conservative dynamics of binary inspirals at the 5PN order. 
Our results cover all the relevant sectors, and are obtained for generic compact binaries in generic 
kinematic settings, via the EFT of spinning gravitating objects introduced in \cite{Levi:2015msa}. Our 
independent framework, including the public \texttt{EFTofPNG} code \cite{Levi:2017kzq}, 
enables to study PN theory in any generic sector, which is critical to 
advance the present high precision frontier in both high-loop and higher-spin orders, such as in the 
present sectors. Such generic theoretical input that our framework provides, improves the ability to 
mine unknown physics from real-world GW data. 
Moreover, the unique treatment of the present sectors, also at the $5$PN order, via the EFT framework, 
provides even more invaluable new information on the UV physics of strong gravity and extreme QCD.  

The N$^3$LO quadratic-in-spin corrections consist of $4$ independent sectors, one of which is new, first 
appearing in the present $5$PN order, similar to the related tidal effect in the point-mass sector, 
which also enters first at this order. The $3$ other quadratic-in-spin sectors here scale as the single 
N$^3$LO spin-orbit sector. 
The overall computational load of the Feynman evaluation is similar to the N$^3$LO spin-orbit sector 
\cite{Kim:2022pou}, with an imperative use of advanced multi-loop methods, and a generic overall 
dimensional treatment due to the appearance of DimReg poles across all loop orders. 
Yet, the procedure of redefinitions of the position and rotational variables is somewhat more intricate 
and demanding in higher-spin sectors such as the present ones, compared to the point-mass or spin-orbit 
sectors, e.g.~at the N$^3$LO \cite{Kim:2022pou}. Nevertheless, the streamlined automated algorithm for 
this procedure, which we set up in \cite{Kim:2022pou}, still performed in an efficient and 
rapid manner.

We presented here the reduced Lagrangians of all relevant sectors for the first time, which can 
be used to directly derive the EOMs for both position and spin \cite{Levi:2015msa}. 
We also presented here the most general Hamiltonians of these sectors for the first time, 
and their consequent useful simplifications. 
Our general Hamiltonians here have been fully verified in \cite{Levi:2022rrq} via the 
completion of the Poincar\'e algebra, which constitutes the most stringent check of validity.
Finally, in these high-order sectors it is especially crucial to derive unambiguous observables, which 
correspond to real-world GW measurements, and we presented here the gauge-invariant relation of the 
angular momentum to the GW frequency. We also showed the binding energy of the new tidal effect with 
spins in comparison with the tidal effect from the point-mass sector, which also enters at the present 
$5$PN order. 

We ended with a derivation of all the extrapolated scattering angles that hold for aligned spins, 
which correspond to an extension of our Hamiltonians to the scattering problem. 
Complete agreement was found with the limited available results at lower orders obtained via 
traditional GR, and via EFT and scattering-amplitudes methods. 
Such agreements with \cite{Kosmopoulos:2021zoq,Jakobsen:2022fcj} are not surprising, as the latter 
approaches have built closely on our theory presented in detail in \cite{Levi:2015msa}, and on our 
comprehensive results along the years, see e.g.~\cite{Levi:2018nxp}, and thus 
\cite{Kosmopoulos:2021zoq,Jakobsen:2022fcj} are clearly dependent on our framework.
Moreover, all the approaches noted above still provide only very partial 
results, that are below the higher orders of either loop or spin of our present new results, 
and are only in the restricted aligned-spins configuration, which is significantly less 
informative at higher-spin sectors. 
In contrast, our results in this work cover all quadratic-in-spin sectors, are 
at the higher N$^3$LO (or three-loop) level, and are not limited to the aligned-spins 
simplification, but rather hold for any generic spin orientations, and moreover -- in general reference 
frames. 

We found a new independent Wilson coefficient in the present sectors, that is unique to 
spinning objects, yet seems to be related to the long-studied ``Love numbers''. Determining the 
value of such characteristic UV coefficients constitutes one of the most challenging tasks in an EFT 
framework. Fortunately the EFT framework is also well-suited to carry out such so-called matching 
of its unknown coefficients, either via analytical studies, or by comparing to numerical or real-world 
data. Indeed in recent years with the incredible success of GW measurements came a notable thrust in 
the theoretical efforts to study such ``Love numbers'', that have been explored for 
both BHs and NSs, notably extending studies to the case of rotating objects, see 
e.g.~\cite{LeTiec:2020spy,LeTiec:2020bos,Chia:2020yla,Charalambous:2021mea,Charalambous:2021kcz,
Castro:2021wyc}. 

Interestingly, almost all studies to date indicate that such ``Love numbers'' vanish for BHs in 
$4$-dimensional GR.
On the other hand, the Wilson coefficients of tidal effects can be significantly large for NSs, 
and thus tidal effects are important for GW astronomy \cite{Blanchet:2000ub}. 
Despite the long history and enormous body of work that has accumulated of studying such ``Love 
numbers'', we believe that the intriguing studies of such coefficients are only in their infancy.  
There is a huge challenge in thoroughly studying them, using various approaches, 
for either BHs or NSs, and in general candidate-theories  of gravity \cite{Levi:2022youtube}. 
Such studies are immensely important to unveil the new physics that 
is encrypted in the continually increasing influx of GW data.

\acknowledgments

J-WK was supported by the Science and Technology Facilities Council (STFC) 
Consolidated Grant ST/T000686/1 \textit{``Amplitudes, Strings and Duality''}.
ML has been supported by the STFC Rutherford Grant ST/V003895 
``\textit{Harnessing QFT for Gravity}'' %, %a Royal Society award IES/R3/213180, 
and by the Mathematical Institute University of Oxford. 
ZY is supported by the Knut and Alice Wallenberg Foundation under grants 
KAW 2018.0116 and KAW 2018.0162.

\appendix

\section{Reduced Actions}
\label{actionresults}

The action that we obtain after the reduction procedure detailed in section \ref{minimalism} is 
standard in the velocity, with the additional spin variable, and contains the following $4$ sectors:
\bea
\label{resultgreataction}
\hat{V}^{\text{N}^3\text{LO}}_{\text{S}^2} =\hat{V}^{\text{N}^3\text{LO}}_{\text{S}_1 
	\text{S}_2} +\hat{V}^{\text{N}^3\text{LO}}_{\text{S}_1^2} + C_{1\text{ES}^2} 
\hat{V}^{\text{N}^3\text{LO}}_{C_{1\text{ES}^2}} + C_{1\text{E$^2$S$^2$}} 
V^{\text{N}^3\text{LO}}_{C_{1\text{E$^2$S$^2$}}} + (1 \leftrightarrow 2),
\eea
where
\bea
\hat{V}^{\text{N}^3\text{LO}}_{\text{S}_1 \text{S}_2} &= &  	\frac{G}{128 r{}^3} \Big[ 3 \vec{S}_{1}\cdot\vec{n} \vec{S}_{2}\cdot\vec{n} \big( 85 v_{1}^2 \vec{v}_{1}\cdot\vec{v}_{2} v_{2}^2 - 192 v_{1}^2 ( \vec{v}_{1}\cdot\vec{v}_{2})^{2} + 44 ( \vec{v}_{1}\cdot\vec{v}_{2})^{3} + 8 v_{1}^{6} \nn\\ 
&& + 124 \vec{v}_{1}\cdot\vec{v}_{2} v_{1}^{4} - 24 v_{2}^2 v_{1}^{4} - 48 v_{1}^2 v_{2}^{4} + 280 \vec{v}_{1}\cdot\vec{n} v_{1}^2 \vec{v}_{2}\cdot\vec{n} \vec{v}_{1}\cdot\vec{v}_{2} - 60 \vec{v}_{1}\cdot\vec{n} v_{1}^2 \vec{v}_{2}\cdot\vec{n} v_{2}^2 \nn\\ 
&& - 640 v_{1}^2 \vec{v}_{1}\cdot\vec{v}_{2} ( \vec{v}_{1}\cdot\vec{n})^{2} - 120 v_{1}^2 v_{2}^2 ( \vec{v}_{1}\cdot\vec{n})^{2} - 320 v_{1}^2 \vec{v}_{1}\cdot\vec{v}_{2} ( \vec{v}_{2}\cdot\vec{n})^{2} \nn\\ 
&& + 480 ( \vec{v}_{1}\cdot\vec{n})^{2} ( \vec{v}_{1}\cdot\vec{v}_{2})^{2} + 40 \vec{v}_{1}\cdot\vec{n} \vec{v}_{2}\cdot\vec{n} ( \vec{v}_{1}\cdot\vec{v}_{2})^{2} + 480 v_{1}^2 v_{2}^2 ( \vec{v}_{2}\cdot\vec{n})^{2} \nn\\ 
&& - 280 \vec{v}_{1}\cdot\vec{n} \vec{v}_{2}\cdot\vec{n} v_{1}^{4} + 120 ( \vec{v}_{2}\cdot\vec{n})^{2} v_{1}^{4} + 560 \vec{v}_{1}\cdot\vec{v}_{2} ( \vec{v}_{1}\cdot\vec{n})^{4} + 840 v_{1}^2 ( \vec{v}_{1}\cdot\vec{n})^{2} ( \vec{v}_{2}\cdot\vec{n})^{2} \nn\\ 
&& + 280 \vec{v}_{1}\cdot\vec{n} v_{1}^2 ( \vec{v}_{2}\cdot\vec{n})^{3} - 560 v_{1}^2 ( \vec{v}_{2}\cdot\vec{n})^{4} \nn\\ 
&& - 770 \vec{v}_{1}\cdot\vec{v}_{2} ( \vec{v}_{1}\cdot\vec{n})^{2} ( \vec{v}_{2}\cdot\vec{n})^{2} -420 ( \vec{v}_{1}\cdot\vec{n})^{3} ( \vec{v}_{2}\cdot\vec{n})^{3} \big) + \vec{S}_{1}\cdot\vec{S}_{2} \big( 154 v_{1}^2 \vec{v}_{1}\cdot\vec{v}_{2} v_{2}^2 \nn\\ 
&& + 240 v_{1}^2 ( \vec{v}_{1}\cdot\vec{v}_{2})^{2} - 160 ( \vec{v}_{1}\cdot\vec{v}_{2})^{3} - 104 v_{1}^{6} + 48 \vec{v}_{1}\cdot\vec{v}_{2} v_{1}^{4} - 224 v_{2}^2 v_{1}^{4} \nn\\ 
&& + 48 v_{1}^2 v_{2}^{4} -216 \vec{v}_{1}\cdot\vec{n} v_{1}^2 \vec{v}_{2}\cdot\vec{n} \vec{v}_{1}\cdot\vec{v}_{2} - 489 \vec{v}_{1}\cdot\vec{n} v_{1}^2 \vec{v}_{2}\cdot\vec{n} v_{2}^2 + 912 \vec{v}_{1}\cdot\vec{n} \vec{v}_{2}\cdot\vec{n} \vec{v}_{1}\cdot\vec{v}_{2} v_{2}^2 \nn\\ 
&& + 456 v_{1}^2 \vec{v}_{1}\cdot\vec{v}_{2} ( \vec{v}_{1}\cdot\vec{n})^{2} + 1032 v_{1}^2 v_{2}^2 ( \vec{v}_{1}\cdot\vec{n})^{2} - 480 \vec{v}_{1}\cdot\vec{v}_{2} v_{2}^2 ( \vec{v}_{1}\cdot\vec{n})^{2} \nn\\ 
&& - 24 v_{1}^2 \vec{v}_{1}\cdot\vec{v}_{2} ( \vec{v}_{2}\cdot\vec{n})^{2} - 768 ( \vec{v}_{1}\cdot\vec{n})^{2} ( \vec{v}_{1}\cdot\vec{v}_{2})^{2} - 180 \vec{v}_{1}\cdot\vec{n} \vec{v}_{2}\cdot\vec{n} ( \vec{v}_{1}\cdot\vec{v}_{2})^{2} \nn\\ 
&& - 288 ( \vec{v}_{2}\cdot\vec{n})^{2} ( \vec{v}_{1}\cdot\vec{v}_{2})^{2} - 864 v_{1}^2 v_{2}^2 ( \vec{v}_{2}\cdot\vec{n})^{2} + 600 \vec{v}_{1}\cdot\vec{v}_{2} v_{2}^2 ( \vec{v}_{2}\cdot\vec{n})^{2} \nn\\ 
&& + 288 ( \vec{v}_{1}\cdot\vec{n})^{2} v_{1}^{4} + 876 \vec{v}_{1}\cdot\vec{n} \vec{v}_{2}\cdot\vec{n} v_{1}^{4} - 120 ( \vec{v}_{2}\cdot\vec{n})^{2} v_{1}^{4} + 528 ( \vec{v}_{1}\cdot\vec{n})^{2} v_{2}^{4} \nn\\ 
&& - 1320 \vec{v}_{1}\cdot\vec{n} \vec{v}_{2}\cdot\vec{n} v_{2}^{4} -2880 v_{1}^2 \vec{v}_{2}\cdot\vec{n} ( \vec{v}_{1}\cdot\vec{n})^{3} - 720 \vec{v}_{1}\cdot\vec{v}_{2} ( \vec{v}_{1}\cdot\vec{n})^{4} \nn\\ 
&& + 240 \vec{v}_{2}\cdot\vec{n} \vec{v}_{1}\cdot\vec{v}_{2} ( \vec{v}_{1}\cdot\vec{n})^{3} - 960 v_{2}^2 ( \vec{v}_{1}\cdot\vec{n})^{4} + 1080 \vec{v}_{2}\cdot\vec{n} v_{2}^2 ( \vec{v}_{1}\cdot\vec{n})^{3} \nn\\ 
&& - 1680 v_{1}^2 ( \vec{v}_{1}\cdot\vec{n})^{2} ( \vec{v}_{2}\cdot\vec{n})^{2} + 240 \vec{v}_{1}\cdot\vec{n} v_{1}^2 ( \vec{v}_{2}\cdot\vec{n})^{3} + 1200 v_{1}^2 ( \vec{v}_{2}\cdot\vec{n})^{4} \nn\\ 
&& + 3120 \vec{v}_{1}\cdot\vec{v}_{2} ( \vec{v}_{1}\cdot\vec{n})^{2} ( \vec{v}_{2}\cdot\vec{n})^{2} - 1200 \vec{v}_{1}\cdot\vec{n} \vec{v}_{1}\cdot\vec{v}_{2} ( \vec{v}_{2}\cdot\vec{n})^{3} - 480 \vec{v}_{1}\cdot\vec{v}_{2} ( \vec{v}_{2}\cdot\vec{n})^{4} \nn\\ 
&& - 720 v_{2}^2 ( \vec{v}_{1}\cdot\vec{n})^{2} ( \vec{v}_{2}\cdot\vec{n})^{2} + 2880 \vec{v}_{1}\cdot\vec{n} v_{2}^2 ( \vec{v}_{2}\cdot\vec{n})^{3} + 1680 \vec{v}_{2}\cdot\vec{n} ( \vec{v}_{1}\cdot\vec{n})^{5} \nn\\ 
&& + 3360 ( \vec{v}_{1}\cdot\vec{n})^{4} ( \vec{v}_{2}\cdot\vec{n})^{2} - 3430 ( \vec{v}_{1}\cdot\vec{n})^{3} ( \vec{v}_{2}\cdot\vec{n})^{3} - 1680 \vec{v}_{1}\cdot\vec{n} ( \vec{v}_{2}\cdot\vec{n})^{5} \big) \nn\\ 
&& - 24 \vec{S}_{2}\cdot\vec{n} \vec{S}_{1}\cdot\vec{v}_{1} \big( 7 \vec{v}_{1}\cdot\vec{n} v_{1}^2 \vec{v}_{1}\cdot\vec{v}_{2} - 3 \vec{v}_{1}\cdot\vec{n} v_{1}^2 v_{2}^2 - 4 v_{1}^2 \vec{v}_{2}\cdot\vec{n} v_{2}^2 + 6 \vec{v}_{1}\cdot\vec{n} \vec{v}_{1}\cdot\vec{v}_{2} v_{2}^2 \nn\\ 
&& - 2 \vec{v}_{1}\cdot\vec{n} ( \vec{v}_{1}\cdot\vec{v}_{2})^{2} + \vec{v}_{1}\cdot\vec{n} v_{1}^{4} - 6 \vec{v}_{1}\cdot\vec{n} v_{2}^{4} -35 v_{1}^2 \vec{v}_{2}\cdot\vec{n} ( \vec{v}_{1}\cdot\vec{n})^{2} \nn\\ 
&& - 20 \vec{v}_{2}\cdot\vec{n} \vec{v}_{1}\cdot\vec{v}_{2} ( \vec{v}_{1}\cdot\vec{n})^{2} - 15 v_{2}^2 ( \vec{v}_{1}\cdot\vec{n})^{3} + 15 \vec{v}_{2}\cdot\vec{n} v_{2}^2 ( \vec{v}_{1}\cdot\vec{n})^{2} \nn\\ 
&& + 15 \vec{v}_{1}\cdot\vec{n} v_{1}^2 ( \vec{v}_{2}\cdot\vec{n})^{2} + 15 v_{1}^2 ( \vec{v}_{2}\cdot\vec{n})^{3} - 10 \vec{v}_{1}\cdot\vec{n} \vec{v}_{1}\cdot\vec{v}_{2} ( \vec{v}_{2}\cdot\vec{n})^{2} \nn\\ 
&& + 60 \vec{v}_{1}\cdot\vec{n} v_{2}^2 ( \vec{v}_{2}\cdot\vec{n})^{2} + 105 ( \vec{v}_{1}\cdot\vec{n})^{3} ( \vec{v}_{2}\cdot\vec{n})^{2} - 70 ( \vec{v}_{1}\cdot\vec{n})^{2} ( \vec{v}_{2}\cdot\vec{n})^{3} \nn\\ 
&& - 70 \vec{v}_{1}\cdot\vec{n} ( \vec{v}_{2}\cdot\vec{n})^{4} \big) + 3 \vec{S}_{1}\cdot\vec{n} \vec{S}_{2}\cdot\vec{v}_{1} \big( 96 \vec{v}_{1}\cdot\vec{n} v_{1}^2 \vec{v}_{1}\cdot\vec{v}_{2} - 48 v_{1}^2 \vec{v}_{2}\cdot\vec{n} \vec{v}_{1}\cdot\vec{v}_{2} \nn\\ 
&& - 176 \vec{v}_{1}\cdot\vec{n} v_{1}^2 v_{2}^2 + 191 v_{1}^2 \vec{v}_{2}\cdot\vec{n} v_{2}^2 + 144 \vec{v}_{1}\cdot\vec{n} \vec{v}_{1}\cdot\vec{v}_{2} v_{2}^2 - 480 \vec{v}_{2}\cdot\vec{n} \vec{v}_{1}\cdot\vec{v}_{2} v_{2}^2 \nn\\ 
&& + 64 \vec{v}_{1}\cdot\vec{n} ( \vec{v}_{1}\cdot\vec{v}_{2})^{2} - 20 \vec{v}_{2}\cdot\vec{n} ( \vec{v}_{1}\cdot\vec{v}_{2})^{2} - 96 \vec{v}_{1}\cdot\vec{n} v_{1}^{4} + 4 \vec{v}_{2}\cdot\vec{n} v_{1}^{4} - 176 \vec{v}_{1}\cdot\vec{n} v_{2}^{4} \nn\\ 
&& + 440 \vec{v}_{2}\cdot\vec{n} v_{2}^{4} + 960 v_{1}^2 \vec{v}_{2}\cdot\vec{n} ( \vec{v}_{1}\cdot\vec{n})^{2} - 160 \vec{v}_{2}\cdot\vec{n} \vec{v}_{1}\cdot\vec{v}_{2} ( \vec{v}_{1}\cdot\vec{n})^{2} + 320 v_{2}^2 ( \vec{v}_{1}\cdot\vec{n})^{3} \nn\\ 
&& - 240 \vec{v}_{2}\cdot\vec{n} v_{2}^2 ( \vec{v}_{1}\cdot\vec{n})^{2} - 200 \vec{v}_{1}\cdot\vec{n} v_{1}^2 ( \vec{v}_{2}\cdot\vec{n})^{2} - 240 v_{1}^2 ( \vec{v}_{2}\cdot\vec{n})^{3} \nn\\ 
&& - 80 \vec{v}_{1}\cdot\vec{n} \vec{v}_{1}\cdot\vec{v}_{2} ( \vec{v}_{2}\cdot\vec{n})^{2} + 640 \vec{v}_{1}\cdot\vec{v}_{2} ( \vec{v}_{2}\cdot\vec{n})^{3} + 240 \vec{v}_{1}\cdot\vec{n} v_{2}^2 ( \vec{v}_{2}\cdot\vec{n})^{2} \nn\\ 
&& - 960 v_{2}^2 ( \vec{v}_{2}\cdot\vec{n})^{3} -560 \vec{v}_{2}\cdot\vec{n} ( \vec{v}_{1}\cdot\vec{n})^{4} - 1120 ( \vec{v}_{1}\cdot\vec{n})^{3} ( \vec{v}_{2}\cdot\vec{n})^{2} \nn\\ 
&& + 910 ( \vec{v}_{1}\cdot\vec{n})^{2} ( \vec{v}_{2}\cdot\vec{n})^{3} + 560 ( \vec{v}_{2}\cdot\vec{n})^{5} \big) + 8 \vec{S}_{1}\cdot\vec{v}_{1} \vec{S}_{2}\cdot\vec{v}_{1} \big( 2 v_{1}^2 \vec{v}_{1}\cdot\vec{v}_{2} + 28 v_{1}^2 v_{2}^2 \nn\\ 
&& - 8 \vec{v}_{1}\cdot\vec{v}_{2} v_{2}^2 - 6 ( \vec{v}_{1}\cdot\vec{v}_{2})^{2} + 13 v_{1}^{4} - 6 v_{2}^{4} -111 \vec{v}_{1}\cdot\vec{n} v_{1}^2 \vec{v}_{2}\cdot\vec{n} - 72 \vec{v}_{1}\cdot\vec{n} \vec{v}_{2}\cdot\vec{n} \vec{v}_{1}\cdot\vec{v}_{2} \nn\\ 
&& - 9 \vec{v}_{1}\cdot\vec{n} \vec{v}_{2}\cdot\vec{n} v_{2}^2 - 63 v_{2}^2 ( \vec{v}_{1}\cdot\vec{n})^{2} + 15 v_{1}^2 ( \vec{v}_{2}\cdot\vec{n})^{2} - 42 \vec{v}_{1}\cdot\vec{v}_{2} ( \vec{v}_{2}\cdot\vec{n})^{2} \nn\\ 
&& + 108 v_{2}^2 ( \vec{v}_{2}\cdot\vec{n})^{2} + 285 ( \vec{v}_{1}\cdot\vec{n})^{2} ( \vec{v}_{2}\cdot\vec{n})^{2} + 30 \vec{v}_{1}\cdot\vec{n} ( \vec{v}_{2}\cdot\vec{n})^{3} - 150 ( \vec{v}_{2}\cdot\vec{n})^{4} \big) \nn\\ 
&& + 3 \vec{S}_{2}\cdot\vec{n} \vec{S}_{1}\cdot\vec{v}_{2} \big( 176 \vec{v}_{1}\cdot\vec{n} v_{1}^2 \vec{v}_{1}\cdot\vec{v}_{2} - 25 \vec{v}_{1}\cdot\vec{n} v_{1}^2 v_{2}^2 - 68 \vec{v}_{1}\cdot\vec{n} ( \vec{v}_{1}\cdot\vec{v}_{2})^{2} - 68 \vec{v}_{1}\cdot\vec{n} v_{1}^{4} \nn\\ 
&& + 640 v_{1}^2 ( \vec{v}_{1}\cdot\vec{n})^{3} - 440 v_{1}^2 \vec{v}_{2}\cdot\vec{n} ( \vec{v}_{1}\cdot\vec{n})^{2} - 480 \vec{v}_{1}\cdot\vec{v}_{2} ( \vec{v}_{1}\cdot\vec{n})^{3} \nn\\ 
&& + 160 \vec{v}_{2}\cdot\vec{n} \vec{v}_{1}\cdot\vec{v}_{2} ( \vec{v}_{1}\cdot\vec{n})^{2} + 40 \vec{v}_{1}\cdot\vec{n} v_{1}^2 ( \vec{v}_{2}\cdot\vec{n})^{2} -560 ( \vec{v}_{1}\cdot\vec{n})^{5} \nn\\ 
&& + 630 ( \vec{v}_{1}\cdot\vec{n})^{3} ( \vec{v}_{2}\cdot\vec{n})^{2} \big) - 2 \vec{S}_{2}\cdot\vec{v}_{1} \vec{S}_{1}\cdot\vec{v}_{2} \big( 96 v_{1}^2 \vec{v}_{1}\cdot\vec{v}_{2} + 33 v_{1}^2 v_{2}^2 - 96 ( \vec{v}_{1}\cdot\vec{v}_{2})^{2} \nn\\ 
&& + 32 v_{1}^{4} -468 \vec{v}_{1}\cdot\vec{n} v_{1}^2 \vec{v}_{2}\cdot\vec{n} + 132 \vec{v}_{1}\cdot\vec{n} \vec{v}_{2}\cdot\vec{n} \vec{v}_{1}\cdot\vec{v}_{2} - 264 \vec{v}_{1}\cdot\vec{n} \vec{v}_{2}\cdot\vec{n} v_{2}^2 \nn\\ 
&& + 372 v_{1}^2 ( \vec{v}_{1}\cdot\vec{n})^{2} - 288 \vec{v}_{1}\cdot\vec{v}_{2} ( \vec{v}_{1}\cdot\vec{n})^{2} - 84 v_{1}^2 ( \vec{v}_{2}\cdot\vec{n})^{2} - 144 \vec{v}_{1}\cdot\vec{v}_{2} ( \vec{v}_{2}\cdot\vec{n})^{2} \nn\\ 
&& + 300 v_{2}^2 ( \vec{v}_{2}\cdot\vec{n})^{2} -360 ( \vec{v}_{1}\cdot\vec{n})^{4} - 120 \vec{v}_{2}\cdot\vec{n} ( \vec{v}_{1}\cdot\vec{n})^{3} + 780 ( \vec{v}_{1}\cdot\vec{n})^{2} ( \vec{v}_{2}\cdot\vec{n})^{2} \nn\\ 
&& + 360 \vec{v}_{1}\cdot\vec{n} ( \vec{v}_{2}\cdot\vec{n})^{3} - 240 ( \vec{v}_{2}\cdot\vec{n})^{4} \big) + 24 \vec{S}_{1}\cdot\vec{n} \vec{S}_{2}\cdot\vec{v}_{2} \big( 12 \vec{v}_{1}\cdot\vec{n} v_{1}^2 \vec{v}_{1}\cdot\vec{v}_{2} \nn\\ 
&& + 2 v_{1}^2 \vec{v}_{2}\cdot\vec{n} \vec{v}_{1}\cdot\vec{v}_{2} - 3 \vec{v}_{1}\cdot\vec{n} v_{1}^2 v_{2}^2 + 22 \vec{v}_{1}\cdot\vec{n} \vec{v}_{1}\cdot\vec{v}_{2} v_{2}^2 - 25 \vec{v}_{2}\cdot\vec{n} \vec{v}_{1}\cdot\vec{v}_{2} v_{2}^2 \nn\\ 
&& - 10 \vec{v}_{1}\cdot\vec{n} ( \vec{v}_{1}\cdot\vec{v}_{2})^{2} + 12 \vec{v}_{2}\cdot\vec{n} ( \vec{v}_{1}\cdot\vec{v}_{2})^{2} - 7 \vec{v}_{1}\cdot\vec{n} v_{1}^{4} \nn\\ 
&& + 2 \vec{v}_{2}\cdot\vec{n} v_{1}^{4} -10 v_{1}^2 \vec{v}_{2}\cdot\vec{n} ( \vec{v}_{1}\cdot\vec{n})^{2} - 20 \vec{v}_{1}\cdot\vec{v}_{2} ( \vec{v}_{1}\cdot\vec{n})^{3} - 5 \vec{v}_{2}\cdot\vec{n} \vec{v}_{1}\cdot\vec{v}_{2} ( \vec{v}_{1}\cdot\vec{n})^{2} \nn\\ 
&& + 10 \vec{v}_{1}\cdot\vec{n} v_{1}^2 ( \vec{v}_{2}\cdot\vec{n})^{2} - 30 \vec{v}_{1}\cdot\vec{n} \vec{v}_{1}\cdot\vec{v}_{2} ( \vec{v}_{2}\cdot\vec{n})^{2} + 20 \vec{v}_{1}\cdot\vec{v}_{2} ( \vec{v}_{2}\cdot\vec{n})^{3} \nn\\ 
&& + 35 ( \vec{v}_{1}\cdot\vec{n})^{3} ( \vec{v}_{2}\cdot\vec{n})^{2} \big) - 4 \vec{S}_{1}\cdot\vec{v}_{1} \vec{S}_{2}\cdot\vec{v}_{2} \big( 50 v_{1}^2 \vec{v}_{1}\cdot\vec{v}_{2} + 7 v_{1}^2 v_{2}^2 - 22 ( \vec{v}_{1}\cdot\vec{v}_{2})^{2} \nn\\ 
&& + 12 \vec{v}_{1}\cdot\vec{n} v_{1}^2 \vec{v}_{2}\cdot\vec{n} - 90 \vec{v}_{1}\cdot\vec{n} \vec{v}_{2}\cdot\vec{n} \vec{v}_{1}\cdot\vec{v}_{2} - 42 v_{1}^2 ( \vec{v}_{1}\cdot\vec{n})^{2} - 60 \vec{v}_{1}\cdot\vec{v}_{2} ( \vec{v}_{1}\cdot\vec{n})^{2} \nn\\ 
&& - 24 v_{2}^2 ( \vec{v}_{1}\cdot\vec{n})^{2} - 66 v_{1}^2 ( \vec{v}_{2}\cdot\vec{n})^{2} -60 \vec{v}_{2}\cdot\vec{n} ( \vec{v}_{1}\cdot\vec{n})^{3} + 345 ( \vec{v}_{1}\cdot\vec{n})^{2} ( \vec{v}_{2}\cdot\vec{n})^{2} \big) \nn\\ 
&& - 8 \vec{S}_{1}\cdot\vec{v}_{2} \vec{S}_{2}\cdot\vec{v}_{2} \big( 14 v_{1}^2 \vec{v}_{1}\cdot\vec{v}_{2} - 25 v_{1}^{4} -3 \vec{v}_{1}\cdot\vec{n} v_{1}^2 \vec{v}_{2}\cdot\vec{n} + 36 \vec{v}_{1}\cdot\vec{n} \vec{v}_{2}\cdot\vec{n} \vec{v}_{1}\cdot\vec{v}_{2} \nn\\ 
&& - 75 \vec{v}_{1}\cdot\vec{n} \vec{v}_{2}\cdot\vec{n} v_{2}^2 + 66 v_{1}^2 ( \vec{v}_{1}\cdot\vec{n})^{2} - 36 \vec{v}_{1}\cdot\vec{v}_{2} ( \vec{v}_{1}\cdot\vec{n})^{2} \nn\\ 
&& + 66 v_{2}^2 ( \vec{v}_{1}\cdot\vec{n})^{2} -60 ( \vec{v}_{1}\cdot\vec{n})^{4} + 45 \vec{v}_{2}\cdot\vec{n} ( \vec{v}_{1}\cdot\vec{n})^{3} - 90 ( \vec{v}_{1}\cdot\vec{n})^{2} ( \vec{v}_{2}\cdot\vec{n})^{2} \nn\\ 
&& + 60 \vec{v}_{1}\cdot\vec{n} ( \vec{v}_{2}\cdot\vec{n})^{3} \big) \Big]\nn\\ && +  	\frac{G^2 m_{1}}{96 r{}^4} \Big[ \vec{S}_{1}\cdot\vec{n} \vec{S}_{2}\cdot\vec{n} \big( 11154 v_{1}^2 \vec{v}_{1}\cdot\vec{v}_{2} + 1076 v_{1}^2 v_{2}^2 - 4544 ( \vec{v}_{1}\cdot\vec{v}_{2})^{2} \nn\\ 
&& - 5859 v_{1}^{4} -32220 \vec{v}_{1}\cdot\vec{n} v_{1}^2 \vec{v}_{2}\cdot\vec{n} + 26142 \vec{v}_{1}\cdot\vec{n} \vec{v}_{2}\cdot\vec{n} \vec{v}_{1}\cdot\vec{v}_{2} + 41211 v_{1}^2 ( \vec{v}_{1}\cdot\vec{n})^{2} \nn\\ 
&& - 37152 \vec{v}_{1}\cdot\vec{v}_{2} ( \vec{v}_{1}\cdot\vec{n})^{2} - 5367 v_{1}^2 ( \vec{v}_{2}\cdot\vec{n})^{2} -2208 ( \vec{v}_{1}\cdot\vec{n})^{4} - 12882 \vec{v}_{2}\cdot\vec{n} ( \vec{v}_{1}\cdot\vec{n})^{3} \nn\\ 
&& + 6738 ( \vec{v}_{1}\cdot\vec{n})^{2} ( \vec{v}_{2}\cdot\vec{n})^{2} \big) - \vec{S}_{1}\cdot\vec{S}_{2} \big( 4995 v_{1}^2 \vec{v}_{1}\cdot\vec{v}_{2} + 2349 v_{1}^2 v_{2}^2 - 2702 ( \vec{v}_{1}\cdot\vec{v}_{2})^{2} \nn\\ 
&& - 2595 v_{1}^{4} -11223 \vec{v}_{1}\cdot\vec{n} v_{1}^2 \vec{v}_{2}\cdot\vec{n} + 12480 \vec{v}_{1}\cdot\vec{n} \vec{v}_{2}\cdot\vec{n} \vec{v}_{1}\cdot\vec{v}_{2} + 3048 \vec{v}_{1}\cdot\vec{n} \vec{v}_{2}\cdot\vec{n} v_{2}^2 \nn\\ 
&& + 24009 v_{1}^2 ( \vec{v}_{1}\cdot\vec{n})^{2} - 35685 \vec{v}_{1}\cdot\vec{v}_{2} ( \vec{v}_{1}\cdot\vec{n})^{2} + 1898 v_{2}^2 ( \vec{v}_{1}\cdot\vec{n})^{2} - 11200 v_{1}^2 ( \vec{v}_{2}\cdot\vec{n})^{2} \nn\\ 
&& - 198 \vec{v}_{1}\cdot\vec{v}_{2} ( \vec{v}_{2}\cdot\vec{n})^{2} -26268 ( \vec{v}_{1}\cdot\vec{n})^{4} + 46683 \vec{v}_{2}\cdot\vec{n} ( \vec{v}_{1}\cdot\vec{n})^{3} \nn\\ 
&& - 2712 ( \vec{v}_{1}\cdot\vec{n})^{2} ( \vec{v}_{2}\cdot\vec{n})^{2} - 6489 \vec{v}_{1}\cdot\vec{n} ( \vec{v}_{2}\cdot\vec{n})^{3} \big) + \vec{S}_{2}\cdot\vec{n} \vec{S}_{1}\cdot\vec{v}_{1} \big( 2211 \vec{v}_{1}\cdot\vec{n} v_{1}^2 \nn\\ 
&& - 294 v_{1}^2 \vec{v}_{2}\cdot\vec{n} - 4698 \vec{v}_{1}\cdot\vec{n} \vec{v}_{1}\cdot\vec{v}_{2} + 320 \vec{v}_{2}\cdot\vec{n} \vec{v}_{1}\cdot\vec{v}_{2} + 250 \vec{v}_{1}\cdot\vec{n} v_{2}^2 \nn\\ 
&& - 576 \vec{v}_{2}\cdot\vec{n} v_{2}^2 -37083 ( \vec{v}_{1}\cdot\vec{n})^{3} + 44856 \vec{v}_{2}\cdot\vec{n} ( \vec{v}_{1}\cdot\vec{n})^{2} + 3477 \vec{v}_{1}\cdot\vec{n} ( \vec{v}_{2}\cdot\vec{n})^{2} \big) \nn\\ 
&& + \vec{S}_{1}\cdot\vec{n} \vec{S}_{2}\cdot\vec{v}_{1} \big( 1062 \vec{v}_{1}\cdot\vec{n} v_{1}^2 - 4071 v_{1}^2 \vec{v}_{2}\cdot\vec{n} - 8106 \vec{v}_{1}\cdot\vec{n} \vec{v}_{1}\cdot\vec{v}_{2} + 4718 \vec{v}_{2}\cdot\vec{n} \vec{v}_{1}\cdot\vec{v}_{2} \nn\\ 
&& + 670 \vec{v}_{1}\cdot\vec{n} v_{2}^2 + 3624 \vec{v}_{2}\cdot\vec{n} v_{2}^2 -27012 ( \vec{v}_{1}\cdot\vec{n})^{3} + 51435 \vec{v}_{2}\cdot\vec{n} ( \vec{v}_{1}\cdot\vec{n})^{2} \nn\\ 
&& - 9762 \vec{v}_{1}\cdot\vec{n} ( \vec{v}_{2}\cdot\vec{n})^{2} - 6489 ( \vec{v}_{2}\cdot\vec{n})^{3} \big) - \vec{S}_{1}\cdot\vec{v}_{1} \vec{S}_{2}\cdot\vec{v}_{1} \big( 2211 v_{1}^2 - 1176 \vec{v}_{1}\cdot\vec{v}_{2} \nn\\ 
&& - 2387 v_{2}^2 + 11076 \vec{v}_{1}\cdot\vec{n} \vec{v}_{2}\cdot\vec{n} - 24747 ( \vec{v}_{1}\cdot\vec{n})^{2} + 9590 ( \vec{v}_{2}\cdot\vec{n})^{2} \big) \nn\\ 
&& - 6 \vec{S}_{2}\cdot\vec{n} \vec{S}_{1}\cdot\vec{v}_{2} \big( 1164 \vec{v}_{1}\cdot\vec{n} v_{1}^2 - 928 \vec{v}_{1}\cdot\vec{n} \vec{v}_{1}\cdot\vec{v}_{2} -7494 ( \vec{v}_{1}\cdot\vec{n})^{3} \nn\\ 
&& + 6999 \vec{v}_{2}\cdot\vec{n} ( \vec{v}_{1}\cdot\vec{n})^{2} \big) + \vec{S}_{2}\cdot\vec{v}_{1} \vec{S}_{1}\cdot\vec{v}_{2} \big( 3561 v_{1}^2 - 3086 \vec{v}_{1}\cdot\vec{v}_{2} + 17354 \vec{v}_{1}\cdot\vec{n} \vec{v}_{2}\cdot\vec{n} \nn\\ 
&& - 31449 ( \vec{v}_{1}\cdot\vec{n})^{2} - 198 ( \vec{v}_{2}\cdot\vec{n})^{2} \big) + 6 \vec{S}_{1}\cdot\vec{n} \vec{S}_{2}\cdot\vec{v}_{2} \big( 1528 \vec{v}_{1}\cdot\vec{n} v_{1}^2 - 590 v_{1}^2 \vec{v}_{2}\cdot\vec{n} \nn\\ 
&& - 1169 \vec{v}_{1}\cdot\vec{n} \vec{v}_{1}\cdot\vec{v}_{2} - 33 \vec{v}_{2}\cdot\vec{n} \vec{v}_{1}\cdot\vec{v}_{2} -243 ( \vec{v}_{1}\cdot\vec{n})^{3} + 942 \vec{v}_{2}\cdot\vec{n} ( \vec{v}_{1}\cdot\vec{n})^{2} \big) \nn\\ 
&& + 2 \vec{S}_{1}\cdot\vec{v}_{1} \vec{S}_{2}\cdot\vec{v}_{2} \big( 174 v_{1}^2 - 346 \vec{v}_{1}\cdot\vec{v}_{2} + 562 \vec{v}_{1}\cdot\vec{n} \vec{v}_{2}\cdot\vec{n} - 5247 ( \vec{v}_{1}\cdot\vec{n})^{2} \big) \nn\\ 
&& + 6 \vec{S}_{1}\cdot\vec{v}_{2} \vec{S}_{2}\cdot\vec{v}_{2} \big( 118 v_{1}^2 + 33 \vec{v}_{1}\cdot\vec{n} \vec{v}_{2}\cdot\vec{n} + 1145 ( \vec{v}_{1}\cdot\vec{n})^{2} \big) \Big] \nn\\ 
&& - 	\frac{G^2 m_{2}}{96 r{}^4} \Big[ 6 \vec{S}_{1}\cdot\vec{n} \vec{S}_{2}\cdot\vec{n} \big( 59 v_{1}^2 \vec{v}_{1}\cdot\vec{v}_{2} + 226 v_{1}^2 v_{2}^2 - 12 v_{1}^{4} + 222 \vec{v}_{1}\cdot\vec{n} v_{1}^2 \vec{v}_{2}\cdot\vec{n} \nn\\ 
&& - 885 \vec{v}_{1}\cdot\vec{v}_{2} ( \vec{v}_{1}\cdot\vec{n})^{2} - 451 v_{1}^2 ( \vec{v}_{2}\cdot\vec{n})^{2} \big) - \vec{S}_{1}\cdot\vec{S}_{2} \big( 927 v_{1}^2 \vec{v}_{1}\cdot\vec{v}_{2} + 1078 v_{1}^2 v_{2}^2 \nn\\ 
&& - 48 v_{1}^{4} -330 \vec{v}_{1}\cdot\vec{n} v_{1}^2 \vec{v}_{2}\cdot\vec{n} - 6568 \vec{v}_{1}\cdot\vec{n} \vec{v}_{2}\cdot\vec{n} \vec{v}_{1}\cdot\vec{v}_{2} + 3474 \vec{v}_{1}\cdot\vec{n} \vec{v}_{2}\cdot\vec{n} v_{2}^2 \nn\\ 
&& - 240 v_{1}^2 ( \vec{v}_{1}\cdot\vec{n})^{2} - 2082 \vec{v}_{1}\cdot\vec{v}_{2} ( \vec{v}_{1}\cdot\vec{n})^{2} - 5018 v_{2}^2 ( \vec{v}_{1}\cdot\vec{n})^{2} - 668 v_{1}^2 ( \vec{v}_{2}\cdot\vec{n})^{2} \nn\\ 
&& - 3678 \vec{v}_{1}\cdot\vec{v}_{2} ( \vec{v}_{2}\cdot\vec{n})^{2} -2250 \vec{v}_{2}\cdot\vec{n} ( \vec{v}_{1}\cdot\vec{n})^{3} + 17382 ( \vec{v}_{1}\cdot\vec{n})^{2} ( \vec{v}_{2}\cdot\vec{n})^{2} \nn\\ 
&& - 5412 \vec{v}_{1}\cdot\vec{n} ( \vec{v}_{2}\cdot\vec{n})^{3} \big) + 2 \vec{S}_{2}\cdot\vec{n} \vec{S}_{1}\cdot\vec{v}_{1} \big( 36 \vec{v}_{1}\cdot\vec{n} v_{1}^2 + 66 v_{1}^2 \vec{v}_{2}\cdot\vec{n} + 408 \vec{v}_{1}\cdot\vec{n} \vec{v}_{1}\cdot\vec{v}_{2} \nn\\ 
&& - 868 \vec{v}_{2}\cdot\vec{n} \vec{v}_{1}\cdot\vec{v}_{2} - 1067 \vec{v}_{1}\cdot\vec{n} v_{2}^2 + 384 \vec{v}_{2}\cdot\vec{n} v_{2}^2 -720 \vec{v}_{2}\cdot\vec{n} ( \vec{v}_{1}\cdot\vec{n})^{2} \nn\\ 
&& + 2199 \vec{v}_{1}\cdot\vec{n} ( \vec{v}_{2}\cdot\vec{n})^{2} - 696 ( \vec{v}_{2}\cdot\vec{n})^{3} \big) - 2 \vec{S}_{1}\cdot\vec{n} \vec{S}_{2}\cdot\vec{v}_{1} \big( 120 \vec{v}_{1}\cdot\vec{n} v_{1}^2 + 273 v_{1}^2 \vec{v}_{2}\cdot\vec{n} \nn\\ 
&& - 3084 \vec{v}_{1}\cdot\vec{n} \vec{v}_{1}\cdot\vec{v}_{2} + 3784 \vec{v}_{2}\cdot\vec{n} \vec{v}_{1}\cdot\vec{v}_{2} + 2120 \vec{v}_{1}\cdot\vec{n} v_{2}^2 - 1353 \vec{v}_{2}\cdot\vec{n} v_{2}^2 \nn\\ 
&& + 1125 \vec{v}_{2}\cdot\vec{n} ( \vec{v}_{1}\cdot\vec{n})^{2} - 8022 \vec{v}_{1}\cdot\vec{n} ( \vec{v}_{2}\cdot\vec{n})^{2} + 2010 ( \vec{v}_{2}\cdot\vec{n})^{3} \big) - 2 \vec{S}_{1}\cdot\vec{v}_{1} \vec{S}_{2}\cdot\vec{v}_{1} \big( 24 v_{1}^2 \nn\\ 
&& + 504 \vec{v}_{1}\cdot\vec{v}_{2} - 539 v_{2}^2 + 48 \vec{v}_{1}\cdot\vec{n} \vec{v}_{2}\cdot\vec{n} + 491 ( \vec{v}_{2}\cdot\vec{n})^{2} \big) \nn\\ 
&& - 6 \vec{S}_{2}\cdot\vec{n} \vec{S}_{1}\cdot\vec{v}_{2} \big( 229 \vec{v}_{1}\cdot\vec{n} v_{1}^2 -885 ( \vec{v}_{1}\cdot\vec{n})^{3} \big) + 3 \vec{S}_{2}\cdot\vec{v}_{1} \vec{S}_{1}\cdot\vec{v}_{2} \big( 713 v_{1}^2 \nn\\ 
&& + 912 \vec{v}_{1}\cdot\vec{n} \vec{v}_{2}\cdot\vec{n} - 2750 ( \vec{v}_{1}\cdot\vec{n})^{2} - 1226 ( \vec{v}_{2}\cdot\vec{n})^{2} \big) - 6 \vec{S}_{1}\cdot\vec{n} \vec{S}_{2}\cdot\vec{v}_{2} \big( 640 \vec{v}_{1}\cdot\vec{n} v_{1}^2 \nn\\ 
&& - 746 v_{1}^2 \vec{v}_{2}\cdot\vec{n} - 456 \vec{v}_{1}\cdot\vec{n} \vec{v}_{1}\cdot\vec{v}_{2} + 613 \vec{v}_{2}\cdot\vec{n} \vec{v}_{1}\cdot\vec{v}_{2} \big) \nn\\ 
&& - 12 \vec{S}_{1}\cdot\vec{v}_{1} \vec{S}_{2}\cdot\vec{v}_{2} \big( 373 \vec{v}_{1}\cdot\vec{n} \vec{v}_{2}\cdot\vec{n} - 320 ( \vec{v}_{1}\cdot\vec{n})^{2} \big) + 6 \vec{S}_{1}\cdot\vec{v}_{2} \vec{S}_{2}\cdot\vec{v}_{2} \big( 613 \vec{v}_{1}\cdot\vec{n} \vec{v}_{2}\cdot\vec{n} \nn\\ 
&& - 456 ( \vec{v}_{1}\cdot\vec{n})^{2} \big) \Big]\nn\\ && +  	\frac{G^3 m_{2}{}^2}{300 r{}^5} \Big[ 75 \vec{S}_{1}\cdot\vec{n} \vec{S}_{2}\cdot\vec{n} \big( 16 v_{1}^2 + 31 ( \vec{v}_{1}\cdot\vec{n})^{2} \big) - \vec{S}_{1}\cdot\vec{S}_{2} \big( 1536 v_{1}^2 -36485 \vec{v}_{1}\cdot\vec{n} \vec{v}_{2}\cdot\vec{n} \nn\\ 
&& + 1395 ( \vec{v}_{1}\cdot\vec{n})^{2} \big) - 5 \vec{S}_{2}\cdot\vec{n} \vec{S}_{1}\cdot\vec{v}_{1} \big( 99 \vec{v}_{1}\cdot\vec{n} + 3236 \vec{v}_{2}\cdot\vec{n} \big) - 5 \vec{S}_{1}\cdot\vec{n} \vec{S}_{2}\cdot\vec{v}_{1} \big( 174 \vec{v}_{1}\cdot\vec{n} \nn\\ 
&& + 4061 \vec{v}_{2}\cdot\vec{n} \big) + 843 \vec{S}_{1}\cdot\vec{v}_{1} \vec{S}_{2}\cdot\vec{v}_{1} \Big] - 	\frac{G^3 m_{1}{}^2}{600 r{}^5} \Big[ 50 \vec{S}_{1}\cdot\vec{n} \vec{S}_{2}\cdot\vec{n} \big( 352 v_{1}^2 \nn\\ 
&& + 451 \vec{v}_{1}\cdot\vec{v}_{2} -4453 \vec{v}_{1}\cdot\vec{n} \vec{v}_{2}\cdot\vec{n} + 4298 ( \vec{v}_{1}\cdot\vec{n})^{2} \big) - \vec{S}_{1}\cdot\vec{S}_{2} \big( 20368 v_{1}^2 \nn\\ 
&& + 9529 \vec{v}_{1}\cdot\vec{v}_{2} -117090 \vec{v}_{1}\cdot\vec{n} \vec{v}_{2}\cdot\vec{n} + 21760 ( \vec{v}_{1}\cdot\vec{n})^{2} \big) - 20 \vec{S}_{2}\cdot\vec{n} \vec{S}_{1}\cdot\vec{v}_{1} \big( 5758 \vec{v}_{1}\cdot\vec{n} \nn\\ 
&& - 1303 \vec{v}_{2}\cdot\vec{n} \big) - 20 \vec{S}_{1}\cdot\vec{n} \vec{S}_{2}\cdot\vec{v}_{1} \big( 3718 \vec{v}_{1}\cdot\vec{n} - 1693 \vec{v}_{2}\cdot\vec{n} \big) + 24300 \vec{S}_{2}\cdot\vec{n} \vec{S}_{1}\cdot\vec{v}_{2} \vec{v}_{1}\cdot\vec{n} \nn\\ 
&& - 13200 \vec{S}_{1}\cdot\vec{n} \vec{S}_{2}\cdot\vec{v}_{2} \vec{v}_{1}\cdot\vec{n} + 40984 \vec{S}_{1}\cdot\vec{v}_{1} \vec{S}_{2}\cdot\vec{v}_{1} - 19099 \vec{S}_{2}\cdot\vec{v}_{1} \vec{S}_{1}\cdot\vec{v}_{2} \nn\\ 
&& + 3626 \vec{S}_{1}\cdot\vec{v}_{1} \vec{S}_{2}\cdot\vec{v}_{2} \Big] + 	\frac{G^3 m_{1} m_{2}}{192 r{}^5} \Big[ \vec{S}_{1}\cdot\vec{n} \vec{S}_{2}\cdot\vec{n} \big( {(33424 - 2115 \pi^2)} v_{1}^2 \nn\\ 
&& - {(47208 - 2115 \pi^2)} \vec{v}_{1}\cdot\vec{v}_{2} + {(81252 - 14805 \pi^2)} \vec{v}_{1}\cdot\vec{n} \vec{v}_{2}\cdot\vec{n} \nn\\ 
&& - {(84526 - 14805 \pi^2)} ( \vec{v}_{1}\cdot\vec{n})^{2} \big) - \vec{S}_{1}\cdot\vec{S}_{2} \big( {(13736 - 423 \pi^2)} v_{1}^2 \nn\\ 
&& - {(24672 - 423 \pi^2)} \vec{v}_{1}\cdot\vec{v}_{2} + {(46992 - 2115 \pi^2)} \vec{v}_{1}\cdot\vec{n} \vec{v}_{2}\cdot\vec{n} \nn\\ 
&& - {(38338 - 2115 \pi^2)} ( \vec{v}_{1}\cdot\vec{n})^{2} \big) + 2 \vec{S}_{2}\cdot\vec{n} \vec{S}_{1}\cdot\vec{v}_{1} \big( {(6316 - 2115 \pi^2)} \vec{v}_{1}\cdot\vec{n} \nn\\ 
&& - {(7149 - 2115 \pi^2)} \vec{v}_{2}\cdot\vec{n} \big) - 2 \vec{S}_{1}\cdot\vec{n} \vec{S}_{2}\cdot\vec{v}_{1} \big( {(5774 + 2115 \pi^2)} \vec{v}_{1}\cdot\vec{n} \nn\\ 
&& - {(4299 + 2115 \pi^2)} \vec{v}_{2}\cdot\vec{n} \big) + 14238 \vec{S}_{2}\cdot\vec{n} \vec{S}_{1}\cdot\vec{v}_{2} \vec{v}_{1}\cdot\vec{n} + 11142 \vec{S}_{1}\cdot\vec{n} \vec{S}_{2}\cdot\vec{v}_{2} \vec{v}_{1}\cdot\vec{n} \nn\\ 
&& + 2 {(2596 + 423 \pi^2)} \vec{S}_{1}\cdot\vec{v}_{1} \vec{S}_{2}\cdot\vec{v}_{1} - 3 {(3088 + 141 \pi^2)} \vec{S}_{2}\cdot\vec{v}_{1} \vec{S}_{1}\cdot\vec{v}_{2} \nn\\ 
&& - 141 {(44 + 3 \pi^2)} \vec{S}_{1}\cdot\vec{v}_{1} \vec{S}_{2}\cdot\vec{v}_{2} \Big]\nn\\ && - 	\frac{G^4 m_{1}{}^3}{100 r{}^6} \Big[ 5131 \vec{S}_{1}\cdot\vec{n} \vec{S}_{2}\cdot\vec{n} - 2137 \vec{S}_{1}\cdot\vec{S}_{2} \Big] \nn\\ 
&& - 	\frac{G^4 m_{1}{}^2 m_{2}}{1200 r{}^6} \Big[ 9 {(86423 - 4050 \pi^2)} \vec{S}_{1}\cdot\vec{n} \vec{S}_{2}\cdot\vec{n} - {(286139 - 12150 \pi^2)} \vec{S}_{1}\cdot\vec{S}_{2} \Big],
\eea
%%%%%%%%%%%%%

%%%%%%%%%%%%%
\bea
\hat{V}^{\text{N}^3\text{LO}}_{\text{S}_1^2} &= &- 	\frac{G m_{2}}{128 m_{1} r{}^3} \Big[ S_{1}^2 \big( 1336 v_{1}^2 \vec{v}_{1}\cdot\vec{v}_{2} v_{2}^2 - 1044 v_{1}^2 ( \vec{v}_{1}\cdot\vec{v}_{2})^{2} + 624 ( \vec{v}_{1}\cdot\vec{v}_{2})^{3} \nn\\ 
&& - 1056 v_{2}^2 ( \vec{v}_{1}\cdot\vec{v}_{2})^{2} - 74 v_{1}^{6} + 488 \vec{v}_{1}\cdot\vec{v}_{2} v_{1}^{4} - 352 v_{2}^2 v_{1}^{4} - 380 v_{1}^2 v_{2}^{4} + 456 \vec{v}_{1}\cdot\vec{v}_{2} v_{2}^{4} \nn\\ 
&& + 4536 \vec{v}_{1}\cdot\vec{n} v_{1}^2 \vec{v}_{2}\cdot\vec{n} \vec{v}_{1}\cdot\vec{v}_{2} - 2280 \vec{v}_{1}\cdot\vec{n} v_{1}^2 \vec{v}_{2}\cdot\vec{n} v_{2}^2 + 2112 \vec{v}_{1}\cdot\vec{n} \vec{v}_{2}\cdot\vec{n} \vec{v}_{1}\cdot\vec{v}_{2} v_{2}^2 \nn\\ 
&& - 876 v_{1}^2 \vec{v}_{1}\cdot\vec{v}_{2} ( \vec{v}_{1}\cdot\vec{n})^{2} + 594 v_{1}^2 v_{2}^2 ( \vec{v}_{1}\cdot\vec{n})^{2} - 516 \vec{v}_{1}\cdot\vec{v}_{2} v_{2}^2 ( \vec{v}_{1}\cdot\vec{n})^{2} \nn\\ 
&& - 3996 v_{1}^2 \vec{v}_{1}\cdot\vec{v}_{2} ( \vec{v}_{2}\cdot\vec{n})^{2} + 984 ( \vec{v}_{1}\cdot\vec{n})^{2} ( \vec{v}_{1}\cdot\vec{v}_{2})^{2} - 3312 \vec{v}_{1}\cdot\vec{n} \vec{v}_{2}\cdot\vec{n} ( \vec{v}_{1}\cdot\vec{v}_{2})^{2} \nn\\ 
&& + 3072 ( \vec{v}_{2}\cdot\vec{n})^{2} ( \vec{v}_{1}\cdot\vec{v}_{2})^{2} + 2256 v_{1}^2 v_{2}^2 ( \vec{v}_{2}\cdot\vec{n})^{2} - 2544 \vec{v}_{1}\cdot\vec{v}_{2} v_{2}^2 ( \vec{v}_{2}\cdot\vec{n})^{2} \nn\\ 
&& + 75 ( \vec{v}_{1}\cdot\vec{n})^{2} v_{1}^{4} - 1224 \vec{v}_{1}\cdot\vec{n} \vec{v}_{2}\cdot\vec{n} v_{1}^{4} + 1116 ( \vec{v}_{2}\cdot\vec{n})^{2} v_{1}^{4} - 258 ( \vec{v}_{1}\cdot\vec{n})^{2} v_{2}^{4} \nn\\ 
&& + 168 \vec{v}_{1}\cdot\vec{n} \vec{v}_{2}\cdot\vec{n} v_{2}^{4} + 1620 v_{1}^2 \vec{v}_{2}\cdot\vec{n} ( \vec{v}_{1}\cdot\vec{n})^{3} - 1200 \vec{v}_{2}\cdot\vec{n} \vec{v}_{1}\cdot\vec{v}_{2} ( \vec{v}_{1}\cdot\vec{n})^{3} \nn\\ 
&& + 150 v_{2}^2 ( \vec{v}_{1}\cdot\vec{n})^{4} - 900 \vec{v}_{2}\cdot\vec{n} v_{2}^2 ( \vec{v}_{1}\cdot\vec{n})^{3} - 3750 v_{1}^2 ( \vec{v}_{1}\cdot\vec{n})^{2} ( \vec{v}_{2}\cdot\vec{n})^{2} \nn\\ 
&& + 4020 \vec{v}_{1}\cdot\vec{n} v_{1}^2 ( \vec{v}_{2}\cdot\vec{n})^{3} - 1920 v_{1}^2 ( \vec{v}_{2}\cdot\vec{n})^{4} + 2520 \vec{v}_{1}\cdot\vec{v}_{2} ( \vec{v}_{1}\cdot\vec{n})^{2} ( \vec{v}_{2}\cdot\vec{n})^{2} \nn\\ 
&& - 3840 \vec{v}_{1}\cdot\vec{n} \vec{v}_{1}\cdot\vec{v}_{2} ( \vec{v}_{2}\cdot\vec{n})^{3} + 2160 \vec{v}_{1}\cdot\vec{v}_{2} ( \vec{v}_{2}\cdot\vec{n})^{4} + 1680 v_{2}^2 ( \vec{v}_{1}\cdot\vec{n})^{2} ( \vec{v}_{2}\cdot\vec{n})^{2} \nn\\ 
&& - 240 \vec{v}_{1}\cdot\vec{n} v_{2}^2 ( \vec{v}_{2}\cdot\vec{n})^{3} -1050 ( \vec{v}_{1}\cdot\vec{n})^{4} ( \vec{v}_{2}\cdot\vec{n})^{2} + 2520 ( \vec{v}_{1}\cdot\vec{n})^{3} ( \vec{v}_{2}\cdot\vec{n})^{3} \nn\\ 
&& - 1680 ( \vec{v}_{1}\cdot\vec{n})^{2} ( \vec{v}_{2}\cdot\vec{n})^{4} \big) + 6 \vec{S}_{1}\cdot\vec{n} \vec{S}_{1}\cdot\vec{v}_{1} \big( 80 \vec{v}_{1}\cdot\vec{n} v_{1}^2 \vec{v}_{1}\cdot\vec{v}_{2} - 56 v_{1}^2 \vec{v}_{2}\cdot\vec{n} \vec{v}_{1}\cdot\vec{v}_{2} \nn\\ 
&& - 70 \vec{v}_{1}\cdot\vec{n} v_{1}^2 v_{2}^2 + 8 v_{1}^2 \vec{v}_{2}\cdot\vec{n} v_{2}^2 + 48 \vec{v}_{1}\cdot\vec{n} \vec{v}_{1}\cdot\vec{v}_{2} v_{2}^2 + 64 \vec{v}_{2}\cdot\vec{n} \vec{v}_{1}\cdot\vec{v}_{2} v_{2}^2 \nn\\ 
&& - 28 \vec{v}_{1}\cdot\vec{n} ( \vec{v}_{1}\cdot\vec{v}_{2})^{2} - 8 \vec{v}_{2}\cdot\vec{n} ( \vec{v}_{1}\cdot\vec{v}_{2})^{2} - 43 \vec{v}_{1}\cdot\vec{n} v_{1}^{4} + 18 \vec{v}_{2}\cdot\vec{n} v_{1}^{4} - 14 \vec{v}_{1}\cdot\vec{n} v_{2}^{4} \nn\\ 
&& - 28 \vec{v}_{2}\cdot\vec{n} v_{2}^{4} -140 v_{1}^2 \vec{v}_{2}\cdot\vec{n} ( \vec{v}_{1}\cdot\vec{n})^{2} + 60 \vec{v}_{2}\cdot\vec{n} \vec{v}_{1}\cdot\vec{v}_{2} ( \vec{v}_{1}\cdot\vec{n})^{2} - 30 v_{2}^2 ( \vec{v}_{1}\cdot\vec{n})^{3} \nn\\ 
&& + 120 \vec{v}_{2}\cdot\vec{n} v_{2}^2 ( \vec{v}_{1}\cdot\vec{n})^{2} + 280 \vec{v}_{1}\cdot\vec{n} v_{1}^2 ( \vec{v}_{2}\cdot\vec{n})^{2} - 30 v_{1}^2 ( \vec{v}_{2}\cdot\vec{n})^{3} \nn\\ 
&& - 40 \vec{v}_{1}\cdot\vec{n} \vec{v}_{1}\cdot\vec{v}_{2} ( \vec{v}_{2}\cdot\vec{n})^{2} - 80 \vec{v}_{1}\cdot\vec{v}_{2} ( \vec{v}_{2}\cdot\vec{n})^{3} + 40 \vec{v}_{1}\cdot\vec{n} v_{2}^2 ( \vec{v}_{2}\cdot\vec{n})^{2} + 40 v_{2}^2 ( \vec{v}_{2}\cdot\vec{n})^{3} \nn\\ 
&& + 210 ( \vec{v}_{1}\cdot\vec{n})^{3} ( \vec{v}_{2}\cdot\vec{n})^{2} - 420 ( \vec{v}_{1}\cdot\vec{n})^{2} ( \vec{v}_{2}\cdot\vec{n})^{3} \big) - 24 \vec{S}_{1}\cdot\vec{n} \vec{S}_{1}\cdot\vec{v}_{2} \big( 18 \vec{v}_{1}\cdot\vec{n} v_{1}^2 \vec{v}_{1}\cdot\vec{v}_{2} \nn\\ 
&& - 25 \vec{v}_{1}\cdot\vec{n} v_{1}^2 v_{2}^2 + 32 \vec{v}_{1}\cdot\vec{n} \vec{v}_{1}\cdot\vec{v}_{2} v_{2}^2 - 15 \vec{v}_{1}\cdot\vec{n} v_{1}^{4} - \vec{v}_{2}\cdot\vec{n} v_{1}^{4} \nn\\ 
&& - 16 \vec{v}_{1}\cdot\vec{n} v_{2}^{4} -5 v_{1}^2 \vec{v}_{2}\cdot\vec{n} ( \vec{v}_{1}\cdot\vec{n})^{2} + 70 \vec{v}_{1}\cdot\vec{n} v_{1}^2 ( \vec{v}_{2}\cdot\vec{n})^{2} - 80 \vec{v}_{1}\cdot\vec{n} \vec{v}_{1}\cdot\vec{v}_{2} ( \vec{v}_{2}\cdot\vec{n})^{2} \nn\\ 
&& + 80 \vec{v}_{1}\cdot\vec{n} v_{2}^2 ( \vec{v}_{2}\cdot\vec{n})^{2} -70 \vec{v}_{1}\cdot\vec{n} ( \vec{v}_{2}\cdot\vec{n})^{4} \big) + 4 \vec{S}_{1}\cdot\vec{v}_{1} \vec{S}_{1}\cdot\vec{v}_{2} \big( 64 v_{1}^2 \vec{v}_{1}\cdot\vec{v}_{2} - 68 v_{1}^2 v_{2}^2 \nn\\ 
&& + 264 \vec{v}_{1}\cdot\vec{v}_{2} v_{2}^2 - 156 ( \vec{v}_{1}\cdot\vec{v}_{2})^{2} + 15 v_{1}^{4} - 114 v_{2}^{4} -132 \vec{v}_{1}\cdot\vec{n} v_{1}^2 \vec{v}_{2}\cdot\vec{n} \nn\\ 
&& + 840 \vec{v}_{1}\cdot\vec{n} \vec{v}_{2}\cdot\vec{n} \vec{v}_{1}\cdot\vec{v}_{2} - 624 \vec{v}_{1}\cdot\vec{n} \vec{v}_{2}\cdot\vec{n} v_{2}^2 - 138 v_{1}^2 ( \vec{v}_{1}\cdot\vec{n})^{2} - 234 \vec{v}_{1}\cdot\vec{v}_{2} ( \vec{v}_{1}\cdot\vec{n})^{2} \nn\\ 
&& + 60 v_{2}^2 ( \vec{v}_{1}\cdot\vec{n})^{2} + 207 v_{1}^2 ( \vec{v}_{2}\cdot\vec{n})^{2} - 768 \vec{v}_{1}\cdot\vec{v}_{2} ( \vec{v}_{2}\cdot\vec{n})^{2} + 636 v_{2}^2 ( \vec{v}_{2}\cdot\vec{n})^{2} \nn\\ 
&& + 270 \vec{v}_{2}\cdot\vec{n} ( \vec{v}_{1}\cdot\vec{n})^{3} - 570 ( \vec{v}_{1}\cdot\vec{n})^{2} ( \vec{v}_{2}\cdot\vec{n})^{2} + 1080 \vec{v}_{1}\cdot\vec{n} ( \vec{v}_{2}\cdot\vec{n})^{3} - 540 ( \vec{v}_{2}\cdot\vec{n})^{4} \big) \nn\\ 
&& - 3 \big( 204 v_{1}^2 \vec{v}_{1}\cdot\vec{v}_{2} v_{2}^2 - 92 v_{1}^2 ( \vec{v}_{1}\cdot\vec{v}_{2})^{2} - 128 v_{2}^2 ( \vec{v}_{1}\cdot\vec{v}_{2})^{2} - 49 v_{1}^{6} + 124 \vec{v}_{1}\cdot\vec{v}_{2} v_{1}^{4} \nn\\ 
&& - 74 v_{2}^2 v_{1}^{4} - 114 v_{1}^2 v_{2}^{4} + 128 \vec{v}_{1}\cdot\vec{v}_{2} v_{2}^{4} + 120 \vec{v}_{1}\cdot\vec{n} v_{1}^2 \vec{v}_{2}\cdot\vec{n} \vec{v}_{1}\cdot\vec{v}_{2} - 60 \vec{v}_{1}\cdot\vec{n} v_{1}^2 \vec{v}_{2}\cdot\vec{n} v_{2}^2 \nn\\ 
&& - 10 v_{1}^2 v_{2}^2 ( \vec{v}_{1}\cdot\vec{n})^{2} - 560 v_{1}^2 \vec{v}_{1}\cdot\vec{v}_{2} ( \vec{v}_{2}\cdot\vec{n})^{2} + 320 ( \vec{v}_{2}\cdot\vec{n})^{2} ( \vec{v}_{1}\cdot\vec{v}_{2})^{2} \nn\\ 
&& + 640 v_{1}^2 v_{2}^2 ( \vec{v}_{2}\cdot\vec{n})^{2} - 640 \vec{v}_{1}\cdot\vec{v}_{2} v_{2}^2 ( \vec{v}_{2}\cdot\vec{n})^{2} - 60 \vec{v}_{1}\cdot\vec{n} \vec{v}_{2}\cdot\vec{n} v_{1}^{4} + 210 ( \vec{v}_{2}\cdot\vec{n})^{2} v_{1}^{4} \nn\\ 
&& + 70 v_{1}^2 ( \vec{v}_{1}\cdot\vec{n})^{2} ( \vec{v}_{2}\cdot\vec{n})^{2} - 560 v_{1}^2 ( \vec{v}_{2}\cdot\vec{n})^{4} + 560 \vec{v}_{1}\cdot\vec{v}_{2} ( \vec{v}_{2}\cdot\vec{n})^{4} \big) ( \vec{S}_{1}\cdot\vec{n})^{2} \nn\\ 
&& - 2 \big( 274 v_{1}^2 \vec{v}_{1}\cdot\vec{v}_{2} - 176 v_{1}^2 v_{2}^2 + 532 \vec{v}_{1}\cdot\vec{v}_{2} v_{2}^2 - 404 ( \vec{v}_{1}\cdot\vec{v}_{2})^{2} - 37 v_{1}^{4} \nn\\ 
&& - 190 v_{2}^{4} -558 \vec{v}_{1}\cdot\vec{n} v_{1}^2 \vec{v}_{2}\cdot\vec{n} + 1848 \vec{v}_{1}\cdot\vec{n} \vec{v}_{2}\cdot\vec{n} \vec{v}_{1}\cdot\vec{v}_{2} - 1116 \vec{v}_{1}\cdot\vec{n} \vec{v}_{2}\cdot\vec{n} v_{2}^2 \nn\\ 
&& - 18 v_{1}^2 ( \vec{v}_{1}\cdot\vec{n})^{2} - 480 \vec{v}_{1}\cdot\vec{v}_{2} ( \vec{v}_{1}\cdot\vec{n})^{2} + 198 v_{2}^2 ( \vec{v}_{1}\cdot\vec{n})^{2} + 558 v_{1}^2 ( \vec{v}_{2}\cdot\vec{n})^{2} \nn\\ 
&& - 1584 \vec{v}_{1}\cdot\vec{v}_{2} ( \vec{v}_{2}\cdot\vec{n})^{2} + 1128 v_{2}^2 ( \vec{v}_{2}\cdot\vec{n})^{2} + 480 \vec{v}_{2}\cdot\vec{n} ( \vec{v}_{1}\cdot\vec{n})^{3} \nn\\ 
&& - 1350 ( \vec{v}_{1}\cdot\vec{n})^{2} ( \vec{v}_{2}\cdot\vec{n})^{2} + 1920 \vec{v}_{1}\cdot\vec{n} ( \vec{v}_{2}\cdot\vec{n})^{3} - 960 ( \vec{v}_{2}\cdot\vec{n})^{4} \big) ( \vec{S}_{1}\cdot\vec{v}_{1})^{2} \nn\\ 
&& - 4 \big( 5 v_{1}^{4} -69 v_{1}^2 ( \vec{v}_{1}\cdot\vec{n})^{2} - 96 v_{2}^2 ( \vec{v}_{1}\cdot\vec{n})^{2} + 240 ( \vec{v}_{1}\cdot\vec{n})^{2} ( \vec{v}_{2}\cdot\vec{n})^{2} \big) ( \vec{S}_{1}\cdot\vec{v}_{2})^{2} \Big]\nn\\ && - 	\frac{G^2 m_{2}}{96 r{}^4} \Big[ S_{1}^2 \big( 6 v_{1}^2 \vec{v}_{1}\cdot\vec{v}_{2} - 230 v_{1}^2 v_{2}^2 - 1632 \vec{v}_{1}\cdot\vec{v}_{2} v_{2}^2 + 1775 ( \vec{v}_{1}\cdot\vec{v}_{2})^{2} - 84 v_{1}^{4} \nn\\ 
&& + 156 v_{2}^{4} -7524 \vec{v}_{1}\cdot\vec{n} v_{1}^2 \vec{v}_{2}\cdot\vec{n} + 1772 \vec{v}_{1}\cdot\vec{n} \vec{v}_{2}\cdot\vec{n} \vec{v}_{1}\cdot\vec{v}_{2} - 432 \vec{v}_{1}\cdot\vec{n} \vec{v}_{2}\cdot\vec{n} v_{2}^2 \nn\\ 
&& + 4728 v_{1}^2 ( \vec{v}_{1}\cdot\vec{n})^{2} - 5376 \vec{v}_{1}\cdot\vec{v}_{2} ( \vec{v}_{1}\cdot\vec{n})^{2} + 2678 v_{2}^2 ( \vec{v}_{1}\cdot\vec{n})^{2} + 620 v_{1}^2 ( \vec{v}_{2}\cdot\vec{n})^{2} \nn\\ 
&& + 4752 \vec{v}_{1}\cdot\vec{v}_{2} ( \vec{v}_{2}\cdot\vec{n})^{2} - 36 v_{2}^2 ( \vec{v}_{2}\cdot\vec{n})^{2} -4845 ( \vec{v}_{1}\cdot\vec{n})^{4} + 15000 \vec{v}_{2}\cdot\vec{n} ( \vec{v}_{1}\cdot\vec{n})^{3} \nn\\ 
&& - 9732 ( \vec{v}_{1}\cdot\vec{n})^{2} ( \vec{v}_{2}\cdot\vec{n})^{2} - 1800 \vec{v}_{1}\cdot\vec{n} ( \vec{v}_{2}\cdot\vec{n})^{3} \big) + 2 \vec{S}_{1}\cdot\vec{n} \vec{S}_{1}\cdot\vec{v}_{1} \big( 258 \vec{v}_{1}\cdot\vec{n} v_{1}^2 \nn\\ 
&& - 888 v_{1}^2 \vec{v}_{2}\cdot\vec{n} + 636 \vec{v}_{1}\cdot\vec{n} \vec{v}_{1}\cdot\vec{v}_{2} + 2306 \vec{v}_{2}\cdot\vec{n} \vec{v}_{1}\cdot\vec{v}_{2} - 1340 \vec{v}_{1}\cdot\vec{n} v_{2}^2 + 84 \vec{v}_{2}\cdot\vec{n} v_{2}^2 \nn\\ 
&& + 12345 ( \vec{v}_{1}\cdot\vec{n})^{3} - 24318 \vec{v}_{2}\cdot\vec{n} ( \vec{v}_{1}\cdot\vec{n})^{2} + 8778 \vec{v}_{1}\cdot\vec{n} ( \vec{v}_{2}\cdot\vec{n})^{2} + 900 ( \vec{v}_{2}\cdot\vec{n})^{3} \big) \nn\\ 
&& - 4 \vec{S}_{1}\cdot\vec{n} \vec{S}_{1}\cdot\vec{v}_{2} \big( 120 \vec{v}_{1}\cdot\vec{n} v_{1}^2 - 419 v_{1}^2 \vec{v}_{2}\cdot\vec{n} - 952 \vec{v}_{1}\cdot\vec{n} \vec{v}_{1}\cdot\vec{v}_{2} + 1692 \vec{v}_{2}\cdot\vec{n} \vec{v}_{1}\cdot\vec{v}_{2} \nn\\ 
&& + 114 \vec{v}_{1}\cdot\vec{n} v_{2}^2 - 36 \vec{v}_{2}\cdot\vec{n} v_{2}^2 + 5190 ( \vec{v}_{1}\cdot\vec{n})^{3} - 8016 \vec{v}_{2}\cdot\vec{n} ( \vec{v}_{1}\cdot\vec{n})^{2} \nn\\ 
&& + 1026 \vec{v}_{1}\cdot\vec{n} ( \vec{v}_{2}\cdot\vec{n})^{2} \big) - 2 \vec{S}_{1}\cdot\vec{v}_{1} \vec{S}_{1}\cdot\vec{v}_{2} \big( 162 v_{1}^2 + 1127 \vec{v}_{1}\cdot\vec{v}_{2} - 588 v_{2}^2 \nn\\ 
&& + 4918 \vec{v}_{1}\cdot\vec{n} \vec{v}_{2}\cdot\vec{n} - 6858 ( \vec{v}_{1}\cdot\vec{n})^{2} + 1800 ( \vec{v}_{2}\cdot\vec{n})^{2} \big) - \big( 4812 v_{1}^2 \vec{v}_{1}\cdot\vec{v}_{2} - 1010 v_{1}^2 v_{2}^2 \nn\\ 
&& + 576 \vec{v}_{1}\cdot\vec{v}_{2} v_{2}^2 - 2848 ( \vec{v}_{1}\cdot\vec{v}_{2})^{2} - 1668 v_{1}^{4} + 144 v_{2}^{4} -22788 \vec{v}_{1}\cdot\vec{n} v_{1}^2 \vec{v}_{2}\cdot\vec{n} \nn\\ 
&& + 28848 \vec{v}_{1}\cdot\vec{n} \vec{v}_{2}\cdot\vec{n} \vec{v}_{1}\cdot\vec{v}_{2} - 1944 \vec{v}_{1}\cdot\vec{n} \vec{v}_{2}\cdot\vec{n} v_{2}^2 + 14301 v_{1}^2 ( \vec{v}_{1}\cdot\vec{n})^{2} \nn\\ 
&& - 15768 \vec{v}_{1}\cdot\vec{v}_{2} ( \vec{v}_{1}\cdot\vec{n})^{2} + 1596 v_{2}^2 ( \vec{v}_{1}\cdot\vec{n})^{2} + 5376 v_{1}^2 ( \vec{v}_{2}\cdot\vec{n})^{2} - 7848 \vec{v}_{1}\cdot\vec{v}_{2} ( \vec{v}_{2}\cdot\vec{n})^{2} \nn\\ 
&& - 108 v_{2}^2 ( \vec{v}_{2}\cdot\vec{n})^{2} + 3360 ( \vec{v}_{1}\cdot\vec{n})^{4} - 11328 \vec{v}_{2}\cdot\vec{n} ( \vec{v}_{1}\cdot\vec{n})^{3} + 6240 ( \vec{v}_{1}\cdot\vec{n})^{2} ( \vec{v}_{2}\cdot\vec{n})^{2} \nn\\ 
&& + 1152 \vec{v}_{1}\cdot\vec{n} ( \vec{v}_{2}\cdot\vec{n})^{3} \big) ( \vec{S}_{1}\cdot\vec{n})^{2} + 2 \big( 126 v_{1}^2 + 15 \vec{v}_{1}\cdot\vec{v}_{2} + 73 v_{2}^2 + 5466 \vec{v}_{1}\cdot\vec{n} \vec{v}_{2}\cdot\vec{n} \nn\\ 
&& - 4728 ( \vec{v}_{1}\cdot\vec{n})^{2} - 166 ( \vec{v}_{2}\cdot\vec{n})^{2} \big) ( \vec{S}_{1}\cdot\vec{v}_{1})^{2} + \big( 599 v_{1}^2 + 528 \vec{v}_{1}\cdot\vec{v}_{2} - 144 v_{2}^2 \nn\\ 
&& + 4128 \vec{v}_{1}\cdot\vec{n} \vec{v}_{2}\cdot\vec{n} - 6728 ( \vec{v}_{1}\cdot\vec{n})^{2} \big) ( \vec{S}_{1}\cdot\vec{v}_{2})^{2} \Big] - 	\frac{G^2 m_{2}{}^2}{128 m_{1} r{}^4} \Big[ S_{1}^2 \big( 1784 v_{1}^2 \vec{v}_{1}\cdot\vec{v}_{2} \nn\\ 
&& - 1396 v_{1}^2 v_{2}^2 + 3180 \vec{v}_{1}\cdot\vec{v}_{2} v_{2}^2 - 1524 ( \vec{v}_{1}\cdot\vec{v}_{2})^{2} - 409 v_{1}^{4} \nn\\ 
&& - 1632 v_{2}^{4} -1368 \vec{v}_{1}\cdot\vec{n} v_{1}^2 \vec{v}_{2}\cdot\vec{n} + 1944 \vec{v}_{1}\cdot\vec{n} \vec{v}_{2}\cdot\vec{n} \vec{v}_{1}\cdot\vec{v}_{2} - 2568 \vec{v}_{1}\cdot\vec{n} \vec{v}_{2}\cdot\vec{n} v_{2}^2 \nn\\ 
&& + 720 v_{1}^2 ( \vec{v}_{1}\cdot\vec{n})^{2} - 3104 \vec{v}_{1}\cdot\vec{v}_{2} ( \vec{v}_{1}\cdot\vec{n})^{2} + 3436 v_{2}^2 ( \vec{v}_{1}\cdot\vec{n})^{2} + 1412 v_{1}^2 ( \vec{v}_{2}\cdot\vec{n})^{2} \nn\\ 
&& - 2340 \vec{v}_{1}\cdot\vec{v}_{2} ( \vec{v}_{2}\cdot\vec{n})^{2} + 3264 v_{2}^2 ( \vec{v}_{2}\cdot\vec{n})^{2} -192 ( \vec{v}_{1}\cdot\vec{n})^{4} + 2304 \vec{v}_{2}\cdot\vec{n} ( \vec{v}_{1}\cdot\vec{n})^{3} \nn\\ 
&& - 3492 ( \vec{v}_{1}\cdot\vec{n})^{2} ( \vec{v}_{2}\cdot\vec{n})^{2} - 24 \vec{v}_{1}\cdot\vec{n} ( \vec{v}_{2}\cdot\vec{n})^{3} - 432 ( \vec{v}_{2}\cdot\vec{n})^{4} \big) \nn\\ 
&& - 8 \vec{S}_{1}\cdot\vec{n} \vec{S}_{1}\cdot\vec{v}_{1} \big( 171 \vec{v}_{1}\cdot\vec{n} v_{1}^2 - 399 v_{1}^2 \vec{v}_{2}\cdot\vec{n} - 295 \vec{v}_{1}\cdot\vec{n} \vec{v}_{1}\cdot\vec{v}_{2} + 735 \vec{v}_{2}\cdot\vec{n} \vec{v}_{1}\cdot\vec{v}_{2} \nn\\ 
&& + 109 \vec{v}_{1}\cdot\vec{n} v_{2}^2 - 517 \vec{v}_{2}\cdot\vec{n} v_{2}^2 -48 ( \vec{v}_{1}\cdot\vec{n})^{3} + 156 \vec{v}_{2}\cdot\vec{n} ( \vec{v}_{1}\cdot\vec{n})^{2} + 186 \vec{v}_{1}\cdot\vec{n} ( \vec{v}_{2}\cdot\vec{n})^{2} \nn\\ 
&& + 573 ( \vec{v}_{2}\cdot\vec{n})^{3} \big) + 8 \vec{S}_{1}\cdot\vec{n} \vec{S}_{1}\cdot\vec{v}_{2} \big( 189 \vec{v}_{1}\cdot\vec{n} v_{1}^2 - 405 v_{1}^2 \vec{v}_{2}\cdot\vec{n} - 75 \vec{v}_{1}\cdot\vec{n} \vec{v}_{1}\cdot\vec{v}_{2} \nn\\ 
&& + 521 \vec{v}_{2}\cdot\vec{n} \vec{v}_{1}\cdot\vec{v}_{2} - 33 \vec{v}_{1}\cdot\vec{n} v_{2}^2 - 396 \vec{v}_{2}\cdot\vec{n} v_{2}^2 -240 \vec{v}_{2}\cdot\vec{n} ( \vec{v}_{1}\cdot\vec{n})^{2} \nn\\ 
&& + 918 \vec{v}_{1}\cdot\vec{n} ( \vec{v}_{2}\cdot\vec{n})^{2} + 648 ( \vec{v}_{2}\cdot\vec{n})^{3} \big) - 4 \vec{S}_{1}\cdot\vec{v}_{1} \vec{S}_{1}\cdot\vec{v}_{2} \big( 436 v_{1}^2 - 318 \vec{v}_{1}\cdot\vec{v}_{2} \nn\\ 
&& + 299 v_{2}^2 -282 \vec{v}_{1}\cdot\vec{n} \vec{v}_{2}\cdot\vec{n} - 386 ( \vec{v}_{1}\cdot\vec{n})^{2} - 717 ( \vec{v}_{2}\cdot\vec{n})^{2} \big) - \big( 2920 v_{1}^2 \vec{v}_{1}\cdot\vec{v}_{2} \nn\\ 
&& - 1492 v_{1}^2 v_{2}^2 + 5752 \vec{v}_{1}\cdot\vec{v}_{2} v_{2}^2 - 3224 ( \vec{v}_{1}\cdot\vec{v}_{2})^{2} - 801 v_{1}^{4} - 3152 v_{2}^{4} \nn\\ 
&& + 1296 \vec{v}_{1}\cdot\vec{n} v_{1}^2 \vec{v}_{2}\cdot\vec{n} - 5904 \vec{v}_{1}\cdot\vec{n} \vec{v}_{2}\cdot\vec{n} \vec{v}_{1}\cdot\vec{v}_{2} + 3600 \vec{v}_{1}\cdot\vec{n} \vec{v}_{2}\cdot\vec{n} v_{2}^2 + 192 v_{1}^2 ( \vec{v}_{1}\cdot\vec{n})^{2} \nn\\ 
&& + 960 v_{2}^2 ( \vec{v}_{1}\cdot\vec{n})^{2} - 1356 v_{1}^2 ( \vec{v}_{2}\cdot\vec{n})^{2} + 576 \vec{v}_{1}\cdot\vec{v}_{2} ( \vec{v}_{2}\cdot\vec{n})^{2} \nn\\ 
&& + 2352 v_{2}^2 ( \vec{v}_{2}\cdot\vec{n})^{2} -1008 ( \vec{v}_{1}\cdot\vec{n})^{2} ( \vec{v}_{2}\cdot\vec{n})^{2} - 288 \vec{v}_{1}\cdot\vec{n} ( \vec{v}_{2}\cdot\vec{n})^{3} \nn\\ 
&& + 1392 ( \vec{v}_{2}\cdot\vec{n})^{4} \big) ( \vec{S}_{1}\cdot\vec{n})^{2} + 4 \big( 102 v_{1}^2 - 6 \vec{v}_{1}\cdot\vec{v}_{2} + 71 v_{2}^2 -248 \vec{v}_{1}\cdot\vec{n} \vec{v}_{2}\cdot\vec{n} \nn\\ 
&& - 38 ( \vec{v}_{1}\cdot\vec{n})^{2} - 27 ( \vec{v}_{2}\cdot\vec{n})^{2} \big) ( \vec{S}_{1}\cdot\vec{v}_{1})^{2} + 4 \big( 341 v_{1}^2 - 512 \vec{v}_{1}\cdot\vec{v}_{2} + 420 v_{2}^2 \nn\\ 
&& + 242 \vec{v}_{1}\cdot\vec{n} \vec{v}_{2}\cdot\vec{n} - 586 ( \vec{v}_{1}\cdot\vec{n})^{2} - 1076 ( \vec{v}_{2}\cdot\vec{n})^{2} \big) ( \vec{S}_{1}\cdot\vec{v}_{2})^{2} \Big]\nn\\ && - 	\frac{G^3 m_{2}{}^3}{16 m_{1} r{}^5} \Big[ S_{1}^2 \big( 16 v_{1}^2 - 127 \vec{v}_{1}\cdot\vec{v}_{2} + 320 v_{2}^2 -257 \vec{v}_{1}\cdot\vec{n} \vec{v}_{2}\cdot\vec{n} - ( \vec{v}_{1}\cdot\vec{n})^{2} \nn\\ 
&& - 146 ( \vec{v}_{2}\cdot\vec{n})^{2} \big) - 2 \vec{S}_{1}\cdot\vec{n} \vec{S}_{1}\cdot\vec{v}_{1} \big( 39 \vec{v}_{1}\cdot\vec{n} + 322 \vec{v}_{2}\cdot\vec{n} \big) + 2 \vec{S}_{1}\cdot\vec{n} \vec{S}_{1}\cdot\vec{v}_{2} \big( 121 \vec{v}_{1}\cdot\vec{n} \nn\\ 
&& + 621 \vec{v}_{2}\cdot\vec{n} \big) + 144 \vec{S}_{1}\cdot\vec{v}_{1} \vec{S}_{1}\cdot\vec{v}_{2} - \big( 5 v_{1}^2 - 357 \vec{v}_{1}\cdot\vec{v}_{2} + 686 v_{2}^2 -285 \vec{v}_{1}\cdot\vec{n} \vec{v}_{2}\cdot\vec{n} \nn\\ 
&& - 80 ( \vec{v}_{1}\cdot\vec{n})^{2} + 395 ( \vec{v}_{2}\cdot\vec{n})^{2} \big) ( \vec{S}_{1}\cdot\vec{n})^{2} - 12 ( \vec{S}_{1}\cdot\vec{v}_{1})^{2} - 335 ( \vec{S}_{1}\cdot\vec{v}_{2})^{2} \Big] \nn\\ 
&& - 	\frac{G^3 m_{1} m_{2}}{117600 r{}^5} \Big[ S_{1}^2 \big( 1389116 v_{1}^2 + 2980008 \vec{v}_{1}\cdot\vec{v}_{2} - 489024 v_{2}^2 -3207240 \vec{v}_{1}\cdot\vec{n} \vec{v}_{2}\cdot\vec{n} \nn\\ 
&& - 4741105 ( \vec{v}_{1}\cdot\vec{n})^{2} + 3087720 ( \vec{v}_{2}\cdot\vec{n})^{2} \big) - 10 \vec{S}_{1}\cdot\vec{n} \vec{S}_{1}\cdot\vec{v}_{1} \big( 364471 \vec{v}_{1}\cdot\vec{n} \nn\\ 
&& - 346476 \vec{v}_{2}\cdot\vec{n} \big) + 480 \vec{S}_{1}\cdot\vec{n} \vec{S}_{1}\cdot\vec{v}_{2} \big( 11287 \vec{v}_{1}\cdot\vec{n} + 7888 \vec{v}_{2}\cdot\vec{n} \big) \nn\\ 
&& - 3287544 \vec{S}_{1}\cdot\vec{v}_{1} \vec{S}_{1}\cdot\vec{v}_{2} - 25 \big( 61561 v_{1}^2 + 143376 \vec{v}_{1}\cdot\vec{v}_{2} - 49152 v_{2}^2 \nn\\ 
&& + 17136 \vec{v}_{1}\cdot\vec{n} \vec{v}_{2}\cdot\vec{n} - 406672 ( \vec{v}_{1}\cdot\vec{n})^{2} + 326424 ( \vec{v}_{2}\cdot\vec{n})^{2} \big) ( \vec{S}_{1}\cdot\vec{n})^{2} \nn\\ 
&& - 48788 ( \vec{S}_{1}\cdot\vec{v}_{1})^{2} - 245568 ( \vec{S}_{1}\cdot\vec{v}_{2})^{2} \Big] - 	\frac{G^3 m_{2}{}^2}{6144 r{}^5} \Big[ S_{1}^2 \big( {(39936 - 189 \pi^2)} v_{1}^2 \nn\\ 
&& + {(242432 - 3942 \pi^2)} \vec{v}_{1}\cdot\vec{v}_{2} + {(22272 + 4131 \pi^2)} v_{2}^2 \nn\\ 
&& - {(497536 - 19710 \pi^2)} \vec{v}_{1}\cdot\vec{n} \vec{v}_{2}\cdot\vec{n} - {(206016 - 945 \pi^2)} ( \vec{v}_{1}\cdot\vec{n})^{2} \nn\\ 
&& + {(441600 - 20655 \pi^2)} ( \vec{v}_{2}\cdot\vec{n})^{2} \big) - 4 \vec{S}_{1}\cdot\vec{n} \vec{S}_{1}\cdot\vec{v}_{1} \big( {(86016 - 945 \pi^2)} \vec{v}_{1}\cdot\vec{n} \nn\\ 
&& + {(36992 - 9855 \pi^2)} \vec{v}_{2}\cdot\vec{n} \big) + 4 \vec{S}_{1}\cdot\vec{n} \vec{S}_{1}\cdot\vec{v}_{2} \big( {(74176 + 9855 \pi^2)} \vec{v}_{1}\cdot\vec{n} \nn\\ 
&& + {(225792 - 20655 \pi^2)} \vec{v}_{2}\cdot\vec{n} \big) - 4 {(8672 + 1971 \pi^2)} \vec{S}_{1}\cdot\vec{v}_{1} \vec{S}_{1}\cdot\vec{v}_{2} \nn\\ 
&& - \big( {(65856 - 945 \pi^2)} v_{1}^2 + {(276736 - 19710 \pi^2)} \vec{v}_{1}\cdot\vec{v}_{2} + {(28416 + 20655 \pi^2)} v_{2}^2 \nn\\ 
&& - {(765952 - 137970 \pi^2)} \vec{v}_{1}\cdot\vec{n} \vec{v}_{2}\cdot\vec{n} - {(677760 - 6615 \pi^2)} ( \vec{v}_{1}\cdot\vec{n})^{2} \nn\\ 
&& + {(1389312 - 144585 \pi^2)} ( \vec{v}_{2}\cdot\vec{n})^{2} \big) ( \vec{S}_{1}\cdot\vec{n})^{2} - 54 {(64 + 7 \pi^2)} ( \vec{S}_{1}\cdot\vec{v}_{1})^{2} \nn\\ 
&& - 6 {(42752 - 1377 \pi^2)} ( \vec{S}_{1}\cdot\vec{v}_{2})^{2} \Big]\nn\\ && - 	\frac{G^4 m_{1}{}^2 m_{2}}{2450 r{}^6} \Big[ 5813 S_{1}^2 - 3684 ( \vec{S}_{1}\cdot\vec{n})^{2} \Big] - 	\frac{29 G^4 m_{2}{}^4}{16 m_{1} r{}^6} \Big[ S_{1}^2 - ( \vec{S}_{1}\cdot\vec{n})^{2} \Big] \nn\\ 
&& - 	\frac{3 G^4 m_{2}{}^3}{4 r{}^6} \Big[ 29 S_{1}^2 + 11 ( \vec{S}_{1}\cdot\vec{n})^{2} \Big] - 	\frac{G^4 m_{1} m_{2}{}^2}{78400 r{}^6} \Big[ {(2549096 - 55125 \pi^2)} S_{1}^2 \nn\\ 
&& - {(958088 - 165375 \pi^2)} ( \vec{S}_{1}\cdot\vec{n})^{2} \Big] ,
\eea
%%%%%%%%%%%%%

%%%%%%%%%%%%%
\bea
\hat{V}^{\text{N}^3\text{LO}}_{C_{1\text{ES}^2}} &= &  	\frac{G m_{2}}{32 m_{1} r{}^3} \Big[ S_{1}^2 \big( 5 v_{1}^2 \vec{v}_{1}\cdot\vec{v}_{2} v_{2}^2 + 26 v_{1}^2 ( \vec{v}_{1}\cdot\vec{v}_{2})^{2} - 22 ( \vec{v}_{1}\cdot\vec{v}_{2})^{3} - 14 v_{2}^2 ( \vec{v}_{1}\cdot\vec{v}_{2})^{2} - 25 v_{1}^{6} \nn\\ 
&& + 27 \vec{v}_{1}\cdot\vec{v}_{2} v_{1}^{4} - 6 v_{2}^2 v_{1}^{4} - 10 v_{1}^2 v_{2}^{4} + 31 \vec{v}_{1}\cdot\vec{v}_{2} v_{2}^{4} - 11 v_{2}^{6} -204 \vec{v}_{1}\cdot\vec{n} v_{1}^2 \vec{v}_{2}\cdot\vec{n} \vec{v}_{1}\cdot\vec{v}_{2} \nn\\ 
&& + 57 \vec{v}_{1}\cdot\vec{n} v_{1}^2 \vec{v}_{2}\cdot\vec{n} v_{2}^2 - 60 \vec{v}_{1}\cdot\vec{n} \vec{v}_{2}\cdot\vec{n} \vec{v}_{1}\cdot\vec{v}_{2} v_{2}^2 + 12 v_{1}^2 \vec{v}_{1}\cdot\vec{v}_{2} ( \vec{v}_{1}\cdot\vec{n})^{2} \nn\\ 
&& + 21 v_{1}^2 v_{2}^2 ( \vec{v}_{1}\cdot\vec{n})^{2} + 39 \vec{v}_{1}\cdot\vec{v}_{2} v_{2}^2 ( \vec{v}_{1}\cdot\vec{n})^{2} + 27 v_{1}^2 \vec{v}_{1}\cdot\vec{v}_{2} ( \vec{v}_{2}\cdot\vec{n})^{2} \nn\\ 
&& - 120 ( \vec{v}_{1}\cdot\vec{n})^{2} ( \vec{v}_{1}\cdot\vec{v}_{2})^{2} + 150 \vec{v}_{1}\cdot\vec{n} \vec{v}_{2}\cdot\vec{n} ( \vec{v}_{1}\cdot\vec{v}_{2})^{2} - 27 v_{1}^2 v_{2}^2 ( \vec{v}_{2}\cdot\vec{n})^{2} \nn\\ 
&& + 42 ( \vec{v}_{1}\cdot\vec{n})^{2} v_{1}^{4} + 87 \vec{v}_{1}\cdot\vec{n} \vec{v}_{2}\cdot\vec{n} v_{1}^{4} - 9 ( \vec{v}_{2}\cdot\vec{n})^{2} v_{1}^{4} + 15 ( \vec{v}_{1}\cdot\vec{n})^{2} v_{2}^{4} \nn\\ 
&& - 21 \vec{v}_{1}\cdot\vec{n} \vec{v}_{2}\cdot\vec{n} v_{2}^{4} -180 v_{1}^2 \vec{v}_{2}\cdot\vec{n} ( \vec{v}_{1}\cdot\vec{n})^{3} + 240 \vec{v}_{2}\cdot\vec{n} \vec{v}_{1}\cdot\vec{v}_{2} ( \vec{v}_{1}\cdot\vec{n})^{3} - 30 v_{2}^2 ( \vec{v}_{1}\cdot\vec{n})^{4} \nn\\ 
&& - 105 \vec{v}_{2}\cdot\vec{n} v_{2}^2 ( \vec{v}_{1}\cdot\vec{n})^{3} - 45 v_{1}^2 ( \vec{v}_{1}\cdot\vec{n})^{2} ( \vec{v}_{2}\cdot\vec{n})^{2} + 75 \vec{v}_{1}\cdot\vec{n} v_{1}^2 ( \vec{v}_{2}\cdot\vec{n})^{3} \nn\\ 
&& - 135 \vec{v}_{1}\cdot\vec{v}_{2} ( \vec{v}_{1}\cdot\vec{n})^{2} ( \vec{v}_{2}\cdot\vec{n})^{2} + 135 v_{2}^2 ( \vec{v}_{1}\cdot\vec{n})^{2} ( \vec{v}_{2}\cdot\vec{n})^{2} + 210 ( \vec{v}_{1}\cdot\vec{n})^{4} ( \vec{v}_{2}\cdot\vec{n})^{2} \nn\\ 
&& - 175 ( \vec{v}_{1}\cdot\vec{n})^{3} ( \vec{v}_{2}\cdot\vec{n})^{3} \big) + 6 \vec{S}_{1}\cdot\vec{n} \vec{S}_{1}\cdot\vec{v}_{1} \big( 10 \vec{v}_{1}\cdot\vec{n} v_{1}^2 \vec{v}_{1}\cdot\vec{v}_{2} + 6 v_{1}^2 \vec{v}_{2}\cdot\vec{n} \vec{v}_{1}\cdot\vec{v}_{2} \nn\\ 
&& - 2 \vec{v}_{1}\cdot\vec{n} v_{1}^2 v_{2}^2 - v_{1}^2 \vec{v}_{2}\cdot\vec{n} v_{2}^2 - 6 \vec{v}_{1}\cdot\vec{n} \vec{v}_{1}\cdot\vec{v}_{2} v_{2}^2 - 4 \vec{v}_{2}\cdot\vec{n} \vec{v}_{1}\cdot\vec{v}_{2} v_{2}^2 + 10 \vec{v}_{1}\cdot\vec{n} ( \vec{v}_{1}\cdot\vec{v}_{2})^{2} \nn\\ 
&& - 10 \vec{v}_{2}\cdot\vec{n} ( \vec{v}_{1}\cdot\vec{v}_{2})^{2} - 11 \vec{v}_{1}\cdot\vec{n} v_{1}^{4} - \vec{v}_{1}\cdot\vec{n} v_{2}^{4} + 7 \vec{v}_{2}\cdot\vec{n} v_{2}^{4} + 35 v_{1}^2 \vec{v}_{2}\cdot\vec{n} ( \vec{v}_{1}\cdot\vec{n})^{2} \nn\\ 
&& - 50 \vec{v}_{2}\cdot\vec{n} \vec{v}_{1}\cdot\vec{v}_{2} ( \vec{v}_{1}\cdot\vec{n})^{2} + 5 v_{2}^2 ( \vec{v}_{1}\cdot\vec{n})^{3} + 15 \vec{v}_{2}\cdot\vec{n} v_{2}^2 ( \vec{v}_{1}\cdot\vec{n})^{2} - 15 \vec{v}_{1}\cdot\vec{n} v_{1}^2 ( \vec{v}_{2}\cdot\vec{n})^{2} \nn\\ 
&& - 5 v_{1}^2 ( \vec{v}_{2}\cdot\vec{n})^{3} + 50 \vec{v}_{1}\cdot\vec{n} \vec{v}_{1}\cdot\vec{v}_{2} ( \vec{v}_{2}\cdot\vec{n})^{2} - 30 \vec{v}_{1}\cdot\vec{n} v_{2}^2 ( \vec{v}_{2}\cdot\vec{n})^{2} -35 ( \vec{v}_{1}\cdot\vec{n})^{3} ( \vec{v}_{2}\cdot\vec{n})^{2} \nn\\ 
&& + 35 ( \vec{v}_{1}\cdot\vec{n})^{2} ( \vec{v}_{2}\cdot\vec{n})^{3} \big) - 6 \vec{S}_{1}\cdot\vec{n} \vec{S}_{1}\cdot\vec{v}_{2} \big( 4 \vec{v}_{1}\cdot\vec{n} v_{1}^2 \vec{v}_{1}\cdot\vec{v}_{2} + 10 v_{1}^2 \vec{v}_{2}\cdot\vec{n} \vec{v}_{1}\cdot\vec{v}_{2} \nn\\ 
&& - 5 \vec{v}_{1}\cdot\vec{n} v_{1}^2 v_{2}^2 - 6 v_{1}^2 \vec{v}_{2}\cdot\vec{n} v_{2}^2 + 4 \vec{v}_{1}\cdot\vec{n} \vec{v}_{1}\cdot\vec{v}_{2} v_{2}^2 + 10 \vec{v}_{1}\cdot\vec{n} ( \vec{v}_{1}\cdot\vec{v}_{2})^{2} - 7 \vec{v}_{1}\cdot\vec{n} v_{1}^{4} \nn\\ 
&& - 6 \vec{v}_{2}\cdot\vec{n} v_{1}^{4} - 7 \vec{v}_{1}\cdot\vec{n} v_{2}^{4} + 30 v_{1}^2 \vec{v}_{2}\cdot\vec{n} ( \vec{v}_{1}\cdot\vec{n})^{2} - 50 \vec{v}_{2}\cdot\vec{n} \vec{v}_{1}\cdot\vec{v}_{2} ( \vec{v}_{1}\cdot\vec{n})^{2} \nn\\ 
&& + 5 v_{2}^2 ( \vec{v}_{1}\cdot\vec{n})^{3} + 30 \vec{v}_{2}\cdot\vec{n} v_{2}^2 ( \vec{v}_{1}\cdot\vec{n})^{2} + 15 \vec{v}_{1}\cdot\vec{n} v_{1}^2 ( \vec{v}_{2}\cdot\vec{n})^{2} -35 ( \vec{v}_{1}\cdot\vec{n})^{3} ( \vec{v}_{2}\cdot\vec{n})^{2} \big) \nn\\ 
&& - 2 \vec{S}_{1}\cdot\vec{v}_{1} \vec{S}_{1}\cdot\vec{v}_{2} \big( 6 v_{1}^2 \vec{v}_{1}\cdot\vec{v}_{2} - v_{1}^2 v_{2}^2 - 4 \vec{v}_{1}\cdot\vec{v}_{2} v_{2}^2 - 10 ( \vec{v}_{1}\cdot\vec{v}_{2})^{2} \nn\\ 
&& + 7 v_{2}^{4} -18 \vec{v}_{1}\cdot\vec{n} v_{1}^2 \vec{v}_{2}\cdot\vec{n} + 60 \vec{v}_{1}\cdot\vec{n} \vec{v}_{2}\cdot\vec{n} \vec{v}_{1}\cdot\vec{v}_{2} - 36 \vec{v}_{1}\cdot\vec{n} \vec{v}_{2}\cdot\vec{n} v_{2}^2 + 21 v_{1}^2 ( \vec{v}_{1}\cdot\vec{n})^{2} \nn\\ 
&& - 30 \vec{v}_{1}\cdot\vec{v}_{2} ( \vec{v}_{1}\cdot\vec{n})^{2} + 9 v_{2}^2 ( \vec{v}_{1}\cdot\vec{n})^{2} - 9 v_{1}^2 ( \vec{v}_{2}\cdot\vec{n})^{2} -30 \vec{v}_{2}\cdot\vec{n} ( \vec{v}_{1}\cdot\vec{n})^{3} \nn\\ 
&& + 45 ( \vec{v}_{1}\cdot\vec{n})^{2} ( \vec{v}_{2}\cdot\vec{n})^{2} \big) - 3 \big( 19 v_{1}^2 \vec{v}_{1}\cdot\vec{v}_{2} v_{2}^2 - 6 v_{1}^2 ( \vec{v}_{1}\cdot\vec{v}_{2})^{2} - 2 ( \vec{v}_{1}\cdot\vec{v}_{2})^{3} \nn\\ 
&& - 6 v_{2}^2 ( \vec{v}_{1}\cdot\vec{v}_{2})^{2} - 11 v_{1}^{6} + 17 \vec{v}_{1}\cdot\vec{v}_{2} v_{1}^{4} - 8 v_{2}^2 v_{1}^{4} - 8 v_{1}^2 v_{2}^{4} + 17 \vec{v}_{1}\cdot\vec{v}_{2} v_{2}^{4} \nn\\ 
&& - 11 v_{2}^{6} -20 \vec{v}_{1}\cdot\vec{n} v_{1}^2 \vec{v}_{2}\cdot\vec{n} \vec{v}_{1}\cdot\vec{v}_{2} + 25 \vec{v}_{1}\cdot\vec{n} v_{1}^2 \vec{v}_{2}\cdot\vec{n} v_{2}^2 - 20 \vec{v}_{1}\cdot\vec{n} \vec{v}_{2}\cdot\vec{n} \vec{v}_{1}\cdot\vec{v}_{2} v_{2}^2 \nn\\ 
&& + 15 v_{1}^2 v_{2}^2 ( \vec{v}_{1}\cdot\vec{n})^{2} - 25 \vec{v}_{1}\cdot\vec{v}_{2} v_{2}^2 ( \vec{v}_{1}\cdot\vec{n})^{2} - 25 v_{1}^2 \vec{v}_{1}\cdot\vec{v}_{2} ( \vec{v}_{2}\cdot\vec{n})^{2} \nn\\ 
&& - 50 \vec{v}_{1}\cdot\vec{n} \vec{v}_{2}\cdot\vec{n} ( \vec{v}_{1}\cdot\vec{v}_{2})^{2} + 15 v_{1}^2 v_{2}^2 ( \vec{v}_{2}\cdot\vec{n})^{2} + 35 \vec{v}_{1}\cdot\vec{n} \vec{v}_{2}\cdot\vec{n} v_{1}^{4} + 15 ( \vec{v}_{2}\cdot\vec{n})^{2} v_{1}^{4} \nn\\ 
&& + 15 ( \vec{v}_{1}\cdot\vec{n})^{2} v_{2}^{4} + 35 \vec{v}_{1}\cdot\vec{n} \vec{v}_{2}\cdot\vec{n} v_{2}^{4} -35 \vec{v}_{2}\cdot\vec{n} v_{2}^2 ( \vec{v}_{1}\cdot\vec{n})^{3} - 105 v_{1}^2 ( \vec{v}_{1}\cdot\vec{n})^{2} ( \vec{v}_{2}\cdot\vec{n})^{2} \nn\\ 
&& - 35 \vec{v}_{1}\cdot\vec{n} v_{1}^2 ( \vec{v}_{2}\cdot\vec{n})^{3} + 175 \vec{v}_{1}\cdot\vec{v}_{2} ( \vec{v}_{1}\cdot\vec{n})^{2} ( \vec{v}_{2}\cdot\vec{n})^{2} - 105 v_{2}^2 ( \vec{v}_{1}\cdot\vec{n})^{2} ( \vec{v}_{2}\cdot\vec{n})^{2} \nn\\ 
&& + 105 ( \vec{v}_{1}\cdot\vec{n})^{3} ( \vec{v}_{2}\cdot\vec{n})^{3} \big) ( \vec{S}_{1}\cdot\vec{n})^{2} - 2 \big( 5 v_{1}^2 \vec{v}_{1}\cdot\vec{v}_{2} - 2 v_{1}^2 v_{2}^2 - \vec{v}_{1}\cdot\vec{v}_{2} v_{2}^2 + 10 ( \vec{v}_{1}\cdot\vec{v}_{2})^{2} \nn\\ 
&& - 7 v_{1}^{4} - 4 v_{2}^{4} + 33 \vec{v}_{1}\cdot\vec{n} v_{1}^2 \vec{v}_{2}\cdot\vec{n} - 60 \vec{v}_{1}\cdot\vec{n} \vec{v}_{2}\cdot\vec{n} \vec{v}_{1}\cdot\vec{v}_{2} + 27 \vec{v}_{1}\cdot\vec{n} \vec{v}_{2}\cdot\vec{n} v_{2}^2 \nn\\ 
&& - 12 v_{1}^2 ( \vec{v}_{1}\cdot\vec{n})^{2} + 15 \vec{v}_{1}\cdot\vec{v}_{2} ( \vec{v}_{1}\cdot\vec{n})^{2} + 15 \vec{v}_{1}\cdot\vec{v}_{2} ( \vec{v}_{2}\cdot\vec{n})^{2} - 9 v_{2}^2 ( \vec{v}_{2}\cdot\vec{n})^{2} \nn\\ 
&& + 15 \vec{v}_{2}\cdot\vec{n} ( \vec{v}_{1}\cdot\vec{n})^{3} - 45 ( \vec{v}_{1}\cdot\vec{n})^{2} ( \vec{v}_{2}\cdot\vec{n})^{2} + 15 \vec{v}_{1}\cdot\vec{n} ( \vec{v}_{2}\cdot\vec{n})^{3} \big) ( \vec{S}_{1}\cdot\vec{v}_{1})^{2} \nn\\ 
&& + 2 \big( 5 v_{1}^2 \vec{v}_{1}\cdot\vec{v}_{2} - 3 v_{1}^2 v_{2}^2 - 3 v_{1}^{4} + 9 \vec{v}_{1}\cdot\vec{n} v_{1}^2 \vec{v}_{2}\cdot\vec{n} + 9 v_{1}^2 ( \vec{v}_{1}\cdot\vec{n})^{2} - 15 \vec{v}_{1}\cdot\vec{v}_{2} ( \vec{v}_{1}\cdot\vec{n})^{2} \nn\\ 
&& + 9 v_{2}^2 ( \vec{v}_{1}\cdot\vec{n})^{2} -15 \vec{v}_{2}\cdot\vec{n} ( \vec{v}_{1}\cdot\vec{n})^{3} \big) ( \vec{S}_{1}\cdot\vec{v}_{2})^{2} \Big]\nn\\ && +  	\frac{G^2 m_{2}}{16 r{}^4} \Big[ S_{1}^2 \big( 35 v_{1}^2 \vec{v}_{1}\cdot\vec{v}_{2} - 64 v_{1}^2 v_{2}^2 + 152 \vec{v}_{1}\cdot\vec{v}_{2} v_{2}^2 - 92 ( \vec{v}_{1}\cdot\vec{v}_{2})^{2} + 15 v_{1}^{4} - 47 v_{2}^{4} \nn\\ 
&& + 308 \vec{v}_{1}\cdot\vec{n} v_{1}^2 \vec{v}_{2}\cdot\vec{n} - 192 \vec{v}_{1}\cdot\vec{n} \vec{v}_{2}\cdot\vec{n} \vec{v}_{1}\cdot\vec{v}_{2} - 255 v_{1}^2 ( \vec{v}_{1}\cdot\vec{n})^{2} + 95 \vec{v}_{1}\cdot\vec{v}_{2} ( \vec{v}_{1}\cdot\vec{n})^{2} \nn\\ 
&& + 77 v_{2}^2 ( \vec{v}_{1}\cdot\vec{n})^{2} - 44 v_{1}^2 ( \vec{v}_{2}\cdot\vec{n})^{2} + 4 \vec{v}_{1}\cdot\vec{v}_{2} ( \vec{v}_{2}\cdot\vec{n})^{2} - 2 v_{2}^2 ( \vec{v}_{2}\cdot\vec{n})^{2} + 92 ( \vec{v}_{1}\cdot\vec{n})^{4} \nn\\ 
&& - 84 \vec{v}_{2}\cdot\vec{n} ( \vec{v}_{1}\cdot\vec{n})^{3} + 78 ( \vec{v}_{1}\cdot\vec{n})^{2} ( \vec{v}_{2}\cdot\vec{n})^{2} - 56 \vec{v}_{1}\cdot\vec{n} ( \vec{v}_{2}\cdot\vec{n})^{3} \big) \nn\\ 
&& - 2 \vec{S}_{1}\cdot\vec{n} \vec{S}_{1}\cdot\vec{v}_{1} \big( 879 \vec{v}_{1}\cdot\vec{n} v_{1}^2 - 418 v_{1}^2 \vec{v}_{2}\cdot\vec{n} - 851 \vec{v}_{1}\cdot\vec{n} \vec{v}_{1}\cdot\vec{v}_{2} + 388 \vec{v}_{2}\cdot\vec{n} \vec{v}_{1}\cdot\vec{v}_{2} \nn\\ 
&& + 68 \vec{v}_{1}\cdot\vec{n} v_{2}^2 - 50 \vec{v}_{2}\cdot\vec{n} v_{2}^2 -164 ( \vec{v}_{1}\cdot\vec{n})^{3} - 33 \vec{v}_{2}\cdot\vec{n} ( \vec{v}_{1}\cdot\vec{n})^{2} + 108 \vec{v}_{1}\cdot\vec{n} ( \vec{v}_{2}\cdot\vec{n})^{2} \nn\\ 
&& - 28 ( \vec{v}_{2}\cdot\vec{n})^{3} \big) + 4 \vec{S}_{1}\cdot\vec{n} \vec{S}_{1}\cdot\vec{v}_{2} \big( 206 \vec{v}_{1}\cdot\vec{n} v_{1}^2 + 13 v_{1}^2 \vec{v}_{2}\cdot\vec{n} - 180 \vec{v}_{1}\cdot\vec{n} \vec{v}_{1}\cdot\vec{v}_{2} \nn\\ 
&& - 46 \vec{v}_{2}\cdot\vec{n} \vec{v}_{1}\cdot\vec{v}_{2} + 17 \vec{v}_{1}\cdot\vec{n} v_{2}^2 - 6 \vec{v}_{2}\cdot\vec{n} v_{2}^2 -6 ( \vec{v}_{1}\cdot\vec{n})^{3} - 63 \vec{v}_{2}\cdot\vec{n} ( \vec{v}_{1}\cdot\vec{n})^{2} \nn\\ 
&& + 60 \vec{v}_{1}\cdot\vec{n} ( \vec{v}_{2}\cdot\vec{n})^{2} \big) - \vec{S}_{1}\cdot\vec{v}_{1} \vec{S}_{1}\cdot\vec{v}_{2} \big( 221 v_{1}^2 - 224 \vec{v}_{1}\cdot\vec{v}_{2} + 40 v_{2}^2 -56 \vec{v}_{1}\cdot\vec{n} \vec{v}_{2}\cdot\vec{n} \nn\\ 
&& + 13 ( \vec{v}_{1}\cdot\vec{n})^{2} + 20 ( \vec{v}_{2}\cdot\vec{n})^{2} \big) + \big( 540 v_{1}^2 \vec{v}_{1}\cdot\vec{v}_{2} + 111 v_{1}^2 v_{2}^2 - 400 \vec{v}_{1}\cdot\vec{v}_{2} v_{2}^2 \nn\\ 
&& - 44 ( \vec{v}_{1}\cdot\vec{v}_{2})^{2} - 336 v_{1}^{4} + 137 v_{2}^{4} -2454 \vec{v}_{1}\cdot\vec{n} v_{1}^2 \vec{v}_{2}\cdot\vec{n} + 2112 \vec{v}_{1}\cdot\vec{n} \vec{v}_{2}\cdot\vec{n} \vec{v}_{1}\cdot\vec{v}_{2} \nn\\ 
&& - 192 \vec{v}_{1}\cdot\vec{n} \vec{v}_{2}\cdot\vec{n} v_{2}^2 + 2646 v_{1}^2 ( \vec{v}_{1}\cdot\vec{n})^{2} - 2322 \vec{v}_{1}\cdot\vec{v}_{2} ( \vec{v}_{1}\cdot\vec{n})^{2} + 51 v_{2}^2 ( \vec{v}_{1}\cdot\vec{n})^{2} \nn\\ 
&& - 66 v_{1}^2 ( \vec{v}_{2}\cdot\vec{n})^{2} + 288 \vec{v}_{1}\cdot\vec{v}_{2} ( \vec{v}_{2}\cdot\vec{n})^{2} + 30 v_{2}^2 ( \vec{v}_{2}\cdot\vec{n})^{2} -592 ( \vec{v}_{1}\cdot\vec{n})^{4} \nn\\ 
&& + 6 \vec{v}_{2}\cdot\vec{n} ( \vec{v}_{1}\cdot\vec{n})^{3} + 546 ( \vec{v}_{1}\cdot\vec{n})^{2} ( \vec{v}_{2}\cdot\vec{n})^{2} - 320 \vec{v}_{1}\cdot\vec{n} ( \vec{v}_{2}\cdot\vec{n})^{3} \big) ( \vec{S}_{1}\cdot\vec{n})^{2} + \big( 231 v_{1}^2 \nn\\ 
&& - 252 \vec{v}_{1}\cdot\vec{v}_{2} + 39 v_{2}^2 -38 \vec{v}_{1}\cdot\vec{n} \vec{v}_{2}\cdot\vec{n} - 11 ( \vec{v}_{1}\cdot\vec{n})^{2} + 26 ( \vec{v}_{2}\cdot\vec{n})^{2} \big) ( \vec{S}_{1}\cdot\vec{v}_{1})^{2} - 2 \big( 5 v_{1}^2 \nn\\ 
&& - 12 \vec{v}_{1}\cdot\vec{v}_{2} - 2 v_{2}^2 + 24 \vec{v}_{1}\cdot\vec{n} \vec{v}_{2}\cdot\vec{n} - 13 ( \vec{v}_{1}\cdot\vec{n})^{2} \big) ( \vec{S}_{1}\cdot\vec{v}_{2})^{2} \Big] \nn\\ 
&& + 	\frac{G^2 m_{2}{}^2}{16 m_{1} r{}^4} \Big[ S_{1}^2 \big( 124 v_{1}^2 \vec{v}_{1}\cdot\vec{v}_{2} - 174 v_{1}^2 v_{2}^2 + 407 \vec{v}_{1}\cdot\vec{v}_{2} v_{2}^2 - 370 ( \vec{v}_{1}\cdot\vec{v}_{2})^{2} - 8 v_{1}^{4} \nn\\ 
&& + 17 v_{2}^{4} + 208 \vec{v}_{1}\cdot\vec{n} v_{1}^2 \vec{v}_{2}\cdot\vec{n} + 2680 \vec{v}_{1}\cdot\vec{n} \vec{v}_{2}\cdot\vec{n} \vec{v}_{1}\cdot\vec{v}_{2} - 2156 \vec{v}_{1}\cdot\vec{n} \vec{v}_{2}\cdot\vec{n} v_{2}^2 \nn\\ 
&& - 160 v_{1}^2 ( \vec{v}_{1}\cdot\vec{n})^{2} - 256 \vec{v}_{1}\cdot\vec{v}_{2} ( \vec{v}_{1}\cdot\vec{n})^{2} + 710 v_{2}^2 ( \vec{v}_{1}\cdot\vec{n})^{2} + 457 v_{1}^2 ( \vec{v}_{2}\cdot\vec{n})^{2} \nn\\ 
&& - 2517 \vec{v}_{1}\cdot\vec{v}_{2} ( \vec{v}_{2}\cdot\vec{n})^{2} + 989 v_{2}^2 ( \vec{v}_{2}\cdot\vec{n})^{2} + 96 ( \vec{v}_{1}\cdot\vec{n})^{4} - 128 \vec{v}_{2}\cdot\vec{n} ( \vec{v}_{1}\cdot\vec{n})^{3} \nn\\ 
&& - 3768 ( \vec{v}_{1}\cdot\vec{n})^{2} ( \vec{v}_{2}\cdot\vec{n})^{2} + 5940 \vec{v}_{1}\cdot\vec{n} ( \vec{v}_{2}\cdot\vec{n})^{3} - 1966 ( \vec{v}_{2}\cdot\vec{n})^{4} \big) \nn\\ 
&& - \vec{S}_{1}\cdot\vec{n} \vec{S}_{1}\cdot\vec{v}_{1} \big( 148 \vec{v}_{1}\cdot\vec{n} v_{1}^2 - 112 v_{1}^2 \vec{v}_{2}\cdot\vec{n} - 36 \vec{v}_{1}\cdot\vec{n} \vec{v}_{1}\cdot\vec{v}_{2} + 1362 \vec{v}_{2}\cdot\vec{n} \vec{v}_{1}\cdot\vec{v}_{2} \nn\\ 
&& - 293 \vec{v}_{1}\cdot\vec{n} v_{2}^2 - 1122 \vec{v}_{2}\cdot\vec{n} v_{2}^2 + 96 ( \vec{v}_{1}\cdot\vec{n})^{3} - 624 \vec{v}_{2}\cdot\vec{n} ( \vec{v}_{1}\cdot\vec{n})^{2} \nn\\ 
&& - 357 \vec{v}_{1}\cdot\vec{n} ( \vec{v}_{2}\cdot\vec{n})^{2} + 1002 ( \vec{v}_{2}\cdot\vec{n})^{3} \big) - \vec{S}_{1}\cdot\vec{n} \vec{S}_{1}\cdot\vec{v}_{2} \big( 156 \vec{v}_{1}\cdot\vec{n} v_{1}^2 + 578 v_{1}^2 \vec{v}_{2}\cdot\vec{n} \nn\\ 
&& + 1672 \vec{v}_{1}\cdot\vec{n} \vec{v}_{1}\cdot\vec{v}_{2} - 3692 \vec{v}_{2}\cdot\vec{n} \vec{v}_{1}\cdot\vec{v}_{2} - 613 \vec{v}_{1}\cdot\vec{n} v_{2}^2 + 1982 \vec{v}_{2}\cdot\vec{n} v_{2}^2 -128 ( \vec{v}_{1}\cdot\vec{n})^{3} \nn\\ 
&& - 4086 \vec{v}_{2}\cdot\vec{n} ( \vec{v}_{1}\cdot\vec{n})^{2} + 6933 \vec{v}_{1}\cdot\vec{n} ( \vec{v}_{2}\cdot\vec{n})^{2} - 3130 ( \vec{v}_{2}\cdot\vec{n})^{3} \big) - 2 \vec{S}_{1}\cdot\vec{v}_{1} \vec{S}_{1}\cdot\vec{v}_{2} \big( 32 v_{1}^2 \nn\\ 
&& - 241 \vec{v}_{1}\cdot\vec{v}_{2} + 49 v_{2}^2 + 355 \vec{v}_{1}\cdot\vec{n} \vec{v}_{2}\cdot\vec{n} + 22 ( \vec{v}_{1}\cdot\vec{n})^{2} - 391 ( \vec{v}_{2}\cdot\vec{n})^{2} \big) - \big( 364 v_{1}^2 \vec{v}_{1}\cdot\vec{v}_{2} \nn\\ 
&& - 156 v_{1}^2 v_{2}^2 - 721 \vec{v}_{1}\cdot\vec{v}_{2} v_{2}^2 + 104 ( \vec{v}_{1}\cdot\vec{v}_{2})^{2} - 74 v_{1}^{4} + 468 v_{2}^{4} -516 \vec{v}_{1}\cdot\vec{n} v_{1}^2 \vec{v}_{2}\cdot\vec{n} \nn\\ 
&& - 1656 \vec{v}_{1}\cdot\vec{n} \vec{v}_{2}\cdot\vec{n} \vec{v}_{1}\cdot\vec{v}_{2} + 2391 \vec{v}_{1}\cdot\vec{n} \vec{v}_{2}\cdot\vec{n} v_{2}^2 - 48 v_{1}^2 ( \vec{v}_{1}\cdot\vec{n})^{2} - 384 \vec{v}_{1}\cdot\vec{v}_{2} ( \vec{v}_{1}\cdot\vec{n})^{2} \nn\\ 
&& + 249 v_{2}^2 ( \vec{v}_{1}\cdot\vec{n})^{2} + 126 v_{1}^2 ( \vec{v}_{2}\cdot\vec{n})^{2} + 2703 \vec{v}_{1}\cdot\vec{v}_{2} ( \vec{v}_{2}\cdot\vec{n})^{2} - 3042 v_{2}^2 ( \vec{v}_{2}\cdot\vec{n})^{2} \nn\\ 
&& + 512 \vec{v}_{2}\cdot\vec{n} ( \vec{v}_{1}\cdot\vec{n})^{3} + 735 ( \vec{v}_{1}\cdot\vec{n})^{2} ( \vec{v}_{2}\cdot\vec{n})^{2} - 1623 \vec{v}_{1}\cdot\vec{n} ( \vec{v}_{2}\cdot\vec{n})^{3} \nn\\ 
&& + 1072 ( \vec{v}_{2}\cdot\vec{n})^{4} \big) ( \vec{S}_{1}\cdot\vec{n})^{2} + \big( 12 v_{1}^2 + 40 \vec{v}_{1}\cdot\vec{v}_{2} - 89 v_{2}^2 -600 \vec{v}_{1}\cdot\vec{n} \vec{v}_{2}\cdot\vec{n} \nn\\ 
&& + 204 ( \vec{v}_{1}\cdot\vec{n})^{2} + 221 ( \vec{v}_{2}\cdot\vec{n})^{2} \big) ( \vec{S}_{1}\cdot\vec{v}_{1})^{2} + \big( 211 v_{1}^2 - 752 \vec{v}_{1}\cdot\vec{v}_{2} + 257 v_{2}^2 \nn\\ 
&& + 2240 \vec{v}_{1}\cdot\vec{n} \vec{v}_{2}\cdot\vec{n} - 850 ( \vec{v}_{1}\cdot\vec{n})^{2} - 1267 ( \vec{v}_{2}\cdot\vec{n})^{2} \big) ( \vec{S}_{1}\cdot\vec{v}_{2})^{2} \Big]\nn\\ && - 	\frac{G^3 m_{1} m_{2}}{58800 r{}^5} \Big[ S_{1}^2 \big( 468263 v_{1}^2 - 731426 \vec{v}_{1}\cdot\vec{v}_{2} + 323538 v_{2}^2 + 2235430 \vec{v}_{1}\cdot\vec{n} \vec{v}_{2}\cdot\vec{n} \nn\\ 
&& - 744265 ( \vec{v}_{1}\cdot\vec{n})^{2} - 726240 ( \vec{v}_{2}\cdot\vec{n})^{2} \big) - 20 \vec{S}_{1}\cdot\vec{n} \vec{S}_{1}\cdot\vec{v}_{1} \big( 100574 \vec{v}_{1}\cdot\vec{n} \nn\\ 
&& - 128539 \vec{v}_{2}\cdot\vec{n} \big) + 20 \vec{S}_{1}\cdot\vec{n} \vec{S}_{1}\cdot\vec{v}_{2} \big( 133789 \vec{v}_{1}\cdot\vec{n} - 83004 \vec{v}_{2}\cdot\vec{n} \big) \nn\\ 
&& - 341452 \vec{S}_{1}\cdot\vec{v}_{1} \vec{S}_{1}\cdot\vec{v}_{2} - 25 \big( 59347 v_{1}^2 - 88090 \vec{v}_{1}\cdot\vec{v}_{2} + 40398 v_{2}^2 \nn\\ 
&& + 427882 \vec{v}_{1}\cdot\vec{n} \vec{v}_{2}\cdot\vec{n} - 149947 ( \vec{v}_{1}\cdot\vec{n})^{2} - 140784 ( \vec{v}_{2}\cdot\vec{n})^{2} \big) ( \vec{S}_{1}\cdot\vec{n})^{2} \nn\\ 
&& + 260326 ( \vec{S}_{1}\cdot\vec{v}_{1})^{2} + 103176 ( \vec{S}_{1}\cdot\vec{v}_{2})^{2} \Big] - 	\frac{G^3 m_{2}{}^3}{48 m_{1} r{}^5} \Big[ S_{1}^2 \big( 522 v_{1}^2 - 110 \vec{v}_{1}\cdot\vec{v}_{2} \nn\\ 
&& - 673 v_{2}^2 + 178 \vec{v}_{1}\cdot\vec{n} \vec{v}_{2}\cdot\vec{n} - 1476 ( \vec{v}_{1}\cdot\vec{n})^{2} + 2147 ( \vec{v}_{2}\cdot\vec{n})^{2} \big) \nn\\ 
&& - 4 \vec{S}_{1}\cdot\vec{n} \vec{S}_{1}\cdot\vec{v}_{1} \big( 1419 \vec{v}_{1}\cdot\vec{n} - 1622 \vec{v}_{2}\cdot\vec{n} \big) + 4 \vec{S}_{1}\cdot\vec{n} \vec{S}_{1}\cdot\vec{v}_{2} \big( 1265 \vec{v}_{1}\cdot\vec{n} - 1246 \vec{v}_{2}\cdot\vec{n} \big) \nn\\ 
&& - 880 \vec{S}_{1}\cdot\vec{v}_{1} \vec{S}_{1}\cdot\vec{v}_{2} - \big( 1218 v_{1}^2 - 922 \vec{v}_{1}\cdot\vec{v}_{2} - 185 v_{2}^2 + 12850 \vec{v}_{1}\cdot\vec{n} \vec{v}_{2}\cdot\vec{n} \nn\\ 
&& - 7464 ( \vec{v}_{1}\cdot\vec{n})^{2} - 2887 ( \vec{v}_{2}\cdot\vec{n})^{2} \big) ( \vec{S}_{1}\cdot\vec{n})^{2} + 636 ( \vec{S}_{1}\cdot\vec{v}_{1})^{2} + 490 ( \vec{S}_{1}\cdot\vec{v}_{2})^{2} \Big] \nn\\ 
&& - 	\frac{G^3 m_{2}{}^2}{6144 r{}^5} \Big[ S_{1}^2 \big( {(183680 - 8109 \pi^2)} v_{1}^2 - {(384256 - 16218 \pi^2)} \vec{v}_{1}\cdot\vec{v}_{2} \nn\\ 
&& + {(242816 - 8109 \pi^2)} v_{2}^2 + {(972032 - 81090 \pi^2)} \vec{v}_{1}\cdot\vec{n} \vec{v}_{2}\cdot\vec{n} \nn\\ 
&& - {(324736 - 40545 \pi^2)} ( \vec{v}_{1}\cdot\vec{n})^{2} - {(684928 - 40545 \pi^2)} ( \vec{v}_{2}\cdot\vec{n})^{2} \big) \nn\\ 
&& - 4 \vec{S}_{1}\cdot\vec{n} \vec{S}_{1}\cdot\vec{v}_{1} \big( {(509152 - 40545 \pi^2)} \vec{v}_{1}\cdot\vec{n} - {(672736 - 40545 \pi^2)} \vec{v}_{2}\cdot\vec{n} \big) \nn\\ 
&& + 4 \vec{S}_{1}\cdot\vec{n} \vec{S}_{1}\cdot\vec{v}_{2} \big( {(447712 - 40545 \pi^2)} \vec{v}_{1}\cdot\vec{n} - {(541024 - 40545 \pi^2)} \vec{v}_{2}\cdot\vec{n} \big) \nn\\ 
&& - 4 {(146240 - 8109 \pi^2)} \vec{S}_{1}\cdot\vec{v}_{1} \vec{S}_{1}\cdot\vec{v}_{2} - \big( {(670720 - 40545 \pi^2)} v_{1}^2 \nn\\ 
&& - {(1422848 - 81090 \pi^2)} \vec{v}_{1}\cdot\vec{v}_{2} + {(869632 - 40545 \pi^2)} v_{2}^2 \nn\\ 
&& + {(7089152 - 567630 \pi^2)} \vec{v}_{1}\cdot\vec{n} \vec{v}_{2}\cdot\vec{n} - {(2963200 - 283815 \pi^2)} ( \vec{v}_{1}\cdot\vec{n})^{2} \nn\\ 
&& - {(3748864 - 283815 \pi^2)} ( \vec{v}_{2}\cdot\vec{n})^{2} \big) ( \vec{S}_{1}\cdot\vec{n})^{2} + 2 {(155840 - 8109 \pi^2)} ( \vec{S}_{1}\cdot\vec{v}_{1})^{2} \nn\\ 
&& + 2 {(153536 - 8109 \pi^2)} ( \vec{S}_{1}\cdot\vec{v}_{2})^{2} \Big]\nn\\ && +  	\frac{19 G^4 m_{1}{}^2 m_{2}}{4900 r{}^6} \Big[ 1571 S_{1}^2 - 4433 ( \vec{S}_{1}\cdot\vec{n})^{2} \Big] + 	\frac{G^4 m_{2}{}^4}{4 m_{1} r{}^6} \Big[ 73 S_{1}^2 - 119 ( \vec{S}_{1}\cdot\vec{n})^{2} \Big] \nn\\ 
&& + 	\frac{G^4 m_{1} m_{2}{}^2}{29400 r{}^6} \Big[ {(4275319 - 154350 \pi^2)} S_{1}^2 - 3 {(3162179 - 154350 \pi^2)} ( \vec{S}_{1}\cdot\vec{n})^{2} \Big] \nn\\ 
&& + 	\frac{G^4 m_{2}{}^3}{24 r{}^6} \Big[ {(3175 - 126 \pi^2)} S_{1}^2 - 9 {(785 - 42 \pi^2)} ( \vec{S}_{1}\cdot\vec{n})^{2} \Big],
\eea
and 
\bea 
V^{\text{N}^3\text{LO}}_{C_{1\text{E$^2$S$^2$}}} &=& 	- \frac{G^4 m_{1} m_{2}{}^2}{2 r{}^6} 
\Big[ S_{1}^2 - 3 ( \vec{S}_{1}\cdot\vec{n})^{2} \Big].
\eea
This action is also provided in machine-readable format in the ancillary files to this 
publication.

\section{General Hamiltonians}
\label{genhamresults}

Our general Hamiltonian for the present sectors, comprised of $4$ parts, can be expressed as:
\bea
H^{\text{N}^3\text{LO}}_{\text{S}^2} = H^{\text{N}^3\text{LO}}_{\text{S}_1 \text{S}_2} +H^{\text{N}^3\text{LO}}_{\text{S}_1^2} + C_{1\text{ES}^2} H^{\text{N}^3\text{LO}}_{C_{1\text{ES}^2}} + C_{1\text{E$^2$S$^2$}} H^{\text{N}^3\text{LO}}_{C_{1\text{E$^2$S$^2$}}} + (1 \leftrightarrow 2),
\eea
where
\bea
H^{\text{N}^3\text{LO}}_{\text{S}_1 \text{S}_2} &= &  	\frac{G}{16 m_{1}{}^6 r{}^3} \Big[ 15 \vec{S}_{1}\cdot\vec{n} \vec{S}_{2}\cdot\vec{n} p_{1}^{6} - 15 \vec{p}_{1}\cdot\vec{S}_{1} \vec{S}_{2}\cdot\vec{n} \vec{p}_{1}\cdot\vec{n} p_{1}^{4} - 15 \vec{S}_{1}\cdot\vec{n} \vec{p}_{1}\cdot\vec{S}_{2} \vec{p}_{1}\cdot\vec{n} p_{1}^{4} \nn\\ 
&& + 17 \vec{p}_{1}\cdot\vec{S}_{1} \vec{p}_{1}\cdot\vec{S}_{2} p_{1}^{4} - \vec{S}_{1}\cdot\vec{S}_{2} \big( 17 p_{1}^{6} -15 ( \vec{p}_{1}\cdot\vec{n})^{2} p_{1}^{4} \big) \Big] \nn\\ 
&& - 	\frac{G}{32 m_{1}{}^5 m_{2} r{}^3} \Big[ 3 \vec{S}_{1}\cdot\vec{n} \vec{S}_{2}\cdot\vec{n} \big( 41 \vec{p}_{1}\cdot\vec{p}_{2} p_{1}^{4} + 40 p_{1}^2 \vec{p}_{1}\cdot\vec{p}_{2} ( \vec{p}_{1}\cdot\vec{n})^{2} \nn\\ 
&& - 80 \vec{p}_{1}\cdot\vec{n} \vec{p}_{2}\cdot\vec{n} p_{1}^{4} -140 \vec{p}_{1}\cdot\vec{p}_{2} ( \vec{p}_{1}\cdot\vec{n})^{4} \big) - 6 \vec{p}_{1}\cdot\vec{S}_{1} \vec{S}_{2}\cdot\vec{n} \big( 16 \vec{p}_{1}\cdot\vec{n} p_{1}^2 \vec{p}_{1}\cdot\vec{p}_{2} \nn\\ 
&& + 9 \vec{p}_{2}\cdot\vec{n} p_{1}^{4} -55 p_{1}^2 \vec{p}_{2}\cdot\vec{n} ( \vec{p}_{1}\cdot\vec{n})^{2} + 20 \vec{p}_{1}\cdot\vec{p}_{2} ( \vec{p}_{1}\cdot\vec{n})^{3} \big) + 3 \vec{p}_{2}\cdot\vec{S}_{1} \vec{S}_{2}\cdot\vec{n} \big( 3 \vec{p}_{1}\cdot\vec{n} p_{1}^{4} \nn\\ 
&& + 80 p_{1}^2 ( \vec{p}_{1}\cdot\vec{n})^{3} \big) - 3 \vec{S}_{1}\cdot\vec{n} \vec{p}_{1}\cdot\vec{S}_{2} \big( 22 \vec{p}_{1}\cdot\vec{n} p_{1}^2 \vec{p}_{1}\cdot\vec{p}_{2} \nn\\ 
&& + \vec{p}_{2}\cdot\vec{n} p_{1}^{4} -70 p_{1}^2 \vec{p}_{2}\cdot\vec{n} ( \vec{p}_{1}\cdot\vec{n})^{2} -140 \vec{p}_{2}\cdot\vec{n} ( \vec{p}_{1}\cdot\vec{n})^{4} \big) \nn\\ 
&& + 2 \vec{p}_{1}\cdot\vec{S}_{1} \vec{p}_{1}\cdot\vec{S}_{2} \big( 10 p_{1}^2 \vec{p}_{1}\cdot\vec{p}_{2} -147 \vec{p}_{1}\cdot\vec{n} p_{1}^2 \vec{p}_{2}\cdot\vec{n} + 60 \vec{p}_{2}\cdot\vec{n} ( \vec{p}_{1}\cdot\vec{n})^{3} \big) \nn\\ 
&& + 2 \vec{p}_{2}\cdot\vec{S}_{1} \vec{p}_{1}\cdot\vec{S}_{2} \big( 5 p_{1}^{4} -12 p_{1}^2 ( \vec{p}_{1}\cdot\vec{n})^{2} -150 ( \vec{p}_{1}\cdot\vec{n})^{4} \big) - 36 \vec{S}_{1}\cdot\vec{n} \vec{p}_{2}\cdot\vec{S}_{2} \vec{p}_{1}\cdot\vec{n} p_{1}^{4} \nn\\ 
&& + 30 \vec{p}_{1}\cdot\vec{S}_{1} \vec{p}_{2}\cdot\vec{S}_{2} p_{1}^{4} - 3 \vec{S}_{1}\cdot\vec{S}_{2} \big( 22 \vec{p}_{1}\cdot\vec{p}_{2} p_{1}^{4} -30 p_{1}^2 \vec{p}_{1}\cdot\vec{p}_{2} ( \vec{p}_{1}\cdot\vec{n})^{2} \nn\\ 
&& - 105 \vec{p}_{1}\cdot\vec{n} \vec{p}_{2}\cdot\vec{n} p_{1}^{4} + 190 p_{1}^2 \vec{p}_{2}\cdot\vec{n} ( \vec{p}_{1}\cdot\vec{n})^{3} - 100 \vec{p}_{1}\cdot\vec{p}_{2} ( \vec{p}_{1}\cdot\vec{n})^{4} \big) \Big] \nn\\ 
&& + 	\frac{3 G}{16 m_{1} m_{2}{}^5 r{}^3} \Big[ 14 \vec{p}_{1}\cdot\vec{S}_{1} \vec{S}_{2}\cdot\vec{n} \vec{p}_{2}\cdot\vec{n} p_{2}^{4} - \vec{p}_{2}\cdot\vec{S}_{1} \vec{S}_{2}\cdot\vec{n} \big( 23 \vec{p}_{2}\cdot\vec{n} \vec{p}_{1}\cdot\vec{p}_{2} p_{2}^2 - 6 \vec{p}_{1}\cdot\vec{n} p_{2}^{4} \nn\\ 
&& + 25 \vec{p}_{1}\cdot\vec{n} p_{2}^2 ( \vec{p}_{2}\cdot\vec{n})^{2} \big) - 11 \vec{p}_{1}\cdot\vec{S}_{1} \vec{p}_{2}\cdot\vec{S}_{2} p_{2}^2 ( \vec{p}_{2}\cdot\vec{n})^{2} + 11 \vec{p}_{2}\cdot\vec{S}_{1} \vec{p}_{2}\cdot\vec{S}_{2} \vec{p}_{1}\cdot\vec{n} \vec{p}_{2}\cdot\vec{n} p_{2}^2 \nn\\ 
&& + \vec{S}_{1}\cdot\vec{S}_{2} \big( 23 \vec{p}_{1}\cdot\vec{p}_{2} p_{2}^2 ( \vec{p}_{2}\cdot\vec{n})^{2} - 20 \vec{p}_{1}\cdot\vec{n} \vec{p}_{2}\cdot\vec{n} p_{2}^{4} + 25 \vec{p}_{1}\cdot\vec{n} p_{2}^2 ( \vec{p}_{2}\cdot\vec{n})^{3} \big) \Big] \nn\\ 
&& + 	\frac{G}{16 m_{1}{}^4 m_{2}{}^2 r{}^3} \Big[ 3 \vec{S}_{1}\cdot\vec{n} \vec{S}_{2}\cdot\vec{n} \big( 2 p_{1}^2 ( \vec{p}_{1}\cdot\vec{p}_{2})^{2} + 3 p_{2}^2 p_{1}^{4} -95 \vec{p}_{1}\cdot\vec{n} p_{1}^2 \vec{p}_{2}\cdot\vec{n} \vec{p}_{1}\cdot\vec{p}_{2} \nn\\ 
&& - 5 p_{1}^2 p_{2}^2 ( \vec{p}_{1}\cdot\vec{n})^{2} + 60 ( \vec{p}_{1}\cdot\vec{n})^{2} ( \vec{p}_{1}\cdot\vec{p}_{2})^{2} - 25 ( \vec{p}_{2}\cdot\vec{n})^{2} p_{1}^{4} + 35 p_{1}^2 ( \vec{p}_{1}\cdot\vec{n})^{2} ( \vec{p}_{2}\cdot\vec{n})^{2} \big) \nn\\ 
&& - 3 \vec{p}_{1}\cdot\vec{S}_{1} \vec{S}_{2}\cdot\vec{n} \big( 12 p_{1}^2 \vec{p}_{2}\cdot\vec{n} \vec{p}_{1}\cdot\vec{p}_{2} + 17 \vec{p}_{1}\cdot\vec{n} p_{1}^2 p_{2}^2 - 14 \vec{p}_{1}\cdot\vec{n} ( \vec{p}_{1}\cdot\vec{p}_{2})^{2} \nn\\ 
&& + 10 \vec{p}_{2}\cdot\vec{n} \vec{p}_{1}\cdot\vec{p}_{2} ( \vec{p}_{1}\cdot\vec{n})^{2} - 15 p_{2}^2 ( \vec{p}_{1}\cdot\vec{n})^{3} - 85 \vec{p}_{1}\cdot\vec{n} p_{1}^2 ( \vec{p}_{2}\cdot\vec{n})^{2} \nn\\ 
&& + 105 ( \vec{p}_{1}\cdot\vec{n})^{3} ( \vec{p}_{2}\cdot\vec{n})^{2} \big) - 3 \vec{p}_{2}\cdot\vec{S}_{1} \vec{S}_{2}\cdot\vec{n} \big( 12 \vec{p}_{1}\cdot\vec{n} p_{1}^2 \vec{p}_{1}\cdot\vec{p}_{2} \nn\\ 
&& - 2 \vec{p}_{2}\cdot\vec{n} p_{1}^{4} -105 p_{1}^2 \vec{p}_{2}\cdot\vec{n} ( \vec{p}_{1}\cdot\vec{n})^{2} - 20 \vec{p}_{1}\cdot\vec{p}_{2} ( \vec{p}_{1}\cdot\vec{n})^{3} + 70 \vec{p}_{2}\cdot\vec{n} ( \vec{p}_{1}\cdot\vec{n})^{4} \big) \nn\\ 
&& + 3 \vec{S}_{1}\cdot\vec{n} \vec{p}_{1}\cdot\vec{S}_{2} \big( 8 p_{1}^2 \vec{p}_{2}\cdot\vec{n} \vec{p}_{1}\cdot\vec{p}_{2} - 5 \vec{p}_{1}\cdot\vec{n} p_{1}^2 p_{2}^2 \nn\\ 
&& + 2 \vec{p}_{1}\cdot\vec{n} ( \vec{p}_{1}\cdot\vec{p}_{2})^{2} -40 \vec{p}_{2}\cdot\vec{n} \vec{p}_{1}\cdot\vec{p}_{2} ( \vec{p}_{1}\cdot\vec{n})^{2} + 15 p_{2}^2 ( \vec{p}_{1}\cdot\vec{n})^{3} \nn\\ 
&& + 60 \vec{p}_{1}\cdot\vec{n} p_{1}^2 ( \vec{p}_{2}\cdot\vec{n})^{2} -105 ( \vec{p}_{1}\cdot\vec{n})^{3} ( \vec{p}_{2}\cdot\vec{n})^{2} \big) + \vec{p}_{1}\cdot\vec{S}_{1} \vec{p}_{1}\cdot\vec{S}_{2} \big( 10 p_{1}^2 p_{2}^2 \nn\\ 
&& - 6 ( \vec{p}_{1}\cdot\vec{p}_{2})^{2} -96 \vec{p}_{1}\cdot\vec{n} \vec{p}_{2}\cdot\vec{n} \vec{p}_{1}\cdot\vec{p}_{2} - 63 p_{2}^2 ( \vec{p}_{1}\cdot\vec{n})^{2} - 63 p_{1}^2 ( \vec{p}_{2}\cdot\vec{n})^{2} \nn\\ 
&& + 405 ( \vec{p}_{1}\cdot\vec{n})^{2} ( \vec{p}_{2}\cdot\vec{n})^{2} \big) - \vec{p}_{2}\cdot\vec{S}_{1} \vec{p}_{1}\cdot\vec{S}_{2} \big( 10 p_{1}^2 \vec{p}_{1}\cdot\vec{p}_{2} + 147 \vec{p}_{1}\cdot\vec{n} p_{1}^2 \vec{p}_{2}\cdot\vec{n} \nn\\ 
&& - 108 \vec{p}_{1}\cdot\vec{p}_{2} ( \vec{p}_{1}\cdot\vec{n})^{2} \big) - 24 \vec{S}_{1}\cdot\vec{n} \vec{p}_{2}\cdot\vec{S}_{2} \big( \vec{p}_{1}\cdot\vec{n} p_{1}^2 \vec{p}_{1}\cdot\vec{p}_{2} -5 p_{1}^2 \vec{p}_{2}\cdot\vec{n} ( \vec{p}_{1}\cdot\vec{n})^{2} \big) \nn\\ 
&& + 3 \vec{p}_{1}\cdot\vec{S}_{1} \vec{p}_{2}\cdot\vec{S}_{2} \big( 7 p_{1}^2 \vec{p}_{1}\cdot\vec{p}_{2} -32 \vec{p}_{1}\cdot\vec{n} p_{1}^2 \vec{p}_{2}\cdot\vec{n} + 6 \vec{p}_{1}\cdot\vec{p}_{2} ( \vec{p}_{1}\cdot\vec{n})^{2} \big) \nn\\ 
&& - \vec{p}_{2}\cdot\vec{S}_{1} \vec{p}_{2}\cdot\vec{S}_{2} \big( 5 p_{1}^{4} + 78 p_{1}^2 ( \vec{p}_{1}\cdot\vec{n})^{2} -60 ( \vec{p}_{1}\cdot\vec{n})^{4} \big) - \vec{S}_{1}\cdot\vec{S}_{2} \big( 4 p_{1}^2 ( \vec{p}_{1}\cdot\vec{p}_{2})^{2} \nn\\ 
&& + 8 p_{2}^2 p_{1}^{4} -387 \vec{p}_{1}\cdot\vec{n} p_{1}^2 \vec{p}_{2}\cdot\vec{n} \vec{p}_{1}\cdot\vec{p}_{2} - 132 p_{1}^2 p_{2}^2 ( \vec{p}_{1}\cdot\vec{n})^{2} + 114 ( \vec{p}_{1}\cdot\vec{n})^{2} ( \vec{p}_{1}\cdot\vec{p}_{2})^{2} \nn\\ 
&& - 69 ( \vec{p}_{2}\cdot\vec{n})^{2} p_{1}^{4} + 120 \vec{p}_{2}\cdot\vec{n} \vec{p}_{1}\cdot\vec{p}_{2} ( \vec{p}_{1}\cdot\vec{n})^{3} + 105 p_{2}^2 ( \vec{p}_{1}\cdot\vec{n})^{4} \nn\\ 
&& + 735 p_{1}^2 ( \vec{p}_{1}\cdot\vec{n})^{2} ( \vec{p}_{2}\cdot\vec{n})^{2} -525 ( \vec{p}_{1}\cdot\vec{n})^{4} ( \vec{p}_{2}\cdot\vec{n})^{2} \big) \Big] \nn\\ 
&& - 	\frac{G}{128 m_{1}{}^3 m_{2}{}^3 r{}^3} \Big[ 3 \vec{S}_{1}\cdot\vec{n} \vec{S}_{2}\cdot\vec{n} \big( 11 p_{1}^2 \vec{p}_{1}\cdot\vec{p}_{2} p_{2}^2 - 44 ( \vec{p}_{1}\cdot\vec{p}_{2})^{3} -140 \vec{p}_{1}\cdot\vec{n} p_{1}^2 \vec{p}_{2}\cdot\vec{n} p_{2}^2 \nn\\ 
&& - 160 \vec{p}_{1}\cdot\vec{p}_{2} p_{2}^2 ( \vec{p}_{1}\cdot\vec{n})^{2} + 320 p_{1}^2 \vec{p}_{1}\cdot\vec{p}_{2} ( \vec{p}_{2}\cdot\vec{n})^{2} \nn\\ 
&& - 40 \vec{p}_{1}\cdot\vec{n} \vec{p}_{2}\cdot\vec{n} ( \vec{p}_{1}\cdot\vec{p}_{2})^{2} -280 \vec{p}_{2}\cdot\vec{n} p_{2}^2 ( \vec{p}_{1}\cdot\vec{n})^{3} + 770 \vec{p}_{1}\cdot\vec{p}_{2} ( \vec{p}_{1}\cdot\vec{n})^{2} ( \vec{p}_{2}\cdot\vec{n})^{2} \nn\\ 
&& + 420 ( \vec{p}_{1}\cdot\vec{n})^{3} ( \vec{p}_{2}\cdot\vec{n})^{3} \big) - 24 \vec{p}_{1}\cdot\vec{S}_{1} \vec{S}_{2}\cdot\vec{n} \big( 10 p_{1}^2 \vec{p}_{2}\cdot\vec{n} p_{2}^2 + 12 \vec{p}_{1}\cdot\vec{n} \vec{p}_{1}\cdot\vec{p}_{2} p_{2}^2 \nn\\ 
&& - 12 \vec{p}_{2}\cdot\vec{n} ( \vec{p}_{1}\cdot\vec{p}_{2})^{2} -35 \vec{p}_{2}\cdot\vec{n} p_{2}^2 ( \vec{p}_{1}\cdot\vec{n})^{2} - 5 p_{1}^2 ( \vec{p}_{2}\cdot\vec{n})^{3} + 25 \vec{p}_{1}\cdot\vec{n} \vec{p}_{1}\cdot\vec{p}_{2} ( \vec{p}_{2}\cdot\vec{n})^{2} \nn\\ 
&& + 70 ( \vec{p}_{1}\cdot\vec{n})^{2} ( \vec{p}_{2}\cdot\vec{n})^{3} \big) - 3 \vec{p}_{2}\cdot\vec{S}_{1} \vec{S}_{2}\cdot\vec{n} \big( 96 p_{1}^2 \vec{p}_{2}\cdot\vec{n} \vec{p}_{1}\cdot\vec{p}_{2} + 55 \vec{p}_{1}\cdot\vec{n} p_{1}^2 p_{2}^2 \nn\\ 
&& - 20 \vec{p}_{1}\cdot\vec{n} ( \vec{p}_{1}\cdot\vec{p}_{2})^{2} -400 \vec{p}_{2}\cdot\vec{n} \vec{p}_{1}\cdot\vec{p}_{2} ( \vec{p}_{1}\cdot\vec{n})^{2} - 240 p_{2}^2 ( \vec{p}_{1}\cdot\vec{n})^{3} \nn\\ 
&& - 320 \vec{p}_{1}\cdot\vec{n} p_{1}^2 ( \vec{p}_{2}\cdot\vec{n})^{2} + 910 ( \vec{p}_{1}\cdot\vec{n})^{3} ( \vec{p}_{2}\cdot\vec{n})^{2} \big) + 3 \vec{S}_{1}\cdot\vec{n} \vec{p}_{1}\cdot\vec{S}_{2} \big( p_{1}^2 \vec{p}_{2}\cdot\vec{n} p_{2}^2 \nn\\ 
&& + 16 \vec{p}_{1}\cdot\vec{n} \vec{p}_{1}\cdot\vec{p}_{2} p_{2}^2 + 68 \vec{p}_{2}\cdot\vec{n} ( \vec{p}_{1}\cdot\vec{p}_{2})^{2} + 360 \vec{p}_{2}\cdot\vec{n} p_{2}^2 ( \vec{p}_{1}\cdot\vec{n})^{2} \nn\\ 
&& - 480 \vec{p}_{1}\cdot\vec{n} \vec{p}_{1}\cdot\vec{p}_{2} ( \vec{p}_{2}\cdot\vec{n})^{2} -630 ( \vec{p}_{1}\cdot\vec{n})^{2} ( \vec{p}_{2}\cdot\vec{n})^{3} \big) \nn\\ 
&& + 24 \vec{p}_{1}\cdot\vec{S}_{1} \vec{p}_{1}\cdot\vec{S}_{2} \big( 4 \vec{p}_{1}\cdot\vec{p}_{2} p_{2}^2 -25 \vec{p}_{1}\cdot\vec{n} \vec{p}_{2}\cdot\vec{n} p_{2}^2 + 28 \vec{p}_{1}\cdot\vec{p}_{2} ( \vec{p}_{2}\cdot\vec{n})^{2} \nn\\ 
&& + 15 \vec{p}_{1}\cdot\vec{n} ( \vec{p}_{2}\cdot\vec{n})^{3} \big) + 2 \vec{p}_{2}\cdot\vec{S}_{1} \vec{p}_{1}\cdot\vec{S}_{2} \big( 29 p_{1}^2 p_{2}^2 - 96 ( \vec{p}_{1}\cdot\vec{p}_{2})^{2} + 132 \vec{p}_{1}\cdot\vec{n} \vec{p}_{2}\cdot\vec{n} \vec{p}_{1}\cdot\vec{p}_{2} \nn\\ 
&& - 324 p_{2}^2 ( \vec{p}_{1}\cdot\vec{n})^{2} + 780 ( \vec{p}_{1}\cdot\vec{n})^{2} ( \vec{p}_{2}\cdot\vec{n})^{2} \big) - 24 \vec{S}_{1}\cdot\vec{n} \vec{p}_{2}\cdot\vec{S}_{2} \big( 3 \vec{p}_{1}\cdot\vec{n} p_{1}^2 p_{2}^2 \nn\\ 
&& + 2 \vec{p}_{1}\cdot\vec{n} ( \vec{p}_{1}\cdot\vec{p}_{2})^{2} -20 \vec{p}_{2}\cdot\vec{n} \vec{p}_{1}\cdot\vec{p}_{2} ( \vec{p}_{1}\cdot\vec{n})^{2} - 10 p_{2}^2 ( \vec{p}_{1}\cdot\vec{n})^{3} + 35 ( \vec{p}_{1}\cdot\vec{n})^{3} ( \vec{p}_{2}\cdot\vec{n})^{2} \big) \nn\\ 
&& + 4 \vec{p}_{1}\cdot\vec{S}_{1} \vec{p}_{2}\cdot\vec{S}_{2} \big( 17 p_{1}^2 p_{2}^2 - 22 ( \vec{p}_{1}\cdot\vec{p}_{2})^{2} -90 \vec{p}_{1}\cdot\vec{n} \vec{p}_{2}\cdot\vec{n} \vec{p}_{1}\cdot\vec{p}_{2} - 66 p_{2}^2 ( \vec{p}_{1}\cdot\vec{n})^{2} \nn\\ 
&& + 12 p_{1}^2 ( \vec{p}_{2}\cdot\vec{n})^{2} + 345 ( \vec{p}_{1}\cdot\vec{n})^{2} ( \vec{p}_{2}\cdot\vec{n})^{2} \big) + 8 \vec{p}_{2}\cdot\vec{S}_{1} \vec{p}_{2}\cdot\vec{S}_{2} \big( 14 p_{1}^2 \vec{p}_{1}\cdot\vec{p}_{2} \nn\\ 
&& + 9 \vec{p}_{1}\cdot\vec{n} p_{1}^2 \vec{p}_{2}\cdot\vec{n} - 78 \vec{p}_{1}\cdot\vec{p}_{2} ( \vec{p}_{1}\cdot\vec{n})^{2} -30 \vec{p}_{2}\cdot\vec{n} ( \vec{p}_{1}\cdot\vec{n})^{3} \big) - \vec{S}_{1}\cdot\vec{S}_{2} \big( 226 p_{1}^2 \vec{p}_{1}\cdot\vec{p}_{2} p_{2}^2 \nn\\ 
&& - 160 ( \vec{p}_{1}\cdot\vec{p}_{2})^{3} -921 \vec{p}_{1}\cdot\vec{n} p_{1}^2 \vec{p}_{2}\cdot\vec{n} p_{2}^2 - 1464 \vec{p}_{1}\cdot\vec{p}_{2} p_{2}^2 ( \vec{p}_{1}\cdot\vec{n})^{2} \nn\\ 
&& + 672 p_{1}^2 \vec{p}_{1}\cdot\vec{p}_{2} ( \vec{p}_{2}\cdot\vec{n})^{2} - 180 \vec{p}_{1}\cdot\vec{n} \vec{p}_{2}\cdot\vec{n} ( \vec{p}_{1}\cdot\vec{p}_{2})^{2} + 1200 \vec{p}_{2}\cdot\vec{n} p_{2}^2 ( \vec{p}_{1}\cdot\vec{n})^{3} \nn\\ 
&& + 1080 \vec{p}_{1}\cdot\vec{n} p_{1}^2 ( \vec{p}_{2}\cdot\vec{n})^{3} + 3120 \vec{p}_{1}\cdot\vec{p}_{2} ( \vec{p}_{1}\cdot\vec{n})^{2} ( \vec{p}_{2}\cdot\vec{n})^{2} -3430 ( \vec{p}_{1}\cdot\vec{n})^{3} ( \vec{p}_{2}\cdot\vec{n})^{3} \big) \Big] \nn\\ 
&& + 	\frac{3 G}{16 m_{1}{}^2 m_{2}{}^4 r{}^3} \Big[ 2 \vec{S}_{1}\cdot\vec{n} \vec{S}_{2}\cdot\vec{n} \big( p_{1}^2 p_{2}^{4} + 10 p_{1}^2 p_{2}^2 ( \vec{p}_{2}\cdot\vec{n})^{2} \nn\\ 
&& + 5 ( \vec{p}_{1}\cdot\vec{n})^{2} p_{2}^{4} -35 p_{1}^2 ( \vec{p}_{2}\cdot\vec{n})^{4} \big) - 2 \vec{p}_{1}\cdot\vec{S}_{1} \vec{S}_{2}\cdot\vec{n} \big( 7 \vec{p}_{2}\cdot\vec{n} \vec{p}_{1}\cdot\vec{p}_{2} p_{2}^2 + 6 \vec{p}_{1}\cdot\vec{n} p_{2}^{4} \nn\\ 
&& + 10 \vec{p}_{1}\cdot\vec{p}_{2} ( \vec{p}_{2}\cdot\vec{n})^{3} + 5 \vec{p}_{1}\cdot\vec{n} p_{2}^2 ( \vec{p}_{2}\cdot\vec{n})^{2} -35 \vec{p}_{1}\cdot\vec{n} ( \vec{p}_{2}\cdot\vec{n})^{4} \big) \nn\\ 
&& + \vec{p}_{2}\cdot\vec{S}_{1} \vec{S}_{2}\cdot\vec{n} \big( 5 p_{1}^2 \vec{p}_{2}\cdot\vec{n} p_{2}^2 + 12 \vec{p}_{1}\cdot\vec{n} \vec{p}_{1}\cdot\vec{p}_{2} p_{2}^2 + 6 \vec{p}_{2}\cdot\vec{n} ( \vec{p}_{1}\cdot\vec{p}_{2})^{2} \nn\\ 
&& + 25 \vec{p}_{2}\cdot\vec{n} p_{2}^2 ( \vec{p}_{1}\cdot\vec{n})^{2} + 25 p_{1}^2 ( \vec{p}_{2}\cdot\vec{n})^{3} \nn\\ 
&& + 20 \vec{p}_{1}\cdot\vec{n} \vec{p}_{1}\cdot\vec{p}_{2} ( \vec{p}_{2}\cdot\vec{n})^{2} -35 ( \vec{p}_{1}\cdot\vec{n})^{2} ( \vec{p}_{2}\cdot\vec{n})^{3} \big) - 2 \vec{S}_{1}\cdot\vec{n} \vec{p}_{1}\cdot\vec{S}_{2} \big( \vec{p}_{1}\cdot\vec{n} p_{2}^{4} \nn\\ 
&& + 10 \vec{p}_{1}\cdot\vec{n} p_{2}^2 ( \vec{p}_{2}\cdot\vec{n})^{2} -35 \vec{p}_{1}\cdot\vec{n} ( \vec{p}_{2}\cdot\vec{n})^{4} \big) + 2 \vec{p}_{1}\cdot\vec{S}_{1} \vec{p}_{1}\cdot\vec{S}_{2} \big( p_{2}^{4} \nn\\ 
&& + 15 p_{2}^2 ( \vec{p}_{2}\cdot\vec{n})^{2} -25 ( \vec{p}_{2}\cdot\vec{n})^{4} \big) - 10 \vec{p}_{2}\cdot\vec{S}_{1} \vec{p}_{1}\cdot\vec{S}_{2} \big( \vec{p}_{1}\cdot\vec{n} \vec{p}_{2}\cdot\vec{n} p_{2}^2 + 2 \vec{p}_{1}\cdot\vec{n} ( \vec{p}_{2}\cdot\vec{n})^{3} \big) \nn\\ 
&& + 2 \vec{p}_{1}\cdot\vec{S}_{1} \vec{p}_{2}\cdot\vec{S}_{2} \big( 3 \vec{p}_{1}\cdot\vec{n} \vec{p}_{2}\cdot\vec{n} p_{2}^2 + 2 \vec{p}_{1}\cdot\vec{p}_{2} ( \vec{p}_{2}\cdot\vec{n})^{2} + 5 \vec{p}_{1}\cdot\vec{n} ( \vec{p}_{2}\cdot\vec{n})^{3} \big) \nn\\ 
&& - 2 \vec{p}_{2}\cdot\vec{S}_{1} \vec{p}_{2}\cdot\vec{S}_{2} \big( 2 \vec{p}_{1}\cdot\vec{n} \vec{p}_{2}\cdot\vec{n} \vec{p}_{1}\cdot\vec{p}_{2} + 3 p_{2}^2 ( \vec{p}_{1}\cdot\vec{n})^{2} + 5 ( \vec{p}_{1}\cdot\vec{n})^{2} ( \vec{p}_{2}\cdot\vec{n})^{2} \big) \nn\\ 
&& - \vec{S}_{1}\cdot\vec{S}_{2} \big( 2 p_{1}^2 p_{2}^{4} -2 \vec{p}_{1}\cdot\vec{n} \vec{p}_{2}\cdot\vec{n} \vec{p}_{1}\cdot\vec{p}_{2} p_{2}^2 + 6 ( \vec{p}_{2}\cdot\vec{n})^{2} ( \vec{p}_{1}\cdot\vec{p}_{2})^{2} + 25 p_{1}^2 p_{2}^2 ( \vec{p}_{2}\cdot\vec{n})^{2} \nn\\ 
&& - 6 ( \vec{p}_{1}\cdot\vec{n})^{2} p_{2}^{4} -45 p_{1}^2 ( \vec{p}_{2}\cdot\vec{n})^{4} + 15 p_{2}^2 ( \vec{p}_{1}\cdot\vec{n})^{2} ( \vec{p}_{2}\cdot\vec{n})^{2} + 35 ( \vec{p}_{1}\cdot\vec{n})^{2} ( \vec{p}_{2}\cdot\vec{n})^{4} \big) \Big] \nn\\ 
&& - 	\frac{21 G}{16 m_{2}{}^6 r{}^3} \Big[ \vec{p}_{2}\cdot\vec{S}_{1} \vec{S}_{2}\cdot\vec{n} \vec{p}_{2}\cdot\vec{n} p_{2}^{4} - \vec{S}_{1}\cdot\vec{S}_{2} ( \vec{p}_{2}\cdot\vec{n})^{2} p_{2}^{4} \Big]\nn\\ && +  	\frac{G^2 m_{2}}{8 m_{1}{}^4 r{}^4} \Big[ 75 \vec{S}_{1}\cdot\vec{n} \vec{S}_{2}\cdot\vec{n} p_{1}^{4} - 74 \vec{p}_{1}\cdot\vec{S}_{1} \vec{S}_{2}\cdot\vec{n} \vec{p}_{1}\cdot\vec{n} p_{1}^2 - 74 \vec{S}_{1}\cdot\vec{n} \vec{p}_{1}\cdot\vec{S}_{2} \vec{p}_{1}\cdot\vec{n} p_{1}^2 \nn\\ 
&& + 56 \vec{p}_{1}\cdot\vec{S}_{1} \vec{p}_{1}\cdot\vec{S}_{2} p_{1}^2 - \vec{S}_{1}\cdot\vec{S}_{2} \big( 57 p_{1}^{4} -74 p_{1}^2 ( \vec{p}_{1}\cdot\vec{n})^{2} \big) \Big] \nn\\ 
&& - 	\frac{G^2}{32 m_{1}{}^3 r{}^4} \Big[ \vec{S}_{1}\cdot\vec{n} \vec{S}_{2}\cdot\vec{n} \big( 1924 p_{1}^2 \vec{p}_{1}\cdot\vec{p}_{2} + 1318 p_{1}^{4} -2358 \vec{p}_{1}\cdot\vec{n} p_{1}^2 \vec{p}_{2}\cdot\vec{n} \nn\\ 
&& - 10305 p_{1}^2 ( \vec{p}_{1}\cdot\vec{n})^{2} - 4086 \vec{p}_{1}\cdot\vec{p}_{2} ( \vec{p}_{1}\cdot\vec{n})^{2} + 796 ( \vec{p}_{1}\cdot\vec{n})^{4} \big) \nn\\ 
&& - \vec{p}_{1}\cdot\vec{S}_{1} \vec{S}_{2}\cdot\vec{n} \big( 1218 \vec{p}_{1}\cdot\vec{n} p_{1}^2 + 558 p_{1}^2 \vec{p}_{2}\cdot\vec{n} + 1554 \vec{p}_{1}\cdot\vec{n} \vec{p}_{1}\cdot\vec{p}_{2} -10501 ( \vec{p}_{1}\cdot\vec{n})^{3} \nn\\ 
&& - 2964 \vec{p}_{2}\cdot\vec{n} ( \vec{p}_{1}\cdot\vec{n})^{2} \big) - 3 \vec{p}_{2}\cdot\vec{S}_{1} \vec{S}_{2}\cdot\vec{n} \big( 42 \vec{p}_{1}\cdot\vec{n} p_{1}^2 -1327 ( \vec{p}_{1}\cdot\vec{n})^{3} \big) \nn\\ 
&& - \vec{S}_{1}\cdot\vec{n} \vec{p}_{1}\cdot\vec{S}_{2} \big( 1394 \vec{p}_{1}\cdot\vec{n} p_{1}^2 + 754 p_{1}^2 \vec{p}_{2}\cdot\vec{n} + 452 \vec{p}_{1}\cdot\vec{n} \vec{p}_{1}\cdot\vec{p}_{2} -10149 ( \vec{p}_{1}\cdot\vec{n})^{3} \nn\\ 
&& - 7050 \vec{p}_{2}\cdot\vec{n} ( \vec{p}_{1}\cdot\vec{n})^{2} \big) + \vec{p}_{1}\cdot\vec{S}_{1} \vec{p}_{1}\cdot\vec{S}_{2} \big( 559 p_{1}^2 - 220 \vec{p}_{1}\cdot\vec{p}_{2} -966 \vec{p}_{1}\cdot\vec{n} \vec{p}_{2}\cdot\vec{n} \nn\\ 
&& - 6341 ( \vec{p}_{1}\cdot\vec{n})^{2} \big) + \vec{p}_{2}\cdot\vec{S}_{1} \vec{p}_{1}\cdot\vec{S}_{2} \big( 755 p_{1}^2 -5088 ( \vec{p}_{1}\cdot\vec{n})^{2} \big) \nn\\ 
&& - 470 \vec{S}_{1}\cdot\vec{n} \vec{p}_{2}\cdot\vec{S}_{2} \vec{p}_{1}\cdot\vec{n} p_{1}^2 + 322 \vec{p}_{1}\cdot\vec{S}_{1} \vec{p}_{2}\cdot\vec{S}_{2} p_{1}^2 - \vec{S}_{1}\cdot\vec{S}_{2} \big( 961 p_{1}^2 \vec{p}_{1}\cdot\vec{p}_{2} \nn\\ 
&& + 499 p_{1}^{4} -2274 \vec{p}_{1}\cdot\vec{n} p_{1}^2 \vec{p}_{2}\cdot\vec{n} - 7395 p_{1}^2 ( \vec{p}_{1}\cdot\vec{n})^{2} - 5540 \vec{p}_{1}\cdot\vec{p}_{2} ( \vec{p}_{1}\cdot\vec{n})^{2} \nn\\ 
&& + 10657 ( \vec{p}_{1}\cdot\vec{n})^{4} + 6945 \vec{p}_{2}\cdot\vec{n} ( \vec{p}_{1}\cdot\vec{n})^{3} \big) \Big] + 	\frac{G^2}{96 m_{1}{}^2 m_{2} r{}^4} \Big[ 6 \vec{S}_{1}\cdot\vec{n} \vec{S}_{2}\cdot\vec{n} \big( 631 p_{1}^2 \vec{p}_{1}\cdot\vec{p}_{2} \nn\\ 
&& - 4 p_{1}^2 p_{2}^2 -2646 \vec{p}_{1}\cdot\vec{n} p_{1}^2 \vec{p}_{2}\cdot\vec{n} - 3105 \vec{p}_{1}\cdot\vec{p}_{2} ( \vec{p}_{1}\cdot\vec{n})^{2} - 227 p_{2}^2 ( \vec{p}_{1}\cdot\vec{n})^{2} \nn\\ 
&& - 1092 p_{1}^2 ( \vec{p}_{2}\cdot\vec{n})^{2} -275 \vec{p}_{2}\cdot\vec{n} ( \vec{p}_{1}\cdot\vec{n})^{3} \big) + 2 \vec{p}_{1}\cdot\vec{S}_{1} \vec{S}_{2}\cdot\vec{n} \big( 549 p_{1}^2 \vec{p}_{2}\cdot\vec{n} \nn\\ 
&& + 2748 \vec{p}_{1}\cdot\vec{n} \vec{p}_{1}\cdot\vec{p}_{2} - 5333 \vec{p}_{2}\cdot\vec{n} \vec{p}_{1}\cdot\vec{p}_{2} - 2539 \vec{p}_{1}\cdot\vec{n} p_{2}^2 + 9396 \vec{p}_{2}\cdot\vec{n} ( \vec{p}_{1}\cdot\vec{n})^{2} \nn\\ 
&& + 3741 \vec{p}_{1}\cdot\vec{n} ( \vec{p}_{2}\cdot\vec{n})^{2} \big) - \vec{p}_{2}\cdot\vec{S}_{1} \vec{S}_{2}\cdot\vec{n} \big( 16710 \vec{p}_{1}\cdot\vec{n} p_{1}^2 - 8797 p_{1}^2 \vec{p}_{2}\cdot\vec{n} \nn\\ 
&& - 7418 \vec{p}_{1}\cdot\vec{n} \vec{p}_{1}\cdot\vec{p}_{2} -39246 ( \vec{p}_{1}\cdot\vec{n})^{3} - 20199 \vec{p}_{2}\cdot\vec{n} ( \vec{p}_{1}\cdot\vec{n})^{2} \big) \nn\\ 
&& - 6 \vec{S}_{1}\cdot\vec{n} \vec{p}_{1}\cdot\vec{S}_{2} \big( 917 p_{1}^2 \vec{p}_{2}\cdot\vec{n} + 114 \vec{p}_{1}\cdot\vec{n} \vec{p}_{1}\cdot\vec{p}_{2} - 4 \vec{p}_{1}\cdot\vec{n} p_{2}^2 -8592 \vec{p}_{2}\cdot\vec{n} ( \vec{p}_{1}\cdot\vec{n})^{2} \nn\\ 
&& - 1092 \vec{p}_{1}\cdot\vec{n} ( \vec{p}_{2}\cdot\vec{n})^{2} \big) + 2 \vec{p}_{1}\cdot\vec{S}_{1} \vec{p}_{1}\cdot\vec{S}_{2} \big( 204 \vec{p}_{1}\cdot\vec{p}_{2} - 455 p_{2}^2 -7416 \vec{p}_{1}\cdot\vec{n} \vec{p}_{2}\cdot\vec{n} \nn\\ 
&& + 2739 ( \vec{p}_{2}\cdot\vec{n})^{2} \big) + 3 \vec{p}_{2}\cdot\vec{S}_{1} \vec{p}_{1}\cdot\vec{S}_{2} \big( 1187 p_{1}^2 -4010 \vec{p}_{1}\cdot\vec{n} \vec{p}_{2}\cdot\vec{n} - 4437 ( \vec{p}_{1}\cdot\vec{n})^{2} \big) \nn\\ 
&& + 6 \vec{S}_{1}\cdot\vec{n} \vec{p}_{2}\cdot\vec{S}_{2} \big( 185 \vec{p}_{1}\cdot\vec{n} p_{1}^2 -1191 ( \vec{p}_{1}\cdot\vec{n})^{3} \big) \nn\\ 
&& - 12 \vec{p}_{1}\cdot\vec{S}_{1} \vec{p}_{2}\cdot\vec{S}_{2} \big( 47 p_{1}^2 -456 \vec{p}_{1}\cdot\vec{n} \vec{p}_{2}\cdot\vec{n} - 345 ( \vec{p}_{1}\cdot\vec{n})^{2} \big) \nn\\ 
&& - 5705 \vec{p}_{2}\cdot\vec{S}_{1} \vec{p}_{2}\cdot\vec{S}_{2} ( \vec{p}_{1}\cdot\vec{n})^{2} - \vec{S}_{1}\cdot\vec{S}_{2} \big( 2667 p_{1}^2 \vec{p}_{1}\cdot\vec{p}_{2} \nn\\ 
&& - 910 p_{1}^2 p_{2}^2 -19110 \vec{p}_{1}\cdot\vec{n} p_{1}^2 \vec{p}_{2}\cdot\vec{n} - 3248 \vec{p}_{1}\cdot\vec{n} \vec{p}_{2}\cdot\vec{n} \vec{p}_{1}\cdot\vec{p}_{2} - 6957 \vec{p}_{1}\cdot\vec{p}_{2} ( \vec{p}_{1}\cdot\vec{n})^{2} \nn\\ 
&& - 6129 p_{2}^2 ( \vec{p}_{1}\cdot\vec{n})^{2} + 2245 p_{1}^2 ( \vec{p}_{2}\cdot\vec{n})^{2} + 54132 \vec{p}_{2}\cdot\vec{n} ( \vec{p}_{1}\cdot\vec{n})^{3} \nn\\ 
&& + 27681 ( \vec{p}_{1}\cdot\vec{n})^{2} ( \vec{p}_{2}\cdot\vec{n})^{2} \big) \Big] + 	\frac{G^2}{96 m_{1} m_{2}{}^2 r{}^4} \Big[ \vec{S}_{1}\cdot\vec{n} \vec{S}_{2}\cdot\vec{n} \big( 2399 p_{1}^2 p_{2}^2 \nn\\ 
&& + 1846 ( \vec{p}_{1}\cdot\vec{p}_{2})^{2} -2424 \vec{p}_{1}\cdot\vec{n} \vec{p}_{2}\cdot\vec{n} \vec{p}_{1}\cdot\vec{p}_{2} + 1014 p_{2}^2 ( \vec{p}_{1}\cdot\vec{n})^{2} \nn\\ 
&& - 9765 p_{1}^2 ( \vec{p}_{2}\cdot\vec{n})^{2} -3162 ( \vec{p}_{1}\cdot\vec{n})^{2} ( \vec{p}_{2}\cdot\vec{n})^{2} \big) - \vec{p}_{1}\cdot\vec{S}_{1} \vec{S}_{2}\cdot\vec{n} \big( 8416 \vec{p}_{2}\cdot\vec{n} \vec{p}_{1}\cdot\vec{p}_{2} \nn\\ 
&& + 8705 \vec{p}_{1}\cdot\vec{n} p_{2}^2 - 12882 \vec{p}_{2}\cdot\vec{n} p_{2}^2 -22305 \vec{p}_{1}\cdot\vec{n} ( \vec{p}_{2}\cdot\vec{n})^{2} + 120 ( \vec{p}_{2}\cdot\vec{n})^{3} \big) \nn\\ 
&& + \vec{p}_{2}\cdot\vec{S}_{1} \vec{S}_{2}\cdot\vec{n} \big( 9508 p_{1}^2 \vec{p}_{2}\cdot\vec{n} + 18122 \vec{p}_{1}\cdot\vec{n} \vec{p}_{1}\cdot\vec{p}_{2} - 12030 \vec{p}_{2}\cdot\vec{n} \vec{p}_{1}\cdot\vec{p}_{2} \nn\\ 
&& + 2091 \vec{p}_{1}\cdot\vec{n} p_{2}^2 -31668 \vec{p}_{2}\cdot\vec{n} ( \vec{p}_{1}\cdot\vec{n})^{2} - 33147 \vec{p}_{1}\cdot\vec{n} ( \vec{p}_{2}\cdot\vec{n})^{2} \big) \nn\\ 
&& + 3 \vec{S}_{1}\cdot\vec{n} \vec{p}_{1}\cdot\vec{S}_{2} \big( 666 \vec{p}_{2}\cdot\vec{n} \vec{p}_{1}\cdot\vec{p}_{2} - 827 \vec{p}_{1}\cdot\vec{n} p_{2}^2 -3693 \vec{p}_{1}\cdot\vec{n} ( \vec{p}_{2}\cdot\vec{n})^{2} \big) \nn\\ 
&& + \vec{p}_{1}\cdot\vec{S}_{1} \vec{p}_{1}\cdot\vec{S}_{2} \big( 2333 p_{2}^2 -2409 ( \vec{p}_{2}\cdot\vec{n})^{2} \big) \nn\\ 
&& - 2 \vec{p}_{2}\cdot\vec{S}_{1} \vec{p}_{1}\cdot\vec{S}_{2} \big( 1321 \vec{p}_{1}\cdot\vec{p}_{2} -346 \vec{p}_{1}\cdot\vec{n} \vec{p}_{2}\cdot\vec{n} \big) \nn\\ 
&& - 36 \vec{S}_{1}\cdot\vec{n} \vec{p}_{2}\cdot\vec{S}_{2} \big( 59 \vec{p}_{1}\cdot\vec{n} \vec{p}_{1}\cdot\vec{p}_{2} -385 \vec{p}_{2}\cdot\vec{n} ( \vec{p}_{1}\cdot\vec{n})^{2} \big) \nn\\ 
&& + 2 \vec{p}_{1}\cdot\vec{S}_{1} \vec{p}_{2}\cdot\vec{S}_{2} \big( 458 \vec{p}_{1}\cdot\vec{p}_{2} -1628 \vec{p}_{1}\cdot\vec{n} \vec{p}_{2}\cdot\vec{n} - 6597 ( \vec{p}_{2}\cdot\vec{n})^{2} \big) \nn\\ 
&& + 2 \vec{p}_{2}\cdot\vec{S}_{1} \vec{p}_{2}\cdot\vec{S}_{2} \big( 54 p_{1}^2 + 6597 \vec{p}_{1}\cdot\vec{n} \vec{p}_{2}\cdot\vec{n} - 1633 ( \vec{p}_{1}\cdot\vec{n})^{2} \big) - \vec{S}_{1}\cdot\vec{S}_{2} \big( 2211 p_{1}^2 p_{2}^2 \nn\\ 
&& - 794 ( \vec{p}_{1}\cdot\vec{p}_{2})^{2} + 390 \vec{p}_{1}\cdot\vec{n} \vec{p}_{2}\cdot\vec{n} \vec{p}_{1}\cdot\vec{p}_{2} + 14973 \vec{p}_{1}\cdot\vec{n} \vec{p}_{2}\cdot\vec{n} p_{2}^2 - 4089 p_{2}^2 ( \vec{p}_{1}\cdot\vec{n})^{2} \nn\\ 
&& - 257 p_{1}^2 ( \vec{p}_{2}\cdot\vec{n})^{2} - 12030 \vec{p}_{1}\cdot\vec{p}_{2} ( \vec{p}_{2}\cdot\vec{n})^{2} -10461 ( \vec{p}_{1}\cdot\vec{n})^{2} ( \vec{p}_{2}\cdot\vec{n})^{2} \nn\\ 
&& - 33267 \vec{p}_{1}\cdot\vec{n} ( \vec{p}_{2}\cdot\vec{n})^{3} \big) \Big] + 	\frac{G^2}{32 m_{2}{}^3 r{}^4} \Big[ 1918 \vec{p}_{1}\cdot\vec{S}_{1} \vec{S}_{2}\cdot\vec{n} \vec{p}_{2}\cdot\vec{n} p_{2}^2 \nn\\ 
&& - \vec{p}_{2}\cdot\vec{S}_{1} \vec{S}_{2}\cdot\vec{n} \big( 3756 \vec{p}_{2}\cdot\vec{n} \vec{p}_{1}\cdot\vec{p}_{2} + 576 \vec{p}_{1}\cdot\vec{n} p_{2}^2 + 1630 \vec{p}_{2}\cdot\vec{n} p_{2}^2 -588 \vec{p}_{1}\cdot\vec{n} ( \vec{p}_{2}\cdot\vec{n})^{2} \nn\\ 
&& - 4337 ( \vec{p}_{2}\cdot\vec{n})^{3} \big) - 2088 \vec{p}_{1}\cdot\vec{S}_{1} \vec{p}_{2}\cdot\vec{S}_{2} ( \vec{p}_{2}\cdot\vec{n})^{2} + 2088 \vec{p}_{2}\cdot\vec{S}_{1} \vec{p}_{2}\cdot\vec{S}_{2} \vec{p}_{1}\cdot\vec{n} \vec{p}_{2}\cdot\vec{n} \nn\\ 
&& - \vec{S}_{1}\cdot\vec{S}_{2} \big( 1342 \vec{p}_{1}\cdot\vec{n} \vec{p}_{2}\cdot\vec{n} p_{2}^2 - 3756 \vec{p}_{1}\cdot\vec{p}_{2} ( \vec{p}_{2}\cdot\vec{n})^{2} - 1630 p_{2}^2 ( \vec{p}_{2}\cdot\vec{n})^{2} \nn\\ 
&& + 588 \vec{p}_{1}\cdot\vec{n} ( \vec{p}_{2}\cdot\vec{n})^{3} + 4337 ( \vec{p}_{2}\cdot\vec{n})^{4} \big) \Big] + 	\frac{7 G^2 m_{1}}{4 m_{2}{}^4 r{}^4} \Big[ \vec{p}_{2}\cdot\vec{S}_{1} \vec{S}_{2}\cdot\vec{n} \vec{p}_{2}\cdot\vec{n} p_{2}^2 \nn\\ 
&& - \vec{S}_{1}\cdot\vec{S}_{2} p_{2}^2 ( \vec{p}_{2}\cdot\vec{n})^{2} \Big]\nn\\ && - 	\frac{1393 G^3 m_{1}{}^2}{20 m_{2}{}^2 r{}^5} \Big[ \vec{p}_{2}\cdot\vec{S}_{1} \vec{S}_{2}\cdot\vec{n} \vec{p}_{2}\cdot\vec{n} - \vec{S}_{1}\cdot\vec{S}_{2} ( \vec{p}_{2}\cdot\vec{n})^{2} \Big] \nn\\ 
&& + 	\frac{G^3 m_{2}{}^2}{100 m_{1}{}^2 r{}^5} \Big[ 25 \vec{S}_{1}\cdot\vec{n} \vec{S}_{2}\cdot\vec{n} \big( 193 p_{1}^2 + 25 ( \vec{p}_{1}\cdot\vec{n})^{2} \big) - 4415 \vec{p}_{1}\cdot\vec{S}_{1} \vec{S}_{2}\cdot\vec{n} \vec{p}_{1}\cdot\vec{n} \nn\\ 
&& - 4325 \vec{S}_{1}\cdot\vec{n} \vec{p}_{1}\cdot\vec{S}_{2} \vec{p}_{1}\cdot\vec{n} - \vec{S}_{1}\cdot\vec{S}_{2} \big( 6337 p_{1}^2 -3820 ( \vec{p}_{1}\cdot\vec{n})^{2} \big) + 5831 \vec{p}_{1}\cdot\vec{S}_{1} \vec{p}_{1}\cdot\vec{S}_{2} \Big] \nn\\ 
&& + 	\frac{G^3 m_{2}}{960 m_{1} r{}^5} \Big[ 5 \vec{S}_{1}\cdot\vec{n} \vec{S}_{2}\cdot\vec{n} \big( {(71716 - 2115 \pi^2)} p_{1}^2 - {(122788 - 14805 \pi^2)} ( \vec{p}_{1}\cdot\vec{n})^{2} \big) \nn\\ 
&& - 2 \vec{p}_{1}\cdot\vec{S}_{1} \vec{S}_{2}\cdot\vec{n} \big( {(9970 + 10575 \pi^2)} \vec{p}_{1}\cdot\vec{n} + 23128 \vec{p}_{2}\cdot\vec{n} \big) + 83284 \vec{p}_{2}\cdot\vec{S}_{1} \vec{S}_{2}\cdot\vec{n} \vec{p}_{1}\cdot\vec{n} \nn\\ 
&& - 140640 \vec{S}_{1}\cdot\vec{n} \vec{p}_{1}\cdot\vec{S}_{2} \vec{p}_{1}\cdot\vec{n} - \vec{S}_{1}\cdot\vec{S}_{2} \big( {(290140 - 2115 \pi^2)} p_{1}^2 + 37028 \vec{p}_{1}\cdot\vec{n} \vec{p}_{2}\cdot\vec{n} \nn\\ 
&& - {(423720 - 31725 \pi^2)} ( \vec{p}_{1}\cdot\vec{n})^{2} \big) + 10 {(18220 + 423 \pi^2)} \vec{p}_{1}\cdot\vec{S}_{1} \vec{p}_{1}\cdot\vec{S}_{2} \Big] \nn\\ 
&& - 	\frac{G^3 m_{1}}{2400 m_{2} r{}^5} \Big[ 200 \vec{S}_{1}\cdot\vec{n} \vec{S}_{2}\cdot\vec{n} \big( 1987 \vec{p}_{1}\cdot\vec{p}_{2} -3652 \vec{p}_{1}\cdot\vec{n} \vec{p}_{2}\cdot\vec{n} \big) \nn\\ 
&& - 174160 \vec{p}_{1}\cdot\vec{S}_{1} \vec{S}_{2}\cdot\vec{n} \vec{p}_{2}\cdot\vec{n} + 5 \vec{p}_{2}\cdot\vec{S}_{1} \vec{S}_{2}\cdot\vec{n} \big( 27808 \vec{p}_{1}\cdot\vec{n} \nn\\ 
&& + {(80650 + 10575 \pi^2)} \vec{p}_{2}\cdot\vec{n} \big) + 120000 \vec{S}_{1}\cdot\vec{n} \vec{p}_{1}\cdot\vec{S}_{2} \vec{p}_{2}\cdot\vec{n} \nn\\ 
&& - 254400 \vec{S}_{1}\cdot\vec{n} \vec{p}_{2}\cdot\vec{S}_{2} \vec{p}_{1}\cdot\vec{n} - \vec{S}_{1}\cdot\vec{S}_{2} \big( 395416 \vec{p}_{1}\cdot\vec{p}_{2} -605760 \vec{p}_{1}\cdot\vec{n} \vec{p}_{2}\cdot\vec{n} \nn\\ 
&& + {(403250 + 52875 \pi^2)} ( \vec{p}_{2}\cdot\vec{n})^{2} \big) + 9104 \vec{p}_{2}\cdot\vec{S}_{1} \vec{p}_{1}\cdot\vec{S}_{2} + 199904 \vec{p}_{1}\cdot\vec{S}_{1} \vec{p}_{2}\cdot\vec{S}_{2} \Big] \nn\\ 
&& + 	\frac{G^3}{4800 r{}^5} \Big[ 25 \vec{S}_{1}\cdot\vec{n} \vec{S}_{2}\cdot\vec{n} \big( 17744 p_{1}^2 - {(78366 - 2115 \pi^2)} \vec{p}_{1}\cdot\vec{p}_{2} \nn\\ 
&& + {(123654 - 14805 \pi^2)} \vec{p}_{1}\cdot\vec{n} \vec{p}_{2}\cdot\vec{n} - 58280 ( \vec{p}_{1}\cdot\vec{n})^{2} \big) + 10 \vec{p}_{1}\cdot\vec{S}_{1} \vec{S}_{2}\cdot\vec{n} \big( 50548 \vec{p}_{1}\cdot\vec{n} \nn\\ 
&& - {(930 - 10575 \pi^2)} \vec{p}_{2}\cdot\vec{n} \big) + 10 \vec{p}_{2}\cdot\vec{S}_{1} \vec{S}_{2}\cdot\vec{n} \big( {(75405 + 10575 \pi^2)} \vec{p}_{1}\cdot\vec{n} \nn\\ 
&& - 46622 \vec{p}_{2}\cdot\vec{n} \big) + 150 \vec{S}_{1}\cdot\vec{n} \vec{p}_{1}\cdot\vec{S}_{2} \big( 484 \vec{p}_{1}\cdot\vec{n} + 1445 \vec{p}_{2}\cdot\vec{n} \big) \nn\\ 
&& + 653100 \vec{S}_{1}\cdot\vec{n} \vec{p}_{2}\cdot\vec{S}_{2} \vec{p}_{1}\cdot\vec{n} - \vec{S}_{1}\cdot\vec{S}_{2} \big( 442456 p_{1}^2 - {(1557600 - 10575 \pi^2)} \vec{p}_{1}\cdot\vec{p}_{2} \nn\\ 
&& + {(2795100 - 52875 \pi^2)} \vec{p}_{1}\cdot\vec{n} \vec{p}_{2}\cdot\vec{n} - 638160 ( \vec{p}_{1}\cdot\vec{n})^{2} - 466220 ( \vec{p}_{2}\cdot\vec{n})^{2} \big) \nn\\ 
&& + 96928 \vec{p}_{1}\cdot\vec{S}_{1} \vec{p}_{1}\cdot\vec{S}_{2} - 75 {(5536 + 141 \pi^2)} \vec{p}_{2}\cdot\vec{S}_{1} \vec{p}_{1}\cdot\vec{S}_{2} \nn\\ 
&& - 75 {(7804 + 141 \pi^2)} \vec{p}_{1}\cdot\vec{S}_{1} \vec{p}_{2}\cdot\vec{S}_{2} \Big]\nn\\ && - 	\frac{G^4 m_{1}{}^3}{100 r{}^6} \Big[ 7531 \vec{S}_{1}\cdot\vec{n} \vec{S}_{2}\cdot\vec{n} - 4537 \vec{S}_{1}\cdot\vec{S}_{2} \Big] \nn\\ 
&& - 	\frac{G^4 m_{1}{}^2 m_{2}}{600 r{}^6} \Big[ 3 {(143047 - 6075 \pi^2)} \vec{S}_{1}\cdot\vec{n} \vec{S}_{2}\cdot\vec{n} - {(183307 - 6075 \pi^2)} \vec{S}_{1}\cdot\vec{S}_{2} \Big],
\eea
%%%%%%%%%%%%%%%%

%%%%%%%%%%%%%%%%
\bea
H^{\text{N}^3\text{LO}}_{\text{S}_1^2} &= &   	\frac{3 G}{8 m_{1}{}^2 m_{2}{}^4 r{}^3} \Big[ 4 \vec{S}_{1}\cdot\vec{n} \vec{p}_{1}\cdot\vec{S}_{1} \big( 2 \vec{p}_{2}\cdot\vec{n} p_{2}^{4} -5 p_{2}^2 ( \vec{p}_{2}\cdot\vec{n})^{3} \big) + \vec{S}_{1}\cdot\vec{n} \vec{p}_{2}\cdot\vec{S}_{1} \big( \vec{p}_{1}\cdot\vec{n} p_{2}^{4} \nn\\ 
&& + 10 \vec{p}_{1}\cdot\vec{n} p_{2}^2 ( \vec{p}_{2}\cdot\vec{n})^{2} -35 \vec{p}_{1}\cdot\vec{n} ( \vec{p}_{2}\cdot\vec{n})^{4} \big) + \vec{p}_{1}\cdot\vec{S}_{1} \vec{p}_{2}\cdot\vec{S}_{1} \big( 5 p_{2}^{4} -38 p_{2}^2 ( \vec{p}_{2}\cdot\vec{n})^{2} \nn\\ 
&& + 45 ( \vec{p}_{2}\cdot\vec{n})^{4} \big) - S_{1}^2 \big( 5 \vec{p}_{1}\cdot\vec{p}_{2} p_{2}^{4} -38 \vec{p}_{1}\cdot\vec{p}_{2} p_{2}^2 ( \vec{p}_{2}\cdot\vec{n})^{2} + 8 \vec{p}_{1}\cdot\vec{n} \vec{p}_{2}\cdot\vec{n} p_{2}^{4} \nn\\ 
&& + 45 \vec{p}_{1}\cdot\vec{p}_{2} ( \vec{p}_{2}\cdot\vec{n})^{4} - 20 \vec{p}_{1}\cdot\vec{n} p_{2}^2 ( \vec{p}_{2}\cdot\vec{n})^{3} \big) - \big( \vec{p}_{1}\cdot\vec{p}_{2} p_{2}^{4} \nn\\ 
&& + 10 \vec{p}_{1}\cdot\vec{p}_{2} p_{2}^2 ( \vec{p}_{2}\cdot\vec{n})^{2} -35 \vec{p}_{1}\cdot\vec{p}_{2} ( \vec{p}_{2}\cdot\vec{n})^{4} \big) ( \vec{S}_{1}\cdot\vec{n})^{2} \Big] \nn\\ 
&& + 	\frac{3 G m_{2}}{128 m_{1}{}^7 r{}^3} \Big[ 38 \vec{S}_{1}\cdot\vec{n} \vec{p}_{1}\cdot\vec{S}_{1} \vec{p}_{1}\cdot\vec{n} p_{1}^{4} + 3 S_{1}^2 \big( 10 p_{1}^{6} -19 ( \vec{p}_{1}\cdot\vec{n})^{2} p_{1}^{4} \big) - 33 p_{1}^{6} ( \vec{S}_{1}\cdot\vec{n})^{2} \nn\\ 
&& - 2 \big( 15 p_{1}^{4} -26 p_{1}^2 ( \vec{p}_{1}\cdot\vec{n})^{2} \big) ( \vec{p}_{1}\cdot\vec{S}_{1})^{2} \Big] - 	\frac{G}{32 m_{1}{}^6 r{}^3} \Big[ 12 \vec{S}_{1}\cdot\vec{n} \vec{p}_{1}\cdot\vec{S}_{1} \big( \vec{p}_{1}\cdot\vec{n} p_{1}^2 \vec{p}_{1}\cdot\vec{p}_{2} \nn\\ 
&& + 3 \vec{p}_{2}\cdot\vec{n} p_{1}^{4} -10 p_{1}^2 \vec{p}_{2}\cdot\vec{n} ( \vec{p}_{1}\cdot\vec{n})^{2} \big) - 108 \vec{S}_{1}\cdot\vec{n} \vec{p}_{2}\cdot\vec{S}_{1} \vec{p}_{1}\cdot\vec{n} p_{1}^{4} + 12 \vec{p}_{1}\cdot\vec{S}_{1} \vec{p}_{2}\cdot\vec{S}_{1} \big( p_{1}^{4} \nn\\ 
&& + 5 p_{1}^2 ( \vec{p}_{1}\cdot\vec{n})^{2} \big) + S_{1}^2 \big( 59 \vec{p}_{1}\cdot\vec{p}_{2} p_{1}^{4} -264 p_{1}^2 \vec{p}_{1}\cdot\vec{p}_{2} ( \vec{p}_{1}\cdot\vec{n})^{2} - 189 \vec{p}_{1}\cdot\vec{n} \vec{p}_{2}\cdot\vec{n} p_{1}^{4} \nn\\ 
&& + 360 p_{1}^2 \vec{p}_{2}\cdot\vec{n} ( \vec{p}_{1}\cdot\vec{n})^{3} \big) + 60 \vec{p}_{1}\cdot\vec{p}_{2} p_{1}^{4} ( \vec{S}_{1}\cdot\vec{n})^{2} - \big( 71 p_{1}^2 \vec{p}_{1}\cdot\vec{p}_{2} -153 \vec{p}_{1}\cdot\vec{n} p_{1}^2 \vec{p}_{2}\cdot\vec{n} \nn\\ 
&& - 240 \vec{p}_{1}\cdot\vec{p}_{2} ( \vec{p}_{1}\cdot\vec{n})^{2} + 240 \vec{p}_{2}\cdot\vec{n} ( \vec{p}_{1}\cdot\vec{n})^{3} \big) ( \vec{p}_{1}\cdot\vec{S}_{1})^{2} \Big] \nn\\ 
&& + 	\frac{G}{32 m_{1}{}^4 m_{2}{}^2 r{}^3} \Big[ 3 \vec{S}_{1}\cdot\vec{n} \vec{p}_{1}\cdot\vec{S}_{1} \big( p_{1}^2 \vec{p}_{2}\cdot\vec{n} p_{2}^2 - 12 \vec{p}_{1}\cdot\vec{n} \vec{p}_{1}\cdot\vec{p}_{2} p_{2}^2 \nn\\ 
&& + 4 \vec{p}_{2}\cdot\vec{n} ( \vec{p}_{1}\cdot\vec{p}_{2})^{2} -70 \vec{p}_{2}\cdot\vec{n} p_{2}^2 ( \vec{p}_{1}\cdot\vec{n})^{2} - 5 p_{1}^2 ( \vec{p}_{2}\cdot\vec{n})^{3} + 20 \vec{p}_{1}\cdot\vec{n} \vec{p}_{1}\cdot\vec{p}_{2} ( \vec{p}_{2}\cdot\vec{n})^{2} \nn\\ 
&& + 210 ( \vec{p}_{1}\cdot\vec{n})^{2} ( \vec{p}_{2}\cdot\vec{n})^{3} \big) - 12 \vec{S}_{1}\cdot\vec{n} \vec{p}_{2}\cdot\vec{S}_{1} \big( 4 \vec{p}_{1}\cdot\vec{n} p_{1}^2 p_{2}^2 -25 \vec{p}_{1}\cdot\vec{n} p_{1}^2 ( \vec{p}_{2}\cdot\vec{n})^{2} \big) \nn\\ 
&& + 3 \vec{p}_{1}\cdot\vec{S}_{1} \vec{p}_{2}\cdot\vec{S}_{1} \big( 17 p_{1}^2 p_{2}^2 + 52 ( \vec{p}_{1}\cdot\vec{p}_{2})^{2} -280 \vec{p}_{1}\cdot\vec{n} \vec{p}_{2}\cdot\vec{n} \vec{p}_{1}\cdot\vec{p}_{2} - 42 p_{2}^2 ( \vec{p}_{1}\cdot\vec{n})^{2} \nn\\ 
&& - 49 p_{1}^2 ( \vec{p}_{2}\cdot\vec{n})^{2} + 190 ( \vec{p}_{1}\cdot\vec{n})^{2} ( \vec{p}_{2}\cdot\vec{n})^{2} \big) - S_{1}^2 \big( 295 p_{1}^2 \vec{p}_{1}\cdot\vec{p}_{2} p_{2}^2 \nn\\ 
&& + 156 ( \vec{p}_{1}\cdot\vec{p}_{2})^{3} -513 \vec{p}_{1}\cdot\vec{n} p_{1}^2 \vec{p}_{2}\cdot\vec{n} p_{2}^2 - 144 \vec{p}_{1}\cdot\vec{p}_{2} p_{2}^2 ( \vec{p}_{1}\cdot\vec{n})^{2} \nn\\ 
&& - 939 p_{1}^2 \vec{p}_{1}\cdot\vec{p}_{2} ( \vec{p}_{2}\cdot\vec{n})^{2} - 828 \vec{p}_{1}\cdot\vec{n} \vec{p}_{2}\cdot\vec{n} ( \vec{p}_{1}\cdot\vec{p}_{2})^{2} -240 \vec{p}_{2}\cdot\vec{n} p_{2}^2 ( \vec{p}_{1}\cdot\vec{n})^{3} \nn\\ 
&& + 945 \vec{p}_{1}\cdot\vec{n} p_{1}^2 ( \vec{p}_{2}\cdot\vec{n})^{3} + 630 \vec{p}_{1}\cdot\vec{p}_{2} ( \vec{p}_{1}\cdot\vec{n})^{2} ( \vec{p}_{2}\cdot\vec{n})^{2} + 630 ( \vec{p}_{1}\cdot\vec{n})^{3} ( \vec{p}_{2}\cdot\vec{n})^{3} \big) \nn\\ 
&& + 6 \big( 11 p_{1}^2 \vec{p}_{1}\cdot\vec{p}_{2} p_{2}^2 -5 \vec{p}_{1}\cdot\vec{n} p_{1}^2 \vec{p}_{2}\cdot\vec{n} p_{2}^2 - 50 p_{1}^2 \vec{p}_{1}\cdot\vec{p}_{2} ( \vec{p}_{2}\cdot\vec{n})^{2} \big) ( \vec{S}_{1}\cdot\vec{n})^{2} \nn\\ 
&& + 4 \big( 61 \vec{p}_{1}\cdot\vec{p}_{2} p_{2}^2 -129 \vec{p}_{1}\cdot\vec{n} \vec{p}_{2}\cdot\vec{n} p_{2}^2 - 198 \vec{p}_{1}\cdot\vec{p}_{2} ( \vec{p}_{2}\cdot\vec{n})^{2} \nn\\ 
&& + 240 \vec{p}_{1}\cdot\vec{n} ( \vec{p}_{2}\cdot\vec{n})^{3} \big) ( \vec{p}_{1}\cdot\vec{S}_{1})^{2} \Big] + 	\frac{G}{64 m_{1}{}^5 m_{2} r{}^3} \Big[ 6 \vec{S}_{1}\cdot\vec{n} \vec{p}_{1}\cdot\vec{S}_{1} \big( 12 p_{1}^2 \vec{p}_{2}\cdot\vec{n} \vec{p}_{1}\cdot\vec{p}_{2} \nn\\ 
&& + 15 \vec{p}_{1}\cdot\vec{n} p_{1}^2 p_{2}^2 + 14 \vec{p}_{1}\cdot\vec{n} ( \vec{p}_{1}\cdot\vec{p}_{2})^{2} -30 \vec{p}_{2}\cdot\vec{n} \vec{p}_{1}\cdot\vec{p}_{2} ( \vec{p}_{1}\cdot\vec{n})^{2} + 15 p_{2}^2 ( \vec{p}_{1}\cdot\vec{n})^{3} \nn\\ 
&& - 20 \vec{p}_{1}\cdot\vec{n} p_{1}^2 ( \vec{p}_{2}\cdot\vec{n})^{2} -105 ( \vec{p}_{1}\cdot\vec{n})^{3} ( \vec{p}_{2}\cdot\vec{n})^{2} \big) - 12 \vec{S}_{1}\cdot\vec{n} \vec{p}_{2}\cdot\vec{S}_{1} \big( 14 \vec{p}_{1}\cdot\vec{n} p_{1}^2 \vec{p}_{1}\cdot\vec{p}_{2} \nn\\ 
&& + \vec{p}_{2}\cdot\vec{n} p_{1}^{4} + 5 p_{1}^2 \vec{p}_{2}\cdot\vec{n} ( \vec{p}_{1}\cdot\vec{n})^{2} \big) - 4 \vec{p}_{1}\cdot\vec{S}_{1} \vec{p}_{2}\cdot\vec{S}_{1} \big( 8 p_{1}^2 \vec{p}_{1}\cdot\vec{p}_{2} -42 \vec{p}_{1}\cdot\vec{n} p_{1}^2 \vec{p}_{2}\cdot\vec{n} \nn\\ 
&& - 117 \vec{p}_{1}\cdot\vec{p}_{2} ( \vec{p}_{1}\cdot\vec{n})^{2} + 135 \vec{p}_{2}\cdot\vec{n} ( \vec{p}_{1}\cdot\vec{n})^{3} \big) + S_{1}^2 \big( 426 p_{1}^2 ( \vec{p}_{1}\cdot\vec{p}_{2})^{2} \nn\\ 
&& + 104 p_{2}^2 p_{1}^{4} -2076 \vec{p}_{1}\cdot\vec{n} p_{1}^2 \vec{p}_{2}\cdot\vec{n} \vec{p}_{1}\cdot\vec{p}_{2} - 333 p_{1}^2 p_{2}^2 ( \vec{p}_{1}\cdot\vec{n})^{2} - 492 ( \vec{p}_{1}\cdot\vec{n})^{2} ( \vec{p}_{1}\cdot\vec{p}_{2})^{2} \nn\\ 
&& - 318 ( \vec{p}_{2}\cdot\vec{n})^{2} p_{1}^{4} + 600 \vec{p}_{2}\cdot\vec{n} \vec{p}_{1}\cdot\vec{p}_{2} ( \vec{p}_{1}\cdot\vec{n})^{3} - 75 p_{2}^2 ( \vec{p}_{1}\cdot\vec{n})^{4} \nn\\ 
&& + 1635 p_{1}^2 ( \vec{p}_{1}\cdot\vec{n})^{2} ( \vec{p}_{2}\cdot\vec{n})^{2} + 525 ( \vec{p}_{1}\cdot\vec{n})^{4} ( \vec{p}_{2}\cdot\vec{n})^{2} \big) + 3 \big( 18 p_{1}^2 ( \vec{p}_{1}\cdot\vec{p}_{2})^{2} + 15 p_{2}^2 p_{1}^{4} \nn\\ 
&& + 60 \vec{p}_{1}\cdot\vec{n} p_{1}^2 \vec{p}_{2}\cdot\vec{n} \vec{p}_{1}\cdot\vec{p}_{2} - 5 p_{1}^2 p_{2}^2 ( \vec{p}_{1}\cdot\vec{n})^{2} - 55 ( \vec{p}_{2}\cdot\vec{n})^{2} p_{1}^{4} \nn\\ 
&& + 35 p_{1}^2 ( \vec{p}_{1}\cdot\vec{n})^{2} ( \vec{p}_{2}\cdot\vec{n})^{2} \big) ( \vec{S}_{1}\cdot\vec{n})^{2} - 2 \big( 52 p_{1}^2 p_{2}^2 \nn\\ 
&& + 202 ( \vec{p}_{1}\cdot\vec{p}_{2})^{2} -924 \vec{p}_{1}\cdot\vec{n} \vec{p}_{2}\cdot\vec{n} \vec{p}_{1}\cdot\vec{p}_{2} - 99 p_{2}^2 ( \vec{p}_{1}\cdot\vec{n})^{2} - 159 p_{1}^2 ( \vec{p}_{2}\cdot\vec{n})^{2} \nn\\ 
&& + 675 ( \vec{p}_{1}\cdot\vec{n})^{2} ( \vec{p}_{2}\cdot\vec{n})^{2} \big) ( \vec{p}_{1}\cdot\vec{S}_{1})^{2} + 2 \big( 5 p_{1}^{4} + 27 p_{1}^2 ( \vec{p}_{1}\cdot\vec{n})^{2} \big) ( \vec{p}_{2}\cdot\vec{S}_{1})^{2} \Big] \nn\\ 
&& - 	\frac{G}{64 m_{1}{}^3 m_{2}{}^3 r{}^3} \Big[ 6 \vec{S}_{1}\cdot\vec{n} \vec{p}_{1}\cdot\vec{S}_{1} \big( 48 \vec{p}_{2}\cdot\vec{n} \vec{p}_{1}\cdot\vec{p}_{2} p_{2}^2 + 13 \vec{p}_{1}\cdot\vec{n} p_{2}^{4} -40 \vec{p}_{1}\cdot\vec{p}_{2} ( \vec{p}_{2}\cdot\vec{n})^{3} \nn\\ 
&& - 100 \vec{p}_{1}\cdot\vec{n} p_{2}^2 ( \vec{p}_{2}\cdot\vec{n})^{2} \big) + 960 \vec{S}_{1}\cdot\vec{n} \vec{p}_{2}\cdot\vec{S}_{1} \vec{p}_{1}\cdot\vec{n} \vec{p}_{1}\cdot\vec{p}_{2} ( \vec{p}_{2}\cdot\vec{n})^{2} \nn\\ 
&& + 48 \vec{p}_{1}\cdot\vec{S}_{1} \vec{p}_{2}\cdot\vec{S}_{1} \big( 9 \vec{p}_{1}\cdot\vec{p}_{2} p_{2}^2 -24 \vec{p}_{1}\cdot\vec{n} \vec{p}_{2}\cdot\vec{n} p_{2}^2 - 32 \vec{p}_{1}\cdot\vec{p}_{2} ( \vec{p}_{2}\cdot\vec{n})^{2} \nn\\ 
&& + 45 \vec{p}_{1}\cdot\vec{n} ( \vec{p}_{2}\cdot\vec{n})^{3} \big) - S_{1}^2 \big( 432 p_{2}^2 ( \vec{p}_{1}\cdot\vec{p}_{2})^{2} + 118 p_{1}^2 p_{2}^{4} -864 \vec{p}_{1}\cdot\vec{n} \vec{p}_{2}\cdot\vec{n} \vec{p}_{1}\cdot\vec{p}_{2} p_{2}^2 \nn\\ 
&& - 1536 ( \vec{p}_{2}\cdot\vec{n})^{2} ( \vec{p}_{1}\cdot\vec{p}_{2})^{2} - 888 p_{1}^2 p_{2}^2 ( \vec{p}_{2}\cdot\vec{n})^{2} + 93 ( \vec{p}_{1}\cdot\vec{n})^{2} p_{2}^{4} + 960 p_{1}^2 ( \vec{p}_{2}\cdot\vec{n})^{4} \nn\\ 
&& + 1920 \vec{p}_{1}\cdot\vec{n} \vec{p}_{1}\cdot\vec{p}_{2} ( \vec{p}_{2}\cdot\vec{n})^{3} - 1080 p_{2}^2 ( \vec{p}_{1}\cdot\vec{n})^{2} ( \vec{p}_{2}\cdot\vec{n})^{2} + 840 ( \vec{p}_{1}\cdot\vec{n})^{2} ( \vec{p}_{2}\cdot\vec{n})^{4} \big) \nn\\ 
&& + 15 \big( p_{1}^2 p_{2}^{4} -32 ( \vec{p}_{2}\cdot\vec{n})^{2} ( \vec{p}_{1}\cdot\vec{p}_{2})^{2} - 32 p_{1}^2 p_{2}^2 ( \vec{p}_{2}\cdot\vec{n})^{2} + 56 p_{1}^2 ( \vec{p}_{2}\cdot\vec{n})^{4} \big) ( \vec{S}_{1}\cdot\vec{n})^{2} \nn\\ 
&& + 2 \big( 59 p_{2}^{4} -444 p_{2}^2 ( \vec{p}_{2}\cdot\vec{n})^{2} + 480 ( \vec{p}_{2}\cdot\vec{n})^{4} \big) ( \vec{p}_{1}\cdot\vec{S}_{1})^{2} \nn\\ 
&& - 480 ( \vec{p}_{1}\cdot\vec{n})^{2} ( \vec{p}_{2}\cdot\vec{n})^{2} ( \vec{p}_{2}\cdot\vec{S}_{1})^{2} \Big]\nn\\ && +  	\frac{G^2 m_{2}{}^2}{16 m_{1}{}^5 r{}^4} \Big[ 3 \vec{S}_{1}\cdot\vec{n} \vec{p}_{1}\cdot\vec{S}_{1} \big( 79 \vec{p}_{1}\cdot\vec{n} p_{1}^2 -16 ( \vec{p}_{1}\cdot\vec{n})^{3} \big) + 4 S_{1}^2 \big( 31 p_{1}^{4} -59 p_{1}^2 ( \vec{p}_{1}\cdot\vec{n})^{2} \nn\\ 
&& + 6 ( \vec{p}_{1}\cdot\vec{n})^{4} \big) - \big( 167 p_{1}^{4} -24 p_{1}^2 ( \vec{p}_{1}\cdot\vec{n})^{2} \big) ( \vec{S}_{1}\cdot\vec{n})^{2} \nn\\ 
&& - \big( 121 p_{1}^2 -163 ( \vec{p}_{1}\cdot\vec{n})^{2} \big) ( \vec{p}_{1}\cdot\vec{S}_{1})^{2} \Big] + 	\frac{G^2 m_{2}}{32 m_{1}{}^4 r{}^4} \Big[ \vec{S}_{1}\cdot\vec{n} \vec{p}_{1}\cdot\vec{S}_{1} \big( 79 \vec{p}_{1}\cdot\vec{n} p_{1}^2 \nn\\ 
&& - 574 p_{1}^2 \vec{p}_{2}\cdot\vec{n} + 452 \vec{p}_{1}\cdot\vec{n} \vec{p}_{1}\cdot\vec{p}_{2} -7288 ( \vec{p}_{1}\cdot\vec{n})^{3} + 258 \vec{p}_{2}\cdot\vec{n} ( \vec{p}_{1}\cdot\vec{n})^{2} \big) \nn\\ 
&& + 731 \vec{S}_{1}\cdot\vec{n} \vec{p}_{2}\cdot\vec{S}_{1} \vec{p}_{1}\cdot\vec{n} p_{1}^2 + 2 \vec{p}_{1}\cdot\vec{S}_{1} \vec{p}_{2}\cdot\vec{S}_{1} \big( 217 p_{1}^2 -851 ( \vec{p}_{1}\cdot\vec{n})^{2} \big) \nn\\ 
&& + S_{1}^2 \big( 26 p_{1}^2 \vec{p}_{1}\cdot\vec{p}_{2} + 119 p_{1}^{4} + 148 \vec{p}_{1}\cdot\vec{n} p_{1}^2 \vec{p}_{2}\cdot\vec{n} - 1260 p_{1}^2 ( \vec{p}_{1}\cdot\vec{n})^{2} \nn\\ 
&& + 1012 \vec{p}_{1}\cdot\vec{p}_{2} ( \vec{p}_{1}\cdot\vec{n})^{2} + 1144 ( \vec{p}_{1}\cdot\vec{n})^{4} - 480 \vec{p}_{2}\cdot\vec{n} ( \vec{p}_{1}\cdot\vec{n})^{3} \big) - \big( 552 p_{1}^2 \vec{p}_{1}\cdot\vec{p}_{2} \nn\\ 
&& + 631 p_{1}^{4} -129 \vec{p}_{1}\cdot\vec{n} p_{1}^2 \vec{p}_{2}\cdot\vec{n} - 3888 p_{1}^2 ( \vec{p}_{1}\cdot\vec{n})^{2} -1120 ( \vec{p}_{1}\cdot\vec{n})^{4} \big) ( \vec{S}_{1}\cdot\vec{n})^{2} - \big( 199 p_{1}^2 \nn\\ 
&& + 460 \vec{p}_{1}\cdot\vec{p}_{2} -578 \vec{p}_{1}\cdot\vec{n} \vec{p}_{2}\cdot\vec{n} - 2900 ( \vec{p}_{1}\cdot\vec{n})^{2} \big) ( \vec{p}_{1}\cdot\vec{S}_{1})^{2} \Big] \nn\\ 
&& + 	\frac{G^2}{32 m_{1}{}^3 r{}^4} \Big[ 2 \vec{S}_{1}\cdot\vec{n} \vec{p}_{1}\cdot\vec{S}_{1} \big( 303 p_{1}^2 \vec{p}_{2}\cdot\vec{n} + 362 \vec{p}_{1}\cdot\vec{n} \vec{p}_{1}\cdot\vec{p}_{2} + 420 \vec{p}_{2}\cdot\vec{n} \vec{p}_{1}\cdot\vec{p}_{2} \nn\\ 
&& - 310 \vec{p}_{1}\cdot\vec{n} p_{2}^2 + 5820 \vec{p}_{2}\cdot\vec{n} ( \vec{p}_{1}\cdot\vec{n})^{2} + 1392 \vec{p}_{1}\cdot\vec{n} ( \vec{p}_{2}\cdot\vec{n})^{2} \big) \nn\\ 
&& + 2 \vec{S}_{1}\cdot\vec{n} \vec{p}_{2}\cdot\vec{S}_{1} \big( 750 \vec{p}_{1}\cdot\vec{n} p_{1}^2 + 23 p_{1}^2 \vec{p}_{2}\cdot\vec{n} - 1231 \vec{p}_{1}\cdot\vec{n} \vec{p}_{1}\cdot\vec{p}_{2} + 3208 ( \vec{p}_{1}\cdot\vec{n})^{3} \nn\\ 
&& - 54 \vec{p}_{2}\cdot\vec{n} ( \vec{p}_{1}\cdot\vec{n})^{2} \big) + 2 \vec{p}_{1}\cdot\vec{S}_{1} \vec{p}_{2}\cdot\vec{S}_{1} \big( 29 p_{1}^2 + 302 \vec{p}_{1}\cdot\vec{p}_{2} + 262 \vec{p}_{1}\cdot\vec{n} \vec{p}_{2}\cdot\vec{n} \nn\\ 
&& - 2738 ( \vec{p}_{1}\cdot\vec{n})^{2} \big) + S_{1}^2 \big( 792 p_{1}^2 \vec{p}_{1}\cdot\vec{p}_{2} - 48 p_{1}^2 p_{2}^2 - 525 ( \vec{p}_{1}\cdot\vec{p}_{2})^{2} + 30 \vec{p}_{1}\cdot\vec{n} p_{1}^2 \vec{p}_{2}\cdot\vec{n} \nn\\ 
&& + 66 \vec{p}_{1}\cdot\vec{n} \vec{p}_{2}\cdot\vec{n} \vec{p}_{1}\cdot\vec{p}_{2} + 1684 \vec{p}_{1}\cdot\vec{p}_{2} ( \vec{p}_{1}\cdot\vec{n})^{2} - 932 p_{2}^2 ( \vec{p}_{1}\cdot\vec{n})^{2} \nn\\ 
&& + 1028 p_{1}^2 ( \vec{p}_{2}\cdot\vec{n})^{2} -2792 \vec{p}_{2}\cdot\vec{n} ( \vec{p}_{1}\cdot\vec{n})^{3} + 588 ( \vec{p}_{1}\cdot\vec{n})^{2} ( \vec{p}_{2}\cdot\vec{n})^{2} \big) + \big( 56 p_{1}^2 \vec{p}_{1}\cdot\vec{p}_{2} \nn\\ 
&& + 634 p_{1}^2 p_{2}^2 + 824 ( \vec{p}_{1}\cdot\vec{p}_{2})^{2} -5208 \vec{p}_{1}\cdot\vec{n} p_{1}^2 \vec{p}_{2}\cdot\vec{n} - 1008 \vec{p}_{1}\cdot\vec{n} \vec{p}_{2}\cdot\vec{n} \vec{p}_{1}\cdot\vec{p}_{2} \nn\\ 
&& - 4752 \vec{p}_{1}\cdot\vec{p}_{2} ( \vec{p}_{1}\cdot\vec{n})^{2} + 264 p_{2}^2 ( \vec{p}_{1}\cdot\vec{n})^{2} - 2037 p_{1}^2 ( \vec{p}_{2}\cdot\vec{n})^{2} -3776 \vec{p}_{2}\cdot\vec{n} ( \vec{p}_{1}\cdot\vec{n})^{3} \nn\\ 
&& - 540 ( \vec{p}_{1}\cdot\vec{n})^{2} ( \vec{p}_{2}\cdot\vec{n})^{2} \big) ( \vec{S}_{1}\cdot\vec{n})^{2} - \big( 714 \vec{p}_{1}\cdot\vec{p}_{2} - 295 p_{2}^2 + 1140 \vec{p}_{1}\cdot\vec{n} \vec{p}_{2}\cdot\vec{n} \nn\\ 
&& + 1347 ( \vec{p}_{2}\cdot\vec{n})^{2} \big) ( \vec{p}_{1}\cdot\vec{S}_{1})^{2} - 2 \big( 187 p_{1}^2 -752 ( \vec{p}_{1}\cdot\vec{n})^{2} \big) ( \vec{p}_{2}\cdot\vec{S}_{1})^{2} \Big] \nn\\ 
&& - 	\frac{G^2}{96 m_{1}{}^2 m_{2} r{}^4} \Big[ 2 \vec{S}_{1}\cdot\vec{n} \vec{p}_{1}\cdot\vec{S}_{1} \big( 2426 \vec{p}_{2}\cdot\vec{n} \vec{p}_{1}\cdot\vec{p}_{2} - 872 \vec{p}_{1}\cdot\vec{n} p_{2}^2 + 777 \vec{p}_{2}\cdot\vec{n} p_{2}^2 \nn\\ 
&& + 12 \vec{p}_{1}\cdot\vec{n} ( \vec{p}_{2}\cdot\vec{n})^{2} + 1107 ( \vec{p}_{2}\cdot\vec{n})^{3} \big) + 2 \vec{S}_{1}\cdot\vec{n} \vec{p}_{2}\cdot\vec{S}_{1} \big( 1366 p_{1}^2 \vec{p}_{2}\cdot\vec{n} \nn\\ 
&& + 5504 \vec{p}_{1}\cdot\vec{n} \vec{p}_{1}\cdot\vec{p}_{2} - 597 \vec{p}_{2}\cdot\vec{n} \vec{p}_{1}\cdot\vec{p}_{2} - 1707 \vec{p}_{1}\cdot\vec{n} p_{2}^2 + 11928 \vec{p}_{2}\cdot\vec{n} ( \vec{p}_{1}\cdot\vec{n})^{2} \nn\\ 
&& + 4914 \vec{p}_{1}\cdot\vec{n} ( \vec{p}_{2}\cdot\vec{n})^{2} \big) - \vec{p}_{1}\cdot\vec{S}_{1} \vec{p}_{2}\cdot\vec{S}_{1} \big( 5062 \vec{p}_{1}\cdot\vec{p}_{2} - 861 p_{2}^2 + 4028 \vec{p}_{1}\cdot\vec{n} \vec{p}_{2}\cdot\vec{n} \nn\\ 
&& + 3213 ( \vec{p}_{2}\cdot\vec{n})^{2} \big) + S_{1}^2 \big( 550 p_{1}^2 p_{2}^2 - 213 \vec{p}_{1}\cdot\vec{p}_{2} p_{2}^2 \nn\\ 
&& + 4331 ( \vec{p}_{1}\cdot\vec{p}_{2})^{2} -5908 \vec{p}_{1}\cdot\vec{n} \vec{p}_{2}\cdot\vec{n} \vec{p}_{1}\cdot\vec{p}_{2} - 1866 \vec{p}_{1}\cdot\vec{n} \vec{p}_{2}\cdot\vec{n} p_{2}^2 + 3914 p_{2}^2 ( \vec{p}_{1}\cdot\vec{n})^{2} \nn\\ 
&& - 5200 p_{1}^2 ( \vec{p}_{2}\cdot\vec{n})^{2} + 5049 \vec{p}_{1}\cdot\vec{p}_{2} ( \vec{p}_{2}\cdot\vec{n})^{2} -804 ( \vec{p}_{1}\cdot\vec{n})^{2} ( \vec{p}_{2}\cdot\vec{n})^{2} \nn\\ 
&& - 2790 \vec{p}_{1}\cdot\vec{n} ( \vec{p}_{2}\cdot\vec{n})^{3} \big) - 2 \big( 629 p_{1}^2 p_{2}^2 - 441 \vec{p}_{1}\cdot\vec{p}_{2} p_{2}^2 + 568 ( \vec{p}_{1}\cdot\vec{p}_{2})^{2} \nn\\ 
&& + 10608 \vec{p}_{1}\cdot\vec{n} \vec{p}_{2}\cdot\vec{n} \vec{p}_{1}\cdot\vec{p}_{2} + 792 \vec{p}_{1}\cdot\vec{n} \vec{p}_{2}\cdot\vec{n} p_{2}^2 + 654 p_{2}^2 ( \vec{p}_{1}\cdot\vec{n})^{2} - 1182 p_{1}^2 ( \vec{p}_{2}\cdot\vec{n})^{2} \nn\\ 
&& + 4176 \vec{p}_{1}\cdot\vec{p}_{2} ( \vec{p}_{2}\cdot\vec{n})^{2} + 2544 ( \vec{p}_{1}\cdot\vec{n})^{2} ( \vec{p}_{2}\cdot\vec{n})^{2} + 36 \vec{p}_{1}\cdot\vec{n} ( \vec{p}_{2}\cdot\vec{n})^{3} \big) ( \vec{S}_{1}\cdot\vec{n})^{2} \nn\\ 
&& - 2 \big( 413 p_{2}^2 -2888 ( \vec{p}_{2}\cdot\vec{n})^{2} \big) ( \vec{p}_{1}\cdot\vec{S}_{1})^{2} + \big( 1043 p_{1}^2 - 816 \vec{p}_{1}\cdot\vec{p}_{2} + 3126 \vec{p}_{1}\cdot\vec{n} \vec{p}_{2}\cdot\vec{n} \nn\\ 
&& - 9560 ( \vec{p}_{1}\cdot\vec{n})^{2} \big) ( \vec{p}_{2}\cdot\vec{S}_{1})^{2} \Big] - 	\frac{G^2}{8 m_{1} m_{2}{}^2 r{}^4} \Big[ 24 \vec{S}_{1}\cdot\vec{n} \vec{p}_{1}\cdot\vec{S}_{1} \big( 4 \vec{p}_{2}\cdot\vec{n} p_{2}^2 + 19 ( \vec{p}_{2}\cdot\vec{n})^{3} \big) \nn\\ 
&& - 2 \vec{S}_{1}\cdot\vec{n} \vec{p}_{2}\cdot\vec{S}_{1} \big( 314 \vec{p}_{2}\cdot\vec{n} \vec{p}_{1}\cdot\vec{p}_{2} + 73 \vec{p}_{1}\cdot\vec{n} p_{2}^2 - 34 \vec{p}_{2}\cdot\vec{n} p_{2}^2 -183 \vec{p}_{1}\cdot\vec{n} ( \vec{p}_{2}\cdot\vec{n})^{2} \nn\\ 
&& + 15 ( \vec{p}_{2}\cdot\vec{n})^{3} \big) + 6 \vec{p}_{1}\cdot\vec{S}_{1} \vec{p}_{2}\cdot\vec{S}_{1} \big( 26 p_{2}^2 -123 ( \vec{p}_{2}\cdot\vec{n})^{2} \big) - S_{1}^2 \big( 180 \vec{p}_{1}\cdot\vec{p}_{2} p_{2}^2 + 26 p_{2}^{4} \nn\\ 
&& + 56 \vec{p}_{1}\cdot\vec{n} \vec{p}_{2}\cdot\vec{n} p_{2}^2 - 818 \vec{p}_{1}\cdot\vec{p}_{2} ( \vec{p}_{2}\cdot\vec{n})^{2} - 8 p_{2}^2 ( \vec{p}_{2}\cdot\vec{n})^{2} + 504 \vec{p}_{1}\cdot\vec{n} ( \vec{p}_{2}\cdot\vec{n})^{3} \nn\\ 
&& - 105 ( \vec{p}_{2}\cdot\vec{n})^{4} \big) + \big( 54 \vec{p}_{1}\cdot\vec{p}_{2} p_{2}^2 + 22 p_{2}^{4} + 84 \vec{p}_{1}\cdot\vec{n} \vec{p}_{2}\cdot\vec{n} p_{2}^2 + 42 \vec{p}_{1}\cdot\vec{p}_{2} ( \vec{p}_{2}\cdot\vec{n})^{2} \nn\\ 
&& + 39 p_{2}^2 ( \vec{p}_{2}\cdot\vec{n})^{2} -96 \vec{p}_{1}\cdot\vec{n} ( \vec{p}_{2}\cdot\vec{n})^{3} - 81 ( \vec{p}_{2}\cdot\vec{n})^{4} \big) ( \vec{S}_{1}\cdot\vec{n})^{2} + 7 \big( 4 \vec{p}_{1}\cdot\vec{p}_{2} + 4 p_{2}^2 \nn\\ 
&& + 56 \vec{p}_{1}\cdot\vec{n} \vec{p}_{2}\cdot\vec{n} - 19 ( \vec{p}_{2}\cdot\vec{n})^{2} \big) ( \vec{p}_{2}\cdot\vec{S}_{1})^{2} \Big] - 	\frac{G^2}{8 m_{2}{}^3 r{}^4} \Big[ 36 \vec{S}_{1}\cdot\vec{n} \vec{p}_{2}\cdot\vec{S}_{1} \vec{p}_{2}\cdot\vec{n} p_{2}^2 \nn\\ 
&& + S_{1}^2 \big( 13 p_{2}^{4} -21 p_{2}^2 ( \vec{p}_{2}\cdot\vec{n})^{2} \big) - 3 \big( 4 p_{2}^{4} + 3 p_{2}^2 ( \vec{p}_{2}\cdot\vec{n})^{2} \big) ( \vec{S}_{1}\cdot\vec{n})^{2} - 12 p_{2}^2 ( \vec{p}_{2}\cdot\vec{S}_{1})^{2} \Big]\nn\\ && +  	\frac{G^3 m_{2}{}^3}{16 m_{1}{}^3 r{}^5} \Big[ 1458 \vec{S}_{1}\cdot\vec{n} \vec{p}_{1}\cdot\vec{S}_{1} \vec{p}_{1}\cdot\vec{n} + S_{1}^2 \big( 441 p_{1}^2 -457 ( \vec{p}_{1}\cdot\vec{n})^{2} \big) - 16 \big( 55 p_{1}^2 \nn\\ 
&& + 8 ( \vec{p}_{1}\cdot\vec{n})^{2} \big) ( \vec{S}_{1}\cdot\vec{n})^{2} - 434 ( \vec{p}_{1}\cdot\vec{S}_{1})^{2} \Big] \nn\\ 
&& + 	\frac{G^3 m_{2}{}^2}{2048 m_{1}{}^2 r{}^5} \Big[ 12 \vec{S}_{1}\cdot\vec{n} \vec{p}_{1}\cdot\vec{S}_{1} \big( {(42112 - 105 \pi^2)} \vec{p}_{1}\cdot\vec{n} + 960 \vec{p}_{2}\cdot\vec{n} \big) \nn\\ 
&& - 231040 \vec{S}_{1}\cdot\vec{n} \vec{p}_{2}\cdot\vec{S}_{1} \vec{p}_{1}\cdot\vec{n} + S_{1}^2 \big( {(92032 + 63 \pi^2)} p_{1}^2 - 60032 \vec{p}_{1}\cdot\vec{p}_{2} \nn\\ 
&& + 260480 \vec{p}_{1}\cdot\vec{n} \vec{p}_{2}\cdot\vec{n} - {(17216 + 315 \pi^2)} ( \vec{p}_{1}\cdot\vec{n})^{2} \big) + 59904 \vec{p}_{1}\cdot\vec{S}_{1} \vec{p}_{2}\cdot\vec{S}_{1} \nn\\ 
&& - \big( {(167872 + 315 \pi^2)} p_{1}^2 - 87424 \vec{p}_{1}\cdot\vec{p}_{2} + 128256 \vec{p}_{1}\cdot\vec{n} \vec{p}_{2}\cdot\vec{n} \nn\\ 
&& + {(400512 - 2205 \pi^2)} ( \vec{p}_{1}\cdot\vec{n})^{2} \big) ( \vec{S}_{1}\cdot\vec{n})^{2} - 42 {(2816 - 3 \pi^2)} ( \vec{p}_{1}\cdot\vec{S}_{1})^{2} \Big] \nn\\ 
&& - 	\frac{G^3 m_{1}}{4900 m_{2} r{}^5} \Big[ 402760 \vec{S}_{1}\cdot\vec{n} \vec{p}_{2}\cdot\vec{S}_{1} \vec{p}_{2}\cdot\vec{n} + 187 S_{1}^2 \big( 402 p_{2}^2 -85 ( \vec{p}_{2}\cdot\vec{n})^{2} \big) \nn\\ 
&& - 25 \big( 1774 p_{2}^2 + 16247 ( \vec{p}_{2}\cdot\vec{n})^{2} \big) ( \vec{S}_{1}\cdot\vec{n})^{2} - 88632 ( \vec{p}_{2}\cdot\vec{S}_{1})^{2} \Big] \nn\\ 
&& + 	\frac{G^3 m_{2}}{3763200 m_{1} r{}^5} \Big[ 10 \vec{S}_{1}\cdot\vec{n} \vec{p}_{1}\cdot\vec{S}_{1} \big( 51835232 \vec{p}_{1}\cdot\vec{n} - {(6554240 + 2414475 \pi^2)} \vec{p}_{2}\cdot\vec{n} \big) \nn\\ 
&& - 2450 \vec{S}_{1}\cdot\vec{n} \vec{p}_{2}\cdot\vec{S}_{1} \big( {(287584 + 9855 \pi^2)} \vec{p}_{1}\cdot\vec{n} + 134688 \vec{p}_{2}\cdot\vec{n} \big) + S_{1}^2 \big( 95727488 p_{1}^2 \nn\\ 
&& - {(152723200 - 2414475 \pi^2)} \vec{p}_{1}\cdot\vec{p}_{2} - 129595200 p_{2}^2 \nn\\ 
&& + {(694467200 - 12072375 \pi^2)} \vec{p}_{1}\cdot\vec{n} \vec{p}_{2}\cdot\vec{n} + 144188960 ( \vec{p}_{1}\cdot\vec{n})^{2} \nn\\ 
&& - 128184000 ( \vec{p}_{2}\cdot\vec{n})^{2} \big) + 2450 {(53600 + 1971 \pi^2)} \vec{p}_{1}\cdot\vec{S}_{1} \vec{p}_{2}\cdot\vec{S}_{1} - 25 \big( 7852000 p_{1}^2 \nn\\ 
&& - {(9715328 - 482895 \pi^2)} \vec{p}_{1}\cdot\vec{p}_{2} - 9384480 p_{2}^2 + {(6234368 - 3380265 \pi^2)} \vec{p}_{1}\cdot\vec{n} \vec{p}_{2}\cdot\vec{n} \nn\\ 
&& + 18470144 ( \vec{p}_{1}\cdot\vec{n})^{2} - 9215136 ( \vec{p}_{2}\cdot\vec{n})^{2} \big) ( \vec{S}_{1}\cdot\vec{n})^{2} - 180953984 ( \vec{p}_{1}\cdot\vec{S}_{1})^{2} \nn\\ 
&& + 122774400 ( \vec{p}_{2}\cdot\vec{S}_{1})^{2} \Big] - 	\frac{G^3}{2508800 r{}^5} \Big[ 85204480 \vec{S}_{1}\cdot\vec{n} \vec{p}_{1}\cdot\vec{S}_{1} \vec{p}_{2}\cdot\vec{n} \nn\\ 
&& + 20 \vec{S}_{1}\cdot\vec{n} \vec{p}_{2}\cdot\vec{S}_{1} \big( 12050944 \vec{p}_{1}\cdot\vec{n} + {(29282400 - 1686825 \pi^2)} \vec{p}_{2}\cdot\vec{n} \big) \nn\\ 
&& + S_{1}^2 \big( 106223104 \vec{p}_{1}\cdot\vec{p}_{2} + {(197097600 + 1686825 \pi^2)} p_{2}^2 -69048320 \vec{p}_{1}\cdot\vec{n} \vec{p}_{2}\cdot\vec{n} \nn\\ 
&& + {(95491200 - 8434125 \pi^2)} ( \vec{p}_{2}\cdot\vec{n})^{2} \big) - 152297472 \vec{p}_{1}\cdot\vec{S}_{1} \vec{p}_{2}\cdot\vec{S}_{1} \nn\\ 
&& - 25 \big( 7097856 \vec{p}_{1}\cdot\vec{p}_{2} + {(9916032 + 337365 \pi^2)} p_{2}^2 -412160 \vec{p}_{1}\cdot\vec{n} \vec{p}_{2}\cdot\vec{n} \nn\\ 
&& + {(19769344 - 2361555 \pi^2)} ( \vec{p}_{2}\cdot\vec{n})^{2} \big) ( \vec{S}_{1}\cdot\vec{n})^{2} - 2450 {(98560 - 1377 \pi^2)} ( \vec{p}_{2}\cdot\vec{S}_{1})^{2} \Big]\nn\\ && +  	\frac{9 G^4 m_{2}{}^4}{m_{1} r{}^6} \Big[ S_{1}^2 - ( \vec{S}_{1}\cdot\vec{n})^{2} \Big] + 	\frac{G^4 m_{2}{}^3}{4 r{}^6} \Big[ 97 S_{1}^2 - 217 ( \vec{S}_{1}\cdot\vec{n})^{2} \Big] + 	\frac{G^4 m_{1}{}^2 m_{2}}{2450 r{}^6} \Big[ 26037 S_{1}^2 \nn\\ 
&& - 28166 ( \vec{S}_{1}\cdot\vec{n})^{2} \Big] + 	\frac{G^4 m_{1} m_{2}{}^2}{78400 r{}^6} \Big[ {(1895204 + 55125 \pi^2)} S_{1}^2 \nn\\ 
&& - {(3486212 + 165375 \pi^2)} ( \vec{S}_{1}\cdot\vec{n})^{2} \Big],
\eea
%%%%%%%%%%%%%%%%

%%%%%%%%%%%%%%%%
\bea
H^{\text{N}^3\text{LO}}_{C_{1\text{ES}^2}} &= &  - \frac{G}{4 m_{1}{}^2 m_{2}{}^4 r{}^3} \Big[ 6 \vec{S}_{1}\cdot\vec{n} \vec{p}_{1}\cdot\vec{S}_{1} \vec{p}_{2}\cdot\vec{n} p_{2}^{4} + 6 \vec{S}_{1}\cdot\vec{n} \vec{p}_{2}\cdot\vec{S}_{1} \vec{p}_{1}\cdot\vec{n} p_{2}^{4} - 2 \vec{p}_{1}\cdot\vec{S}_{1} \vec{p}_{2}\cdot\vec{S}_{1} p_{2}^{4} \nn\\ 
&& + S_{1}^2 \big( \vec{p}_{1}\cdot\vec{p}_{2} p_{2}^{4} -3 \vec{p}_{1}\cdot\vec{n} \vec{p}_{2}\cdot\vec{n} p_{2}^{4} \big) + 3 \big( \vec{p}_{1}\cdot\vec{p}_{2} p_{2}^{4} -5 \vec{p}_{1}\cdot\vec{n} \vec{p}_{2}\cdot\vec{n} p_{2}^{4} \big) ( \vec{S}_{1}\cdot\vec{n})^{2} \Big] \nn\\ 
&& - 	\frac{7 G}{32 m_{1} m_{2}{}^5 r{}^3} \Big[ S_{1}^2 p_{2}^{6} - 3 p_{2}^{6} ( \vec{S}_{1}\cdot\vec{n})^{2} \Big] - 	\frac{G}{16 m_{1}{}^6 r{}^3} \Big[ 3 \vec{S}_{1}\cdot\vec{n} \vec{p}_{1}\cdot\vec{S}_{1} \big( 26 \vec{p}_{1}\cdot\vec{n} p_{1}^2 \vec{p}_{1}\cdot\vec{p}_{2} \nn\\ 
&& + 9 \vec{p}_{2}\cdot\vec{n} p_{1}^{4} -5 p_{1}^2 \vec{p}_{2}\cdot\vec{n} ( \vec{p}_{1}\cdot\vec{n})^{2} \big) + 24 \vec{S}_{1}\cdot\vec{n} \vec{p}_{2}\cdot\vec{S}_{1} \vec{p}_{1}\cdot\vec{n} p_{1}^{4} - 3 \vec{p}_{1}\cdot\vec{S}_{1} \vec{p}_{2}\cdot\vec{S}_{1} \big( 3 p_{1}^{4} \nn\\ 
&& - p_{1}^2 ( \vec{p}_{1}\cdot\vec{n})^{2} \big) + 12 S_{1}^2 \big( 3 \vec{p}_{1}\cdot\vec{p}_{2} p_{1}^{4} -8 p_{1}^2 \vec{p}_{1}\cdot\vec{p}_{2} ( \vec{p}_{1}\cdot\vec{n})^{2} - \vec{p}_{1}\cdot\vec{n} \vec{p}_{2}\cdot\vec{n} p_{1}^{4} \big) \nn\\ 
&& + 12 \big( \vec{p}_{1}\cdot\vec{p}_{2} p_{1}^{4} -5 \vec{p}_{1}\cdot\vec{n} \vec{p}_{2}\cdot\vec{n} p_{1}^{4} \big) ( \vec{S}_{1}\cdot\vec{n})^{2} - \big( 31 p_{1}^2 \vec{p}_{1}\cdot\vec{p}_{2} + 3 \vec{p}_{1}\cdot\vec{n} p_{1}^2 \vec{p}_{2}\cdot\vec{n} \nn\\ 
&& - 15 \vec{p}_{1}\cdot\vec{p}_{2} ( \vec{p}_{1}\cdot\vec{n})^{2} -15 \vec{p}_{2}\cdot\vec{n} ( \vec{p}_{1}\cdot\vec{n})^{3} \big) ( \vec{p}_{1}\cdot\vec{S}_{1})^{2} \Big] \nn\\ 
&& + 	\frac{G m_{2}}{32 m_{1}{}^7 r{}^3} \Big[ 54 \vec{S}_{1}\cdot\vec{n} \vec{p}_{1}\cdot\vec{S}_{1} \vec{p}_{1}\cdot\vec{n} p_{1}^{4} + S_{1}^2 \big( 11 p_{1}^{6} -54 ( \vec{p}_{1}\cdot\vec{n})^{2} p_{1}^{4} \big) + 21 p_{1}^{6} ( \vec{S}_{1}\cdot\vec{n})^{2} \nn\\ 
&& - 18 p_{1}^{4} ( \vec{p}_{1}\cdot\vec{S}_{1})^{2} \Big] + 	\frac{G}{32 m_{1}{}^5 m_{2} r{}^3} \Big[ 6 \vec{S}_{1}\cdot\vec{n} \vec{p}_{1}\cdot\vec{S}_{1} \big( 30 p_{1}^2 \vec{p}_{2}\cdot\vec{n} \vec{p}_{1}\cdot\vec{p}_{2} + 6 \vec{p}_{1}\cdot\vec{n} p_{1}^2 p_{2}^2 \nn\\ 
&& + 10 \vec{p}_{1}\cdot\vec{n} ( \vec{p}_{1}\cdot\vec{p}_{2})^{2} -50 \vec{p}_{2}\cdot\vec{n} \vec{p}_{1}\cdot\vec{p}_{2} ( \vec{p}_{1}\cdot\vec{n})^{2} + 5 p_{2}^2 ( \vec{p}_{1}\cdot\vec{n})^{3} \nn\\ 
&& + 5 \vec{p}_{1}\cdot\vec{n} p_{1}^2 ( \vec{p}_{2}\cdot\vec{n})^{2} -35 ( \vec{p}_{1}\cdot\vec{n})^{3} ( \vec{p}_{2}\cdot\vec{n})^{2} \big) + 12 \vec{S}_{1}\cdot\vec{n} \vec{p}_{2}\cdot\vec{S}_{1} \big( 10 \vec{p}_{1}\cdot\vec{n} p_{1}^2 \vec{p}_{1}\cdot\vec{p}_{2} \nn\\ 
&& + \vec{p}_{2}\cdot\vec{n} p_{1}^{4} -5 p_{1}^2 \vec{p}_{2}\cdot\vec{n} ( \vec{p}_{1}\cdot\vec{n})^{2} \big) - 12 \vec{p}_{1}\cdot\vec{S}_{1} \vec{p}_{2}\cdot\vec{S}_{1} \big( 5 p_{1}^2 \vec{p}_{1}\cdot\vec{p}_{2} + \vec{p}_{1}\cdot\vec{n} p_{1}^2 \vec{p}_{2}\cdot\vec{n} \nn\\ 
&& - 5 \vec{p}_{1}\cdot\vec{p}_{2} ( \vec{p}_{1}\cdot\vec{n})^{2} -5 \vec{p}_{2}\cdot\vec{n} ( \vec{p}_{1}\cdot\vec{n})^{3} \big) + S_{1}^2 \big( 78 p_{1}^2 ( \vec{p}_{1}\cdot\vec{p}_{2})^{2} \nn\\ 
&& + 16 p_{2}^2 p_{1}^{4} -228 \vec{p}_{1}\cdot\vec{n} p_{1}^2 \vec{p}_{2}\cdot\vec{n} \vec{p}_{1}\cdot\vec{p}_{2} - 33 p_{1}^2 p_{2}^2 ( \vec{p}_{1}\cdot\vec{n})^{2} - 120 ( \vec{p}_{1}\cdot\vec{n})^{2} ( \vec{p}_{1}\cdot\vec{p}_{2})^{2} \nn\\ 
&& + 9 ( \vec{p}_{2}\cdot\vec{n})^{2} p_{1}^{4} + 240 \vec{p}_{2}\cdot\vec{n} \vec{p}_{1}\cdot\vec{p}_{2} ( \vec{p}_{1}\cdot\vec{n})^{3} - 30 p_{2}^2 ( \vec{p}_{1}\cdot\vec{n})^{4} - 135 p_{1}^2 ( \vec{p}_{1}\cdot\vec{n})^{2} ( \vec{p}_{2}\cdot\vec{n})^{2} \nn\\ 
&& + 210 ( \vec{p}_{1}\cdot\vec{n})^{4} ( \vec{p}_{2}\cdot\vec{n})^{2} \big) + 3 \big( 2 p_{1}^2 ( \vec{p}_{1}\cdot\vec{p}_{2})^{2} + 2 p_{2}^2 p_{1}^{4} -100 \vec{p}_{1}\cdot\vec{n} p_{1}^2 \vec{p}_{2}\cdot\vec{n} \vec{p}_{1}\cdot\vec{p}_{2} \nn\\ 
&& - 5 p_{1}^2 p_{2}^2 ( \vec{p}_{1}\cdot\vec{n})^{2} - 5 ( \vec{p}_{2}\cdot\vec{n})^{2} p_{1}^{4} + 35 p_{1}^2 ( \vec{p}_{1}\cdot\vec{n})^{2} ( \vec{p}_{2}\cdot\vec{n})^{2} \big) ( \vec{S}_{1}\cdot\vec{n})^{2} - 2 \big( 8 p_{1}^2 p_{2}^2 \nn\\ 
&& + 10 ( \vec{p}_{1}\cdot\vec{p}_{2})^{2} -60 \vec{p}_{1}\cdot\vec{n} \vec{p}_{2}\cdot\vec{n} \vec{p}_{1}\cdot\vec{p}_{2} \nn\\ 
&& + 6 p_{1}^2 ( \vec{p}_{2}\cdot\vec{n})^{2} -45 ( \vec{p}_{1}\cdot\vec{n})^{2} ( \vec{p}_{2}\cdot\vec{n})^{2} \big) ( \vec{p}_{1}\cdot\vec{S}_{1})^{2} - 2 \big( p_{1}^{4} -3 p_{1}^2 ( \vec{p}_{1}\cdot\vec{n})^{2} \big) ( \vec{p}_{2}\cdot\vec{S}_{1})^{2} \Big] \nn\\ 
&& + 	\frac{G}{32 m_{1}{}^3 m_{2}{}^3 r{}^3} \Big[ 6 \vec{S}_{1}\cdot\vec{n} \vec{p}_{1}\cdot\vec{S}_{1} \big( 20 \vec{p}_{2}\cdot\vec{n} \vec{p}_{1}\cdot\vec{p}_{2} p_{2}^2 + 7 \vec{p}_{1}\cdot\vec{n} p_{2}^{4} -10 \vec{p}_{1}\cdot\vec{n} p_{2}^2 ( \vec{p}_{2}\cdot\vec{n})^{2} \big) \nn\\ 
&& + 12 \vec{S}_{1}\cdot\vec{n} \vec{p}_{2}\cdot\vec{S}_{1} \big( p_{1}^2 \vec{p}_{2}\cdot\vec{n} p_{2}^2 + 10 \vec{p}_{1}\cdot\vec{n} \vec{p}_{1}\cdot\vec{p}_{2} p_{2}^2 -5 \vec{p}_{2}\cdot\vec{n} p_{2}^2 ( \vec{p}_{1}\cdot\vec{n})^{2} \big) \nn\\ 
&& - 8 \vec{p}_{1}\cdot\vec{S}_{1} \vec{p}_{2}\cdot\vec{S}_{1} \big( 5 \vec{p}_{1}\cdot\vec{p}_{2} p_{2}^2 -3 \vec{p}_{1}\cdot\vec{n} \vec{p}_{2}\cdot\vec{n} p_{2}^2 \big) + S_{1}^2 \big( 38 p_{2}^2 ( \vec{p}_{1}\cdot\vec{p}_{2})^{2} \nn\\ 
&& + 12 p_{1}^2 p_{2}^{4} -84 \vec{p}_{1}\cdot\vec{n} \vec{p}_{2}\cdot\vec{n} \vec{p}_{1}\cdot\vec{p}_{2} p_{2}^2 - 9 p_{1}^2 p_{2}^2 ( \vec{p}_{2}\cdot\vec{n})^{2} - 39 ( \vec{p}_{1}\cdot\vec{n})^{2} p_{2}^{4} \nn\\ 
&& + 45 p_{2}^2 ( \vec{p}_{1}\cdot\vec{n})^{2} ( \vec{p}_{2}\cdot\vec{n})^{2} \big) + 3 \big( 2 p_{2}^2 ( \vec{p}_{1}\cdot\vec{p}_{2})^{2} + 2 p_{1}^2 p_{2}^{4} -100 \vec{p}_{1}\cdot\vec{n} \vec{p}_{2}\cdot\vec{n} \vec{p}_{1}\cdot\vec{p}_{2} p_{2}^2 \nn\\ 
&& - 5 p_{1}^2 p_{2}^2 ( \vec{p}_{2}\cdot\vec{n})^{2} - 5 ( \vec{p}_{1}\cdot\vec{n})^{2} p_{2}^{4} + 35 p_{2}^2 ( \vec{p}_{1}\cdot\vec{n})^{2} ( \vec{p}_{2}\cdot\vec{n})^{2} \big) ( \vec{S}_{1}\cdot\vec{n})^{2} - 6 \big( 2 p_{2}^{4} \nn\\ 
&& - p_{2}^2 ( \vec{p}_{2}\cdot\vec{n})^{2} \big) ( \vec{p}_{1}\cdot\vec{S}_{1})^{2} - 2 \big( p_{1}^2 p_{2}^2 -3 p_{2}^2 ( \vec{p}_{1}\cdot\vec{n})^{2} \big) ( \vec{p}_{2}\cdot\vec{S}_{1})^{2} \Big] \nn\\ 
&& - 	\frac{G}{32 m_{1}{}^4 m_{2}{}^2 r{}^3} \Big[ 6 \vec{S}_{1}\cdot\vec{n} \vec{p}_{1}\cdot\vec{S}_{1} \big( 9 p_{1}^2 \vec{p}_{2}\cdot\vec{n} p_{2}^2 + 18 \vec{p}_{1}\cdot\vec{n} \vec{p}_{1}\cdot\vec{p}_{2} p_{2}^2 \nn\\ 
&& + 10 \vec{p}_{2}\cdot\vec{n} ( \vec{p}_{1}\cdot\vec{p}_{2})^{2} -5 \vec{p}_{2}\cdot\vec{n} p_{2}^2 ( \vec{p}_{1}\cdot\vec{n})^{2} + 5 p_{1}^2 ( \vec{p}_{2}\cdot\vec{n})^{3} \nn\\ 
&& - 50 \vec{p}_{1}\cdot\vec{n} \vec{p}_{1}\cdot\vec{p}_{2} ( \vec{p}_{2}\cdot\vec{n})^{2} -35 ( \vec{p}_{1}\cdot\vec{n})^{2} ( \vec{p}_{2}\cdot\vec{n})^{3} \big) + 30 \vec{S}_{1}\cdot\vec{n} \vec{p}_{2}\cdot\vec{S}_{1} \big( 2 p_{1}^2 \vec{p}_{2}\cdot\vec{n} \vec{p}_{1}\cdot\vec{p}_{2} \nn\\ 
&& + \vec{p}_{1}\cdot\vec{n} p_{1}^2 p_{2}^2 + 2 \vec{p}_{1}\cdot\vec{n} ( \vec{p}_{1}\cdot\vec{p}_{2})^{2} -10 \vec{p}_{2}\cdot\vec{n} \vec{p}_{1}\cdot\vec{p}_{2} ( \vec{p}_{1}\cdot\vec{n})^{2} + p_{2}^2 ( \vec{p}_{1}\cdot\vec{n})^{3} \nn\\ 
&& + 3 \vec{p}_{1}\cdot\vec{n} p_{1}^2 ( \vec{p}_{2}\cdot\vec{n})^{2} -7 ( \vec{p}_{1}\cdot\vec{n})^{3} ( \vec{p}_{2}\cdot\vec{n})^{2} \big) - 2 \vec{p}_{1}\cdot\vec{S}_{1} \vec{p}_{2}\cdot\vec{S}_{1} \big( 9 p_{1}^2 p_{2}^2 \nn\\ 
&& + 10 ( \vec{p}_{1}\cdot\vec{p}_{2})^{2} -60 \vec{p}_{1}\cdot\vec{n} \vec{p}_{2}\cdot\vec{n} \vec{p}_{1}\cdot\vec{p}_{2} - 3 p_{2}^2 ( \vec{p}_{1}\cdot\vec{n})^{2} \nn\\ 
&& + 9 p_{1}^2 ( \vec{p}_{2}\cdot\vec{n})^{2} -45 ( \vec{p}_{1}\cdot\vec{n})^{2} ( \vec{p}_{2}\cdot\vec{n})^{2} \big) + S_{1}^2 \big( 41 p_{1}^2 \vec{p}_{1}\cdot\vec{p}_{2} p_{2}^2 \nn\\ 
&& + 22 ( \vec{p}_{1}\cdot\vec{p}_{2})^{3} -51 \vec{p}_{1}\cdot\vec{n} p_{1}^2 \vec{p}_{2}\cdot\vec{n} p_{2}^2 - 99 \vec{p}_{1}\cdot\vec{p}_{2} p_{2}^2 ( \vec{p}_{1}\cdot\vec{n})^{2} - 27 p_{1}^2 \vec{p}_{1}\cdot\vec{p}_{2} ( \vec{p}_{2}\cdot\vec{n})^{2} \nn\\ 
&& - 150 \vec{p}_{1}\cdot\vec{n} \vec{p}_{2}\cdot\vec{n} ( \vec{p}_{1}\cdot\vec{p}_{2})^{2} + 45 \vec{p}_{2}\cdot\vec{n} p_{2}^2 ( \vec{p}_{1}\cdot\vec{n})^{3} - 75 \vec{p}_{1}\cdot\vec{n} p_{1}^2 ( \vec{p}_{2}\cdot\vec{n})^{3} \nn\\ 
&& + 135 \vec{p}_{1}\cdot\vec{p}_{2} ( \vec{p}_{1}\cdot\vec{n})^{2} ( \vec{p}_{2}\cdot\vec{n})^{2} + 175 ( \vec{p}_{1}\cdot\vec{n})^{3} ( \vec{p}_{2}\cdot\vec{n})^{3} \big) + 3 \big( 13 p_{1}^2 \vec{p}_{1}\cdot\vec{p}_{2} p_{2}^2 \nn\\ 
&& - 2 ( \vec{p}_{1}\cdot\vec{p}_{2})^{3} -25 \vec{p}_{1}\cdot\vec{n} p_{1}^2 \vec{p}_{2}\cdot\vec{n} p_{2}^2 - 25 \vec{p}_{1}\cdot\vec{p}_{2} p_{2}^2 ( \vec{p}_{1}\cdot\vec{n})^{2} - 25 p_{1}^2 \vec{p}_{1}\cdot\vec{p}_{2} ( \vec{p}_{2}\cdot\vec{n})^{2} \nn\\ 
&& - 50 \vec{p}_{1}\cdot\vec{n} \vec{p}_{2}\cdot\vec{n} ( \vec{p}_{1}\cdot\vec{p}_{2})^{2} -35 \vec{p}_{2}\cdot\vec{n} p_{2}^2 ( \vec{p}_{1}\cdot\vec{n})^{3} - 35 \vec{p}_{1}\cdot\vec{n} p_{1}^2 ( \vec{p}_{2}\cdot\vec{n})^{3} \nn\\ 
&& + 175 \vec{p}_{1}\cdot\vec{p}_{2} ( \vec{p}_{1}\cdot\vec{n})^{2} ( \vec{p}_{2}\cdot\vec{n})^{2} + 105 ( \vec{p}_{1}\cdot\vec{n})^{3} ( \vec{p}_{2}\cdot\vec{n})^{3} \big) ( \vec{S}_{1}\cdot\vec{n})^{2} \nn\\ 
&& - 2 \big( 13 \vec{p}_{1}\cdot\vec{p}_{2} p_{2}^2 -15 \vec{p}_{1}\cdot\vec{n} \vec{p}_{2}\cdot\vec{n} p_{2}^2 \nn\\ 
&& - 15 \vec{p}_{1}\cdot\vec{p}_{2} ( \vec{p}_{2}\cdot\vec{n})^{2} -15 \vec{p}_{1}\cdot\vec{n} ( \vec{p}_{2}\cdot\vec{n})^{3} \big) ( \vec{p}_{1}\cdot\vec{S}_{1})^{2} - 2 \big( 5 p_{1}^2 \vec{p}_{1}\cdot\vec{p}_{2} + 9 \vec{p}_{1}\cdot\vec{n} p_{1}^2 \vec{p}_{2}\cdot\vec{n} \nn\\ 
&& - 15 \vec{p}_{1}\cdot\vec{p}_{2} ( \vec{p}_{1}\cdot\vec{n})^{2} -15 \vec{p}_{2}\cdot\vec{n} ( \vec{p}_{1}\cdot\vec{n})^{3} \big) ( \vec{p}_{2}\cdot\vec{S}_{1})^{2} \Big]\nn\\ && +  	\frac{G^2 m_{2}{}^2}{8 m_{1}{}^5 r{}^4} \Big[ 2 \vec{S}_{1}\cdot\vec{n} \vec{p}_{1}\cdot\vec{S}_{1} \big( 97 \vec{p}_{1}\cdot\vec{n} p_{1}^2 -24 ( \vec{p}_{1}\cdot\vec{n})^{3} \big) + 2 S_{1}^2 \big( 13 p_{1}^{4} -104 p_{1}^2 ( \vec{p}_{1}\cdot\vec{n})^{2} \nn\\ 
&& + 24 ( \vec{p}_{1}\cdot\vec{n})^{4} \big) + \big( 67 p_{1}^{4} -24 p_{1}^2 ( \vec{p}_{1}\cdot\vec{n})^{2} \big) ( \vec{S}_{1}\cdot\vec{n})^{2} \nn\\ 
&& - 2 \big( 31 p_{1}^2 -33 ( \vec{p}_{1}\cdot\vec{n})^{2} \big) ( \vec{p}_{1}\cdot\vec{S}_{1})^{2} \Big] - 	\frac{G^2 m_{2}}{16 m_{1}{}^4 r{}^4} \Big[ 2 \vec{S}_{1}\cdot\vec{n} \vec{p}_{1}\cdot\vec{S}_{1} \big( 640 \vec{p}_{1}\cdot\vec{n} p_{1}^2 \nn\\ 
&& + 138 p_{1}^2 \vec{p}_{2}\cdot\vec{n} + 429 \vec{p}_{1}\cdot\vec{n} \vec{p}_{1}\cdot\vec{p}_{2} -299 ( \vec{p}_{1}\cdot\vec{n})^{3} \big) \nn\\ 
&& + \vec{S}_{1}\cdot\vec{n} \vec{p}_{2}\cdot\vec{S}_{1} \big( 189 \vec{p}_{1}\cdot\vec{n} p_{1}^2 -128 ( \vec{p}_{1}\cdot\vec{n})^{3} \big) - 2 \vec{p}_{1}\cdot\vec{S}_{1} \vec{p}_{2}\cdot\vec{S}_{1} \big( 2 p_{1}^2 + 53 ( \vec{p}_{1}\cdot\vec{n})^{2} \big) \nn\\ 
&& + S_{1}^2 \big( 208 p_{1}^2 \vec{p}_{1}\cdot\vec{p}_{2} - 109 p_{1}^{4} + 144 \vec{p}_{1}\cdot\vec{n} p_{1}^2 \vec{p}_{2}\cdot\vec{n} + 293 p_{1}^2 ( \vec{p}_{1}\cdot\vec{n})^{2} \nn\\ 
&& - 488 \vec{p}_{1}\cdot\vec{p}_{2} ( \vec{p}_{1}\cdot\vec{n})^{2} + 178 ( \vec{p}_{1}\cdot\vec{n})^{4} - 496 \vec{p}_{2}\cdot\vec{n} ( \vec{p}_{1}\cdot\vec{n})^{3} \big) + \big( 272 p_{1}^2 \vec{p}_{1}\cdot\vec{p}_{2} \nn\\ 
&& + 342 p_{1}^{4} -1221 \vec{p}_{1}\cdot\vec{n} p_{1}^2 \vec{p}_{2}\cdot\vec{n} - 2136 p_{1}^2 ( \vec{p}_{1}\cdot\vec{n})^{2} - 384 \vec{p}_{1}\cdot\vec{p}_{2} ( \vec{p}_{1}\cdot\vec{n})^{2} \nn\\ 
&& + 592 ( \vec{p}_{1}\cdot\vec{n})^{4} + 512 \vec{p}_{2}\cdot\vec{n} ( \vec{p}_{1}\cdot\vec{n})^{3} \big) ( \vec{S}_{1}\cdot\vec{n})^{2} - \big( 135 p_{1}^2 \nn\\ 
&& + 324 \vec{p}_{1}\cdot\vec{p}_{2} -308 \vec{p}_{1}\cdot\vec{n} \vec{p}_{2}\cdot\vec{n} - 161 ( \vec{p}_{1}\cdot\vec{n})^{2} \big) ( \vec{p}_{1}\cdot\vec{S}_{1})^{2} \Big] \nn\\ 
&& + 	\frac{G^2}{16 m_{1}{}^3 r{}^4} \Big[ \vec{S}_{1}\cdot\vec{n} \vec{p}_{1}\cdot\vec{S}_{1} \big( 556 p_{1}^2 \vec{p}_{2}\cdot\vec{n} + 742 \vec{p}_{1}\cdot\vec{n} \vec{p}_{1}\cdot\vec{p}_{2} - 498 \vec{p}_{2}\cdot\vec{n} \vec{p}_{1}\cdot\vec{p}_{2} \nn\\ 
&& + 723 \vec{p}_{1}\cdot\vec{n} p_{2}^2 -930 \vec{p}_{2}\cdot\vec{n} ( \vec{p}_{1}\cdot\vec{n})^{2} + 1287 \vec{p}_{1}\cdot\vec{n} ( \vec{p}_{2}\cdot\vec{n})^{2} \big) \nn\\ 
&& + 2 \vec{S}_{1}\cdot\vec{n} \vec{p}_{2}\cdot\vec{S}_{1} \big( 246 \vec{p}_{1}\cdot\vec{n} p_{1}^2 - 221 p_{1}^2 \vec{p}_{2}\cdot\vec{n} - 374 \vec{p}_{1}\cdot\vec{n} \vec{p}_{1}\cdot\vec{p}_{2} -144 ( \vec{p}_{1}\cdot\vec{n})^{3} \nn\\ 
&& + 2373 \vec{p}_{2}\cdot\vec{n} ( \vec{p}_{1}\cdot\vec{n})^{2} \big) - 3 \vec{p}_{1}\cdot\vec{S}_{1} \vec{p}_{2}\cdot\vec{S}_{1} \big( 43 p_{1}^2 - 38 \vec{p}_{1}\cdot\vec{p}_{2} + 386 \vec{p}_{1}\cdot\vec{n} \vec{p}_{2}\cdot\vec{n} \nn\\ 
&& - 57 ( \vec{p}_{1}\cdot\vec{n})^{2} \big) - S_{1}^2 \big( 357 p_{1}^2 \vec{p}_{1}\cdot\vec{p}_{2} - 51 p_{1}^2 p_{2}^2 - 64 ( \vec{p}_{1}\cdot\vec{p}_{2})^{2} -128 \vec{p}_{1}\cdot\vec{n} p_{1}^2 \vec{p}_{2}\cdot\vec{n} \nn\\ 
&& - 2368 \vec{p}_{1}\cdot\vec{n} \vec{p}_{2}\cdot\vec{n} \vec{p}_{1}\cdot\vec{p}_{2} - 465 \vec{p}_{1}\cdot\vec{p}_{2} ( \vec{p}_{1}\cdot\vec{n})^{2} - 304 p_{2}^2 ( \vec{p}_{1}\cdot\vec{n})^{2} \nn\\ 
&& - 612 p_{1}^2 ( \vec{p}_{2}\cdot\vec{n})^{2} -486 \vec{p}_{2}\cdot\vec{n} ( \vec{p}_{1}\cdot\vec{n})^{3} + 4848 ( \vec{p}_{1}\cdot\vec{n})^{2} ( \vec{p}_{2}\cdot\vec{n})^{2} \big) + \big( 766 p_{1}^2 \vec{p}_{1}\cdot\vec{p}_{2} \nn\\ 
&& - 15 p_{1}^2 p_{2}^2 - 302 ( \vec{p}_{1}\cdot\vec{p}_{2})^{2} -1476 \vec{p}_{1}\cdot\vec{n} p_{1}^2 \vec{p}_{2}\cdot\vec{n} - 252 \vec{p}_{1}\cdot\vec{n} \vec{p}_{2}\cdot\vec{n} \vec{p}_{1}\cdot\vec{p}_{2} \nn\\ 
&& - 1548 \vec{p}_{1}\cdot\vec{p}_{2} ( \vec{p}_{1}\cdot\vec{n})^{2} - 147 p_{2}^2 ( \vec{p}_{1}\cdot\vec{n})^{2} - 411 p_{1}^2 ( \vec{p}_{2}\cdot\vec{n})^{2} + 1056 \vec{p}_{2}\cdot\vec{n} ( \vec{p}_{1}\cdot\vec{n})^{3} \nn\\ 
&& - 1521 ( \vec{p}_{1}\cdot\vec{n})^{2} ( \vec{p}_{2}\cdot\vec{n})^{2} \big) ( \vec{S}_{1}\cdot\vec{n})^{2} - \big( 22 \vec{p}_{1}\cdot\vec{p}_{2} + 245 p_{2}^2 -188 \vec{p}_{1}\cdot\vec{n} \vec{p}_{2}\cdot\vec{n} \nn\\ 
&& - 81 ( \vec{p}_{2}\cdot\vec{n})^{2} \big) ( \vec{p}_{1}\cdot\vec{S}_{1})^{2} + \big( 199 p_{1}^2 -928 ( \vec{p}_{1}\cdot\vec{n})^{2} \big) ( \vec{p}_{2}\cdot\vec{S}_{1})^{2} \Big] \nn\\ 
&& - 	\frac{G^2}{16 m_{1}{}^2 m_{2} r{}^4} \Big[ 2 \vec{S}_{1}\cdot\vec{n} \vec{p}_{1}\cdot\vec{S}_{1} \big( 58 \vec{p}_{2}\cdot\vec{n} \vec{p}_{1}\cdot\vec{p}_{2} - 128 \vec{p}_{1}\cdot\vec{n} p_{2}^2 \nn\\ 
&& - 287 \vec{p}_{2}\cdot\vec{n} p_{2}^2 -102 \vec{p}_{1}\cdot\vec{n} ( \vec{p}_{2}\cdot\vec{n})^{2} + 633 ( \vec{p}_{2}\cdot\vec{n})^{3} \big) + \vec{S}_{1}\cdot\vec{n} \vec{p}_{2}\cdot\vec{S}_{1} \big( 20 p_{1}^2 \vec{p}_{2}\cdot\vec{n} \nn\\ 
&& + 60 \vec{p}_{1}\cdot\vec{n} \vec{p}_{1}\cdot\vec{p}_{2} - 3200 \vec{p}_{2}\cdot\vec{n} \vec{p}_{1}\cdot\vec{p}_{2} - 527 \vec{p}_{1}\cdot\vec{n} p_{2}^2 -168 \vec{p}_{2}\cdot\vec{n} ( \vec{p}_{1}\cdot\vec{n})^{2} \nn\\ 
&& + 7713 \vec{p}_{1}\cdot\vec{n} ( \vec{p}_{2}\cdot\vec{n})^{2} \big) - 2 \vec{p}_{1}\cdot\vec{S}_{1} \vec{p}_{2}\cdot\vec{S}_{1} \big( 2 \vec{p}_{1}\cdot\vec{p}_{2} + 23 p_{2}^2 -54 \vec{p}_{1}\cdot\vec{n} \vec{p}_{2}\cdot\vec{n} \nn\\ 
&& + 465 ( \vec{p}_{2}\cdot\vec{n})^{2} \big) - S_{1}^2 \big( 79 p_{1}^2 p_{2}^2 + 203 \vec{p}_{1}\cdot\vec{p}_{2} p_{2}^2 \nn\\ 
&& + 194 ( \vec{p}_{1}\cdot\vec{p}_{2})^{2} -248 \vec{p}_{1}\cdot\vec{n} \vec{p}_{2}\cdot\vec{n} \vec{p}_{1}\cdot\vec{p}_{2} - 2196 \vec{p}_{1}\cdot\vec{n} \vec{p}_{2}\cdot\vec{n} p_{2}^2 - 134 p_{2}^2 ( \vec{p}_{1}\cdot\vec{n})^{2} \nn\\ 
&& + 4 p_{1}^2 ( \vec{p}_{2}\cdot\vec{n})^{2} - 2579 \vec{p}_{1}\cdot\vec{p}_{2} ( \vec{p}_{2}\cdot\vec{n})^{2} -204 ( \vec{p}_{1}\cdot\vec{n})^{2} ( \vec{p}_{2}\cdot\vec{n})^{2} + 6294 \vec{p}_{1}\cdot\vec{n} ( \vec{p}_{2}\cdot\vec{n})^{3} \big) \nn\\ 
&& + \big( 60 p_{1}^2 p_{2}^2 - 831 \vec{p}_{1}\cdot\vec{p}_{2} p_{2}^2 + 242 ( \vec{p}_{1}\cdot\vec{p}_{2})^{2} -132 \vec{p}_{1}\cdot\vec{n} \vec{p}_{2}\cdot\vec{n} \vec{p}_{1}\cdot\vec{p}_{2} \nn\\ 
&& + 1599 \vec{p}_{1}\cdot\vec{n} \vec{p}_{2}\cdot\vec{n} p_{2}^2 + 228 p_{2}^2 ( \vec{p}_{1}\cdot\vec{n})^{2} - 42 p_{1}^2 ( \vec{p}_{2}\cdot\vec{n})^{2} + 1929 \vec{p}_{1}\cdot\vec{p}_{2} ( \vec{p}_{2}\cdot\vec{n})^{2} \nn\\ 
&& + 288 ( \vec{p}_{1}\cdot\vec{n})^{2} ( \vec{p}_{2}\cdot\vec{n})^{2} - 2673 \vec{p}_{1}\cdot\vec{n} ( \vec{p}_{2}\cdot\vec{n})^{3} \big) ( \vec{S}_{1}\cdot\vec{n})^{2} + \big( 59 p_{2}^2 \nn\\ 
&& + 10 ( \vec{p}_{2}\cdot\vec{n})^{2} \big) ( \vec{p}_{1}\cdot\vec{S}_{1})^{2} - 2 \big( p_{1}^2 - 339 \vec{p}_{1}\cdot\vec{p}_{2} + 1191 \vec{p}_{1}\cdot\vec{n} \vec{p}_{2}\cdot\vec{n} \nn\\ 
&& - 5 ( \vec{p}_{1}\cdot\vec{n})^{2} \big) ( \vec{p}_{2}\cdot\vec{S}_{1})^{2} \Big] - 	\frac{G^2}{16 m_{1} m_{2}{}^2 r{}^4} \Big[ 8 \vec{S}_{1}\cdot\vec{n} \vec{p}_{1}\cdot\vec{S}_{1} \big( 23 \vec{p}_{2}\cdot\vec{n} p_{2}^2 -7 ( \vec{p}_{2}\cdot\vec{n})^{3} \big) \nn\\ 
&& + 2 \vec{S}_{1}\cdot\vec{n} \vec{p}_{2}\cdot\vec{S}_{1} \big( 92 \vec{p}_{2}\cdot\vec{n} \vec{p}_{1}\cdot\vec{p}_{2} + 116 \vec{p}_{1}\cdot\vec{n} p_{2}^2 + 1179 \vec{p}_{2}\cdot\vec{n} p_{2}^2 -120 \vec{p}_{1}\cdot\vec{n} ( \vec{p}_{2}\cdot\vec{n})^{2} \nn\\ 
&& - 1565 ( \vec{p}_{2}\cdot\vec{n})^{3} \big) - 4 \vec{p}_{1}\cdot\vec{S}_{1} \vec{p}_{2}\cdot\vec{S}_{1} \big( 13 p_{2}^2 -5 ( \vec{p}_{2}\cdot\vec{n})^{2} \big) + S_{1}^2 \big( 4 \vec{p}_{1}\cdot\vec{p}_{2} p_{2}^2 \nn\\ 
&& + 147 p_{2}^{4} -72 \vec{p}_{1}\cdot\vec{n} \vec{p}_{2}\cdot\vec{n} p_{2}^2 - 4 \vec{p}_{1}\cdot\vec{p}_{2} ( \vec{p}_{2}\cdot\vec{n})^{2} - 1595 p_{2}^2 ( \vec{p}_{2}\cdot\vec{n})^{2} + 56 \vec{p}_{1}\cdot\vec{n} ( \vec{p}_{2}\cdot\vec{n})^{3} \nn\\ 
&& + 1966 ( \vec{p}_{2}\cdot\vec{n})^{4} \big) + 2 \big( 96 \vec{p}_{1}\cdot\vec{p}_{2} p_{2}^2 + 191 p_{2}^{4} -348 \vec{p}_{1}\cdot\vec{n} \vec{p}_{2}\cdot\vec{n} p_{2}^2 - 144 \vec{p}_{1}\cdot\vec{p}_{2} ( \vec{p}_{2}\cdot\vec{n})^{2} \nn\\ 
&& - 1356 p_{2}^2 ( \vec{p}_{2}\cdot\vec{n})^{2} + 160 \vec{p}_{1}\cdot\vec{n} ( \vec{p}_{2}\cdot\vec{n})^{3} + 536 ( \vec{p}_{2}\cdot\vec{n})^{4} \big) ( \vec{S}_{1}\cdot\vec{n})^{2} - \big( 24 \vec{p}_{1}\cdot\vec{p}_{2} \nn\\ 
&& + 351 p_{2}^2 -48 \vec{p}_{1}\cdot\vec{n} \vec{p}_{2}\cdot\vec{n} - 1267 ( \vec{p}_{2}\cdot\vec{n})^{2} \big) ( \vec{p}_{2}\cdot\vec{S}_{1})^{2} \Big] \nn\\ 
&& + 	\frac{G^2}{16 m_{2}{}^3 r{}^4} \Big[ 24 \vec{S}_{1}\cdot\vec{n} \vec{p}_{2}\cdot\vec{S}_{1} \vec{p}_{2}\cdot\vec{n} p_{2}^2 - S_{1}^2 \big( 27 p_{2}^{4} -2 p_{2}^2 ( \vec{p}_{2}\cdot\vec{n})^{2} \big) \nn\\ 
&& + 5 \big( 17 p_{2}^{4} -6 p_{2}^2 ( \vec{p}_{2}\cdot\vec{n})^{2} \big) ( \vec{S}_{1}\cdot\vec{n})^{2} - 4 p_{2}^2 ( \vec{p}_{2}\cdot\vec{S}_{1})^{2} \Big]\nn\\ && +  	\frac{G^3 m_{2}{}^3}{8 m_{1}{}^3 r{}^5} \Big[ 1382 \vec{S}_{1}\cdot\vec{n} \vec{p}_{1}\cdot\vec{S}_{1} \vec{p}_{1}\cdot\vec{n} - S_{1}^2 \big( 311 p_{1}^2 -434 ( \vec{p}_{1}\cdot\vec{n})^{2} \big) \nn\\ 
&& + \big( 635 p_{1}^2 -1556 ( \vec{p}_{1}\cdot\vec{n})^{2} \big) ( \vec{S}_{1}\cdot\vec{n})^{2} - 194 ( \vec{p}_{1}\cdot\vec{S}_{1})^{2} \Big] \nn\\ 
&& + 	\frac{G^3 m_{2}{}^2}{6144 m_{1}{}^2 r{}^5} \Big[ 20 \vec{S}_{1}\cdot\vec{n} \vec{p}_{1}\cdot\vec{S}_{1} \big( {(67040 - 8109 \pi^2)} \vec{p}_{1}\cdot\vec{n} - 61568 \vec{p}_{2}\cdot\vec{n} \big) \nn\\ 
&& + 3584 \vec{S}_{1}\cdot\vec{n} \vec{p}_{2}\cdot\vec{S}_{1} \vec{p}_{1}\cdot\vec{n} - S_{1}^2 \big( {(738944 - 8109 \pi^2)} p_{1}^2 - 592384 \vec{p}_{1}\cdot\vec{p}_{2} \nn\\ 
&& + 940544 \vec{p}_{1}\cdot\vec{n} \vec{p}_{2}\cdot\vec{n} - {(1429888 - 40545 \pi^2)} ( \vec{p}_{1}\cdot\vec{n})^{2} \big) + 65024 \vec{p}_{1}\cdot\vec{S}_{1} \vec{p}_{2}\cdot\vec{S}_{1} \nn\\ 
&& + \big( {(1338112 - 40545 \pi^2)} p_{1}^2 - 908288 \vec{p}_{1}\cdot\vec{p}_{2} + 1640960 \vec{p}_{1}\cdot\vec{n} \vec{p}_{2}\cdot\vec{n} \nn\\ 
&& - {(3014656 - 283815 \pi^2)} ( \vec{p}_{1}\cdot\vec{n})^{2} \big) ( \vec{S}_{1}\cdot\vec{n})^{2} - 2 {(62144 - 8109 \pi^2)} ( \vec{p}_{1}\cdot\vec{S}_{1})^{2} \Big] \nn\\ 
&& + 	\frac{G^3 m_{1}}{9800 m_{2} r{}^5} \Big[ 453080 \vec{S}_{1}\cdot\vec{n} \vec{p}_{2}\cdot\vec{S}_{1} \vec{p}_{2}\cdot\vec{n} - S_{1}^2 \big( 83323 p_{2}^2 -150440 ( \vec{p}_{2}\cdot\vec{n})^{2} \big) \nn\\ 
&& + 25 \big( 11437 p_{2}^2 -34048 ( \vec{p}_{2}\cdot\vec{n})^{2} \big) ( \vec{S}_{1}\cdot\vec{n})^{2} - 46596 ( \vec{p}_{2}\cdot\vec{S}_{1})^{2} \Big] \nn\\ 
&& + 	\frac{G^3 m_{2}}{3763200 m_{1} r{}^5} \Big[ 10 \vec{S}_{1}\cdot\vec{n} \vec{p}_{1}\cdot\vec{S}_{1} \big( 14566912 \vec{p}_{1}\cdot\vec{n} \nn\\ 
&& - {(125871200 - 9933525 \pi^2)} \vec{p}_{2}\cdot\vec{n} \big) - 2450 \vec{S}_{1}\cdot\vec{n} \vec{p}_{2}\cdot\vec{S}_{1} \big( {(18016 - 40545 \pi^2)} \vec{p}_{1}\cdot\vec{n} \nn\\ 
&& + 62080 \vec{p}_{2}\cdot\vec{n} \big) - S_{1}^2 \big( 77714432 p_{1}^2 - {(972238400 - 9933525 \pi^2)} \vec{p}_{1}\cdot\vec{p}_{2} \nn\\ 
&& + 215600000 p_{2}^2 + {(2404763200 - 49667625 \pi^2)} \vec{p}_{1}\cdot\vec{n} \vec{p}_{2}\cdot\vec{n} - 93496960 ( \vec{p}_{1}\cdot\vec{n})^{2} \nn\\ 
&& - 399212800 ( \vec{p}_{2}\cdot\vec{n})^{2} \big) + 2450 {(12032 - 8109 \pi^2)} \vec{p}_{1}\cdot\vec{S}_{1} \vec{p}_{2}\cdot\vec{S}_{1} + 25 \big( 7683712 p_{1}^2 \nn\\ 
&& - {(65388736 - 1986705 \pi^2)} \vec{p}_{1}\cdot\vec{p}_{2} + 10944640 p_{2}^2 \nn\\ 
&& + {(178040128 - 13906935 \pi^2)} \vec{p}_{1}\cdot\vec{n} \vec{p}_{2}\cdot\vec{n} - 12089728 ( \vec{p}_{1}\cdot\vec{n})^{2} \nn\\ 
&& + 1718528 ( \vec{p}_{2}\cdot\vec{n})^{2} \big) ( \vec{S}_{1}\cdot\vec{n})^{2} - 10075264 ( \vec{p}_{1}\cdot\vec{S}_{1})^{2} + 50489600 ( \vec{p}_{2}\cdot\vec{S}_{1})^{2} \Big] \nn\\ 
&& - 	\frac{G^3}{7526400 r{}^5} \Big[ 409968640 \vec{S}_{1}\cdot\vec{n} \vec{p}_{1}\cdot\vec{S}_{1} \vec{p}_{2}\cdot\vec{n} + 20 \vec{S}_{1}\cdot\vec{n} \vec{p}_{2}\cdot\vec{S}_{1} \big( 23428352 \vec{p}_{1}\cdot\vec{n} \nn\\ 
&& - {(46561760 - 9933525 \pi^2)} \vec{p}_{2}\cdot\vec{n} \big) - S_{1}^2 \big( 199932928 \vec{p}_{1}\cdot\vec{p}_{2} \nn\\ 
&& - {(1049148800 - 9933525 \pi^2)} p_{2}^2 -386800640 \vec{p}_{1}\cdot\vec{n} \vec{p}_{2}\cdot\vec{n} \nn\\ 
&& + {(2783670400 - 49667625 \pi^2)} ( \vec{p}_{2}\cdot\vec{n})^{2} \big) - 60640256 \vec{p}_{1}\cdot\vec{S}_{1} \vec{p}_{2}\cdot\vec{S}_{1} \nn\\ 
&& + 25 \big( 22151168 \vec{p}_{1}\cdot\vec{p}_{2} - {(70422016 - 1986705 \pi^2)} p_{2}^2 -69859328 \vec{p}_{1}\cdot\vec{n} \vec{p}_{2}\cdot\vec{n} \nn\\ 
&& + {(185086720 - 13906935 \pi^2)} ( \vec{p}_{2}\cdot\vec{n})^{2} \big) ( \vec{S}_{1}\cdot\vec{n})^{2} \nn\\ 
&& - 2450 {(3904 + 8109 \pi^2)} ( \vec{p}_{2}\cdot\vec{S}_{1})^{2} \Big]\nn\\ && +  	\frac{19 G^4 m_{1}{}^2 m_{2}}{4900 r{}^6} \Big[ 1571 S_{1}^2 - 4433 ( \vec{S}_{1}\cdot\vec{n})^{2} \Big] + 	\frac{G^4 m_{2}{}^4}{4 m_{1} r{}^6} \Big[ 73 S_{1}^2 - 119 ( \vec{S}_{1}\cdot\vec{n})^{2} \Big] \nn\\ 
&& + 	\frac{G^4 m_{1} m_{2}{}^2}{29400 r{}^6} \Big[ {(4275319 - 154350 \pi^2)} S_{1}^2 - 3 {(3162179 - 154350 \pi^2)} ( \vec{S}_{1}\cdot\vec{n})^{2} \Big] \nn\\ 
&& + 	\frac{G^4 m_{2}{}^3}{24 r{}^6} \Big[ {(3175 - 126 \pi^2)} S_{1}^2 - 9 {(785 - 42 \pi^2)} ( \vec{S}_{1}\cdot\vec{n})^{2} \Big],
\eea
and
\bea
H^{\text{N}^3\text{LO}}_{C_{1\text{E$^2$S$^2$}}}&=& 	- \frac{G^4 m_{1} m_{2}{}^2}{2 r{}^6} 
\Big[ S_{1}^2 - 3 ( \vec{S}_{1}\cdot\vec{n})^{2} \Big].
\eea
These Hamiltonians are also provided in machine-readable format in the ancillary files to 
this publication.

\section{COM Hamiltonians}
\label{comhamresults}

The Hamiltonian of the present sectors in the COM frame is written as:
\bea
\tilde{H}^{\text{N}^3\text{LO}}_{\text{S}^2} =
\tilde{H}^{\text{N}^3\text{LO}}_{\text{S}_1 \text{S}_2} 
+\tilde{H}^{\text{N}^3\text{LO}}_{\text{S}_1^2} 
+ C_{1\text{ES}^2} \tilde{H}^{\text{N}^3\text{LO}}_{C_{1\text{ES}^2}} 
+ C_{1\text{E$^2$S$^2$}} \tilde{H}^{\text{N}^3\text{LO}}_{C_{1\text{E$^2$S$^2$}}} 
+ (1 \leftrightarrow 2),
\eea
with
\bea
\tilde{H}^{\text{N}^3\text{LO}}_{\text{S}_1 \text{S}_2} &=& \frac{\nu \tilde{\vec{S}}_1 \cdot  \tilde{\vec{S}}_2}{\tilde{r}^6}  \left[\left(\frac{101641}{1200}-\frac{81 \pi ^2}{16}\right) \nu +\frac{4537}{200}+ \left( -3 \nu ^2+\left(\frac{423 \pi ^2}{128}-\frac{67947}{800}\right) \nu \right.\right.\nn\\
&&\left.-\frac{253}{100} \right) \frac{\tilde{L}^2}{\tilde{r}} 
+\left( -\frac{\nu ^2}{8}-\frac{9 \nu }{4}-\frac{1}{16} \right) \frac{\tilde{L}^4}{\tilde{r}^2}
+\left( -\frac{5 \nu ^3}{32}+\frac{5 \nu ^2}{16}-\frac{3 \nu }{32} \right) \frac{\tilde{L}^6}{\tilde{r}^3}\nn\\
&&+ \tilde{p}_r^2 \tilde{r} \left(\frac{109 \nu ^2}{16}+\left(\frac{306553}{2400}-\frac{141 \pi ^2}{32}\right) \nu +\frac{556}{25} 
+\left(\frac{1373 \nu ^2}{64}-\frac{93 \nu }{8}+\frac{1}{8} \right) \frac{\tilde{L}^2}{\tilde{r}} \right.\nn\\
&&\left.+\left(-\frac{33 \nu ^3}{128}+\frac{215 \nu ^2}{32}-\frac{237 \nu }{64}+\frac{19}{32} \right) \frac{\tilde{L}^4}{\tilde{r}^2}\right)\nn\\
&&+ \tilde{p}_r^4 \tilde{r}^2 \left(\frac{575 \nu ^2}{64}-\frac{231 \nu }{16}+\frac{3}{16}
+\left(\frac{15 \nu ^3}{64}+\frac{353 \nu ^2}{16}-\frac{369 \nu }{32}+\frac{19}{16} \right) \frac{\tilde{L}^2}{\tilde{r}} \right)\nn\\
&&\left.+ \tilde{p}_r^6 \tilde{r}^3 \left(-\frac{187 \nu ^3}{128}+\frac{153 \nu ^2}{16}-\frac{327 \nu }{64}+\frac{19}{32} \right) \right]\nn\\%%%%%%%
&&+\frac{\nu \tilde{\vec{S}}_1 \cdot \tilde{\vec{L}}  \tilde{\vec{S}}_2\cdot \tilde{\vec{L}}}{\tilde{r}^7}  \left[-\frac{161 \nu ^2}{16}+\left(-\frac{26183}{1200}-\frac{141 \pi ^2}{64}\right) \nu -\frac{5831}{200}
\right.\nn\\
&&+\left( \frac{487 \nu ^2}{32}+\frac{411 \nu }{32}-\frac{7}{2} \right) \frac{\tilde{L}^2}{\tilde{r}} 
+\left( -\frac{15 \nu ^3}{64}-\frac{49 \nu ^2}{32}+\frac{9 \nu }{4}-\frac{17}{32} \right) \frac{\tilde{L}^4}{\tilde{r}^2} \nn\\
&&+ \tilde{p}_r^2 \tilde{r} \left(-\frac{579 \nu ^2}{32}+\frac{535 \nu }{64}-\frac{7}{2}
+\left(-\frac{51 \nu ^3}{32}-\frac{377 \nu ^2}{32}+\frac{303 \nu }{32}-\frac{17}{16} \right) \frac{\tilde{L}^2}{\tilde{r}} \right)\nn\\
&&\left. + \tilde{p}_r^4 \tilde{r}^2 \left(-\frac{237 \nu ^3}{64}-\frac{943 \nu ^2}{32}+\frac{321 \nu }{32}-\frac{17}{32}\right)
\right]\nn\\%%%%%%%
&&+\frac{\nu \tilde{\vec{S}}_1 \cdot \vec{n} \tilde{\vec{S}}_2 \cdot \vec{n} }{\tilde{r}^6}  \left[\left(\frac{243 \pi ^2}{16}-\frac{97861}{400}\right) \nu -\frac{7531}{200}
+ \left( -\frac{11 \nu ^2}{32} \right.\right.\nn\\
&&\left.+\left(\frac{362959}{2400}-\frac{987 \pi ^2}{128}\right) \nu -\frac{503}{100} \right) \frac{\tilde{L}^2}{\tilde{r}} 
+\left( \frac{801 \nu ^2}{64}-\frac{9 \nu }{8}+\frac{19}{16} \right) \frac{\tilde{L}^4}{\tilde{r}^2}\nn\\
&&
+\left( \frac{75 \nu ^3}{128}-\frac{139 \nu ^2}{32}+\frac{87 \nu }{64}-\frac{1}{16} \right) \frac{\tilde{L}^6}{\tilde{r}^3}\nn\\
&&+ \tilde{p}_r^2 \tilde{r} \left(-\frac{53 \nu ^2}{16}+\left(\frac{423 \pi ^2}{32}-\frac{158163}{800}\right) \nu -\frac{553}{25}
+\left(-\frac{1583 \nu ^2}{32}+\frac{987 \nu }{16}+1 \right) \frac{\tilde{L}^2}{\tilde{r}} \right.\nn\\
&&\left.+\left(\frac{231 \nu ^3}{128}-\frac{565 \nu ^2}{32}+\frac{555 \nu }{64}-\frac{23}{32} \right) \frac{\tilde{L}^4}{\tilde{r}^2}\right)\nn\\
&&
+ \tilde{p}_r^4 \tilde{r}^2 \left(-\frac{2943 \nu ^2}{64}+\frac{223 \nu }{16}-\frac{3}{16}
+\left(\frac{255 \nu ^3}{128}-\frac{725 \nu ^2}{32}+\frac{783 \nu }{64}-\frac{5}{4} \right) \frac{\tilde{L}^2}{\tilde{r}} \right) \nn\\
&&\left.+ \tilde{p}_r^6 \tilde{r}^3 \left(\frac{699 \nu ^3}{128}-\frac{217 \nu ^2}{16}+\frac{375 \nu }{64}-\frac{19}{32} \right) \right]\nn\\%%%%%%%
&&+\frac{\nu \tilde{p}_r \tilde{\vec{S}}_1 \cdot \vec{n}  \tilde{\vec{S}}_2\times \tilde{\vec{L}}\cdot \vec{n} }{\tilde{r}^6}  \left[-\frac{1309 \nu ^2}{16}+\left(\frac{31469}{300}-\frac{141 \pi ^2}{8}\right) \nu +\frac{354}{25}
\right.\nn\\
&&+\left(3 \nu ^3+\frac{1605 \nu ^2}{16}-\frac{181 \nu }{16}-\frac{9}{4} \right) \frac{\tilde{L}^2}{\tilde{r}} 
+\left(-\frac{291 \nu ^3}{64}+\frac{17 \nu ^2}{2}-\frac{105 \nu }{32}+\frac{1}{8} \right) \frac{\tilde{L}^4}{\tilde{r}^2} \nn\\
&&+ \tilde{p}_r^2 \tilde{r} \left(\frac{27 \nu ^3}{2}+\frac{3409 \nu ^2}{16}-\frac{3631 \nu }{32}-\frac{9}{4}
+\left(-\frac{669 \nu ^3}{32}+\frac{59 \nu ^2}{4}-\frac{39 \nu }{16}+\frac{1}{4} \right) \frac{\tilde{L}^2}{\tilde{r}} \right)\nn\\
&&\left. + \tilde{p}_r^4 \tilde{r}^2 \left( -\frac{2637 \nu ^3}{64}+25 \nu ^2-\frac{153 \nu }{32}+\frac{1}{8}\right)
\right]\nn\\%%%%%%%
&&+\frac{\nu  \tilde{p}_r \tilde{\vec{S}}_1 \cdot \vec{n}  \tilde{\vec{S}}_2\times \tilde{\vec{L}}\cdot \vec{n} }{q\tilde{r}^6}  \left[-\frac{1221 \nu ^2}{16}-\frac{275 \nu }{4}
\right.\nn\\
&&+\left( 3 \nu ^3+\frac{677 \nu ^2}{16}+\frac{7 \nu }{4} \right) \frac{\tilde{L}^2}{\tilde{r}} 
+\left( -\frac{27 \nu ^3}{8}+\frac{111 \nu ^2}{16}-\frac{21 \nu }{16} \right) \frac{\tilde{L}^4}{\tilde{r}^2} \nn\\
&&+ \tilde{p}_r^2 \tilde{r} \left( \frac{27 \nu ^3}{2}+\frac{695 \nu ^2}{4}+\frac{7 \nu }{4}
+\left( -\frac{129 \nu ^3}{8}+\frac{177 \nu ^2}{16}-\frac{21 \nu }{8} \right) \frac{\tilde{L}^2}{\tilde{r}} \right)\nn\\
&&\left. + \tilde{p}_r^4 \tilde{r}^2 \left(  -\frac{519 \nu ^3}{16}+\frac{69 \nu ^2}{4}-\frac{21 \nu }{16}\right)
\right],%%%%%%%
\eea

\bea
\tilde{H}^{\text{N}^3\text{LO}}_{\text{S}_1^2} &=& \frac{\nu^2 \tilde{S}_1^2}{\tilde{r}^6}  \left[\left(\frac{45 \pi ^2}{64}-\frac{6679}{3920}\right) \nu +\frac{3987  }{2450}
+ \left( \left(\frac{29221}{3920}-\frac{135 \pi ^2}{64}\right) \nu  \right.\right.\nn\\
&&\left.+\frac{45257  }{19600} \right) \frac{\tilde{L}^2}{\tilde{r}} 
+\left(\frac{9 \nu ^2}{16}+\frac{59 \nu }{16}-\frac{5  }{16} \right) \frac{\tilde{L}^4}{\tilde{r}^2}
\nn\\
&&+ \tilde{p}_r^2 \tilde{r} \left(\left(\frac{45 \pi ^2}{16}-\frac{143221}{7840}\right) \nu -\frac{54379  }{4900} 
+\left(\frac{53 \nu ^2}{32}+\frac{179 \nu }{32}-\frac{5  }{2} \right) \frac{\tilde{L}^2}{\tilde{r}} \right.\nn\\
&&\left.+\left(\frac{131 \nu ^2}{128}+\frac{43 \nu }{8}-\frac{75  }{128} \right) \frac{\tilde{L}^4}{\tilde{r}^2}\right)\nn\\
&&+ \tilde{p}_r^4 \tilde{r}^2 \left(\frac{177 \nu ^2}{32}+\frac{43 \nu }{32}+\frac{13  }{2}
+\left(\frac{185 \nu ^2}{64}-\frac{175 \nu }{32}+\frac{3  }{64} \right)\frac{\tilde{L}^2}{\tilde{r}}\right)\nn\\
&&\left.+ \tilde{p}_r^6 \tilde{r}^3 \left(-\frac{31 \nu ^2}{128}-\frac{47 \nu }{32}+\frac{81  }{128} \right) \right]\nn\\%%%%%%%
&&+\frac{\nu \tilde{S}_1^2}{q\tilde{r}^6}  \left[\left(\frac{45 \pi ^2}{64}-\frac{6679}{3920}\right) \nu ^2-\frac{11 \nu }{4}+9
+ \left( \left(\frac{20751}{3920}-\frac{135 \pi ^2}{64}\right) \nu ^2 \right.\right.\nn\\
&&\left.+\left(\frac{189 \pi ^2}{2048}-\frac{29}{2}\right) \nu +\frac{7}{16}\right) \frac{\tilde{L}^2}{\tilde{r}} 
+\left(\frac{9 \nu ^3}{16}+3 \nu ^2-\frac{55 \nu }{16}+\frac{3}{16} \right) \frac{\tilde{L}^4}{\tilde{r}^2}
\nn\\
&&+ \tilde{p}_r^2 \tilde{r} \left(\left(\frac{45 \pi ^2}{16}-\frac{132651}{7840}\right) \nu ^2+\left(-\frac{1835}{32}-\frac{63 \pi ^2}{512}\right) \nu -1\right.\nn\\
&&\left.
+\left(\frac{53 \nu ^3}{32}-\frac{35 \nu ^2}{32}+\frac{1077 \nu }{32}+\frac{27}{8} \right) \frac{\tilde{L}^2}{\tilde{r}} +\left(\frac{29 \nu ^3}{32}+\frac{67 \nu ^2}{16}-\frac{107 \nu }{64}+\frac{75}{128} \right) \frac{\tilde{L}^4}{\tilde{r}^2}\right)\nn\\
&&+ \tilde{p}_r^4 \tilde{r}^2 \left(\frac{177 \nu ^3}{32}+\frac{371 \nu ^2}{8}+\frac{177 \nu }{32}-\frac{11}{2}
+\left(\frac{79 \nu ^3}{32}-\frac{91 \nu ^2}{16}-\frac{103 \nu }{16}-\frac{3}{64} \right)\frac{\tilde{L}^2}{\tilde{r}} \right)\nn\\
&&\left.+ \tilde{p}_r^6 \tilde{r}^3 \left(-\frac{25 \nu ^3}{32}-\frac{137 \nu ^2}{64}+\frac{175 \nu }{64}-\frac{81}{128} \right) \right]\nn\\%%%%%%%
&&+\frac{\nu^2 (\tilde{\vec{S}}_1 \cdot \tilde{\vec{L}} )^2 }{\tilde{r}^7}  \left[\left(\frac{45 \pi ^2}{32}-\frac{7869}{245}\right) \nu -\frac{443089}{9800}
\right.\nn\\
&&+\left( \frac{29 \nu ^2}{16}+\frac{73 \nu }{32}-\frac{145}{16} \right) \frac{\tilde{L}^2}{\tilde{r}} 
+\left(\frac{29 \nu ^2}{16}+\frac{73 \nu }{32}-\frac{145}{16} \right) \frac{\tilde{L}^4}{\tilde{r}^2} \nn\\
&&+ \tilde{p}_r^2 \tilde{r} \left(\frac{115 \nu ^2}{16}+\frac{3971 \nu }{32}+\frac{9}{8}
+\left(-\frac{257 \nu ^2}{32}+\frac{23 \nu }{32}-\frac{3}{16} \right) \frac{\tilde{L}^2}{\tilde{r}} \right)\nn\\
&&\left. + \tilde{p}_r^4 \tilde{r}^2 \left(-\frac{845 \nu ^2}{64}+\frac{29 \nu }{4}+\frac{33}{64}\right)
\right]\nn\\%%%%%%%
&&+\frac{\nu (\tilde{\vec{S}}_1 \cdot \tilde{\vec{L}} )^2 }{q\tilde{r}^7}  \left[\left(\frac{45 \pi ^2}{32}-\frac{94789}{3920}\right) \nu ^2+\left(-\frac{43}{2}-\frac{63 \pi ^2}{1024}\right) \nu +\frac{217}{8}
\right.\nn\\
&&+\left(\frac{29 \nu ^3}{16}-\frac{25 \nu ^2}{4}-\frac{1037 \nu }{32}+\frac{121}{16} \right) \frac{\tilde{L}^2}{\tilde{r}} 
+\left(-\frac{13 \nu ^3}{8}+\frac{127 \nu ^2}{16}-\frac{19 \nu }{8}+\frac{45}{64}\right) \frac{\tilde{L}^4}{\tilde{r}^2} \nn\\
&&+ \tilde{p}_r^2 \tilde{r} \left(\frac{115 \nu ^3}{16}+\frac{2619 \nu ^2}{32}-\frac{3431 \nu }{32}-\frac{21}{8}
+\left(-\frac{109 \nu ^3}{16}+\frac{403 \nu ^2}{32}-\frac{371 \nu }{32}+\frac{3}{16} \right) \frac{\tilde{L}^2}{\tilde{r}} \right)\nn\\
&&\left. + \tilde{p}_r^4 \tilde{r}^2 \left(-\frac{173 \nu ^3}{16}+\frac{67 \nu ^2}{16}-\frac{55 \nu }{32}-\frac{33}{64}\right)
\right]\nn\\%%%%%%%
&&+\frac{\nu^2 (\tilde{\vec{S}}_1 \cdot \vec{n})^2 }{\tilde{r}^6}  \left[\left(\frac{9627}{784}-\frac{135 \pi ^2}{64}\right) \nu -\frac{3058}{1225}
+ \left( \left(\frac{315 \pi ^2}{64}-\frac{1197}{160}\right) \nu  \right.\right.\nn\\
&&\left.+\frac{26373}{1400} \right) \frac{\tilde{L}^2}{\tilde{r}} 
+\left( -\frac{77 \nu ^2}{32}-\frac{433 \nu }{16}+\frac{23}{8}\right) \frac{\tilde{L}^4}{\tilde{r}^2}
+\left(\frac{215 \nu ^2}{128}-\frac{7 \nu }{4}+\frac{9}{128}\right) \frac{\tilde{L}^6}{\tilde{r}^3}  \nn\\
&&+ \tilde{p}_r^2 \tilde{r} \left(\left(\frac{770773}{7840}-\frac{135 \pi ^2}{16}\right) \nu +\frac{131497}{4900}
+\left(
-\frac{121 \nu ^2}{16}-\frac{1955 \nu }{32}+\frac{115}{8}
\right) \frac{\tilde{L}^2}{\tilde{r}} \right.\nn\\
&&\left.+\left(
\frac{707 \nu ^2}{128}-\frac{323 \nu }{32}+\frac{249}{128}
\right) \frac{\tilde{L}^4}{\tilde{r}^2}\right)\nn\\
&&
+ \tilde{p}_r^4 \tilde{r}^2 \left(
-\frac{317 \nu ^2}{32}+\frac{601 \nu }{32}-\frac{47}{8}
+\left(
\frac{823 \nu ^2}{128}-\frac{5 \nu }{4}+\frac{159}{128}
\right) \frac{\tilde{L}^2}{\tilde{r}} \right) \nn\\
&&\left.+ \tilde{p}_r^6 \tilde{r}^3 \left(
\frac{31 \nu ^2}{128}+\frac{47 \nu }{32}-\frac{81}{128}
\right) \right]\nn\\%%%%%%%
&&+\frac{\nu (\tilde{\vec{S}}_1 \cdot \vec{n})^2 }{q\tilde{r}^6}  \left[
\left(\frac{9627}{784}-\frac{135 \pi ^2}{64}\right) \nu ^2-\frac{109 \nu }{4}-9
+ \left( 
\left(\frac{315 \pi ^2}{64}-\frac{1837}{560}\right) \nu ^2
\right.\right.\nn\\
&&\left.
+\left(\frac{2363}{32}-\frac{441 \pi ^2}{2048}\right) \nu -\frac{223}{8}
\right) \frac{\tilde{L}^2}{\tilde{r}} 
+\left( 
-\frac{77 \nu ^3}{32}-\frac{613 \nu ^2}{32}+\frac{277 \nu }{16}-\frac{23}{8}
\right) \frac{\tilde{L}^4}{\tilde{r}^2}\nn\\
&&
+\left(
\frac{25 \nu ^3}{16}-\frac{23 \nu ^2}{4}+\frac{265 \nu }{64}-\frac{9}{128}
\right) \frac{\tilde{L}^6}{\tilde{r}^3}  \nn\\
&&+ \tilde{p}_r^2 \tilde{r} \left(
\left(\frac{677323}{7840}-\frac{135 \pi ^2}{16}\right) \nu ^2+\left(\frac{171}{32}+\frac{189 \pi ^2}{512}\right) \nu +1\right.\nn\\
&&
\left.+\left(
-\frac{121 \nu ^3}{16}-\frac{2301 \nu ^2}{32}+\frac{1407 \nu }{32}-\frac{59}{4}
\right) \frac{\tilde{L}^2}{\tilde{r}}  +\left(
\frac{167 \nu ^3}{32}-\frac{607 \nu ^2}{32}+\frac{199 \nu }{64}-\frac{249}{128}
\right) \frac{\tilde{L}^4}{\tilde{r}^2}\right)\nn\\
&&
+ \tilde{p}_r^4 \tilde{r}^2 \left(
-\frac{317 \nu ^3}{32}-24 \nu ^2-\frac{305 \nu }{32}+\frac{11}{2}
+\left(
\frac{101 \nu ^3}{16}-\frac{139 \nu ^2}{64}+\frac{719 \nu }{64}-\frac{159}{128}
\right) \frac{\tilde{L}^2}{\tilde{r}} \right) \nn\\
&&\left.+ \tilde{p}_r^6 \tilde{r}^3 \left(
\frac{25 \nu ^3}{32}+\frac{137 \nu ^2}{64}-\frac{175 \nu }{64}+\frac{81}{128}
\right) \right]\nn\\%%%%%%%
&&+\frac{\nu^2 \tilde{p}_r \tilde{\vec{S}}_1 \cdot \vec{n}  \tilde{\vec{S}}_1\times \tilde{\vec{L}}\cdot \vec{n} }{\tilde{r}^6}  \left[
\left(\frac{45 \pi ^2}{4}-\frac{91071}{1960}\right) \nu -\frac{812367}{9800}
\right.\nn\\
&&+\left(
\frac{7 \nu ^2}{2}-\frac{373 \nu }{16}-\frac{19}{16}
\right) \frac{\tilde{L}^2}{\tilde{r}} 
+\left(
-\frac{173 \nu ^2}{64}-\frac{29 \nu }{8}+\frac{33}{64}
\right) \frac{\tilde{L}^4}{\tilde{r}^2} \nn\\
&&+ \tilde{p}_r^2 \tilde{r} \left(
\frac{43 \nu ^2}{4}+\frac{531 \nu }{8}-\frac{297}{16}
+\left(
-\frac{151 \nu ^2}{16}+\frac{163 \nu }{16}-\frac{45}{32}
\right) \frac{\tilde{L}^2}{\tilde{r}} \right)\nn\\
&&\left. + \tilde{p}_r^4 \tilde{r}^2 \left( 
-\frac{581 \nu ^2}{64}+\frac{131 \nu }{16}-\frac{123}{64}
\right)
\right]\nn\\%%%%%%%
&&+\frac{\nu  \tilde{p}_r \tilde{\vec{S}}_1 \cdot \vec{n}  \tilde{\vec{S}}_1\times \tilde{\vec{L}}\cdot \vec{n} }{q\tilde{r}^6}  \left[
\left(\frac{45 \pi ^2}{4}-\frac{193027}{3920}\right) \nu ^2+\left(\frac{519}{16}-\frac{63 \pi ^2}{128}\right) \nu +\frac{295}{8}
\right.\nn\\
&&+\left( 
\frac{7 \nu ^3}{2}-\frac{3 \nu ^2}{2}-\frac{413 \nu }{16}-\frac{5}{16}
\right) \frac{\tilde{L}^2}{\tilde{r}} 
+\left( 
-\frac{79 \nu ^3}{32}+\frac{25 \nu ^2}{16}-\frac{79 \nu }{32}-\frac{33}{64}
\right) \frac{\tilde{L}^4}{\tilde{r}^2} \nn\\
&&+ \tilde{p}_r^2 \tilde{r} \left(
\frac{43 \nu ^3}{4}+\frac{249 \nu ^2}{8}-97 \nu +\frac{273}{16}
+\left( 
-\frac{275 \nu ^3}{32}+\frac{655 \nu ^2}{32}+5 \nu +\frac{45}{32}
\right) \frac{\tilde{L}^2}{\tilde{r}} \right)\nn\\
&&\left. + \tilde{p}_r^4 \tilde{r}^2 \left( 
-8 \nu ^3+10 \nu ^2-\frac{241 \nu }{32}+\frac{123}{64}
\right)\right],
\eea

\bea
\tilde{H}^{\text{N}^3\text{LO}}_{C_{1\text{ES}^2}} &=& \frac{\nu^2 \tilde{S}_1^2}{\tilde{r}^6}  \left[
\frac{177 \nu }{7}-\frac{14894}{1225}
+ \left(
-\frac{1399 \nu }{56}+\frac{244353}{4900}
\right) \frac{\tilde{L}^2}{\tilde{r}} 
+\left(
-\frac{9 \nu ^2}{16}-\frac{115 \nu }{16}+\frac{41}{16}
\right) \frac{\tilde{L}^4}{\tilde{r}^2}\right. 
\nn\\
&&+ \tilde{p}_r^2 \tilde{r} \left(
\frac{1445 \nu }{14}-\frac{41779}{4900}
+\left(
-\frac{15 \nu ^2}{8}-\frac{35 \nu }{2}+\frac{31}{2}
\right) \frac{\tilde{L}^2}{\tilde{r}} +\left(
-\frac{9 \nu ^2}{16}-\frac{23 \nu }{16}+\frac{9}{8}
\right) \frac{\tilde{L}^4}{\tilde{r}^2}\right)\nn\\
&&+ \tilde{p}_r^4 \tilde{r}^2 \left(
-\frac{77 \nu ^2}{16}-\frac{269 \nu }{16}+\frac{243}{16}
+\left(
-\frac{45 \nu ^2}{16}-\frac{23 \nu }{8}+\frac{9}{4}
\right)\frac{\tilde{L}^2}{\tilde{r}}\right)\nn\\
&&\left.+ \tilde{p}_r^6 \tilde{r}^3 \left(
-6 \nu ^2-\frac{\nu }{2}+\frac{9}{8}
\right) \right]\nn\\%%%%%%%
&&+\frac{\nu \tilde{S}_1^2}{q\tilde{r}^6}  \left[
\frac{177 \nu ^2}{7}+\left(\frac{1861}{24}-\frac{21 \pi ^2}{4}\right) \nu +\frac{73}{4}
+ \left( 
-\frac{1625 \nu ^2}{112} \right.\right.\nn\\
&&\left.+\left(5+\frac{8109 \pi ^2}{2048}\right) \nu -\frac{505}{8}
\right) \frac{\tilde{L}^2}{\tilde{r}} 
+\left(
-\frac{9 \nu ^3}{16}-\frac{41 \nu ^2}{8}+\frac{121 \nu }{4}-\frac{9}{2}
\right) \frac{\tilde{L}^4}{\tilde{r}^2}
\nn\\
&&+\left(
-\frac{5 \nu ^3}{32}-\frac{35 \nu ^2}{32}+\frac{17 \nu }{16}-\frac{7}{32}
\right) \frac{\tilde{L}^6}{\tilde{r}^3}\nn\\
&&+ \tilde{p}_r^2 \tilde{r} \left(
\frac{3771 \nu ^2}{28}+\left(\frac{861}{8}-\frac{2703 \pi ^2}{512}\right) \nu +\frac{123}{8}
\right.\nn\\
&&\left.
+\left(
-\frac{15 \nu ^3}{8}+\frac{187 \nu ^2}{16}+\frac{1365 \nu }{16}-19
\right) \frac{\tilde{L}^2}{\tilde{r}} 
+\left(
-\frac{15 \nu ^3}{16}-\frac{153 \nu ^2}{32}+\frac{53 \nu }{8}-\frac{57}{32}
\right) \frac{\tilde{L}^4}{\tilde{r}^2}\right)\nn\\
&&+ \tilde{p}_r^4 \tilde{r}^2 \left(
-\frac{77 \nu ^3}{16}-\frac{193 \nu ^2}{4}+\frac{173 \nu }{8}-\frac{67}{4}
+\left(
-3 \nu ^3-\frac{69 \nu ^2}{16}+\frac{161 \nu }{16}-\frac{93}{32}
\right)\frac{\tilde{L}^2}{\tilde{r}} \right)\nn\\
&&\left.+ \tilde{p}_r^6 \tilde{r}^3 \left(
-8 \nu ^3+5 \nu ^2+\frac{57 \nu }{16}-\frac{43}{32}
\right) \right]\nn\\%%%%%%%
&&+\frac{\nu^2 (\tilde{\vec{S}}_1 \cdot \tilde{\vec{L}} )^2 }{\tilde{r}^7}  \left[
\frac{419 \nu }{28}-\frac{95527}{4900}
\right. +\left(
-\frac{3 \nu ^2}{4}-\frac{27 \nu }{4}-\frac{15}{2} 
\right) \frac{\tilde{L}^2}{\tilde{r}} 
+\left(
\frac{15 \nu ^2}{16}+\frac{\nu }{4}-\frac{9}{16}
\right) \frac{\tilde{L}^4}{\tilde{r}^2} \nn\\
&&+ \tilde{p}_r^2 \tilde{r} \left(
-\frac{11 \nu ^2}{4}+46 \nu +\frac{3}{4}
+\left(
\frac{63 \nu ^2}{16}+\frac{23 \nu }{16}-\frac{9}{8}
\right) \frac{\tilde{L}^2}{\tilde{r}} \right)\nn\\
&&\left. + \tilde{p}_r^4 \tilde{r}^2 \left(
\frac{93 \nu ^2}{16}+\frac{17 \nu }{8}-\frac{9}{16}
\right)
\right]\nn\\%%%%%%%
&&+\frac{\nu (\tilde{\vec{S}}_1 \cdot \tilde{\vec{L}} )^2 }{q\tilde{r}^7}  \left[
\frac{97 \nu ^2}{28}+\left(-\frac{1059}{16}-\frac{2703 \pi ^2}{1024}\right) \nu +\frac{97}{4}
\right.\nn\\
&&+\left(
-\frac{3 \nu ^3}{4}-\frac{29 \nu ^2}{8}-\frac{427 \nu }{16}+\frac{31}{4}
\right) \frac{\tilde{L}^2}{\tilde{r}} 
+\left(
\frac{15 \nu ^3}{16}-\frac{21 \nu ^2}{16}-\frac{7 \nu }{8}+\frac{9}{16}
\right) \frac{\tilde{L}^4}{\tilde{r}^2} \nn\\
&&+ \tilde{p}_r^2 \tilde{r} \left(
-\frac{11 \nu ^3}{4}+\frac{1637 \nu ^2}{16}+12 \nu -\frac{1}{2}
+\left(
\frac{57 \nu ^3}{16}-\frac{63 \nu ^2}{16}-\frac{43 \nu }{16}+\frac{9}{8}
\right) \frac{\tilde{L}^2}{\tilde{r}} \right)\nn\\
&&\left. + \tilde{p}_r^4 \tilde{r}^2 \left(
\frac{9 \nu ^3}{2}-\frac{57 \nu ^2}{16}-\frac{11 \nu }{4}+\frac{9}{16}
\right)
\right]\nn\\%%%%%%%
&&+\frac{\nu^2 (\tilde{\vec{S}}_1 \cdot \vec{n})^2 }{\tilde{r}^6}  \left[
\frac{15387}{1225}-\frac{286 \nu }{7}
+ \left(
\frac{3205 \nu }{56}-\frac{24393}{350}  
\right) \frac{\tilde{L}^2}{\tilde{r}} 
+\left(
\frac{9 \nu ^2}{16}-\frac{81 \nu }{16}-\frac{169}{16}
\right) \frac{\tilde{L}^4}{\tilde{r}^2}\right.\nn\\
&&
+\left(
\frac{15 \nu ^2}{16}+\frac{\nu }{4}-\frac{9}{16}
\right) \frac{\tilde{L}^6}{\tilde{r}^3}  \nn\\
&&+ \tilde{p}_r^2 \tilde{r} \left(
-\frac{2823 \nu }{14}-\frac{242933}{4900}
+\left(
\frac{21 \nu ^2}{8}+\frac{333 \nu }{4}-\frac{39}{2}
\right) \frac{\tilde{L}^2}{\tilde{r}}  +\left(
\frac{9 \nu ^2}{2}+\frac{23 \nu }{8}-\frac{9}{4}
\right) \frac{\tilde{L}^4}{\tilde{r}^2}\right)\nn\\
&&
\left.+ \tilde{p}_r^4 \tilde{r}^2 \left(
\frac{217 \nu ^2}{16}+\frac{1129 \nu }{16}-\frac{311}{16}
+\left(
\frac{69 \nu ^2}{8}+5 \nu -\frac{45}{16}
\right) \frac{\tilde{L}^2}{\tilde{r}} \right)  + \tilde{p}_r^6 \tilde{r}^3 \left(
6 \nu ^2+\frac{\nu }{2}-\frac{9}{8}
\right) \right]\nn\\%%%%%%%
&&+\frac{\nu (\tilde{\vec{S}}_1 \cdot \vec{n})^2 }{q\tilde{r}^6}  \left[
-\frac{286 \nu ^2}{7}+\left(\frac{63 \pi ^2}{4}-\frac{1641}{8}\right) \nu -\frac{119}{4}
+ \left( 
\frac{629 \nu ^2}{16}
\right.\right.\nn\\
&&\left.
+\left(-\frac{289}{16}-\frac{18921 \pi ^2}{2048}\right) \nu +\frac{829}{8}
\right) \frac{\tilde{L}^2}{\tilde{r}} 
+\left( 
\frac{9 \nu ^3}{16}-\frac{53 \nu ^2}{4}-\frac{1167 \nu }{16}+\frac{129}{8}
\right) \frac{\tilde{L}^4}{\tilde{r}^2}\nn\\
&&
+\left(
\frac{45 \nu ^3}{32}+\frac{63 \nu ^2}{32}-\frac{65 \nu }{16}+\frac{39}{32}
\right) \frac{\tilde{L}^6}{\tilde{r}^3}  \nn\\
&&+ \tilde{p}_r^2 \tilde{r} \left(
-\frac{6161 \nu ^2}{28}+\left(\frac{8109 \pi ^2}{512}-\frac{1107}{8}\right) \nu +\frac{267}{8}
\right.\nn\\
&&
\left.+\left(
\frac{21 \nu ^3}{8}+\frac{585 \nu ^2}{8}-\frac{1987 \nu }{16}+\frac{119}{4}
\right) \frac{\tilde{L}^2}{\tilde{r}}  +\left(
6 \nu ^3+\frac{297 \nu ^2}{32}-\frac{275 \nu }{16}+\frac{135}{32}
\right) \frac{\tilde{L}^4}{\tilde{r}^2}\right)\nn\\
&&
+ \tilde{p}_r^4 \tilde{r}^2 \left(
\frac{217 \nu ^3}{16}+\frac{115 \nu ^2}{2}-\frac{297 \nu }{4}+\frac{193}{8}
+\left(
\frac{21 \nu ^3}{2}+12 \nu ^2-\frac{355 \nu }{16}+\frac{153}{32}
\right) \frac{\tilde{L}^2}{\tilde{r}} \right) \nn\\
&&\left.+ \tilde{p}_r^6 \tilde{r}^3 \left(
12 \nu ^3-\frac{115 \nu }{16}+\frac{57}{32}
\right) \right]\nn\\%%%%%%%
&&+\frac{\nu^2 \tilde{p}_r \tilde{\vec{S}}_1 \cdot \vec{n}  \tilde{\vec{S}}_1\times \tilde{\vec{L}}\cdot \vec{n} }{\tilde{r}^6}  \left[
\frac{91 \nu }{2}-\frac{428881}{4900}
\right.\nn\\
&&+\left(
-\frac{25 \nu ^2}{16}-\frac{493 \nu }{16}-\frac{31}{4}
\right) \frac{\tilde{L}^2}{\tilde{r}} 
+\left(
-\frac{3 \nu ^2}{8}+\frac{19 \nu }{16}-\frac{9}{16}
\right) \frac{\tilde{L}^4}{\tilde{r}^2} \nn\\
&&+ \tilde{p}_r^2 \tilde{r} \left(
-\frac{129 \nu ^2}{16}-\frac{105 \nu }{16}-\frac{73}{4}
+\left(
-\frac{9 \nu ^2}{8}+\frac{23 \nu }{16}-\frac{9}{8}
\right) \frac{\tilde{L}^2}{\tilde{r}} \right)\nn\\
&&\left. + \tilde{p}_r^4 \tilde{r}^2 \left( 
\frac{3 \nu ^2}{16}-\frac{13 \nu }{8}-\frac{9}{16}
\right)
\right]\nn\\%%%%%%%
&&+\frac{\nu  \tilde{p}_r \tilde{\vec{S}}_1 \cdot \vec{n}  \tilde{\vec{S}}_1\times \tilde{\vec{L}}\cdot \vec{n} }{q\tilde{r}^6}  \left[
-\frac{743 \nu ^2}{28}+\left(-\frac{2249}{16}-\frac{2703 \pi ^2}{128}\right) \nu +\frac{497}{4}
\right.\nn\\
&&+\left( 
-\frac{25 \nu ^3}{16}-29 \nu ^2-\frac{1043 \nu }{16}+\frac{35}{4}
\right) \frac{\tilde{L}^2}{\tilde{r}} 
+\left( 
-\frac{21 \nu ^3}{16}-\frac{5 \nu }{16}+\frac{9}{16}
\right) \frac{\tilde{L}^4}{\tilde{r}^2} \nn\\
&&+ \tilde{p}_r^2 \tilde{r} \left(
-\frac{129 \nu ^3}{16}+\frac{1411 \nu ^2}{16}-\frac{1331 \nu }{16}+\frac{77}{4}
+\left( 
-\frac{81 \nu ^3}{16}-3 \nu ^2+\frac{5 \nu }{16}+\frac{9}{8}
\right) \frac{\tilde{L}^2}{\tilde{r}} \right)\nn\\
&&\left. + \tilde{p}_r^4 \tilde{r}^2 \left( 
-\frac{15 \nu ^3}{2}-\frac{123 \nu ^2}{16}+\frac{5 \nu }{2}+\frac{9}{16}
\right)
\right],%%%%%%%
\eea
\bea
\tilde{H}^{\text{N}^3\text{LO}}_{C_{1\text{E$^2$S$^2$}}} &=& -\frac{\nu^3    }{2\tilde{r}^6}  
\left(1 + \frac{1}{q} \right) \left( \tilde{S}_1^2 - 3 (\tilde{\vec{S}}_1 \cdot \vec{n} 
)^2 \right).
\eea

\bibliographystyle{jhep}
\bibliography{gwbibtex}

\end{document}